\documentclass[12pt,a4paper,twoside,dvips]{article}
\newcommand{\mydirfig}{}
\usepackage{a4p}
\usepackage{cite,mcite}
\usepackage{physics}
\usepackage{l3_titlePH}
\usepackage{ifthen}
\usepackage{graphicx}
\usepackage{amssymb}
\usepackage{amsmath}
\usepackage{dcolumn}
\usepackage{pennames}
\usepackage{subscript}
\usepackage{rotating}
\journalname{Physics Reports}
\date{June 8, 2004}
\preprint{2004-024}
\graphicspath{{.}}
\newlength{\capindent}
\newlength{\capwidth}
\newlength{\figwidth}
\newcommand{\icaption}[2][!*!,!]{\hspace*{\capindent}%
  \begin{minipage}{\capwidth}
    \ifthenelse{\equal{#1}{!*!,!}}%
      {\caption{#2}}%
      {\caption[#1]{#2}}
  \end{minipage}}
\newcommand{\PZ}{\ensuremath{\mathrm{Z}}}
\renewcommand{\Pqb}{\ensuremath{\mathrm{b}}}
\renewcommand{\Paqb}{\ensuremath{\bar{\mathrm{b}}}}
 
\newcommand{\JETSET}{{\scshape Jetset}}

\newcommand{\LEP}{{\scshape lep}}
\newcommand{\LEPone}{{\scshape lep}{\small1}}
\newcommand{\LEPtwo}{{\scshape lep}{\small2}}
\newcommand{\CERN}{{\scshape cern}}
\newcommand{\ALEPH}{{\scshape aleph}}
\newcommand{\OPAL}{{\scshape opal}}
\newcommand{\DELPHI}{{\scshape delphi}}
\newcommand{\Lthree}{{\scshape l}{\small 3}}
\newcommand{\TASSO}{{\scshape tasso}}
\newcommand{\AMY}{{\scshape amy}}

\newcommand{\SRC}{{\scshape src}}
\newcommand{\ECLU}{{\scshape eclu}}
 
\newcommand{\ISR}{{\scshape ISR}}
\newcommand{\FSR}{{\scshape FSR}}

\newcommand{\QCD}{{\scshape QCD}}
\newcommand{\LPHD}{{\scshape LPHD}}

\newcommand{\PSPLT}[1]{{\scshape psplt}{\small(#1)}}
\newcommand{\CLSMR}[1]{{\scshape clsmr}{\small(#1)}}
\newcommand{\CLPOW}{{\scshape clpow}}
\newcommand{\CLMAX}{{\scshape clmax}}
\newcommand{\DECWT}{{\scshape decwt}}
 
\newcommand{\beqa}{\begin{eqnarray}}   \newcommand{\eeqa}{\end{eqnarray}}
\newcommand{\beqan}{\begin{eqnarray*}} \newcommand{\eeqan}{\end{eqnarray*}}
\newcommand{\abs}[1]{\left|#1\right|}
 
\newcommand{\kt}{{\ensuremath{k_\mathrm{t}}}}              
\renewcommand{\pt}{{\ensuremath{{p_\mathrm{t}}}}}          

\newcommand{\chisq}{{\ensuremath{\chi^2}}}                 
 
\newcommand{\dd}[1]{\ensuremath{\,{\mathrm{d}#1}}}%
\newcommand{\aplan}{\ensuremath{A}}
\renewcommand{\as}{\ensuremath{\alpha_\mathrm{s}}}
\newcommand{\asp}{\ensuremath{\overline{\alpha}_{\mathrm{s}}}}
\newcommand{\bb}{{\Pqb\Paqb}}
\newcommand{\BEtt}{\text{BE\textsubscript{32}}}
\newcommand{\bt}{\ensuremath{B_\mathrm{T}}}
\newcommand{\bw}{\ensuremath{B_\mathrm{W}}}
\newcommand{\Bwt}{\ensuremath{B_{n}}}

\newcommand{\dc}{\dd{C}}
\newcommand{\dD}{\dd{D}}

\newcommand{\ds}{\dd{\sigma}}
\newcommand{\dt}{\dd{T}}
\renewcommand{\ee}{\ensuremath{\mathrm{e^{+}e^{-}}}}
\newcommand{\egam}{\ensuremath{E_{\gamma}}}
\newcommand{\epar}{\ensuremath{E_{\parallel}}}
\newcommand{\eperp}{\ensuremath{E_\perp}}
\newcommand{\evis}{\ensuremath{E_{\mathrm{vis}}}}

\newcommand{\fwm}{\ensuremath{H_{4}}}

\newcommand{\xistar}{\ensuremath{\xi^\star}}
\newcommand{\mh}{\ensuremath{M_\mathrm{H}}}
\newcommand{\nch}{\ensuremath{N_\mathrm{ch}}}

\newcommand{\Nf}{\ensuremath{N_{\text{f}}}}        
\newcommand{\Nc}{\ensuremath{N_{\text{c}}}}        

\newcommand{\nt}{\ensuremath{\vec{n}_\mathrm{T}}}
\newcommand{\obla}{\ensuremath{O}}
\newcommand{\onet}{\ensuremath{1-T}}
\newcommand{\paa}{\ensuremath{\vec{p}_{a}}}
\newcommand{\pai}{\ensuremath{\vec{p}_{i}}}
\newcommand{\paii}{\ensuremath{p_{i}}}
\newcommand{\pk}{\ensuremath{\vec{\alpha}}}
\newcommand{\pp}{\ensuremath{p}}
\newcommand{\qq}{\ensuremath{\mathrm{q\bar{q}}}}
\newcommand{\rhoh}{\ensuremath{\rho_{\mathrm{H}}}}
\newcommand{\rhol}{\rho_{\mathrm{L}}}
\newcommand{\rs}{\ensuremath{\sqrt{s}}}
\newcommand{\rsp}{\ensuremath{\sqrt{s'}}}
\newcommand{\spher}{\ensuremath{S}}
\newcommand{\sQ}{\ensuremath{\sigma_{Q}}}
\newcommand{\thr}{\ensuremath{T}}
\newcommand{\tmajor}{\ensuremath{T_\mathrm{major}}}
\newcommand{\tminor}{\ensuremath{T_\mathrm{minor}}}
\newcommand{\tminorns}{\ensuremath{T^\mathrm{NS}_\mathrm{minor}}}

\newcommand{\ycJ}{\ensuremath{y_\mathrm{cut}^\mathrm{J}}}
\newcommand{\ycD}{\ensuremath{y_\mathrm{cut}^\mathrm{D}}}
\newcommand{\yd}{\ensuremath{y_{23}^\mathrm{D}}}
\newcommand{\yj}{\ensuremath{y_{23}^\mathrm{J}}}
\newcommand{\Gfactors}{$G$-factors}
\newcommand{\LambdaAR}{$\Lambda_{\text{\scshape AR}}$}
\newcommand{\LambdaLL}{$\Lambda_{\text{\scshape LL}}$}
\newcommand{\LambdaLLA}{$\Lambda_{\text{\scshape LLA}}$}
\newcommand{\LambdaMLLA}{$\Lambda_{\text{\scshape MLLA}}$}
\newcommand{\ME}{{\textsc{ME}}}

\newcommand{\LLA}{{\textsc{LLA}}}
\newcommand{\DLA}{{\textsc{DLA}}}
\newcommand{\MLLA}{{\textsc{MLLA}}}
\newcommand{\NLLA}{{\textsc{NLLA}}}

\newcommand{\LO}{{\textsc{LO}}}
\newcommand{\NLO}{{\textsc{NLO}}}
\newcommand{\NNLO}{{\textsc{2NLO}}}
\newcommand{\NNNLO}{{\textsc{3NLO}}}
 
\begin{document}
 
\begin{titlepage}
\title{Studies of Hadronic Event Structure \\
       in {\boldmath$\ee$} Annihilation \\
       from {\boldmath$30 \GeV$} to \boldmath{$209 \GeV$} \\
       with the  L3 Detector}
 
\author{ The L3 Collaboration
}
%
 
\begin{abstract}
In this Report, \QCD\ results obtained from a study of hadronic event structure in high
energy \ee\ interactions with the \Lthree\  detector are presented.
 The operation of the \LEP\ collider at many different collision
 energies from 91\,\GeV\ to 209\,\GeV\
 offers a unique  opportunity  to  test \QCD\ by measuring
  the energy dependence  of different observables.
The main results concern the measurement of the strong coupling constant, \as,
from hadronic event shapes  and  the study of effects of
soft gluon coherence in charged particle multiplicity and momentum  distributions.
\end{abstract}
 
\submitted                   
 
\end{titlepage}
 
\normalsize
\pagenumbering{roman}  \setcounter{page}{2}
\tableofcontents
\cleardoublepage
 
\pagenumbering{arabic}

\section{Introduction}\label{sec:intr}
 
Hadronic events produced in \ee\ annihilation offer a good environment to
test the predictions of Quantum Chromodynamics (\QCD)
\cite{qcd_qqgth,qcd_qqgth1,qcd_qqgth2,qcd_qqgth3,qcd_qqgth4,qcd_qqgth5,qcd_qqgth6,qcd_qqgth7,qcd_qqgth8}.
The high energy phase of the \LEP\ collider has given a
unique opportunity to measure \QCD\ observables over a wide energy range and
perform precise tests of the energy dependence of the strong interaction. In
addition, it allows to check the validity of the \QCD\ models very often used
for background modelling in other studies such as electro-weak studies and new particle searches.
 
From 1989 to 1995 \LEP\ operated in the region of the Z pole,
\ie, at centre-of-mass energies, \rs, around 91.2\,\GeV.
During this period,  known as \LEPone,
each of the four \LEP\ experiments (\ALEPH, \DELPHI, \Lthree, and \OPAL) collected about 4 million
hadronic events.
This high statistics, combined with very low background, made it possible
to perform many detailed \QCD\ studies and precise measurements of the hadronic event structure.
Further, events with an observed
high energy photon, which have a lower effective hadronic centre-of-mass energy,
$\rsp<\rs$ due to initial- (\ISR) or final-state radiation (\FSR), enable studies of
energy dependence.
 
In 1995 \LEP\ entered  
a new phase, known as \LEPtwo, of steadily increasing energy.
Data were taken at a number of centre-of-mass energies, listed in Table~\ref{tab:events}, between 130 and
209\,\GeV.
While the total integrated luminosity collected by \Lthree\ at these high energies
(more than 600\,\pb) is much larger than for the Z-pole region (about 140\,\pb),
the number of hadronic events is much less. This is due in first instance to the much lower hadronic
cross section, \eg, about 20\,pb at $\rs = 200\,\GeV$, which is roughly 200 times smaller than at the Z pole.
Secondly, at high energies a large fraction of the
events correspond to hard initial state radiation (\ISR) bringing down the effective hadronic
centre-of-mass energy, \rsp, to the Z pole. When these events with hard \ISR\ are rejected
the data samples have typically a few hundred to a few thousand hadronic events per energy point.
Another experimental challenge
at these high energies
is the treatment of the dominant background, which, above the W-pair production threshold ($\rs >161\,\GeV$),
comes from W pairs decaying into four quarks.
Part of this background can be rejected using topological identification, but the remaining contamination
must be subtracted according to model predictions.
Nevertheless, the availability of a large range of energies 
is very important for testing \QCD,
since the theory predicts, essentially, the energy variation of observables
rather than their  absolute values.
In addition, it is important that the \QCD\ measurements  be performed at each energy
using the same experimental technique and the same theoretical calculation.
The experimental and theoretical uncertainties on the measurement of an observable are then highly
correlated between energies. The measurement of
the energy dependence of the observable is then insensitive to these uncertainties.
We note that the flavour composition changes with the energy away from the Z pole.
For example, when there is no \ISR, the fraction of $\bb$ drops from about 22\%
at the Z pole to
about 16\%
at $\rs\approx209\,\GeV$.
This must be taken into account when measuring the energy dependence of observables which depend on
the quark-flavour composition of the events.
 
The work presented in this report concerns mainly
the variation of hadronic event shapes with centre-of-mass energy
and  the study of soft gluon coherence through charged particle multiplicity and momentum distributions.
The measurements of the event shapes are used to determine the strong coupling constant, \as.
 
Six event-shape distributions are measured, as well as the
charged particle multiplicity and momentum distributions,
using the data collected with the \Lthree\
detector~\cite{l3:det,l3:det1,l3:det2,l3:det3,l3:smd,fbmuon,spaghetti}
at various energies.
At the \PZ\ pole they are measured for b and lighter (udsc) flavours as well as for all flavours.
The measured distributions are compared with predictions from
event generators based on an improved leading logarithmic approximation
(Parton Shower models including \QCD\ coherence effects).
These Monte Carlo programs use different approaches to describe both the
perturbative parton shower evolution and non-perturbative hadronisation
processes. They are tuned to reproduce the global event-shape
distributions and the charged particle multiplicity distribution measured at
91.2\,\GeV.
 
Moments of the event-shape variables are measured between 41.4\,\GeV\ and 206.2\,\GeV.
Perturbative and non-perturbative \QCD\ contributions are obtained from a fit using a
power correction ansatz \cite{qcdpower,qcdpower1,qcdpower2,qcdpower3,milan}.
 
The strong coupling constant is also determined at each of these
centre-of-mass energies by comparing the measured event-shape
distributions with predictions of second order \QCD\ calculations
\cite{kunznas,ert-me} supplemented by resummed leading and next-to-leading order
terms \cite{qcd-t,qcd-hjm,qcd-jbm,qcd-jbm2,qcd-oth,qcd-c}.
 
The mean charged particle multiplicity
and the peak position, $\xistar$, of the distribution of
$\xi= -\ln x$, where $x$ is the
charged particle momentum scaled by the beam energy,
are measured at different centre-of-mass energies.
The energy dependence of these two observables is compared with \QCD\ predictions including
soft gluon coherence.
A study of the differences between udsc-quark, b-quark
 and all flavours is also  presented for the Z-pole data.

The results presented here update and complete
previously published \Lthree\ results on  \QCD\ obtained
from various \ee\ energy studies. The first one was
a study of hadronic event structure at the Z pole \cite{l3qcd91z,l3qcd91,l3qcd91a}.
This study was extended subsequently to high energies \cite{l3qcd133,l3qcdlep2,l3qcdlep2a,l3qcdlep2b,l3qcdlep2c}.
The energy range was also extended to as low as 30\,\GeV\ by exploiting
hadronic events from Z decays with isolated high energy photons, which gives
reduced hadronic centre-of-mass energies \cite{l3qqg}.
In these events the high energy photons are radiated
through initial state radiation or through bremsstrahlung from quarks.
 
\section{QCD  and the Process \boldmath{$\ee \rightarrow \mathrm{hadrons}$ }}\label{sec:qcd}
 
\subsection{Theoretical Framework}  \label{sec:theor_frame}
\QCD\
\cite{qcd_qqgth,qcd_qqgth1,qcd_qqgth2,qcd_qqgth3,qcd_qqgth4,qcd_qqgth5,qcd_qqgth6,qcd_qqgth7,qcd_qqgth8}
is the gauge theory proposed for  the strong interaction.
It describes the interactions between the quarks 
and the neutral vector gauge bosons mediating the strong interactions, the gluons.
Quarks and gluons carry a quantum number, called colour, which allows
the existence of a coupling  between gluons as well as between quarks and gluons.
This gluon self interaction leads
to a fundamental property of \QCD, called asymptotic freedom, predicting the
decrease of the strong coupling constant, \as, with energy scale.
 
From the theoretical point of view, the process of hadron production from a
quark-antiquark pair in \ee\
annihilations may be seen as composed of two different regimes governed by the strong
interactions and referred to as the perturbative and non-perturbative phases.
Asymptotic freedom guarantees that calculational techniques based on
perturbation theory may be
applied to describe quark and gluon production with high momentum transfers. This
defines the first regime corresponding to a parton cascade where primary
quarks split into further partons down to an energy scale of about 1\,\GeV,
where perturbative techniques cease to be valid.
The main perturbative calculations available to describe the hadronic event
structure at the parton level are:
\begin{itemize}
 \item{${\cal{O}}(\as^{2})$ calculations of event-shape variables~\cite{kunznas,ert-me}};
 \item{improved calculations, incorporating the resummation of leading and next-to-leading
   logarithmic terms~\cite{qcd-t,qcd-hjm,qcd-jbm,qcd-jbm2,qcd-oth,qcd-c} matched to ${\cal{O}}(\as^{2})$
   results, for several event shape variables};
 \item{${\cal{O}}(\as^{3})$ calculation (full 1-loop) of 4-parton states} \cite{as3,as3a};
 \item{analytical calculations based on several leading logarithmic
   approximations~\cite{lla,lla1,lla2,lla3,lla4,nlla,nlla1,nlla2,nlla3,mlla1,mlla2,mlla3}}.
\end{itemize}
 
In order to relate the  parton-level  calculations to  final state hadrons,
one approach is to use phenomenological models describing the non-perturbative transition phase.
These models are included in the commonly used \QCD\ Monte Carlo programs.
%
Another, more recent, approach
consists of describing the non-perturbative
effects analytically by means of  power corrections.
These corrections have been calculated for low-order moments
and differential distributions of some \ee\ event-shape observables~\cite{power}.
 
 Finally, in the case of analytical calculations of inclusive quantities (\eg, charged
 particle mean multiplicity or momentum distributions)
 the hypothesis of Local Parton Hadron Duality \cite{lphd,lphd2} is usually invoked.
 It suggests that the calculated parton distributions
 are related to the measured hadron distributions by a simple normalization constant.
 This hypothesis is used here for the study of the energy dependence
 of charged particle distributions.
 
\subsection{Experimental Framework}  \label{sec:expt_frame}
 
In the last 25 years
the study of hadronic events produced  in  \ee\ annihilation
has made a major contribution to demonstrating the validity of \QCD.
This is largely due to the fact that \ee\ interactions offer a very clean
environment to study basic \QCD\ processes.
\QCD\ affects only the final state; there is no contamination from beam remnants; and, apart
from initial and final state electromagnetic radiation,
the hadronic centre-of-mass energy is well defined.
The observed hadronic event structure is directly related to the gluon radiation pattern
produced in the parton (quark and gluon) \QCD\ processes.
 
Direct evidence for the existence of
 the quark  was given by the observation of two-jet structure in hadronic events produced
 in \ee\ collisions at {\scshape spear}~\cite{quark}, and analysis of
 the jet angular distributions established that their spin is 1/2.
The observation of the first three-jet events  at {\scshape petra}
gave the first evidence for the existence of the gluon~\cite{obla,gluonT,gluonP,gluonJ} and its coupling to quarks.
Subsequently, particle production in the region between a quark and anti-quark was found to be
suppressed compared to that between a quark and gluon, the so-called string effect \cite{jadefl,jadefl1}.
The existence of the triple gluon vertex coupling was confirmed in a study of jet angular distributions
in 4-jet events measured at {\scshape tristan}~\cite{amy}.
 
Many quantitative tests of \QCD\ have been performed at various \ee\ colliders.
Some detailed reviews of these studies can be found, \eg, in
References \citen{qcdee,qcdleprev,qcdleprev1,qcdleprev2,qcdleprev3,qcdleprev4}.
 
The \LEP\ experiments have been very active since 1989 in performing quantitative tests
of \QCD\ \cite{qcdleprev,qcdleprev1,qcdleprev2,qcdleprev3,qcdleprev4}.
Due to its large hadronic branching ratio, negligible background from other
processes, and a strong suppression of initial state
radiation, the Z resonance has offered unique conditions for detailed \QCD\ studies.
In addition, the precise micro-vertex detectors of the tracking systems of
the \LEP\ experiments have allowed flavour-dependent \QCD\ studies to
be performed with the high statistics Z-pole data.
The higher centre-of-mass energies of \LEPtwo\    
have allowed studies of the  energy-scale dependence of \QCD\ predictions over a wider range.
The energy-scale dependence has also been observed
in ep deep inelastic scattering at {\scshape hera} \cite{herarunning,herarunning1}
and in \Pp\Pap\ interactions at {\scshape tevatron} \cite{fnalrunning}.

\subsection{Monte Carlo Programs}  \label{sec:MCprogs}
 
Monte Carlo programs simulate the process  $\ee\rightarrow \mathrm{hadrons}$ by factorizing it into
four different phases:
\begin{enumerate}
  \item production of $\qq(\gamma)$ (electroweak),
  \item gluon radiation (perturbative \QCD),
  \item hadronisation of quarks and gluons (non-perturbative \QCD),
  \item decays of unstable particles.
\end{enumerate}
 
Two approaches to the modelling of perturbative \QCD\ exist~\cite{lepgen}. One is
the matrix element method, in which Feynman diagrams are calculated exactly,
order by order. Because of the technical difficulties of the calculation,
matrix elements are only available for
a maximum of four partons in the final state.
 
The other approach is the parton shower method, which
is based on an approximation of the full matrix element expression.
Each parton produced in the initial hard process may split into two partons, as may successive partons.
This results in a description
of multi-jet events, with no explicit upper limit on the number of partons involved.
The parton shower picture is derived within the framework
of the Leading Logarithmic Approximation (\LLA)~\cite{lla,lla1,lla2,lla3,lla4}, in which only the
leading terms in the perturbative expansion are kept,
or within the framework
of the Modified Leading Logarithmic Approximation (\MLLA)~\cite{mlla1,mlla2,mlla3},
in which some interference effects \cite{coh,coh1,coh2} found in the
Next-to-Leading Logarithmic Approximation (\NLLA)~\cite{nlla,nlla1,nlla2,nlla3}
are also included.
In the branching the energy fractions are distributed according to the leading-order
DGLAP splitting functions \cite{dglap,dglap1,dglap2,dglap3}.
There are many ambiguities
in the \LLA\ description, especially in the renormalisation scheme. Therefore,
the parton shower scale parameters extracted from the \LLA\ models through
comparisons with data do not correspond to the \QCD\ scale parameter
$\Lambda_{\overline{\text{MS}}}$.
Note, however, that the parton shower programs impose energy-momentum conservation at each splitting,
a feature which goes beyond these approximation schemes. 
 
Because perturbative \QCD\ calculations are not valid at low energy scales, the fragmentation
of coloured quarks and gluons into colourless hadrons cannot be calculated by
perturbative \QCD. One needs to rely on phenomenological models. The separation
between the perturbative and fragmentation phases is generally characterised by an
energy scale ($Q_0$) with a typical value of a 1--2\,\GeV. Three different
fragmentation models~\cite{lepgen} have been developed:
independent~\cite{if,if1,if2,if3,ifff,ifalihoyer,ifalihoyer1,ifalihoyer2},
string~\cite{sf,sf1,sf2,sf3}, and  cluster~\cite{cf,cf1}.
 
The independent fragmentation model assumes that partons fragment in isolation
from each other. In this scheme, high momentum quarks evolve separately,
splitting into colourless particles and other quarks. It has been shown that
the independent fragmentation model fails to describe some experimental
data~\cite{jadefl,jadefl1}.
 
The string model is derived from the \QCD\ inspired idea that a colour flux tube
(string) is stretched between quark and anti-quark pairs, with gluons
corresponding to kinks in the string. Hadrons are generated in the formalism
of string breaking.
 
In the cluster model, gluons from the perturbative phase are first split into
quark and anti-quark pairs. The quark and anti-quark pairs then form colourless
clusters which, depending on their masses, decay either into lower mass
clusters or directly into particles.
 
These different perturbative \QCD\ approaches and fragmentation models have
been incorporated into several Monte Carlo programs~\cite{lepgen}.
In this Report we compare results with the predictions of the following set of programs:
\textsc{Jetset} 7.4~PS~\cite{jetset-pythia},
\textsc{Ariadne}~4.06~\cite{ariadne},
\textsc{Jetset} 7.4~ME~\cite{jetset-pythia},
\textsc{Herwig}~5.9~\cite{herwig59}, and \textsc{Cojets}~6.23~\cite{cojets6}.
This set of Monte Carlo programs reflects wide differences in the application
of perturbative \QCD\ approaches and fragmentation processes.

\renewcommand{\descriptionlabel}[1]{\hspace\labelsep\normalfont\bfseries\scshape #1}
\begin{description}
  \item[J{\scriptsize ETSET} PS]
  The \textsc{Jetset} parton shower Monte Carlo program \cite{jetset-pythia}
  and its successor \textsc{Pythia} \cite{jetset-pythia,pythiasix}
  simulate $\ee$ annihilation into partons and the subsequent quark and gluon branchings.
  The parton shower is based on the          leading logarithmic approximation using
  as the evolution variable the mass squared of the branching parton.
  Angular ordering, which is a consequence of gluon interference in the next-to-leading logarithmic
  approximation, as well as nearest-neighbour intrajet spin correlations,
  are incorporated in an {\it ad hoc} manner.
  The distribution of the first gluon is modified to match the ${\cal O}(\as)$
  matrix element distribution.
  Initial state radiation is included in \textsc{Jetset} using the lowest order calculation,
  following the   approach presented in References \citen{jetewc} and \citen{jetewc1}.
  In \textsc{Pythia} an `initial-state shower' is used to simulate \ISR.
  The programs contain both string and independent fragmentation options.
  Here we only study string fragmentation. Various fragmentation functions are available.
  They provide the distribution of
  the fraction, $z$, of the light-cone fraction, $E+p_\mathrm{L}$, carried by the resulting hadron,
\begin{equation}
     z=\frac{(E+p_\mathrm{L})^\mathrm{had}}{(E+p_\mathrm{L})^\mathrm{par}}\;.
\end{equation}
  Here  $E$ and $p_\mathrm{L}$ are the energy and longitudinal momentum relative to the primary parton
  direction, and the superscripts (had) and (par) refer, respectively, to the hadron
  after   its creation and the parton before creation of the hadron.
  For  c- and  b-quarks, we use the
  Peterson fragmentation function~\cite{peterson}
\begin{equation}
     f(z) \propto \frac{1}{z\left(1-\frac{1}{z}-\frac{\epsilon}{1-z}\right)^2}
\end{equation}
  where $\epsilon$ is a fragmentation parameter depending on the flavour of the quark.
  The light quarks
  are fragmented according to the Lund symmetric function~\cite{lundsym}
\begin{equation}
    f(z) \propto \frac{1}{z} (1-z)^a \exp\left(-\frac{bm^2_\mathrm{T}}{z}\right) \;,
\end{equation}
  where
  $m_\mathrm{T}=\sqrt{E^2-p_\mathrm{L}^2}= \sqrt{m^2+p_\mathrm{T}^2}$
  is the transverse mass of the system, and $a$ and $b$ are fragmentation parameters.
  The spectrum of the transverse momentum, $p_\mathrm{T}$, of the hadron is described by the Gaussian
  function
\begin{equation}
      f(p_\mathrm{T}) \propto \exp\left(-\frac{p_\mathrm{T}^2}{2\sigma_q^2}\right)
\end{equation}
  with $\sigma_q$ a parameter.
  The parameters that affect hadronic event structure most are the parton
  shower scale \LambdaLL, the parton shower cut-off parameter $Q_0$, and
  the fragmentation parameters $a$, $b$, and $\sigma_q$.
 
  \item[A{\scriptsize RIADNE}]
  The \textsc{Ariadne} program \cite{ariadne} also uses a parton shower algorithm.
  The perturbative \QCD\ cascade in \textsc{Ariadne} is formulated in
  terms of two-parton systems, which form  colour dipoles.  When a gluon is radiated
  from a dipole, the dipole is then converted into two independent dipoles.
  This formulation
  is equivalent, to \MLLA\ 
  accuracy,
  to a parton shower with angular ordering automatically incorporated \cite{azi}.
  The evolution variable is $Q^2=p_\mathrm{t}^2$,
  where $p_\mathrm{t}$ is the transverse momentum of the radiated gluon.
  \textsc{Ariadne} itself does not provide functions for
  fragmentation and decay processes. Instead, it is interfaced to the
  \textsc{Jetset} or \textsc{Pythia}  fragmentation and decay routines. In addition,
  \textsc{Ariadne} uses \textsc{Jetset} or \textsc{Pythia} routines to generate the initial
  $\qq$ system and \ISR.
  Only the string fragmentation is used here.
  In the \textsc{Ariadne} perturbative phase, there are two main parameters that affect the parton
  configuration most: the parton shower scale parameter \LambdaAR\
  and the parton shower cut-off parameter $p^\mathrm{min}_\mathrm{t}$.
  The relevant fragmentation
  parameters are the same as those in the \textsc{Jetset}~PS model.
 
  \item[J{\scriptsize ETSET} ME]
  Besides the parton shower option,
  \textsc{Jetset} also provides for a full ${\cal O}(\alpha^2_\mathrm{s})$  matrix
  element~\cite{ert-me} treatment of perturbative \QCD. In our application,
  we use `optimised perturbation theory'~\cite{opt,opt1} with the renormalisation
  scale, $f$, set to 0.003 and the minimum scaled invariant mass squared of any
  two partons in 3- or 4-jet events, $y_\mathrm{min}$, set to 0.01. The scale
  $f$ is chosen so that $Q^2$ is above the  b-quark mass
  while $y_\mathrm{min}$ is
  close to the minimum allowed value that still gives a positive 2-jet
  production cross section. It has been shown that a small scale $f$ gives
  significantly improved agreement with the data~\cite{scd}. In addition,
  we apply the parameterisation given in Reference \citen{zhu} for the second
  order corrections to the 3-jet rate. The generated partons are subsequently
  fragmented using the string fragmentation model.  As for \textsc{Jetset} PS, we use the Peterson
  function for heavy quark fragmentation
  and the Lund symmetric function for light quark fragmentation.
  The relevant parameters for our study are the \QCD\ scale parameter
  $\Lambda_\mathrm{ME}$
  and the fragmentation parameters $\sigma_q$, $a$ and $b$ of the string model.
 
  \item[H{\scriptsize ERWIG}]
  The \textsc{Herwig} Monte Carlo program \cite{herwig,herwig1,herwig59} is based on
  parton shower simulation using a coherent branching algorithm.  While the energy
  fractions are distributed according to the \LLA,  
  phase space is restricted to an angular-ordered region.
  The choice of evolution variable is $\approx E^2(1-\cos\theta)$,
  where $E$ is the energy of the branching parton and
  $\theta$ is the angle between the two resulting partons.
  This facilitates the inclusion of interference phenomena~\cite{coh,coh1,coh2} in the treatment
  of parton shower development.
  The description of hard gluon emission is improved by matching the parton shower calculation to
  an ${\cal O}(\as)$ matrix element calculation.
  Fragmentation is performed by a cluster model,
  which incorporates the preconfinement property of perturbative \QCD\ \cite{preconfine,lphd,cf,cf1,azi}.
  The event-shape variables are most sensitive to the parton shower scale parameter, \LambdaMLLA,
  the effective gluon mass, $M_\mathrm{g}$,
  and two parameters which control the splitting of clusters:
  the maximum cluster mass, \CLMAX, 
  and the power of the mass, \CLPOW, in the expression for the cluster splitting criterion.
 
  \item[C{\scriptsize OJETS}]
  The Monte Carlo program \textsc{Cojets} \cite{cojets6,cojets,cojets1}
  simulates the multiple gluon radiation in the \LLA. 
  Like \textsc{Jetset PS} it uses the mass squared of the branching parton as evolution
  variable, but with incoherent branching. The parton shower algorithm is
  corrected for single hard gluon emission using an ${\cal O}(\as)$ calculation.
  This simulation is integrated with the independent jet fragmentation
  according to a modified version of the Field-Feynman model~\cite{ifff}.
  \textsc{Cojets} has four free parameters in its longitudinal fragmentation
  function and one free parameter to control the transverse momentum spectra
  in the fragmentation cascade. Since quarks and gluons fragment independently,
  these parameters can have different values for quark and gluon jets.
  As in other parton shower programs, there are also parameters for
  the parton shower scale,
  \LambdaLL, and the parton shower cut-off,  $Q_0$.
\end{description}
 
\section{Hadronic Events in {\scshape L3}}\label{sec:sel}
 
\subsection{Calorimeter Energy Measurement}  \label{sec:energycal}
 
\begin{figure}
 \begin{center}
  \includegraphics[width=.7\figwidth]{\mydirfig 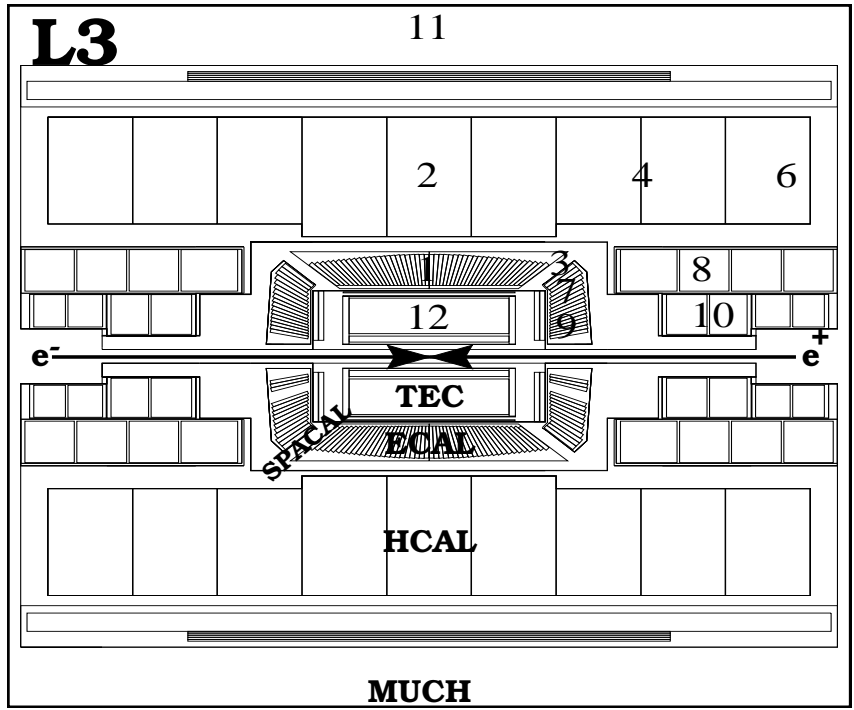}
 \end{center}
 \caption{The eleven regions of L3 detectors as used in the energy
            measurement for the \LEPtwo\ configuration.
          A twelfth region, 5, was present only in earlier set-ups.
            }
 \label{fig:l3region}
\end{figure}

The selection of hadronic events is based on the
energy measured in the electromagnetic and hadron calorimeters.
Two algorithms are developed to estimate the energy of an event
from the raw energy deposits.
In its \LEPtwo\ configuration,
the \Lthree\  detector is divided into eleven broad regions,  nine of which are
calorimeters (regions 1--4 and 6--10, region 5 being no longer present for \LEPtwo).
The other two are the central tracker (region 12) and the muon chambers (region 11).
A particle can deposit its energy in more than one region.
The definition of the regions changed with time depending on the exact detector
configuration. The regions as defined during the \LEPtwo\ runs are shown
in Figure~\ref{fig:l3region}. The main changes with respect to \LEPone\
are the addition of forward/backward muon chambers \cite{fbmuon} and
calorimeters (\textsc{spacal}), constructed using lead and scintillating fibres
between the barrel and endcap electromagnetic calorimeters \cite{spaghetti}.
 
In one of the approaches (linear algorithm), the energy of a particle, detected as the smallest
resolvable calorimeter cluster (\SRC), is expressed as a
linear sum of energy deposits in the calorimeter, $E^\mathrm{c}_i$:
\begin{equation} \label{eq:linG}
    E^\mathrm{c}  =  \sum_{i=1}^{10} G_{i}^\mathrm{L} \cdot E^\mathrm{c}_{i}
\end{equation}
The weighting factors, $G_{i}^\mathrm{L}$, are called \Gfactors.
They compensate for the different calorimeter response to different particle types.
%
The energy of the event is obtained by adding the energy, $E^\mathrm{c}$, of all the \SRC's
in the event and the momenta of the muons.
Since  the  noise levels in different parts of the detectors
can have a wide variation, the energy thresholds for different types of \SRC's
are handled separately.
 
In the second approach (non-linear algorithm), the clusters are redefined to include tracks
in the energy measurement. The new objects, called super-clusters (\ECLU),
are built by associating the different components of a cluster with charged tracks
and muon candidates using angular proximity. The association is carried
out as a four step algorithm:
\begin{enumerate}
\item All possible pairs of constituents, whose angular separation
      is smaller than a given cut, are combined to form seeds.
\item If the angular separation between a constituent and the ones which
      form a seed is smaller than a given cut, it is added to the
      super-cluster associated to the seed. Each constituent can, in
      principle, be included in several super-clusters.
\item The ambiguities are then solved by assigning each constituent to its
      closest super-cluster.
\item The energy of each super-cluster is then calculated.
\end{enumerate}
 
The energy of the super-cluster is given by
\begin{equation}  \label{eq:nlG1}
   \widetilde{E}^\mathrm{sc}_\ell  = \sum_{i=1}^{12} \widetilde{G}_i^\mathrm{NL} \cdot E_{i} +
             \sum_{j,k=1}^{12} \widetilde{A}_{jk}^\mathrm{NL} \cdot C_\ell(E_{j},E_{k}) \;,
\end{equation}
where $E_i$ is the uncorrected energy measured in region $i$.
For the calorimeter regions, $E_i=E^\mathrm{c}_{i}$; for tracks $E_i$ is the momentum of the track.
The correlation function $C_\ell$ introduces a non-linear term in the energy measurement.
Two parametrisations are tried:
\begin{equation}
C_1(E_{j},E_{k})  =  {E_{j}\cdot E_{k}}
     \qquad\text{and}\qquad
C_2(E_{j},E_{k})  =  {\frac{E_{j}\cdot E_{k}}{E_{j} + E_{k}}}
  \; .
\end{equation}
The first parametrisation leads to a better energy resolution while the second
provides a smaller non-linearity in the energy measurement.
The non-linearity is  reduced by scaling the super-cluster energy to obtain
\begin{equation} \label{eq:nlG3}
  {E}^\mathrm{sc}_\ell  =
           \frac{E_\mathrm{tot}}{\sum \widetilde{E}^\mathrm{sc}_\ell}
           \cdot \widetilde{E}^\mathrm{sc}_\ell
   \;.
\end{equation}
The total energy, $E^\mathrm{tot}_\ell$, is given by
\begin{equation}  \label{eq:nlG2}
  E^\mathrm{tot}_\ell  =  \sum_{i=1}^{12} G_{i}^\mathrm{NL} \cdot E^\mathrm{T}_{i} +
              \sum_{j,k=1}^{12} {A}_{jk}^\mathrm{NL} \cdot C_\ell(E^\mathrm{T}_{j},E^\mathrm{T}_{k}) \;,
\end{equation}
an expression analogous to that for a super-cluster.
Here, $E^\mathrm{T}_{i}$ is the sum of the uncorrected energies
of all constituents of the entire event in detector region $i$.
The factors $\widetilde{A}_{jk}^\mathrm{NL}$ and ${A}_{jk}^\mathrm{NL}$ have non-vanishing
values only for connected neighbouring detector regions.

Energies of electrons, photons and muons are accurately measured in the \Lthree\ detector.
To benefit from this, with both algorithms,
active particle identification has been used to identify electrons, photons and muons.
The corresponding clusters, tracks and muon candidates are removed
from the list considered in finding the clusters. The
identified electromagnetic clusters and muons are then added with
their energy measurements from the electromagnetic calorimeter or muon
chamber to the list of reconstructed clusters.
 
For both algorithms, the numerical values of the various coefficients (\Gfactors),
$G^\mathrm{L}$, $\widetilde{G}^\mathrm{NL}$,  $\widetilde{A}^\mathrm{NL}$, $G^\mathrm{NL}$ and $A^\mathrm{NL}$,
are determined by minimising the total energy resolution on hadronic events
while constraining the mean visible energy to the centre-of-mass energy.
This procedure is performed only after precise absolute calibration of each detector component.
The coefficients are re-determined
whenever the detector configuration is modified or the beam energy of \LEP\ is significantly changed.
This is to overcome a certain amount of non-linearity still left in
these energy measurements.  This is more pronounced in the non-linear \Gfactors,
but is somewhat reduced by  a proper choice of the correlation function $C_\ell$
and a better identification algorithm for the final state particles.
 
The non-linear \Gfactors\ are only appropriate for events with small missing energy.
The linear \Gfactors\ are found to be independent of
time variation of detector responses and are nearly energy independent.
The linear algorithm is well suited to comparison of physics measurements over several
centre-of-mass energies.  We have therefore used the linear \Gfactors\ for all our subsequent analyses and
used the non-linear \Gfactors\ for systematic checks.
 
\subsection{Energy and Angular Resolutions of Jets}
 
\begin{figure}[htbp]
 \begin{center}
  \includegraphics[width=.5\figwidth]{\mydirfig 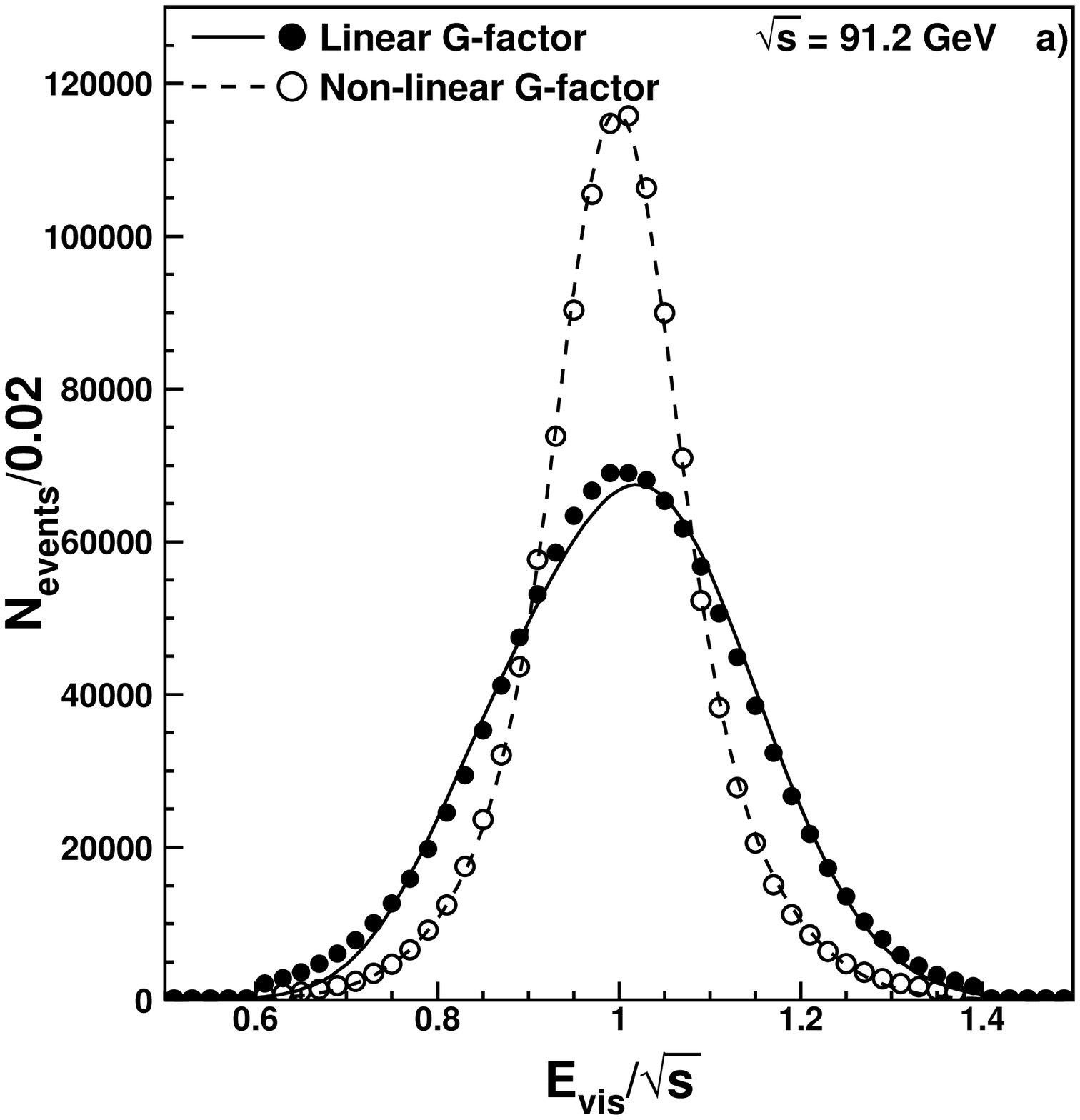}
  \includegraphics[width=.5\figwidth]{\mydirfig 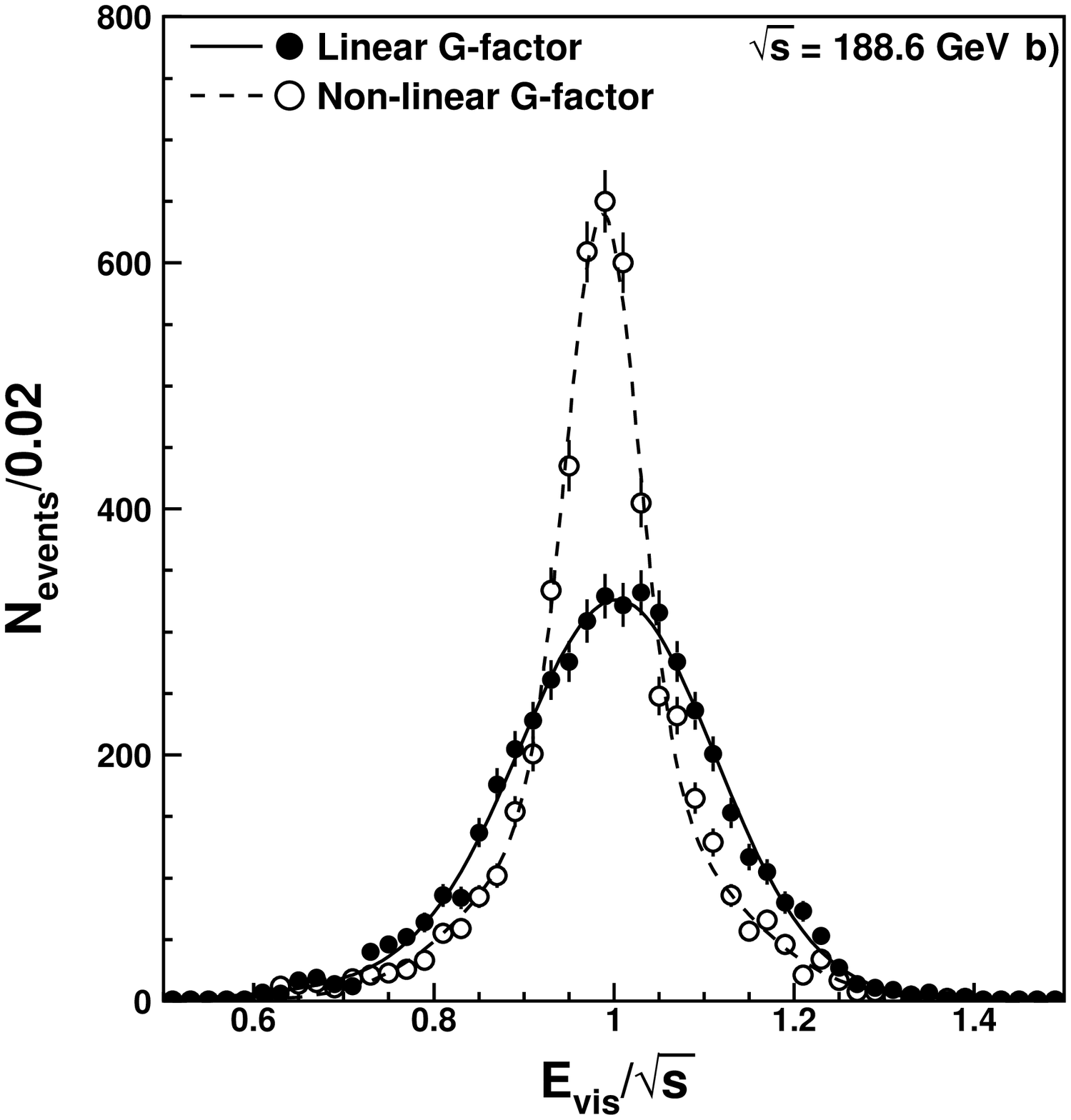}
 \end{center}
 \caption[Distributions of scaled visible energy for clusters with linear
          and non-linear \Gfactors\ in data at (a) $\rs= 91.2\,\GeV$
          and (b) $\rs= 188.6\,\GeV$.]
         {Distributions of scaled visible energy for clusters with linear
          and non-linear \Gfactors\ in data at (a) $\rs = 91.2\,\GeV$
          and (b) $\rs=  188.6\,\GeV$.
          The points correspond to the measurements
          and the smooth curves are from fits of a sum of Gaussian
          distributions as described in the text.}
 \label{fig:evisqq}
\end{figure}

We use energy clusters in the calorimeters with a minimum energy of 100 \MeV.
Figure \ref{fig:evisqq} shows the scaled visible energy ($\evis/\rs$)
distribution at centre-of-mass energies of 91.2 and 188.6\,\GeV\ for the
two different algorithms. The smooth curves shown on the plot
are the results of fits of a sum of two Gaussians to the observed distributions.
%
Table~\ref{tab:angr} summarises the results of the fit as well as the RMS
values from the data at $\rs = 91.2$ and 188.6\,\GeV.
The energy resolution improves substantially with the \ECLU\ algorithm.
 
Jet angular resolutions obtained with both the linear and the non-linear \Gfactors\
are shown in Figure \ref{fig:resth} for
polar angle $\theta$
and
azimuthal angle $\phi$.
They are computed from  the angle between the jets
in selected 2-jet events at $\rs = 91.2\,\GeV$ and 188.6\,\GeV.
The curves correspond to fits with a sum of two Gaussians for each distribution.
The fit results are summarised in Table~\ref{tab:angr} where the
Gaussian widths are denoted $\sigma_{i}$.
The RMS values of the distributions are also given.
Table~\ref{tab:angr} also summarises the resolutions 
obtained using linear  \Gfactors.
There is a slight improvement in $\phi$ resolution with the \ECLU\
algorithm while the $\theta$ resolution is the same.
This difference in improvement is due to a better \Lthree\ track resolution
in $\phi$ than in~$\theta$.

\subsection{Selection of Hadronic Events}
 
The principal variables used to distinguish hadronic events from
backgrounds are the cluster multiplicity and the energy imbalances.
We use energy clusters in the calorimeters to measure the total visible energy,
\evis, and the energy imbalances parallel and perpendicular to the beam direction:
$\epar=\abs{\sum E\cos\theta}$ and
$\eperp=\sqrt{(\sum E\sin\theta\sin\phi)^2+(\sum E\sin\theta\cos\phi)^2}$, respectively,
where $E$ is the energy of a cluster and $\theta$ and $\phi$ are its polar and azimuthal angles
with respect to the beam direction.
Backgrounds are different for hadronic Z decays, hadronic events at reduced
centre-of-mass energies and at high energies. This  results in different selection
cuts for these three types of event.
 
The efficiency of the selection
criteria and purity of the data sample are estimated using
Monte Carlo events.
For the process
$\ee \rightarrow \qq (\gamma)$
Monte Carlo events are generated by the 
programs \textsc{Jetset} 7.3 at the Z pole, \textsc{Pythia} 5.7 for higher energies up
to 189\,\GeV\ and \textsc{KK2}f \cite{kk2f,kk2f1}, which uses \textsc{Pythia} for hadronisation,
for the highest energies.
The generated events are passed through the
\Lthree\  detector simulation, which is based on \textsc{Geant}~\cite{geant}
using the \textsc{Gheisha} program \cite{gheisha} to simulate hadronic interactions.
Background events are simulated with appropriate event generators: \textsc{Pythia}
and \textsc{Phojet} \cite{phojet,phojet1} for hadron production in two-photon interactions,
\textsc{KoralZ} \cite{koralz} for the $\tau^{+}\tau^{-}(\gamma)$ process,
\textsc{Bhagene} \cite{bhagene,bhageneMC} and \textsc{Bhwide} \cite{bhwide} for Bhabha
events, \textsc{KoralW} \cite{koralw,koralw1} for W-pair production and
\textsc{Pythia} for Z-pair production.
 
Hadronic Z decays
are selected \cite{l3qcd91z} by imposing simple cuts on visible energy,
$0.6 < \evis /\rs< 1.4$, relative energy imbalances, $\epar / \evis < 0.4$ and
$\eperp /\evis < 0.4$, and number of clusters $> 12$.
The event-shape distributions for all flavours have been previously published \cite{l3qcd91z}
and are not updated here.  They are based on 8.3\,\pb\ of integrated luminosity, rather than the
full luminosity available (142.4\,\pb).
This is sufficient to
provide an experimental error on \as, which is smaller by a factor 3 than theoretical uncertainties.
 
Events at reduced centre-of-mass energies are obtained from the
entire data collected at the Z pole.
Hadronic events are initially selected
with the same criteria as described above.
In this event sample, isolated photons are selected with energy $E_\gamma> 5\,\GeV$.
The lateral shower profile of the candidate is required to be consistent with an electromagnetic shower and
no other cluster with energy above 250 \MeV\ may lie within 10$^{\circ}$
around the candidate. With these criteria, $1.3\cdot10^5$ 
events are selected.
The centre-of-mass energy of the remaining hadronic system is given by
\begin{equation}
   \rsp = \sqrt{s \left( 1 - \frac{2\egam}{\rs} \right)} \quad.
\end{equation}
Six intervals of \rsp\ are chosen such that each interval has reasonable statistics.
We have studied whether \rsp\ is the correct scale of hadron production by comparing Monte Carlo
hadronic \PZ\ decay events containing isolated final-state photons with  Monte Carlo
\epem\ events generated  without \ISR\ or \FSR\ at $\rs\approx\rsp$.
The distributions of event-shape variables are similar, suggesting that \rsp\ can be used as the \QCD\ scale.
 
The background for the direct photons is dominated by
unresolved $\pi^0$ and $\eta$ decays. To reduce this
background, we require that the shower be isolated and that its
shape be compatible with the electromagnetic shower of a single photon.
We use a shower-shape discriminator based on an artificial neural network to distinguish
multi-photon showers from those of a single photon.
The cut values 
are tuned separately for photon candidates in each of the six different energy ranges
by optimising the efficiency and purity at each energy.
Details of this selection are given in Reference \citen{l3qqg}.
 
\begin{figure}[htbp]
\begin{center}
 \includegraphics[width=.5\figwidth]{\mydirfig 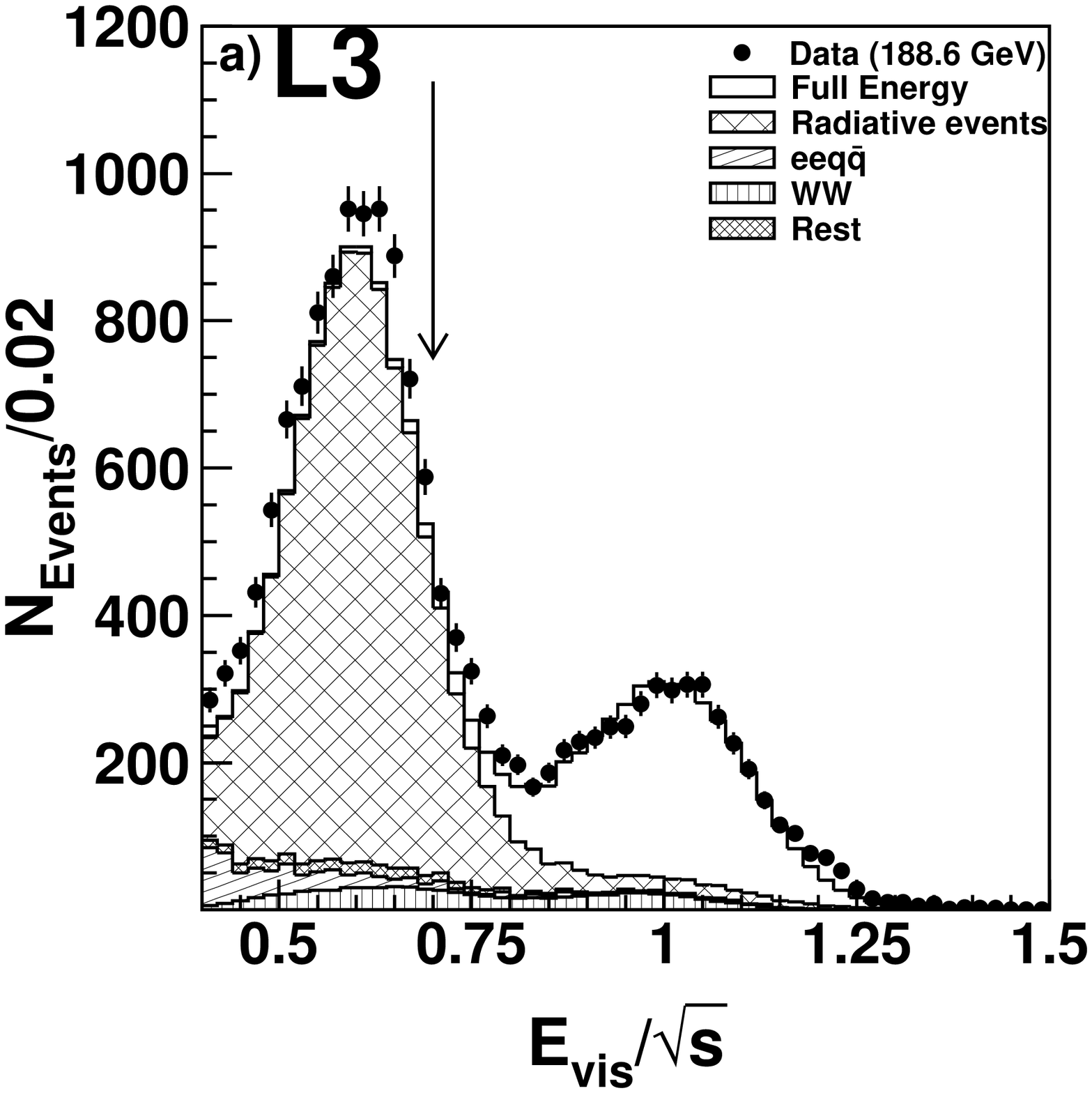}
 \includegraphics[width=.5\figwidth]{\mydirfig 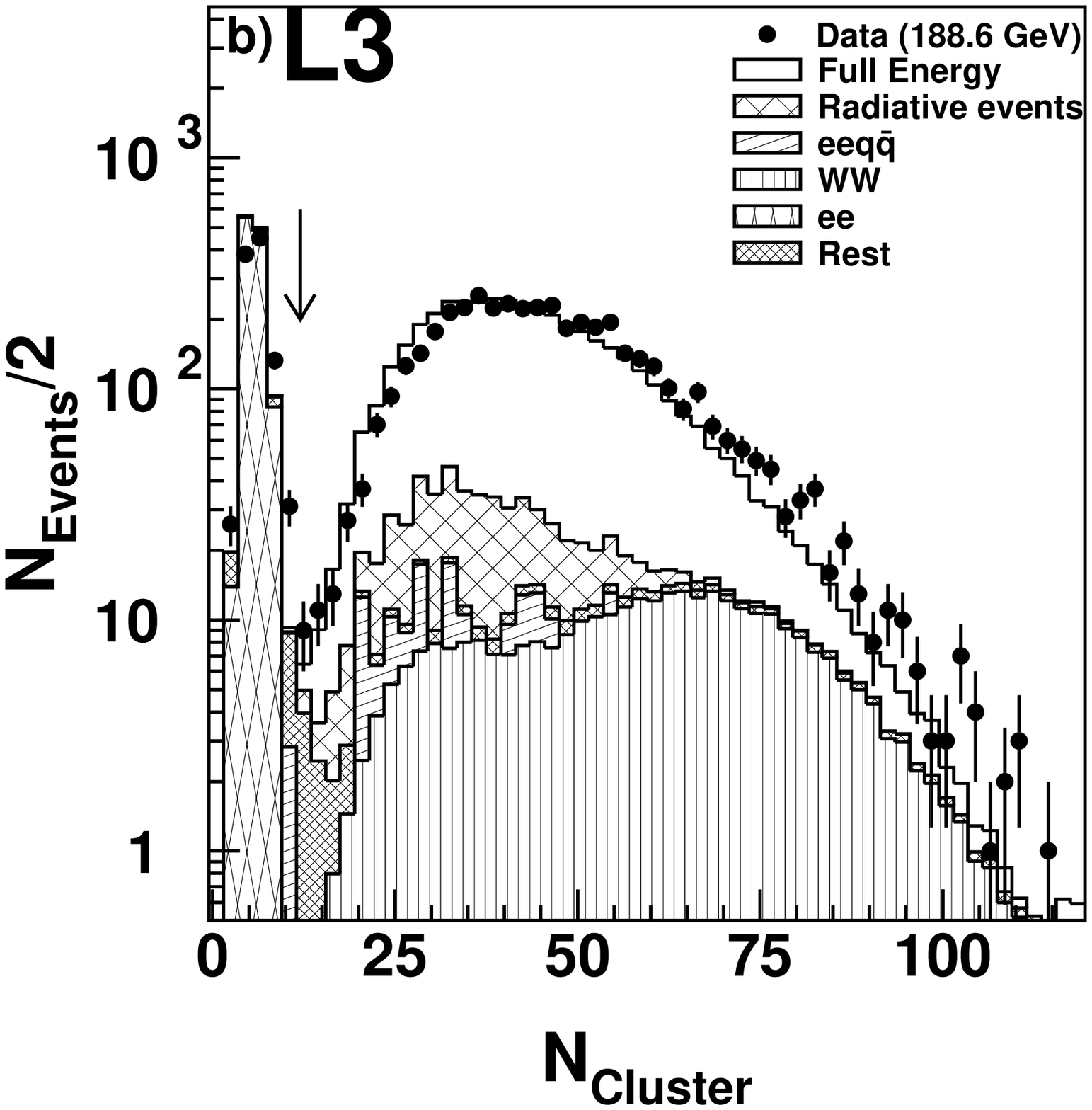}
\end{center}
\caption{Distributions of (a) visible energy and (b) number of calorimetric clusters
    at $\rs=188.6\,\GeV$.
    The arrows indicate the selection cuts.}
\label{fig:selec1}
\end{figure}

\begin{figure}[htbp]
\begin{center}
 \includegraphics[width=.5\figwidth]{\mydirfig 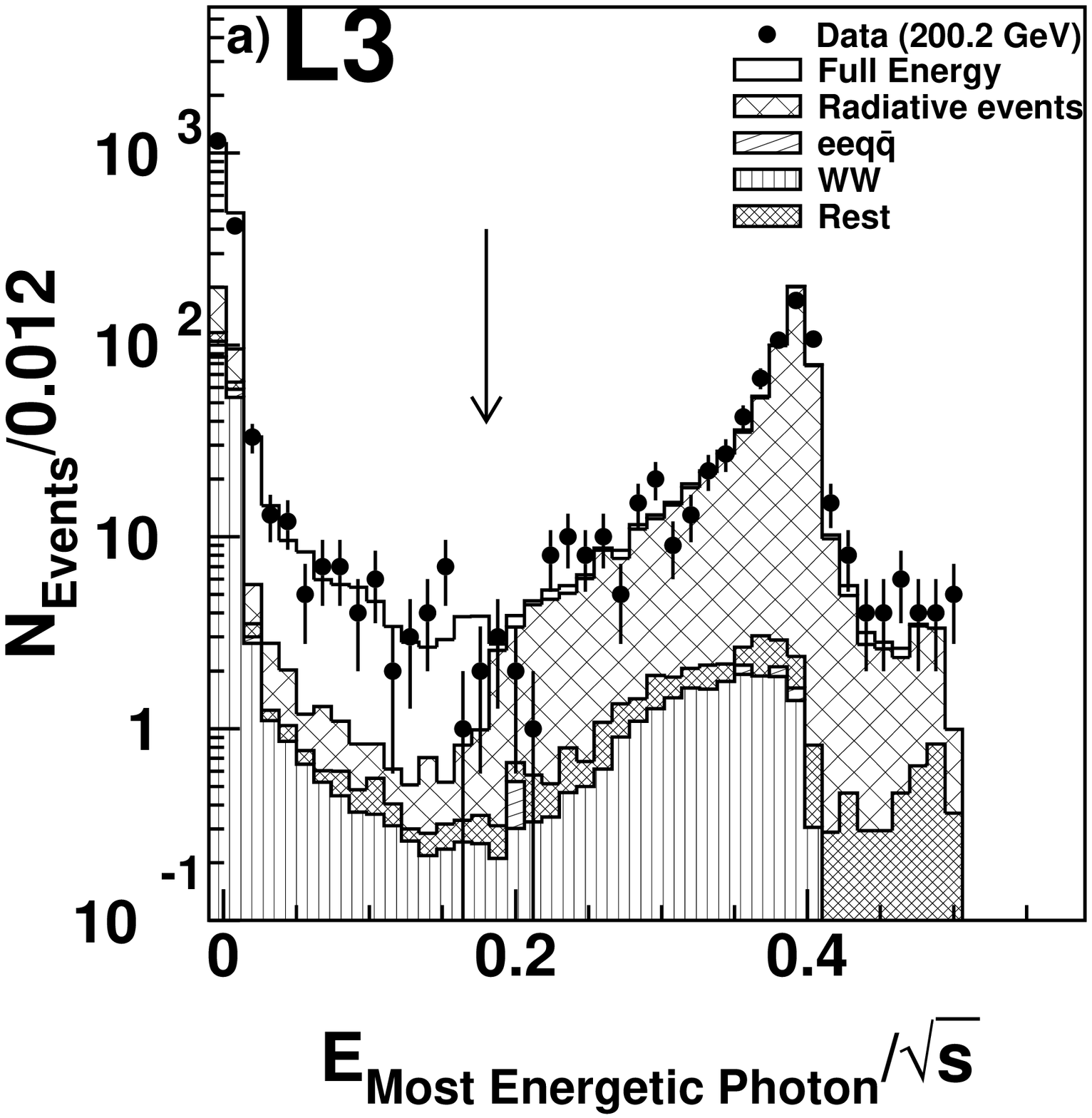}
 \includegraphics[width=.5\figwidth]{\mydirfig 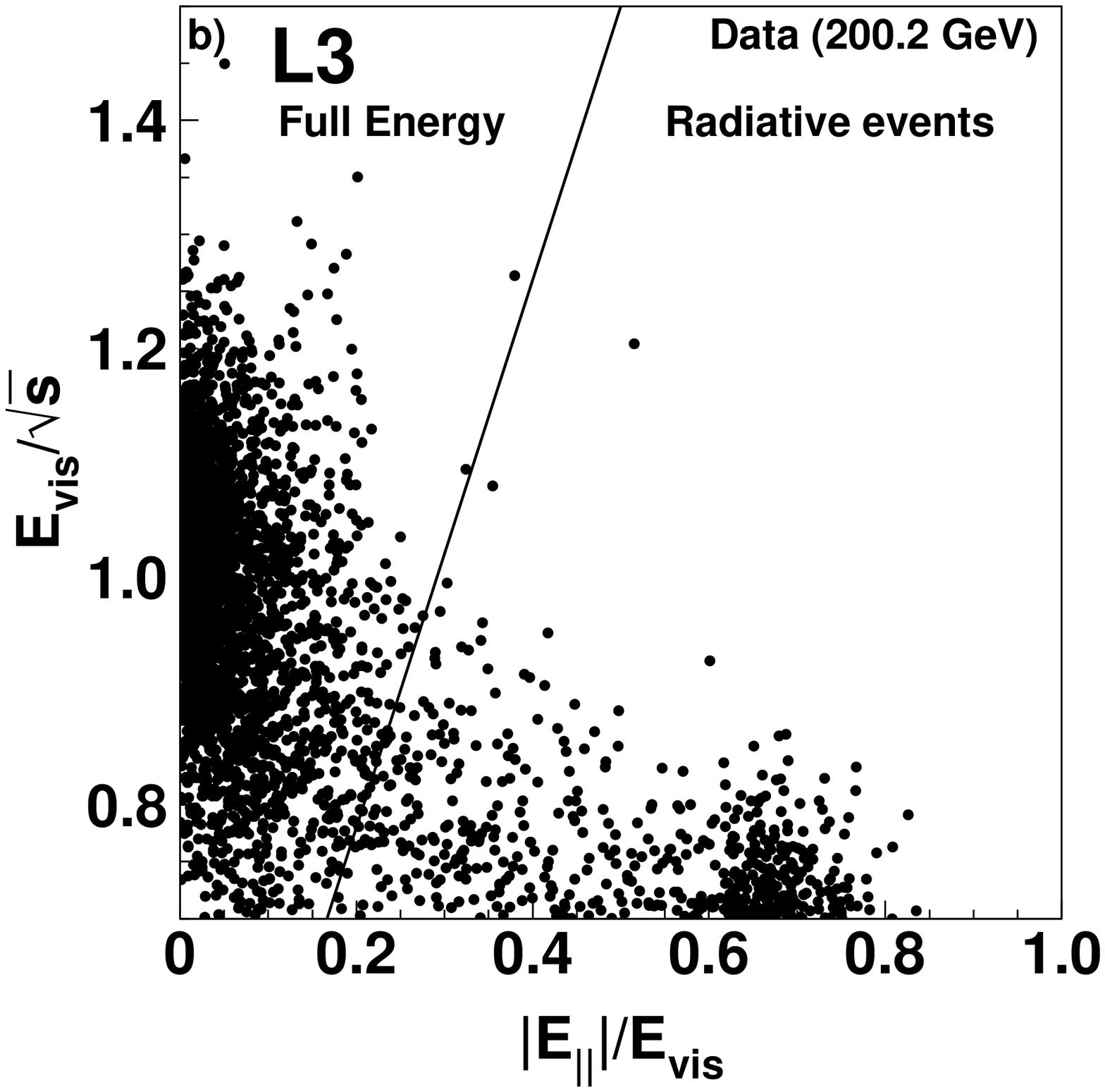}
\end{center}
\caption{(a) Distribution of the energy of the most energetic photon candidate at $\rs=200.2\,\GeV$.
             The arrow indicates the selection cut.
         (b) Plot of visible energy \vs\ energy imbalance along the beam direction for $\rs=200.2\,\GeV$.
             the cut used to remove radiative events is indicated.
         }
\label{fig:selec2}
\end{figure}

At $\rs > 130\,\GeV$,
the main background comes from so-called radiative return events, where
\ISR\ results in a mass of the hadronic system close to that of the Z boson, \MZ.
Events are selected by requiring
$\evis/\rs>0.7$, $\eperp/\evis<0.4$, number of clusters $> 12$, and at least one well measured charged
track. The distributions of  $\evis/\rs$ and the number of clusters are shown, for representative energies,
in Figure~\ref{fig:selec1}. These cuts eliminate a large fraction of the radiative
return events as well as two-photon interactions and other backgrounds.
To further reduce the radiative return background,
events are rejected if they have a high-energy photon candidate, defined as
a cluster in the electromagnetic calorimeter with at least 85\% of its energy within a $15^\circ$ cone
and a total energy greater
than 15\,\GeV\ at $\rs=130.1$ and 136.1\,\GeV\ and greater than $0.18\rs$ at higher \rs.
The distribution of the  energy of the most energetic photon candidate
is shown in Figure~\ref{fig:selec2}a.
Since the \ISR\ photon is often produced at too low an angle to enter the detector,
a cut in the two dimensional plane of $\epar/\evis$ and $\evis/\rs$ is also applied, requiring
$ \evis/\rs > k  \epar/\evis + 0.5 $ where $k$ is 2.5 at $\rs = 130.1$ and 136.1\,\GeV,
 1.5 at $\rs=161.3\,\GeV$, and 2.0 for $\rs\ge172.3\,\GeV$.
This cut is illustrated in Figure~\ref{fig:selec2}b.

Data at $\rs = 130.1$ and 136.1\,\GeV\  were collected in two separate runs
during  1995  \cite{l3qcd133} and 1997.
In the current analysis, data sets from the two years are combined.
 
\begin{figure}[htbp]
\begin{center}
  \includegraphics[width=.5\figwidth]{\mydirfig 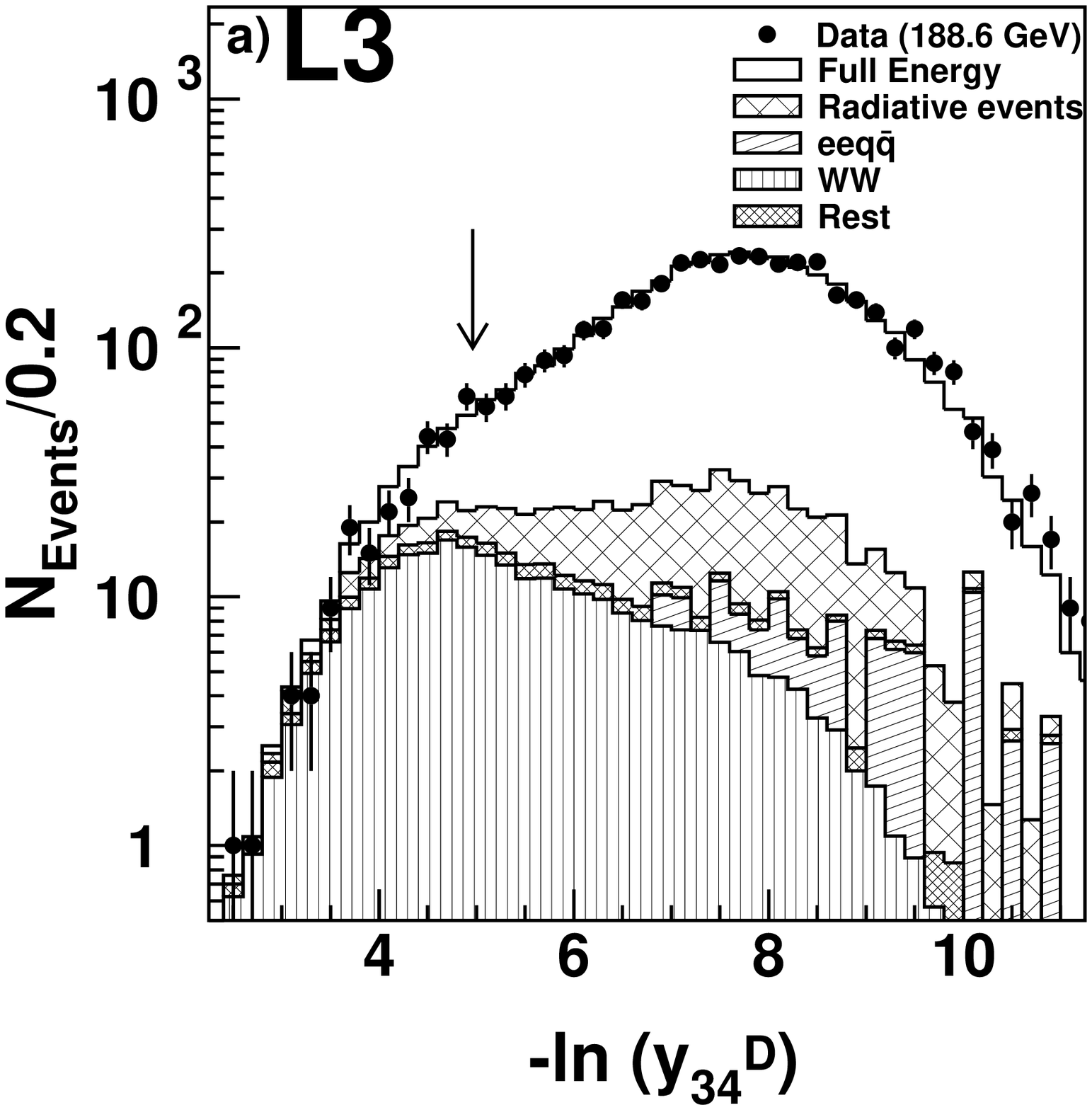}
  \includegraphics[width=.5\figwidth]{\mydirfig 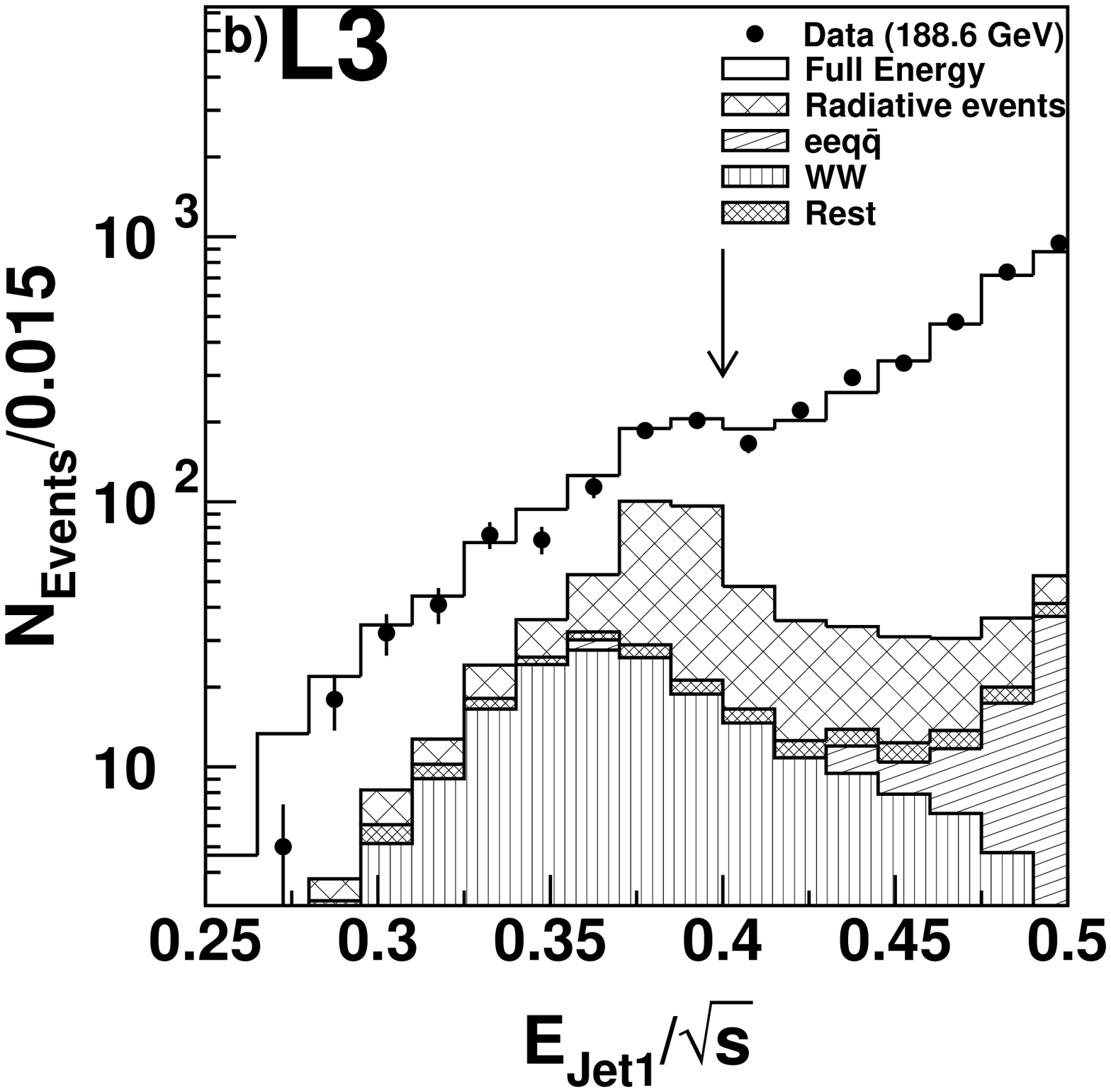}
  \includegraphics[width=.5\figwidth]{\mydirfig 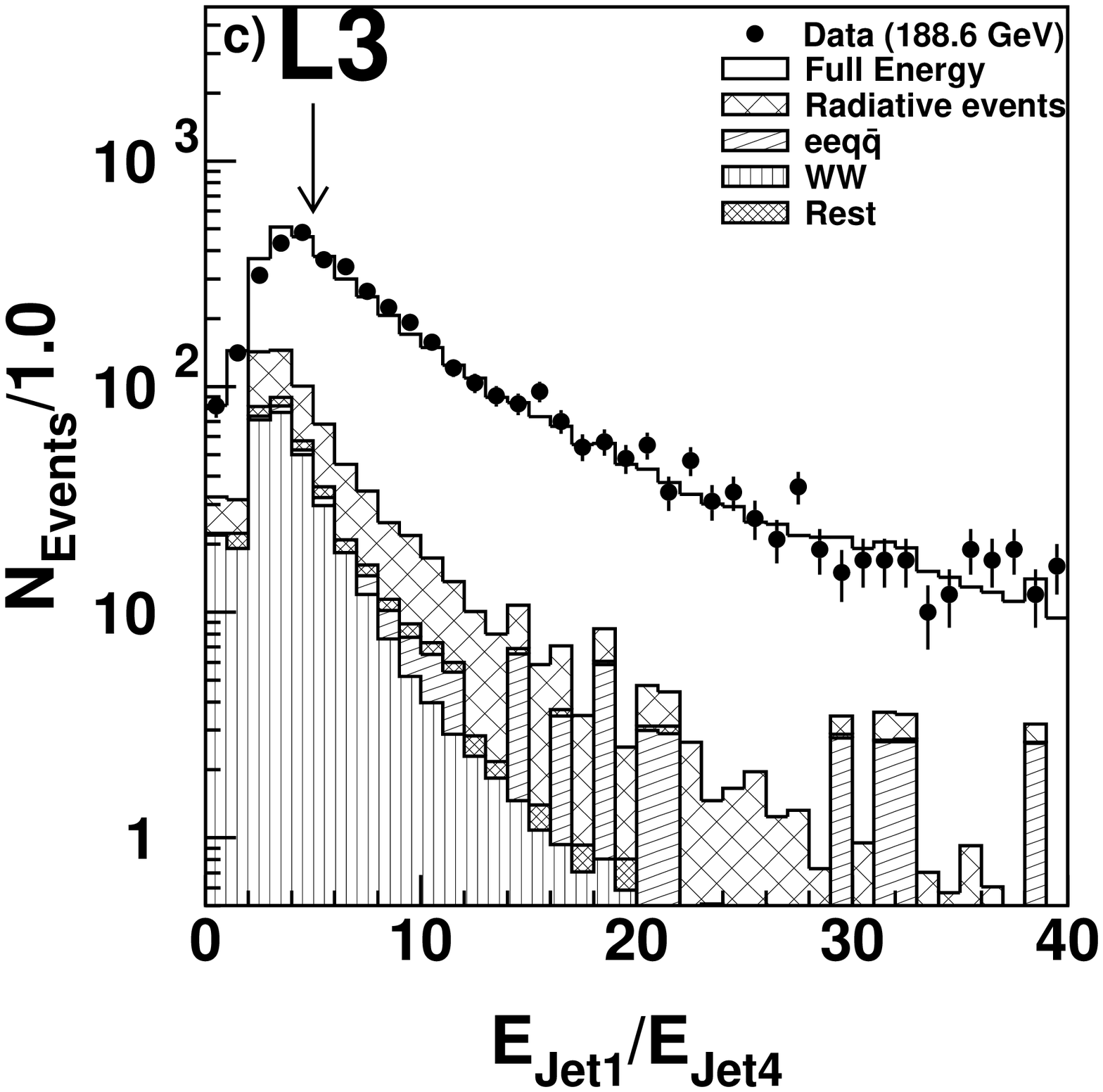}
\end{center}
\caption{For events at $\rs=188.6\,\GeV$,
         (a) distribution of $y_{34}^\mathrm{D}$, the value of the Durham jet resolution parameter at which
             the classification of an event changes from 3-jet to 4-jet.
         (b) distribution of the energy of the most energetic jet after the kinematic fit.
         (c) ratio of energy of the most energetic jet
                    to that of the least energetic jet after the kinematic fit.
         The arrows indicate the selection cuts.}
\label{fig:selec3}
\end{figure}

For the data at $\rs\geq 161.3\,\GeV$, additional backgrounds arise
from W-pair and Z-pair production.
A substantial fraction ($\sim80$\%)
of these events are removed by specific selections \cite{l3qcdlep2,l3qcdlep2a,l3qcdlep2b,l3qcdlep2c}.
To reject events where a W or Z decays into leptons we remove events having an electron or muon
with energy greater than 40\,\GeV.
Fully hadronic decays are rejected by
\begin{itemize}
 \item forcing the event to a 4-jet topology using the Durham algorithm \cite{kt,kt1,kt2,kt3},
 \item performing a kinematic fit imposing the constraints of energy-momentum conservation,
 \item making cuts on the energies of the most- and the least-energetic jets
             and on ${y_{34}^{\mathrm{D}}}$,
             the value of the jet resolution parameter at which the event classification changes from
             3-jet to 4-jet.
           Events are rejected if
           the energy of the most energetic jet is less than $0.4\rs$ (see Figure~\ref{fig:selec3}a),
           the ratio of the energy of the most energetic jet to that of the least energetic jet is smaller
             than 5 (see Figure~\ref{fig:selec3}b),
           ${y_{34}^{\mathrm{D}}}>0.007$ (see Figure~\ref{fig:selec3}c),
           there are more than 40 clusters and more than 15 charged tracks,
           and $\epar < 0.2\evis$ after the kinematic fit.
\end{itemize}
 
These cuts
remove between 3.6\% of signal events at the W-pair threshold
and           10.6\% at the highest centre-of-mass energy.
%
The data collected at high energy are combined into several energy bins.
The integrated luminosity, selection efficiency, purity and number of selected
events for each of the energy points are summarised in Table \ref{tab:events}.
 
\subsection{Flavour Tagging}
 
\begin{figure}[htbp]
\begin{center}
 \includegraphics[width=.98\figwidth]{\mydirfig 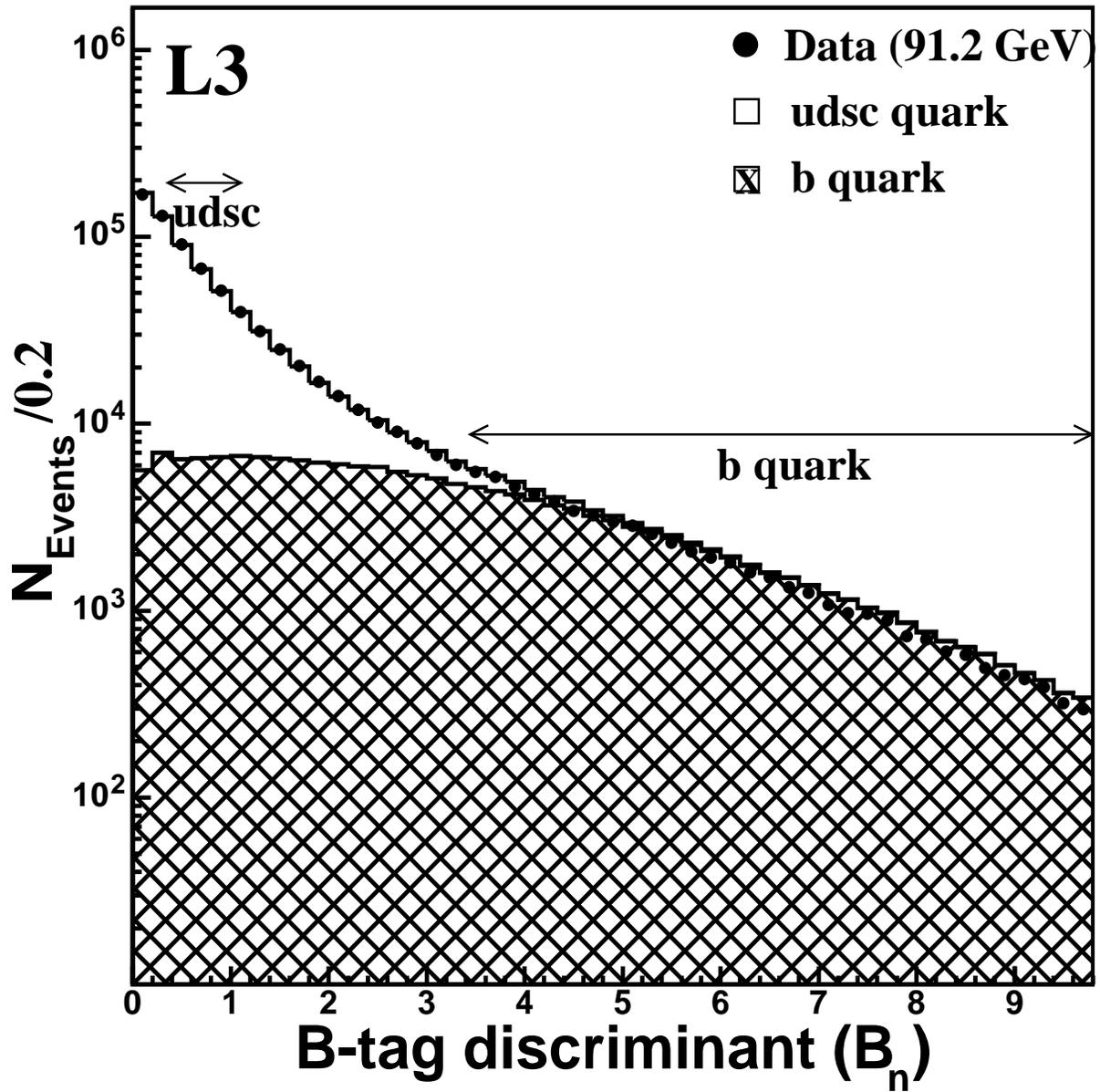}
\end{center}
\caption{Weighted discriminant for b-tagging, $\Bwt$, for the Z-pole data
         compared to the expectation of the \textsc{Jetset} PS Monte Carlo program.
         The cuts used to select udsc- and b-enriched samples are indicated.
         }
\label{fig:btag}
\end{figure}

Events with b-quarks can be separated from events with other flavours at the Z pole
using the characteristic decay properties of the b-hadrons.
As the first step, the interaction vertex 
is estimated by iteratively fitting all of the
good tracks measured in the detector in each beam-storage period.
Measurements of the decay lengths of all $n$ tracks in the event
contribute to a probability, $P^{[n]}$,
which would be flat for zero lifetime but otherwise peak at zero.
Figure \ref{fig:btag} shows the distribution of a weighted discriminant
$\Bwt = -\log\{P^{[n]}\sum_{j=0}^{n-1}\left(-\ln P^{[n]}\right)^j\left/j!\right.\}$
where $P^{[n]}=\prod_{j=1}^n P_j$
and $P_j$ is the probability that track $j$ originates at the primary vertex \cite{l3btag}.
 
A cut on this discriminant is made to distinguish udsc- from b-quark events.
The udsc-flavour events are selected using $0.3 < \Bwt < 1.0$ with
an efficiency of 39.2\% and a purity of 91.0\%.
The b-quark contamination amounts to 8.8\% of the selected events.
The b-flavour events are selected with a cut on $\Bwt> 3.4$ yielding
$6.3\cdot10^4$ b-enriched events with efficiency of 36.2\% and purity of 92.9\%.
The contamination due to udsc-flavour events in the sample is 7.0\%.
Measurement of flavour-tagged quantities uses only data taken after installation
and commissioning of the silicon micro-vertex detector \cite{l3:smd}.

\section{Event-Shape Variables}\label{sec:evshap}
 
\subsection{Choice of Variables}\label{sec:variables}
 
Event-shape variables, constructed from linear sums of measured particle momenta, are
  sensitive to the amount of hard gluon radiation and offer one of the most direct ways to
measure \as\ in \ee\ annihilation. They are insensitive to soft and collinear
radiation (`infra-red safe') and so can be reliably calculated in perturbative \QCD.
We measure six global event-shape variables for which improved analytical \QCD\
calculations \cite{qcd-t,qcd-hjm,qcd-jbm,qcd-jbm2,qcd-oth,qcd-c} are available.
These are thrust (\thr), scaled heavy jet mass (\rhoh), total (\bt) and
wide (\bw) jet broadening variables and the $C$- and $D$-parameters.
 
\renewcommand{\descriptionlabel}[1]{\hspace\labelsep\normalfont\bfseries #1}
\begin{description}
\item[Thrust:] The global event-shape variable thrust, \thr, \cite{thrust,thrust1}
is defined as
\begin{equation}
   \thr\;  =\;  \frac{\sum \, | \pai\cdot \nt |}{\sum \, |\pai |}\; ,
\end{equation}
where $\pai$ is the momentum  vector of particle $i$. The thrust axis, \nt,
is the unit vector which maximises the above expression. The value of the
thrust can vary from 0.5 for spherical events to 1.0 for narrow 2-jet events.
\end{description}
 
The plane normal to \nt\ divides space into two hemispheres,
$S_{\pm}$, which are used in the following definitions.
 
\begin{description}
\item[Scaled heavy jet mass:] The heavy jet mass,  \mh, is defined
\cite{hjm,hjm1,hjm2}
as
\begin{equation}
   \mh  =  \max\left[M_{+}(\nt),M_{-}(\nt)\right] \; ,
\end{equation}
where $M_{\pm}$ are the masses of the system of particles in the two hemispheres,  
\begin{equation}
   M_{\pm}^2 =  \left[ \sum_{i \in {S}_{\pm}} \paii \right]^{\textstyle2}   \; ,
\end{equation}
where  \paii\ is the four-momentum of particle $i$.
The scaled heavy jet mass, \rhoh, is defined as
\begin{equation}
   \rhoh = \left. \mh^2  \right/ E_\mathrm{vis}^2   \;.
\end{equation}
 
\item[Jet broadening variables:] These variables are defined \cite{qcd-jbm,qcd-jbm2} by
computing in each hemisphere the quantity
\begin{equation}
   B_{\pm} = \frac{\sum_{i\in {S}_{\pm}} | \pai\times\nt |}  {2\sum_{i} | \pai |}\;,
\end{equation}
in terms of which
the total jet broadening, \bt, and the wide jet broadening, \bw, are defined as
\begin{equation}
   \bt   =   B_{+} + B_{-} \;\;\; \mathrm{and} \;\;\;
   \bw   =   \max({B}_{+},B_{-})\;.
\end{equation}
 
\item[\boldmath{$C$}- and \boldmath{$D$}-parameters:] The $C$- and $D$-parameters are derived from the
eigenvalues of the linearised momentum tensor~\cite{tmatx,tmatx1}:
\begin{equation}
   \Theta^{ij} = \left. \sum_a \frac{\pp^i_a \pp^j_a}{\abs{\vec{\pp}_a}} \, \right/ \,
                         {\sum_a \abs{\vec{\pp}_a}} \quad i,j=1,2,3,
\end{equation}
where $\pp^i_a$ is the $i^\mathrm{th}$ component of the momentum vector, $\vec{\pp}_a$, of particle $a$.
With  $\lambda_1$, $\lambda_2$, and $\lambda_3$
the eigenvalues of $\Theta$,
the $C$- and $D$-parameters are defined as
\begin{eqnarray}
  C &=& 3 (\lambda_1\lambda_2 + \lambda_2\lambda_3 + \lambda_3\lambda_1) \;  \\
  D &=& 27 \lambda_1\lambda_2\lambda_3 \ .
\end{eqnarray}
\end{description}
 
A few other global event-shape variables are also measured for comparison
with the predictions of Monte Carlo models. These variables have linear or
quadratic dependence on particle momenta. Some of these parameters are particularly
sensitive to the details of fragmentation and hence are used to
tune and test Monte Carlo models.
 
\begin{description}
 \item[Major, minor:] Major (\tmajor)~\cite{obla}
  is defined in the same way as
  thrust but is maximised in the plane perpendicular to the thrust axis.
  The resulting direction is called the major axis, $\vec{n}_\mathrm{major}$.
  The minor axis, $\vec{n}_\mathrm{minor} = \vec{n}_\mathrm{major}\times\nt$,
  is defined to give an orthonormal system.
  Minor (\tminor) is the normalised sum of momenta
  projected onto $\vec{n}_\mathrm{minor}$.
 
 \item[Oblateness:] Oblateness ($\obla$)~\cite{obla} is the difference of the
  major and minor values:
\begin{equation}
              \obla = \tmajor - \tminor \;.
\end{equation}
 
 \item[Minor of the narrow side:] After dividing an event into
  two hemispheres by the plane perpendicular to the thrust axis, the transverse
  momentum fraction
\begin{equation}
    f_\mathrm{t} = \frac{\sum_i \abs{\pai \times \nt}}{\sum_i \abs{\pai}}
\end{equation}
  is calculated for each hemisphere.
  The hemisphere with the smaller $f_\mathrm{t}$ is
  called the narrow side. The minor calculated using only the particles in this
  hemisphere is defined as the minor of the narrow side, \tminorns,~\cite{mns}.
 
  \item[Scaled light jet mass:] This quantity is defined analogously to the scaled heavy jet mass:
\begin{equation}
   \rhol = \left. M_\mathrm{L}^2  \right/ E_\mathrm{vis}^2    \;, \quad
     M_\mathrm{L}  =  \min\left[M_{+}(\nt),M_{-}(\nt)\right] \;.
\end{equation}

  \item[Jet resolution parameters:] Jets are reconstructed using an invariant
  mass (\textsc{Jade}~\cite{jade,jade1}) or scaled transverse momentum (\kt\
  or Durham \cite{kt,kt1,kt2,kt3}) jet algorithm.
  The value of the jet resolution parameter, $y_{ij}$, at which the classification of an event changes
  from 2-jet to 3-jet is called the
  3-jet resolution parameter, $\yj$ and $\yd$ for the \textsc{Jade} and Durham algorithms, respectively.
 
  \item[Fox-Wolfram Moments:] The
  Fox-Wolfram moments~\cite{fwm,fwm1,fwm2} are given by
\begin{equation}
    H_\ell = \sum_{i,j} \frac{\abs{\vec{p}_i}\abs{\vec{p}_j}}{s} \, P_\ell(\cos\alpha_{ij})
\end{equation}
  where $\vec{p}_i$ and $\vec{p}_j$ are the momenta of particles $i$ and $j$,
  respectively, $\alpha_{ij}$ is the angle between these two particles, and
  $P_\ell$ is the Legendre polynomial of order $\ell$.
  The sums run over all particles in the events.
 
 \item[Sphericity, aplanarity:] Sphericity, $\spher$, and aplanarity, $\aplan$,
  are defined using the eigenvalues of the sphericity tensor~\cite{smatx},
\begin{equation}
    s^{ij} = {\frac{\sum_a p^i_a p^j_a}{\sum_a p^2_a}} ~~ i,j=1,2,3,
\end{equation}
  where $p^i_a$ is the $i^\mathrm{th}$ component of the momentum vector $\paa$.
  From the eigenvalues of $s^{ij}$,
  $Q_1 \le Q_2 \le Q_3$,
   the sphericity and aplanarity are defined as
\begin{equation}
                  S = \frac{3}{2}\, (Q_1 + Q_2)\ ; \qquad
                  A = \frac{3}{2}\, Q_1 \;.
\end{equation}
 
 \item[Spherocity:] The global event-shape variable, spherocity ($S^\prime$) \cite{spher,spher1}
   is defined as
\begin{equation}
     S^\prime   = \frac{4}{\pi} \cdot  \frac{\sum \, | \pai\times\vec{n}_{\mathrm{S}} |}
                                            {\sum \, |\pai |}\;,
\end{equation}
   where $\vec{n}_{\mathrm{S}}$, called  the spherocity axis,
   is the unit vector which minimises the above expression.
\end{description}

\subsection{Measurements}\label{sec:measurement}
 
The  distributions of the event-shape variables are measured
over the full energy range, 30--209\,\GeV, which includes the three types of event:
reduced-energy, Z-pole  and high-energy.
For the Z-pole data, they are also measured for b- and udsc-quark samples separately
using an integrated luminosity of 26.3\,\pb\
 
\begin{figure}[htbp]
\begin{center}
  \includegraphics*[width=.5\figwidth]{\mydirfig 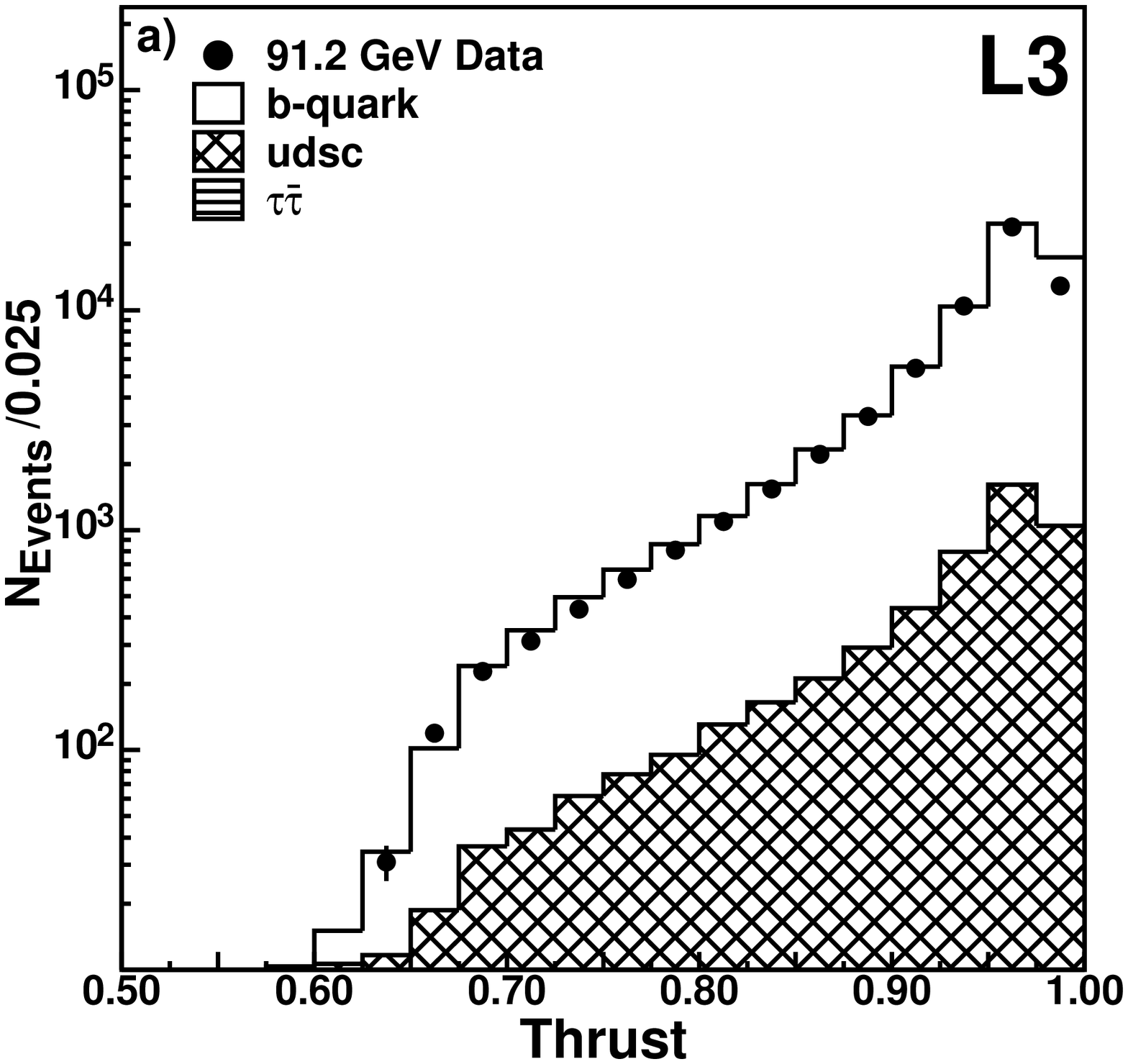}
  \includegraphics*[width=.5\figwidth]{\mydirfig 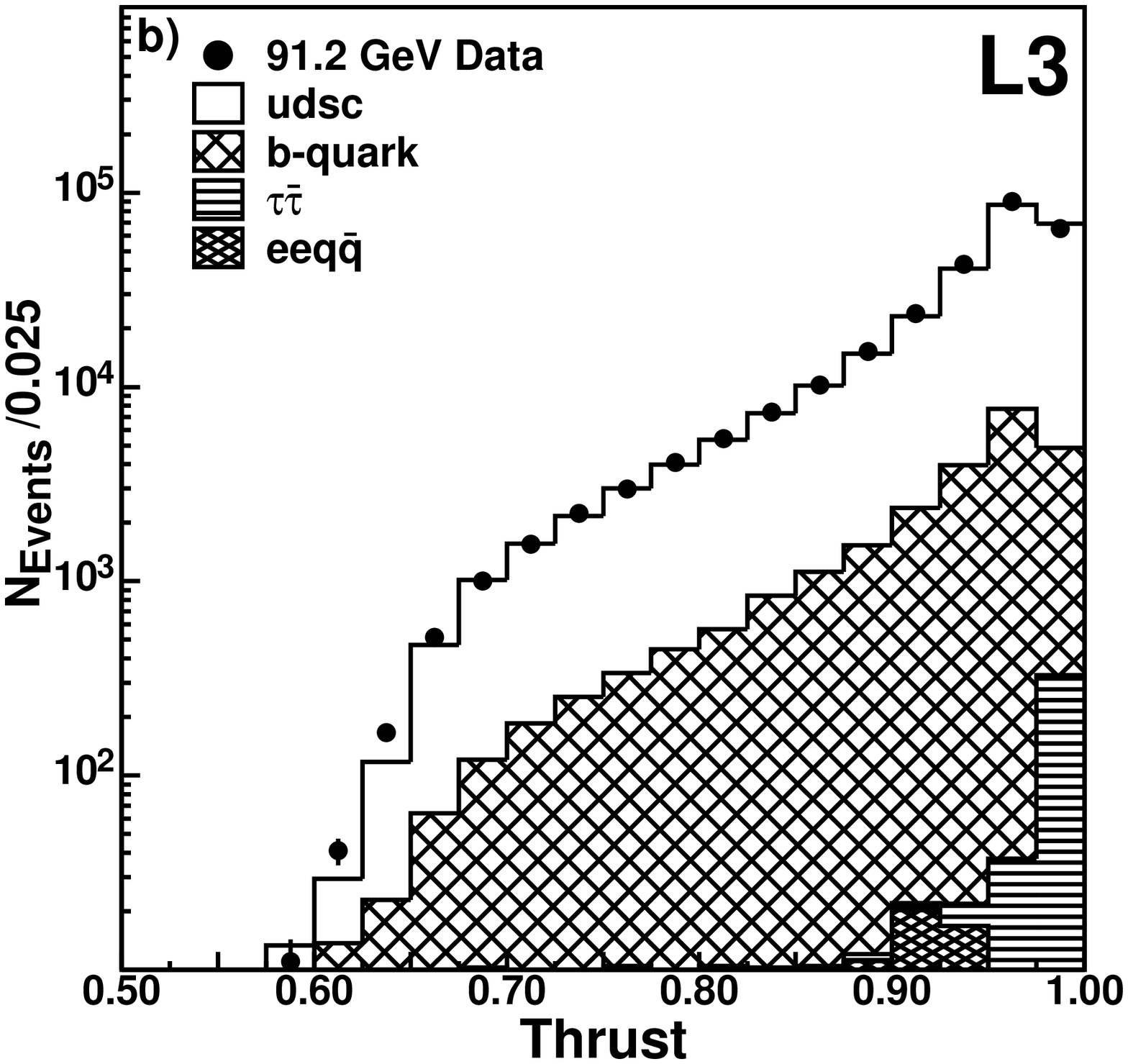}  \\[8mm]
  \includegraphics*[width=.5\figwidth]{\mydirfig 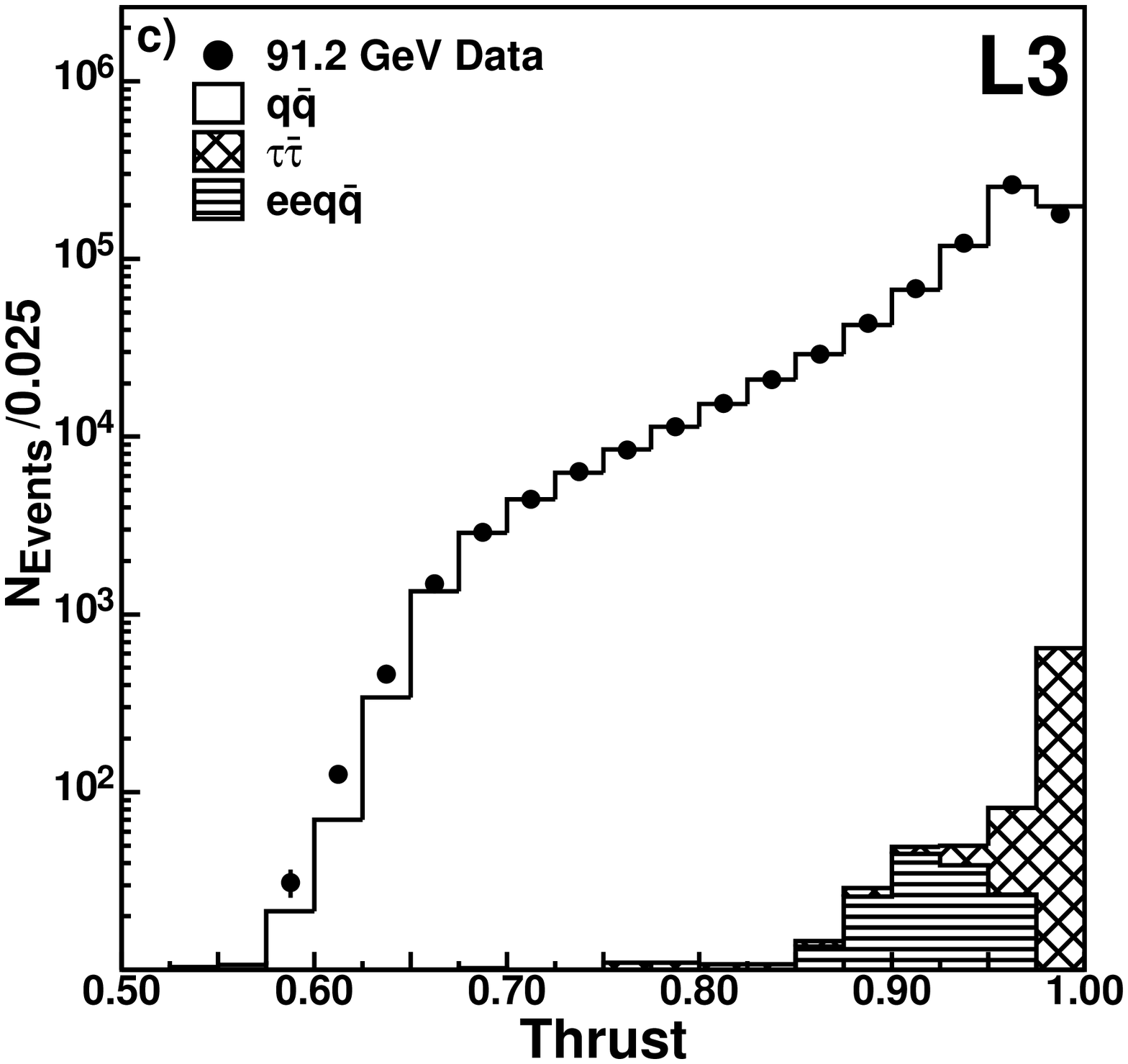}
\end{center}
\caption{Measured thrust distributions at the Z-pole for the (a) b- and (b) udsc-flavour-tagged samples,
         as well as for (c) all events.
           The solid lines correspond to the overall expectations from theory.
           The shaded areas refer to different backgrounds and the clear
           area refers to the signal predicted by \textsc{Jetset}.}
\label{fig:raw2}
\end{figure}

\begin{figure}[htbp]
\begin{center}
  \includegraphics*[width=.5\figwidth]{\mydirfig 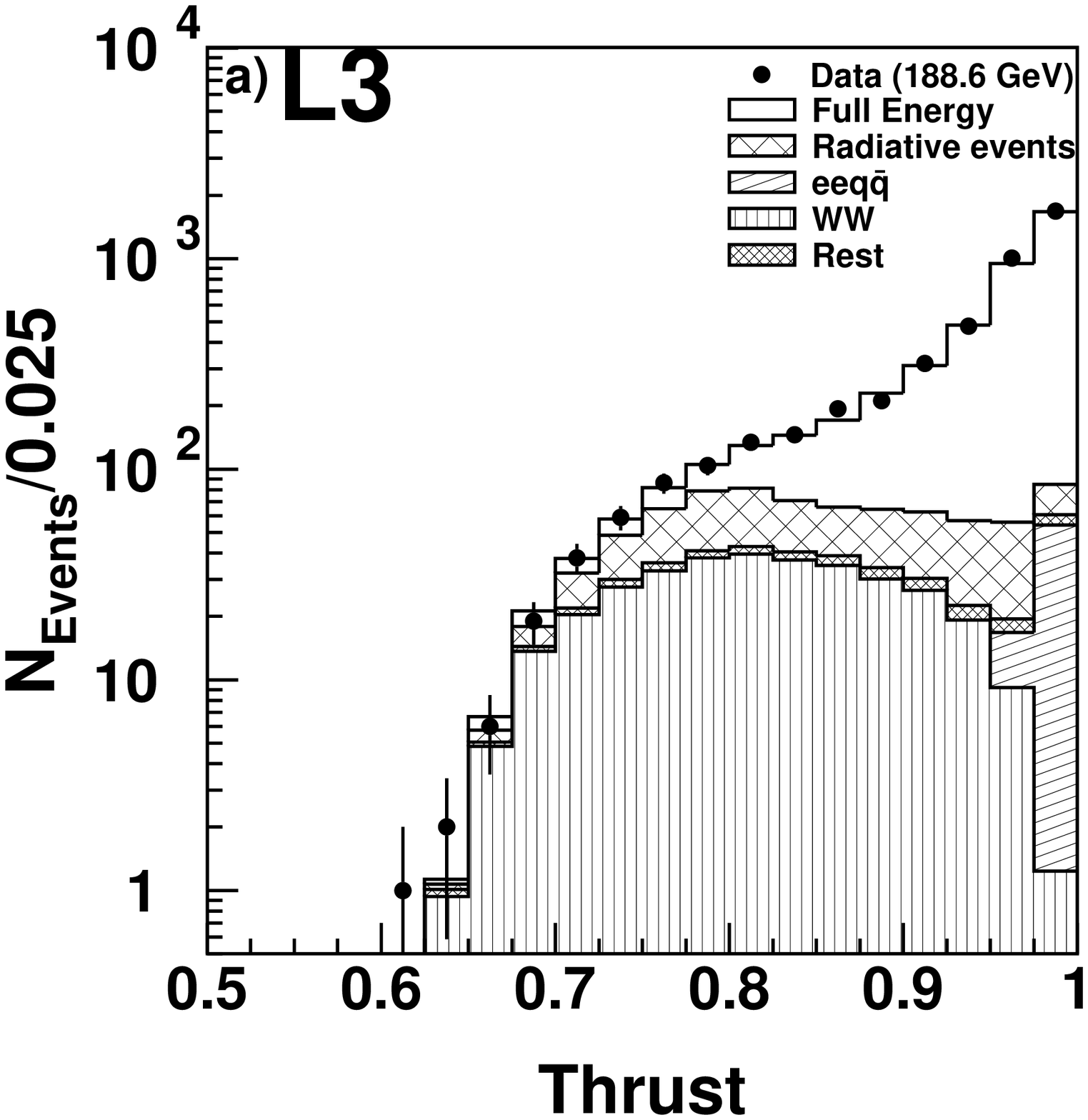}
  \includegraphics*[width=.5\figwidth]{\mydirfig 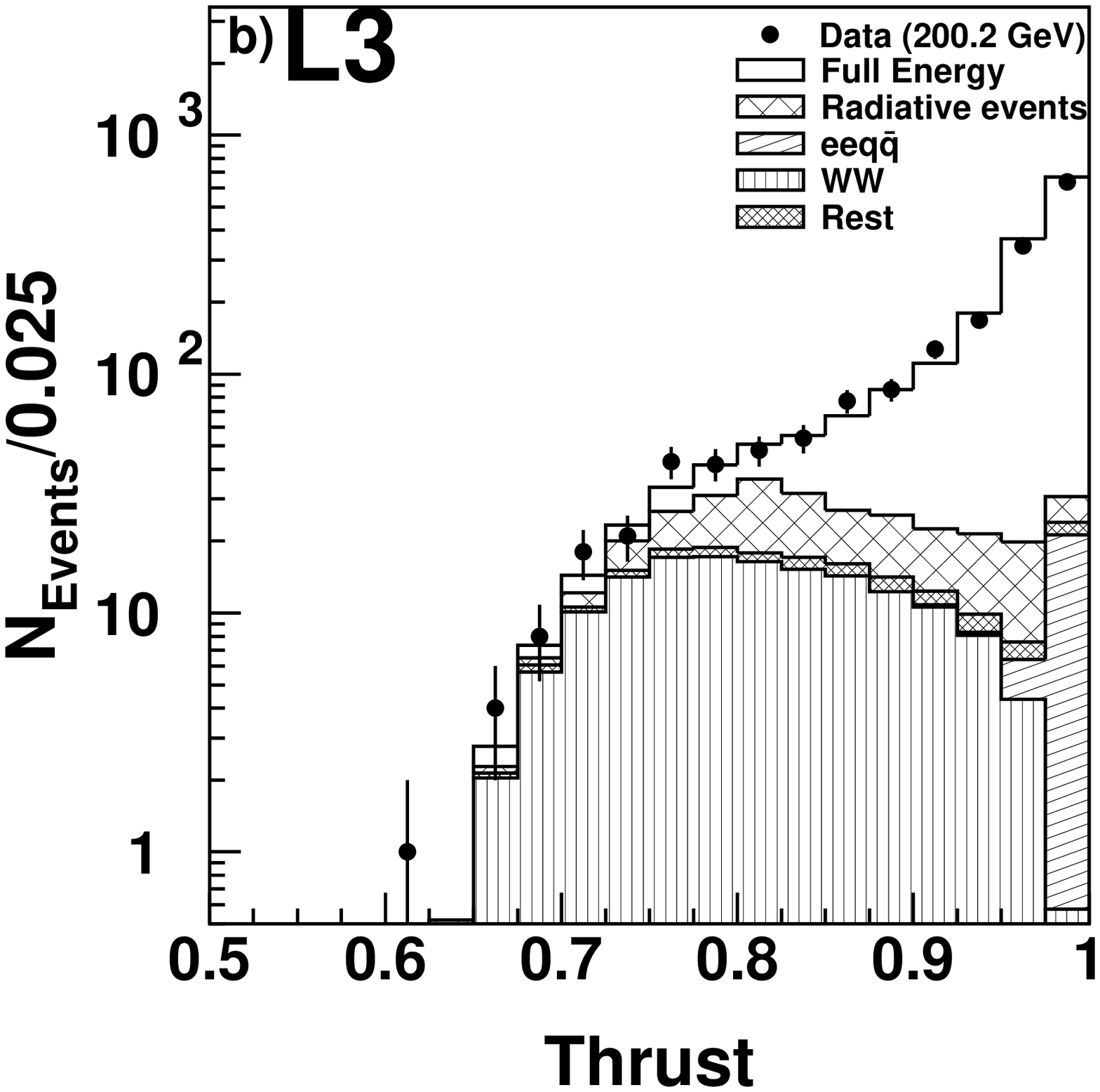}
\end{center}
\caption{Measured thrust distributions at $\rs=188.6\,\GeV$ and $\rs=200.2\,\GeV$
           The solid lines correspond to the overall expectations from theory.
           The shaded areas refer to different backgrounds and the clear
           area refers to the signal predicted by \textsc{Jetset}.}
\label{fig:raw3}
\end{figure}

The data distributions are compared to a combination of the signal and the
different background Monte Carlo distributions obtained using the same
selection procedure and normalised to the integrated luminosity.
Figures~\ref{fig:raw1}--\ref{fig:raw3} show uncorrected thrust
distributions measured in the six energy bins of the reduced centre-of-mass
energy, flavour-tagged Z-pole, and high-energy samples, respectively, compared to Monte Carlo predictions.
The contributions of the shaded areas indicate the various backgrounds.
For the reduced-energy events (Figure~\ref{fig:raw1} the backgrounds considered are
unresolved \Pgpz's and $\eta$'s in the hadronic sample, as well as $\tau$-pair and 2-photon
processes.
The prediction of \textsc{Jetset} has been scaled to account for the lack of isolated
energetic \Pgpz's in the string fragmentation process \cite{scalepi0}.
At the Z pole, background in the flavour tagged samples is
dominated by hadronic events of the other flavour class, but
is negligible for the full sample (Figure~\ref{fig:raw2}).
At high energies (Figure \ref{fig:raw3}) the main backgrounds are radiative events, W-pair production
and 2-photon processes.
The Monte Carlo distributions agree with the data reasonably well at all centre-of-mass energies.
 
The global event-shape variables are calculated
before, `particle level', and after, `detector level', detector simulation.
The calculation before detector simulation takes into account all stable
charged and neutral particles.
The measured distributions at detector level differ from those at
particle level because of detector effects, limited acceptance and
resolution.
The resolution for the thrust varies from about 0.02 at high values to 0.05 at low values.
The resolution is similar for the other shape variables.
After subtracting the background obtained from simulations, the measured
distributions are corrected for detector effects, acceptance and resolution
on a bin-by-bin basis by comparing the detector level results with the
particle level results.
The level of migration is kept at  
an acceptable level by using
a bin size approximately equal to or greater than the experimental resolution.
We also correct the data for initial and final state photon radiation
bin-by-bin using Monte Carlo distributions at particle level with and
without radiation.

\subsection{Systematic Uncertainties}     \label{sec:systematics}
 
The systematic uncertainties in the distributions of event-shape variables
arise mainly from uncertainties in the estimation of detector corrections and
background.
The uncertainty in the detector correction is estimated by several independent checks:
\begin{itemize}
 \item The definition of reconstructed objects used to calculate the
       observables is changed. Instead of using only \SRC\ calorimetric
       clusters, the analysis is repeated using the \ECLU\ objects
       defined in Section \ref{sec:energycal}.
 \item The effect of different particle densities in correcting the measured
       distribution is estimated by using a different signal Monte Carlo program,
       \textsc{Herwig} instead of \textsc{Jetset PS} or \textsc{Pythia}.
 \item The acceptance is reduced by restricting the events to the more precise
       central part of the detector, $\abs{\cos(\theta_\mathrm{T})}< 0.7$, where
       $\theta_\mathrm{T}$ is the polar angle of the thrust axis relative
       to the beam direction.
\end{itemize}
 
The uncertainty on the background composition of the selected event sample
is estimated differently at different centre-of-mass energies.
The systematic uncertainty in the Z-pole flavour-tagged samples is estimated by varying the background from
mis-tagged events by $\pm10\%$.
In addition, the background in the udsc sample from 2-photon processes is varied by $\pm30\%$.
 

At reduced 
energies,
the systematic uncertainties are estimated by varying:
\begin{itemize}
 \item the amount of background from
       misidentified hadrons or non-direct photon production
       by the uncertainty of its estimation from data \cite{scalepi0};
 \item the selection cuts used to select direct photons:
       jet and local isolation angles, energy in the local isolation cone, and
       the neural network probability.
\end{itemize}
The uncertainty at high energies is estimated by repeating the analysis
with:
\begin{itemize}
 \item an alternative criterion to reject the hard initial state photon
       events based on a cut on the kinematically reconstructed effective
       centre-of-mass energy;
 \item a variation of the estimated two-photon interaction background by
       $\pm$ 30\%
       and by using the program
       \textsc{Phojet} instead of \textsc{Pythia} to estimate this background;
 \item a variation of the background estimate by changing the W-pair rejection criteria.
       As an extreme variation, no 4-jet events are rejected from
       the data sample and the number of W-pair events is estimated from
       \textsc{KoralW} Monte Carlo and subtracted from the data.
\end{itemize}
At high energies, uncertainties due to \ISR\ and W-pair background are the most
important. They are roughly equal and are 2--3 times
larger than the uncertainties due to the detector correction.
 
The systematic uncertainties obtained from different sources are combined in
quadrature. Statistical fluctuations
are not negligible in the estimation of systematic effects.
The statistical component of the systematic uncertainty is determined by
splitting the overall Monte Carlo sample into luminosity weighted
sub-samples and treating each of these sub-samples as data.
The statistical component
of the systematic uncertainty is estimated from the differences in these sub-samples.
This component is subtracted in quadrature from the original estimate.

\subsection{Tuning of Monte Carlo Parameters}\label{sec:tuning}
 
The Monte Carlo models involve several parameters.
Particular shape-variable distributions
are especially sensitive to certain parameters and these
distributions are used to tune their values.
To match Monte Carlo with data we proceed as follows.
First, a few 
event-shape variables with special sensitivity to certain parameters
are chosen to be tuning variables
for the comparison of data and Monte Carlo:
\begin{itemize}
   \item the jet resolution parameter in the \textsc{Jade} algorithm which corresponds to
         the transition from 2 to 3 jets ($\yj$).
         This variable is sensitive primarily to the 3-jet rate.
   \item the fourth Fox-Wolfram moment ($\fwm$), which is sensitive to the angles between jets.
   \item the minor of the narrow side ($\tminorns$).
         This variable is sensitive to the lateral size of the quark jet.
   \item the charged particle multiplicity ($\nch$).
\end{itemize}
If a model to describe Bose-Einstein correlations is tuned, the distributions of the four-momentum
difference for like- and unlike-sign charged particle pairs are also used.
 
For a set of values of the parameters, $\pk$,
to be tuned, the Monte Carlo distributions of the tuning variables
are compared to the data distributions. This is quantified by
\begin{equation}
  \chisq(\pk)  = \sum_{i = \mathrm{tun.\,var}} \,  \sum_{j = \mathrm{bins}} \,
      {\frac{[{\text{Data}(i,j)} - {\text{MC}(i,j,\pk)}]^{2}}
            {[\sigma^\mathrm{stat}_{\text{Data}}(i,j)]^{2} +
             [\sigma^\mathrm{syst}_{\text{Data}}(i,j)]^{2} +
             [\sigma^\mathrm{stat}_{\text{MC}}(i,j,\pk)]^{2}}}
\end{equation}
where the individual contributions to \chisq\  are summed
over all bins ($j$) of the distributions of the chosen tuning variables ($i$).
The optimal parameter set is taken to be the one that minimises
the above \chisq\ function, and is found
using the \CERN\ program package \textsc{Minuit} \cite{minuit}.
Bins with insignificant statistics are ignored in the fit.
 
The parameters of a model to be tuned span a continuous multi-dimensional
space, and thus the \chisq\ function is a continuous function of the tuning variables.
However, in any realistic tuning procedure, one starts off with a finite
set of guesses for the optimal parameter set, and generates
Monte Carlo distributions for the event-shape variables only at these
discrete points in the parameter space.
 
More than $10^5$ events are generated for several points on a grid
in the parameter space.
For a grid with $k$-parameters and $n_p$ different values for parameter $p$,
one needs to generate events at $\prod_{p=1}^{k} n_p$ points.
In the subsequent minimisation procedure, \chisq\ values at points
between the grid points are found by  a local multidimensional interpolation,
either linear or  non-linear.
The Monte Carlo distribution
corresponding to the $j^\mathrm{th}$ bin for the $i^\mathrm{th}$ tuning variable,
$\text{MC}(i,j)$, for points in parameter space inside the grid
using a polynomial of given degree is given by 
\begin{equation}
   \mathrm{MC}(i,j,\vec{\alpha}_{0} + \delta \vec{\alpha})  =
      a_0(i,j)
   + \sum_{m=1}^k   a_{1}(i,j)_{m}  \delta\alpha_{m}
   + \sum_{m,n=1}^k a_{2}(i,j)_{mn} \delta\alpha_{m} \delta\alpha_{n}
   + \cdots
\end{equation}
 
These fits are repeated by varying the fit range of the tuning variables,
the degree of the polynomial in the interpolation, and also by changing the choice of grid points.
Each of these systematic variations yields possible sets of optimal values for the tuning parameters.
To decide among them, a new \chisq\ is calculated
using additional global event-shape variables:
$T$, \rhoh, $\rho_\mathrm{L}$, 
\bt, \yd, $\spher$, $\aplan$, $S^\prime$, $C$, $D$, \tmajor, \tminor, $O$ and  $H_3$.
For the tuning of
\textsc{Ariadne} 4.12,  
\textsc{Pythia} 6.2     
and \textsc{Herwig} 6.2 \cite{herwigsix}
the sums of the components of
momentum in and perpendicular to the event plane, as well as $\xi$ were also used.
The set with the smallest value of this \chisq\ is taken as the
tuned parameter set for the Monte Carlo model.
The systematic uncertainties on the parameters are obtained by varying the fit ranges and degree of
polynomials in the interpolation function.

Tuning is carried out with event-shape distributions obtained at the Z pole.
Separate tunings were done for all quark flavours and for udsc flavours.
The results of the tuning are summarised in Table \ref{tab:tunmodel}
for the models which are compared to data in this Report,
except for \textsc{Cojets}, which was previously tuned~\cite{l3qcd91z}.
The cut-off parameter $Q_0$ and
the fragmentation parameter $a$ in the \textsc{Jetset} 7.4 PS model
are fixed at $Q_0 = 1\,\GeV$ and $a=0.5$.
The parameter of the Peterson fragmentation function parameters for charm and bottom quarks
are fixed at $\epsilon_\mathrm{c}=0.03$ and $\epsilon_\mathrm{b}=0.0035$, respectively, which are chosen
to reproduce the mean energies of c and b hadrons \cite{hfwg}.
For the  \textsc{Jetset} 7.4 ME model the parameters kept fixed are: $a=0.5$, $\epsilon_\mathrm{c}=0.10$ and
$\epsilon_\mathrm{b}=0.004$ 
in order to obtain the same mean energies for c and b hadrons as for the PS model.
 
The udsc flavour-tagged data are also used to tune models for precision
studies of W-boson processes. The results of tuning the
\textsc{Pythia} 6.2 parton shower program are summarised in
Table~\ref{tab:tunpythia}. These results refer to the cut-off parameter value $Q_0= 1\,\GeV$
and the fragmentation parameter values $a  = 0.5$, $\epsilon_\mathrm{c}=0.03$ and
$\epsilon_\mathrm{b}=0.002$. The results of tuning
the \textsc{Ariadne} 4.12 and \textsc{Herwig} 6.2 models are summarised in
Tables~\ref{tab:tunariadne} and \ref{tab:tunherwig}, respectively.
The \textsc{Herwig} program is adapted to use the particle decay
and Bose-Einstein routines \cite{lundbe,lundbe1} of \textsc{Pythia} 6.2.

\section{Event-Shape Distributions and {\boldmath $\as$}}\label{sec:evshapd}
 
Since the probability of hard gluon radiation is directly determined by \as,
a direct measurement of \as\ is provided by 
the fraction of events having a specified number of jets.
These so-called jet fractions are measured and their behaviour
as a function of centre-of-mass energy investigated in Section~\ref{sec:jetfrac}.
To determine \as, we  
use the event-shape variables.  Their distributions are measured and compared
to Monte Carlo models in Section~\ref{sec:mcmodel}.
The applicability of power law corrections is investigated in  Section~\ref{sec:power}, and
\as\ is extracted in Section~\ref{sec:alphas}.

\subsection{Jet Fractions}\label{sec:jetfrac}
 
Jets are constructed using the \textsc{Jade} algorithm \cite{jade,jade1}.
The following expression is evaluated for each pair of particles $i$ and $j$:
\begin{equation}
   y_{ij}^\mathrm{J}  =  \frac{2E_{i}E_{j}}{s} \left( 1 - \cos\theta_{ij} \right)
\end{equation}
where $E_{i}$ and $E_{j}$ are their energies and $\theta_{ij}$ is the angle between them.
The pair for which $y_{ij}^\mathrm{J}$ is the smallest is replaced by a pseudo-particle $l$ with
four-momentum
\begin{equation}
   p_{l}  =  p_{i} + p_{j} \; .
\end{equation}
This procedure is repeated until all the $y_{ij}^\mathrm{J}$, calculated using the remaining particles
and pseudo-particles, exceed the jet resolution parameter \ycJ.
These remaining particles and pseudo-particles are called jets.
The jet fraction $f_{i}$ is the fraction of all hadronic events containing $i$ jets
\begin{equation}
   f_{i}  =  \frac{N_{i\;\mathrm{jets}}}{N_\mathrm{tot}} \;.
\end{equation}
 
The observed jet fractions are corrected, on a bin-by-bin basis, for the effects of remaining
background, detector resolution and acceptance using Monte Carlo events
for signal and background processes as described
in the treatment of event-shape variables in section~\ref{sec:measurement}.
 
\begin{figure}[htbp]
\begin{center}
 \includegraphics*[width=0.5\figwidth]{\mydirfig 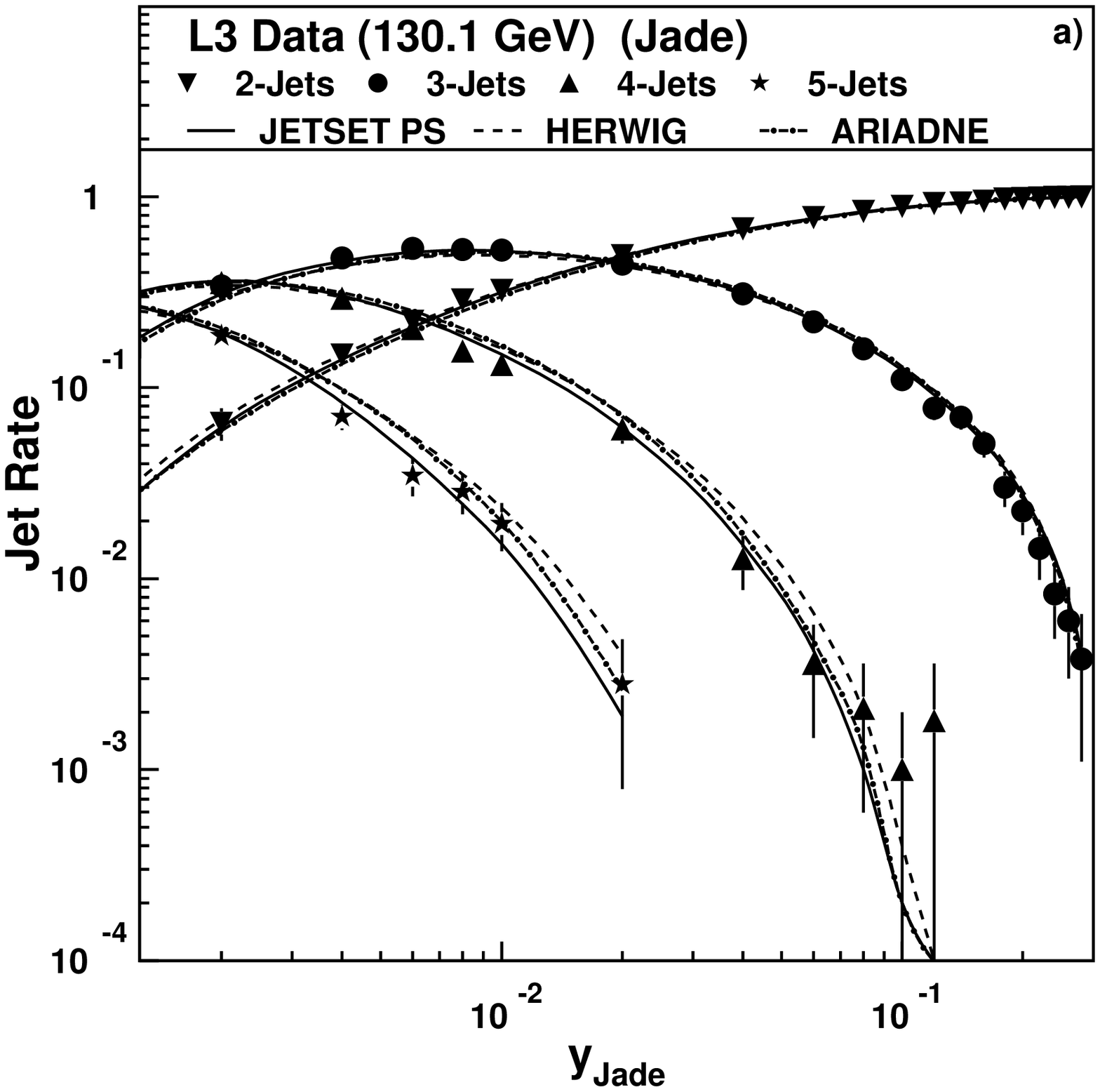}
 \includegraphics*[width=0.5\figwidth]{\mydirfig 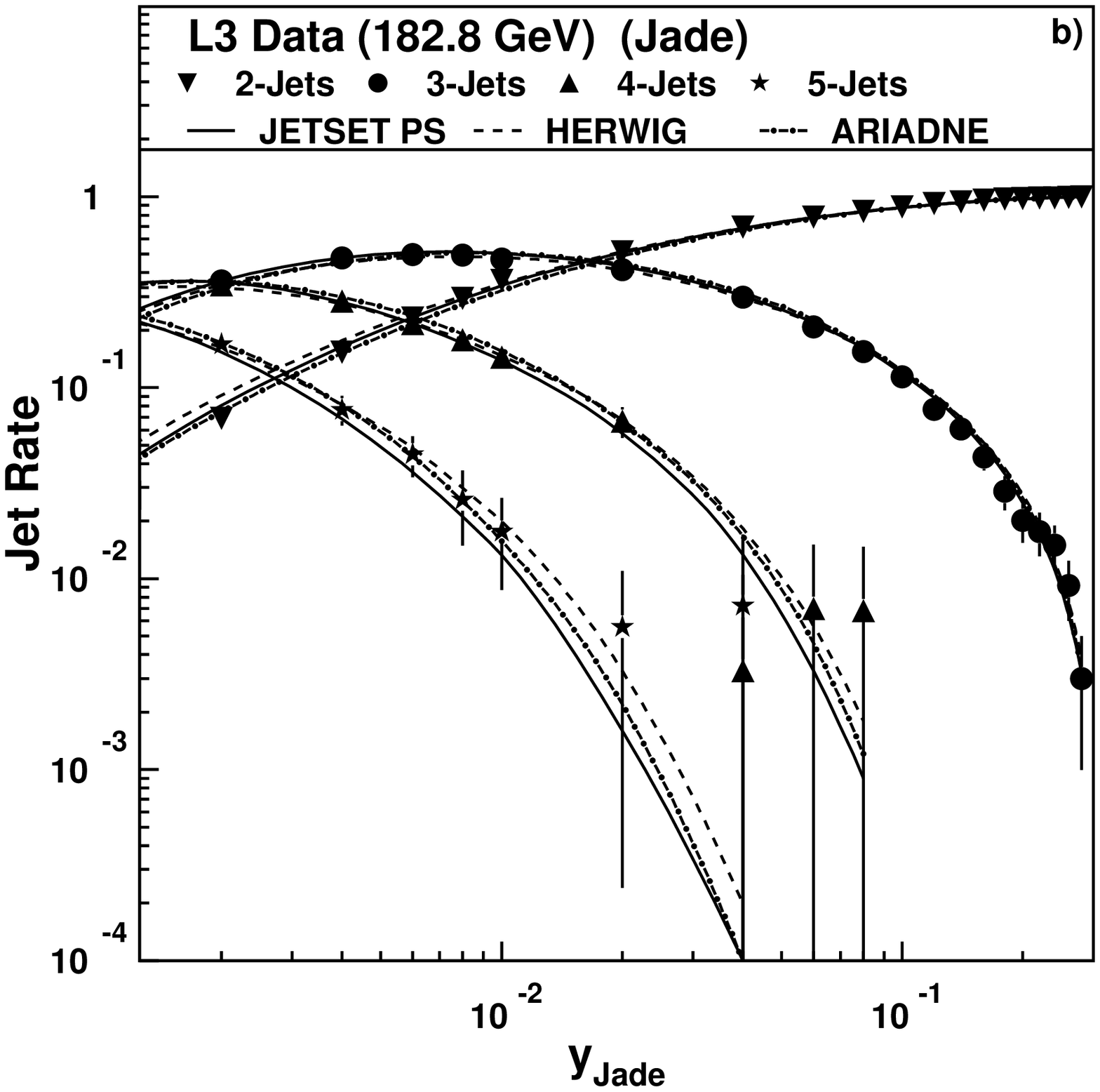}
 \includegraphics*[width=0.5\figwidth]{\mydirfig 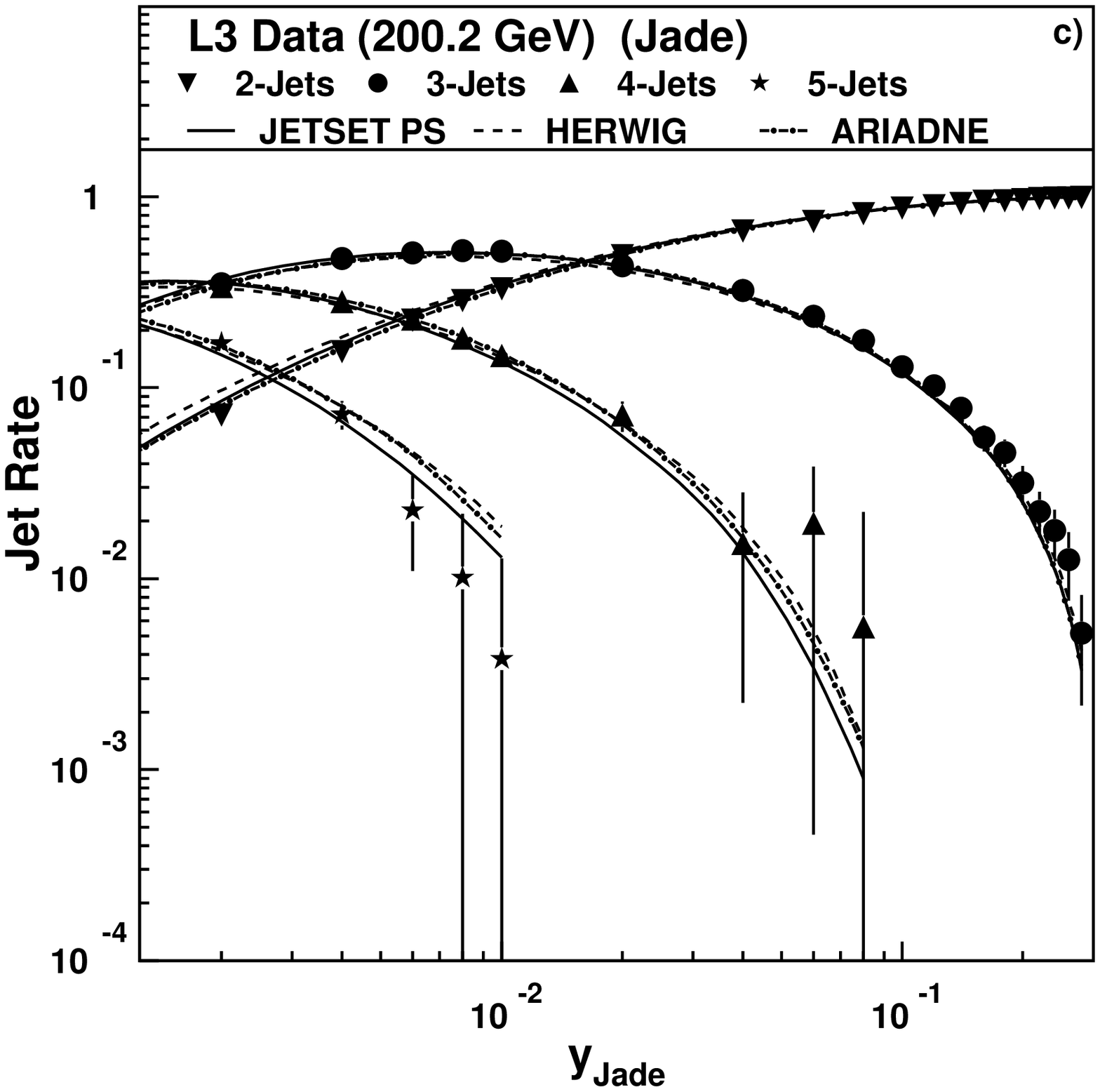}
 \includegraphics*[width=0.5\figwidth]{\mydirfig 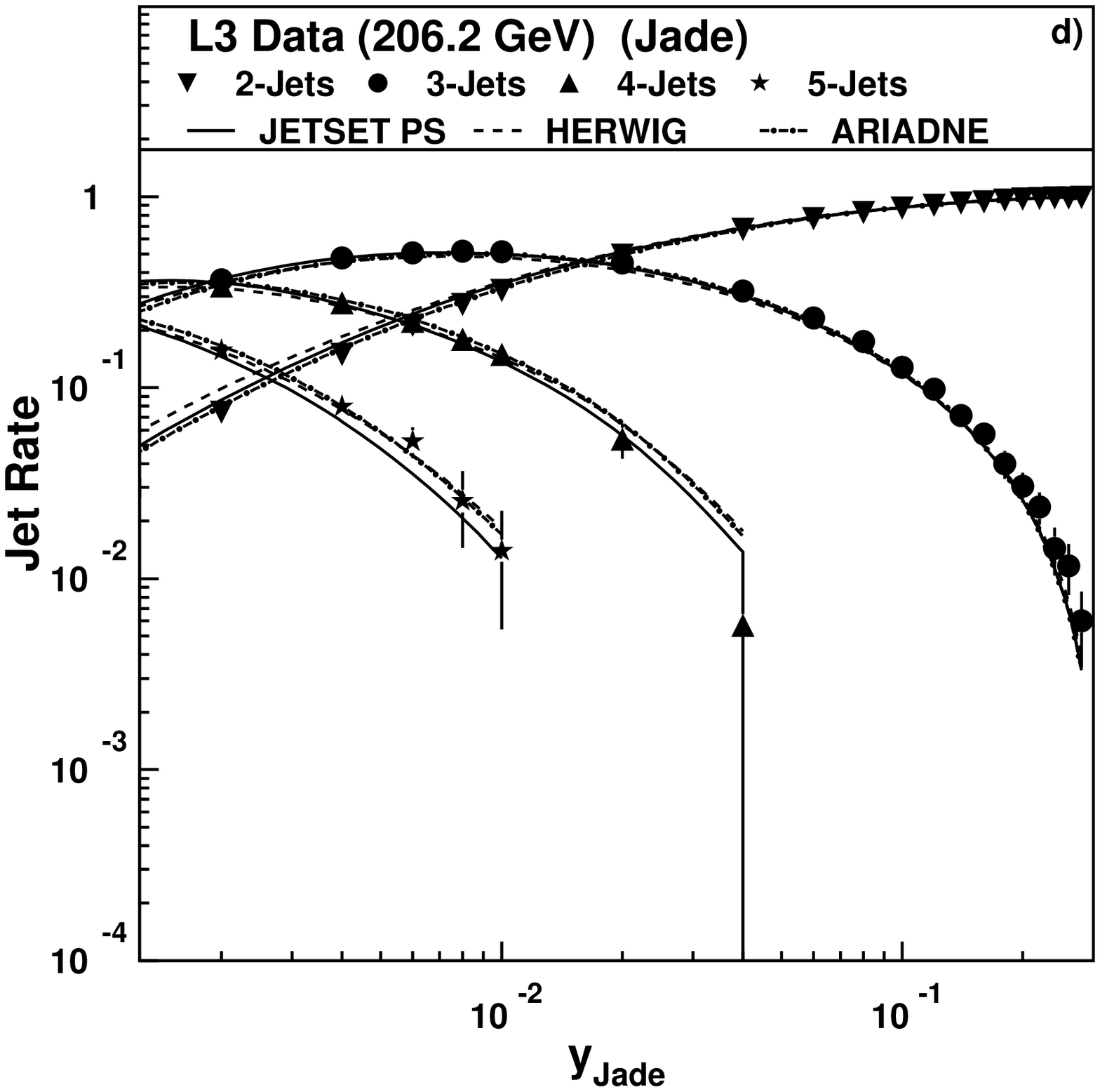}
\end{center}
\caption{Fraction of 2-, 3-, 4- and 5-jet events as a function of jet resolution parameter \ycJ\
           at $\rs = 130.1, 182.8,  200.2\text{ and }206.2\,\GeV$
           for the \textsc{Jade} algorithm.}
\label{fig:jetjd}
\end{figure}

The corrected fractions for 2-, 3-, 4- and 5-jet production at the different
centre-of-mass energies are summarised in
Tables \ref{tab:jade130}--\ref{tab:jade206}.
These fractions are plotted as a function of the jet resolution parameter \ycJ\ in Figure~\ref{fig:jetjd}
at mean centre-of-mass energies of 130.1, 182.8, 200.2 and 206.2\,\GeV.
The data are compared with predictions of various parton shower models,
which are found to describe the data rather well.
 
\begin{figure}[htbp]
\begin{center}
 \includegraphics*[width=0.5\figwidth]{\mydirfig 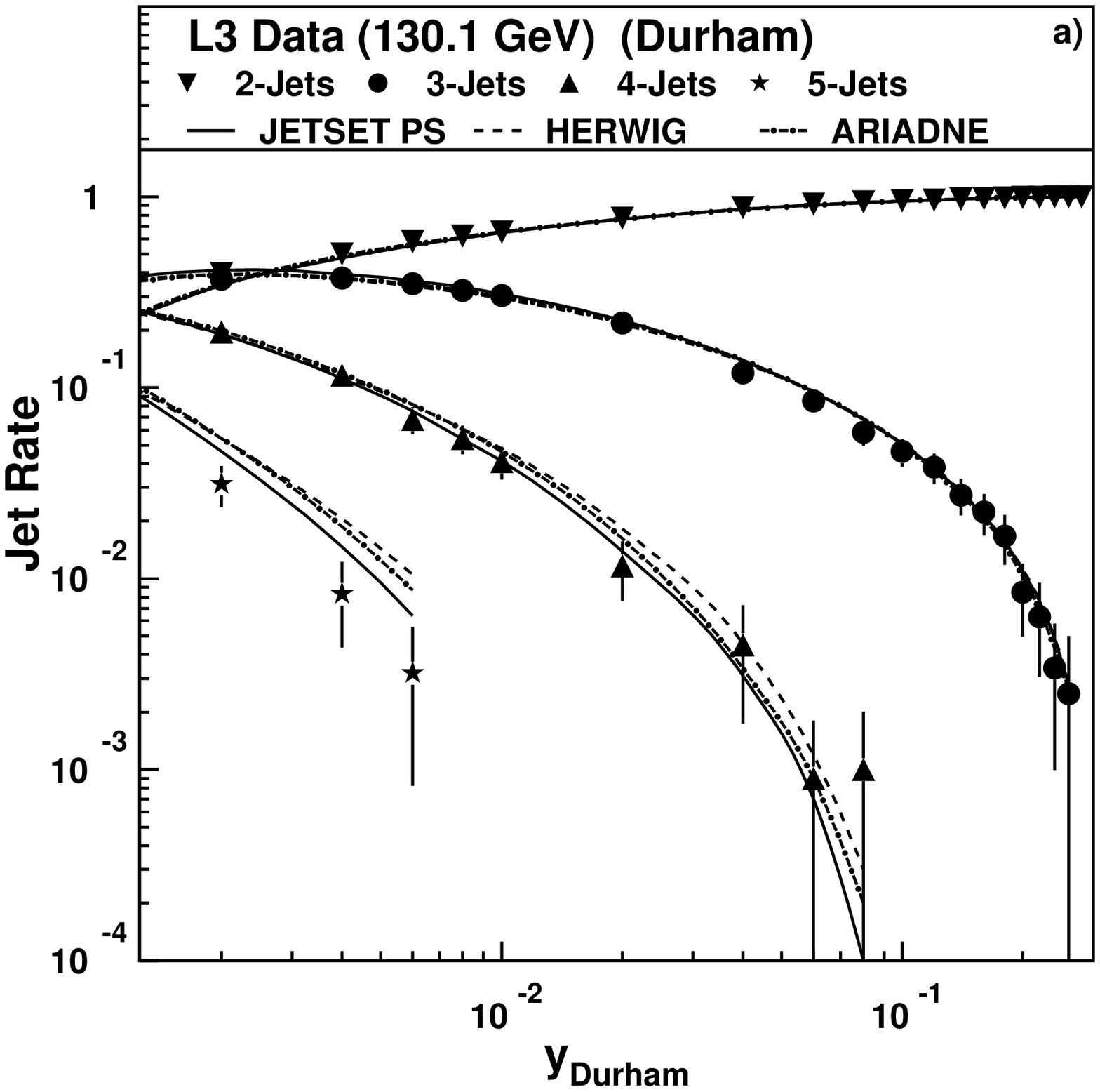}
 \includegraphics*[width=0.5\figwidth]{\mydirfig 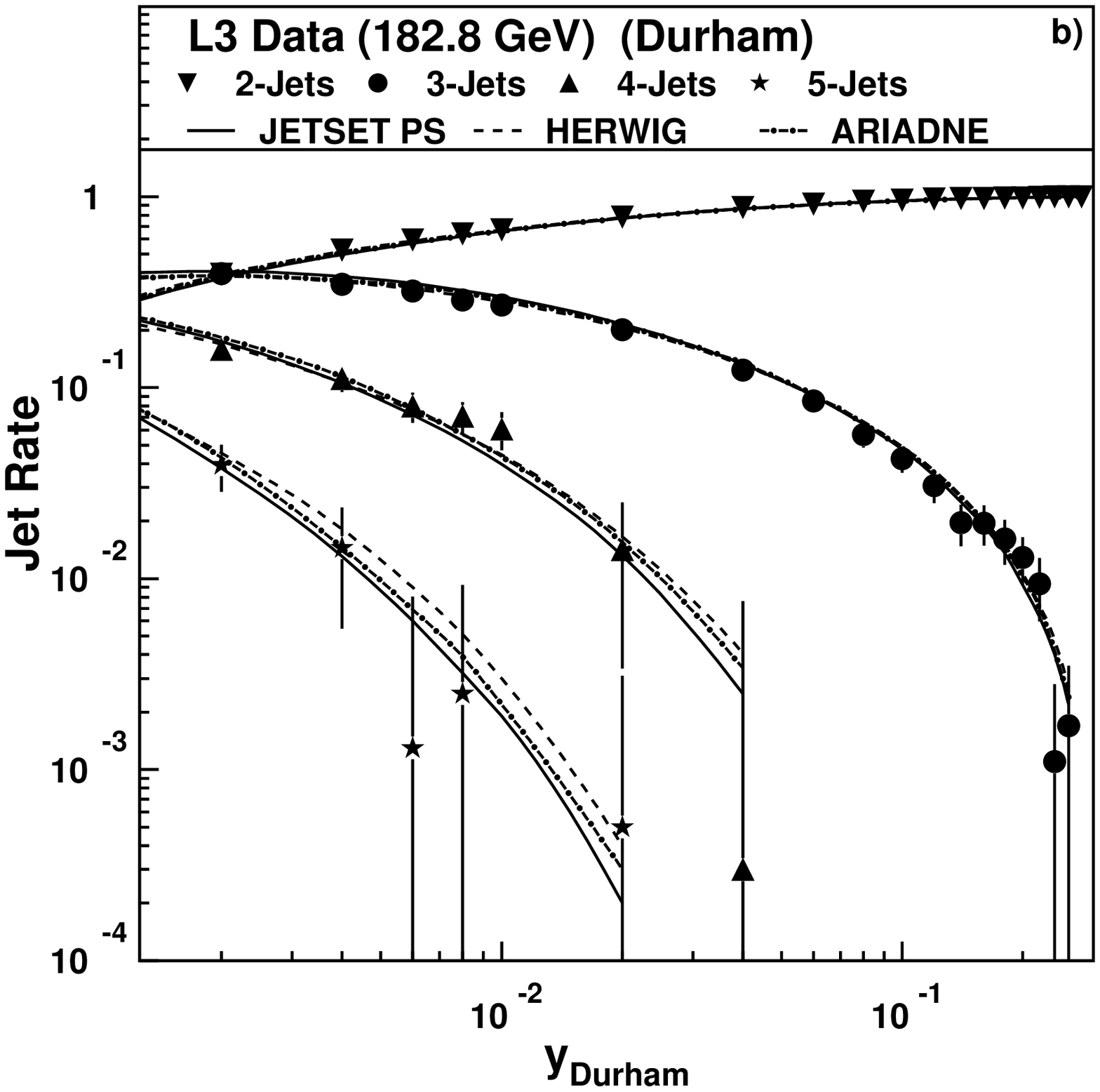}
 \includegraphics*[width=0.5\figwidth]{\mydirfig 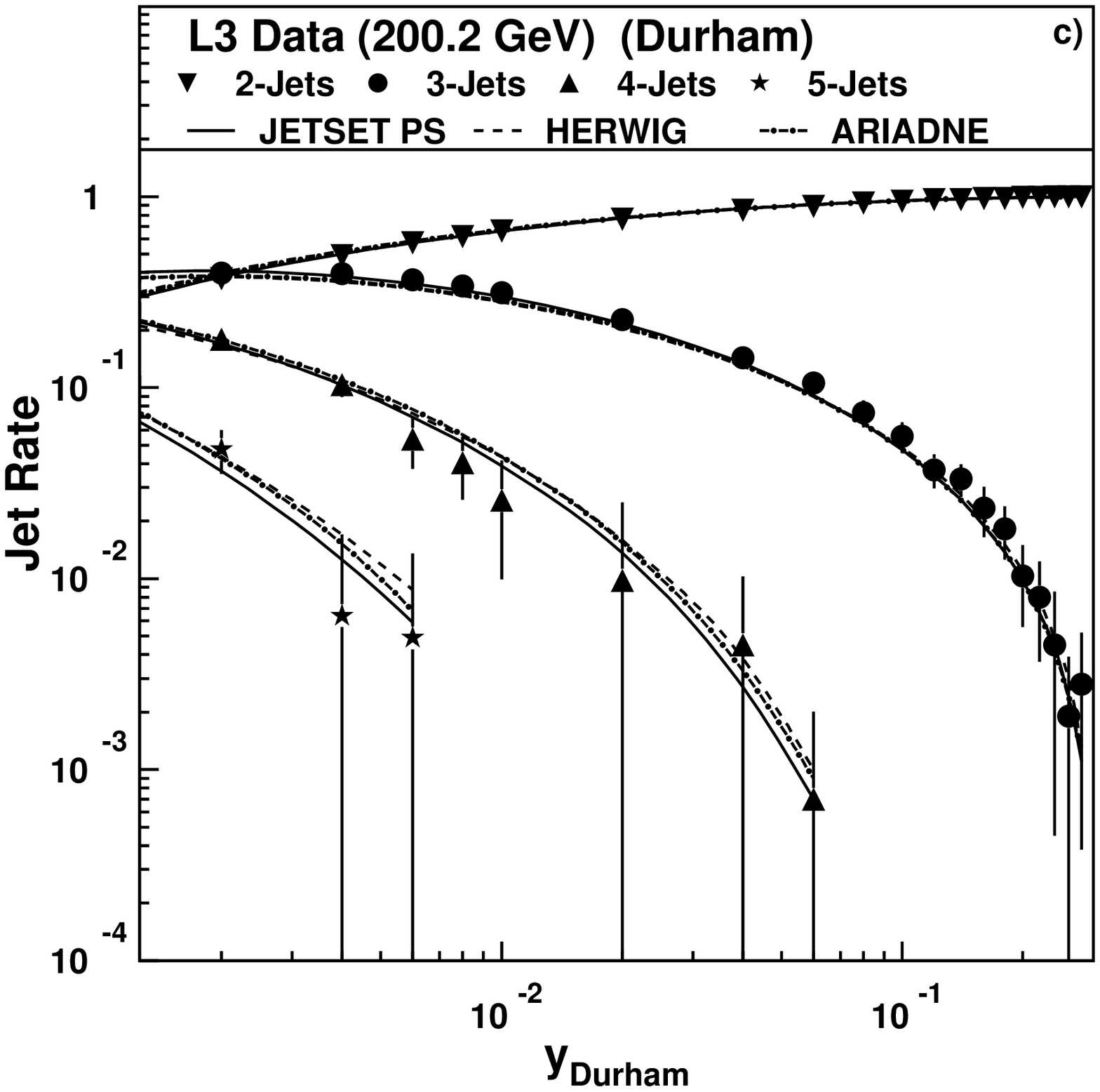}
 \includegraphics*[width=0.5\figwidth]{\mydirfig 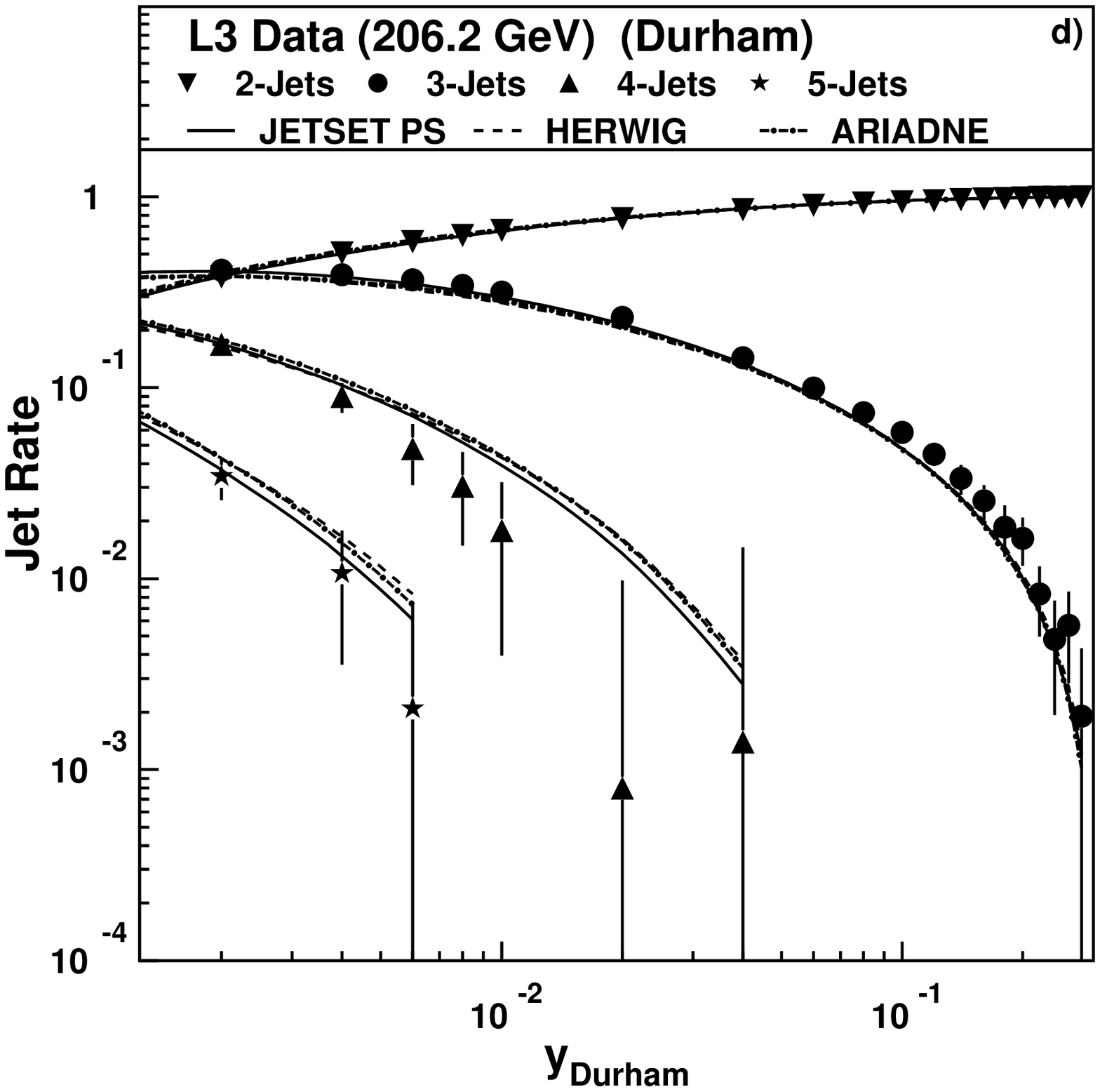}
\end{center}
\caption{Fraction of 2-, 3-, 4- and 5-jet events as a function of jet resolution parameter \ycD\
           at $\rs = 130.1, 182.8,  200.2\text{ and }206.2\,\GeV$
         for the Durham algorithm.}
\label{fig:jetkt}
\end{figure}

Similarly,
Tables \ref{tab:kt130}--\ref{tab:kt206}
show the corrected jet fractions as a
function of 
\ycD\ for 2-, 3-, 4- and 5-jets at
different centre-of-mass energies where the jets are reconstructed using the
\kt\ or Durham algorithm \cite{kt,kt1,kt2,kt3}. This algorithm differs from the \textsc{Jade}
algorithm in the definition of the jet resolution parameter $y_{ij}$
between two particles in order to better treat the summing up of soft gluon emission:
\begin{equation}
    y_{ij}^\mathrm{D}  =  \frac{2 \min (E_{i}^{2}, E_{j}^{2})}{s}
                 \left(1 - \cos\theta_{ij} \right)
\end{equation}
The data are compared with different parton shower models in Figure~\ref{fig:jetkt}
at mean centre-of-mass energies of 130.1, 182.8, 200.2 and 206.2\,\GeV.
Again the data are well described by the different parton shower models.

\begin{figure}[htbp]
\begin{center}
 \includegraphics*[width=0.5\figwidth]{\mydirfig 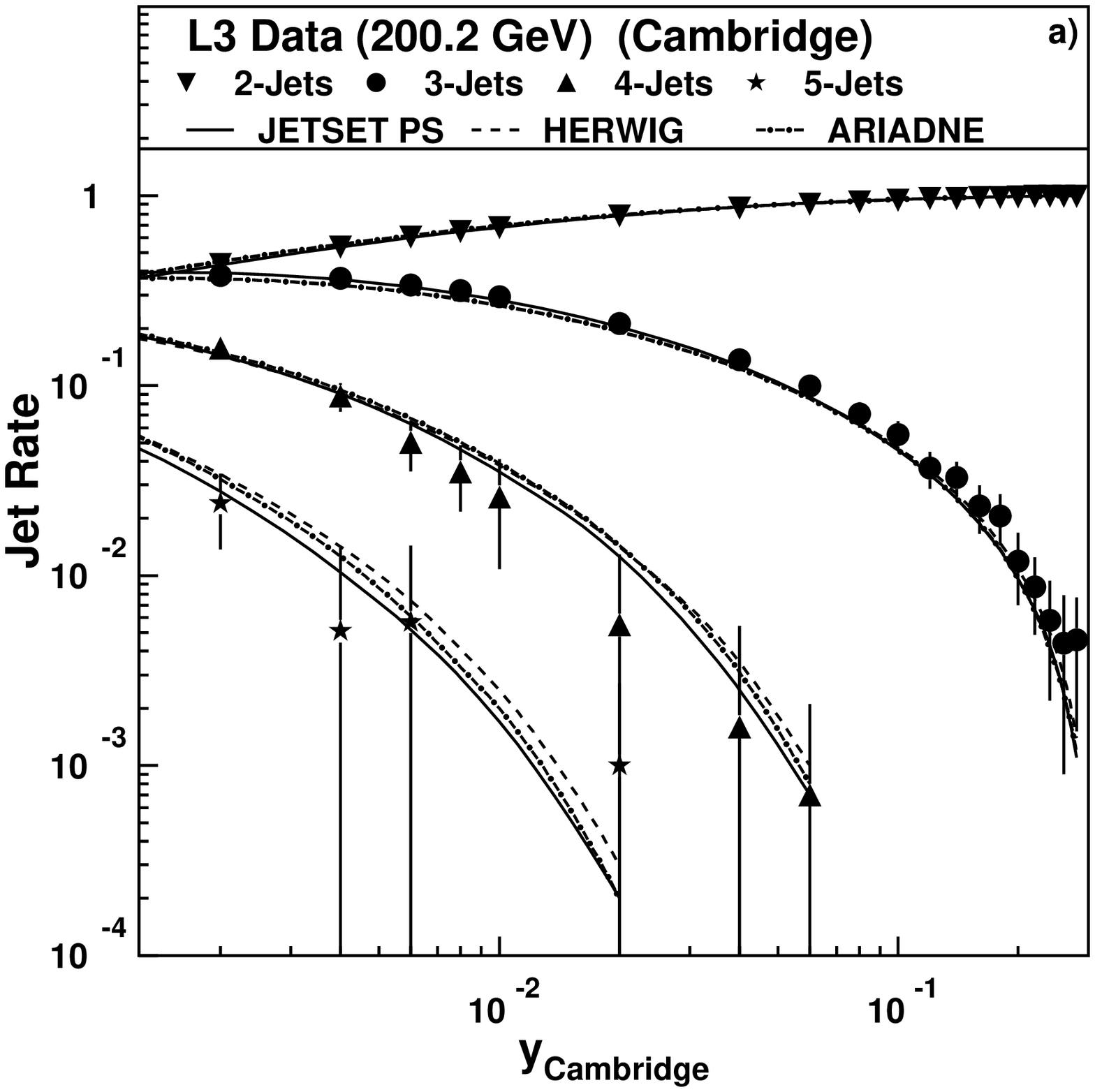}
 \includegraphics*[width=0.5\figwidth]{\mydirfig 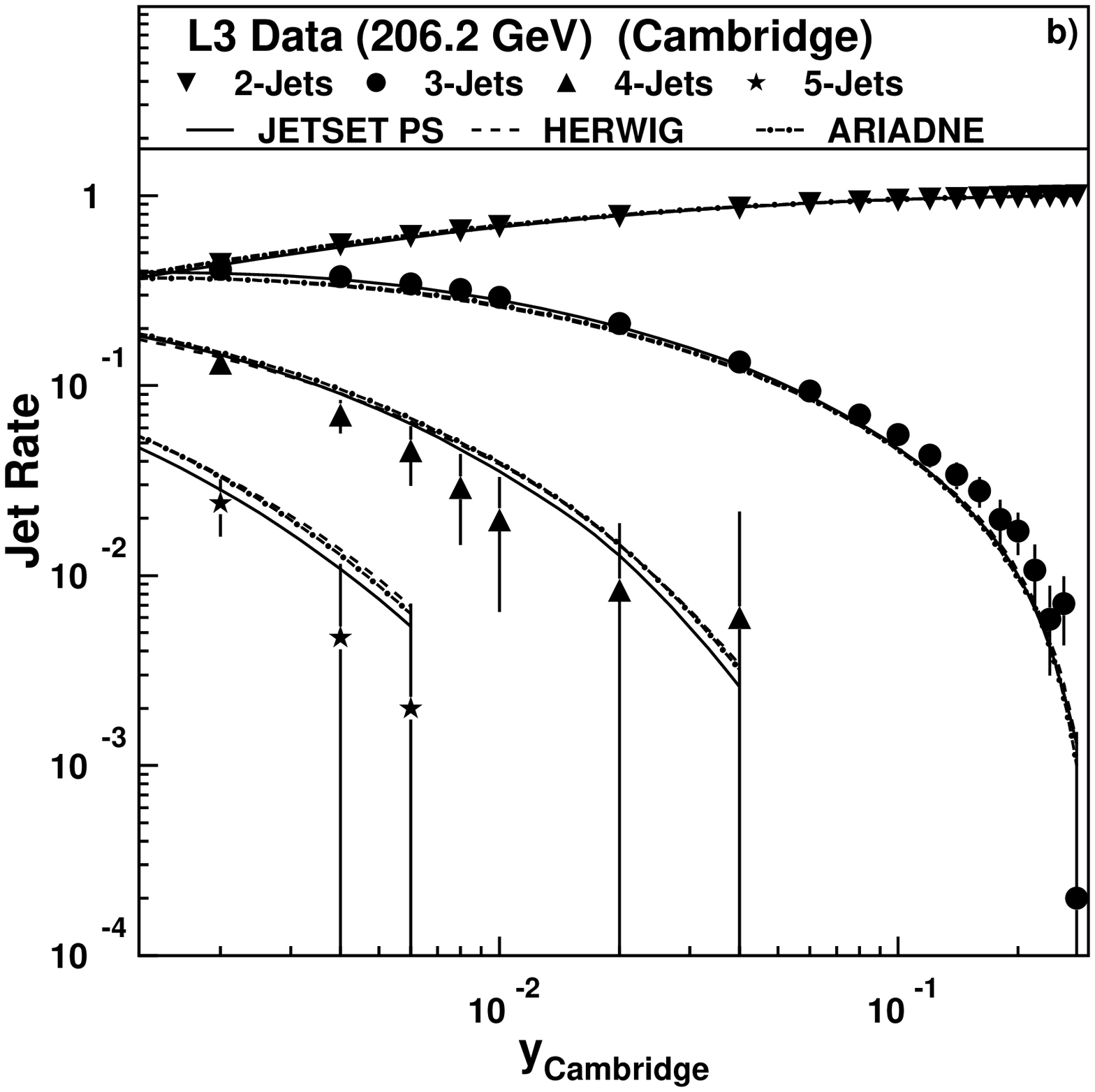}
\end{center}
\caption{Fraction of 2-, 3-, 4- and 5-jet events as a function of the jet resolution parameter \ycD\
         at $\rs = 200.2\text{ and }206.2\,\GeV$
         for the Cambridge algorithm.}
\label{fig:jetcm}
\end{figure}

In the Cambridge algorithm \cite{cambridge} the ordering
parameter for combining particles into pseudo-particles is separated from the
jet resolution parameter and a concept called `soft freezing' is introduced.
In this algorithm, the ordering parameter
$v_{ij}$ is chosen to be
\begin{equation}
   v_{ij}  =  \left(1 - \cos\theta_{ij} \right) \; .
\end{equation}
At each step, the pair having the smallest value of $v_{ij}$ is examined.
If $y_{ij}^\mathrm{D} < \ycD$, particles $i$ and $j$ are combined
to form a pseudo-particle $l$ as in the previous two algorithms, but
if $y_{ij}^\mathrm{D}$ is larger than \ycD, the smaller energy object (between~$i$
and~$j$) is frozen as a jet and is not considered further.
Jet fractions are measured at $\rs = 200.2$ and $206.2\,\GeV$ using this
algorithm and are tabulated in Tables \ref{tab:camb200} and \ref{tab:camb206}.
Figure \ref{fig:jetcm} shows the corrected jet fractions for the
Cambridge algorithm as a function of 
\ycD\ for 2-, 3-, 4- and 5-jets at centre-of-mass energies of 200.2 and 206.2\,\GeV,
respectively. The different \QCD\ models are in good agreement with the data.
 
\begin{figure}[htbp]
\begin{center}
  \includegraphics[width=.95\figwidth]{\mydirfig 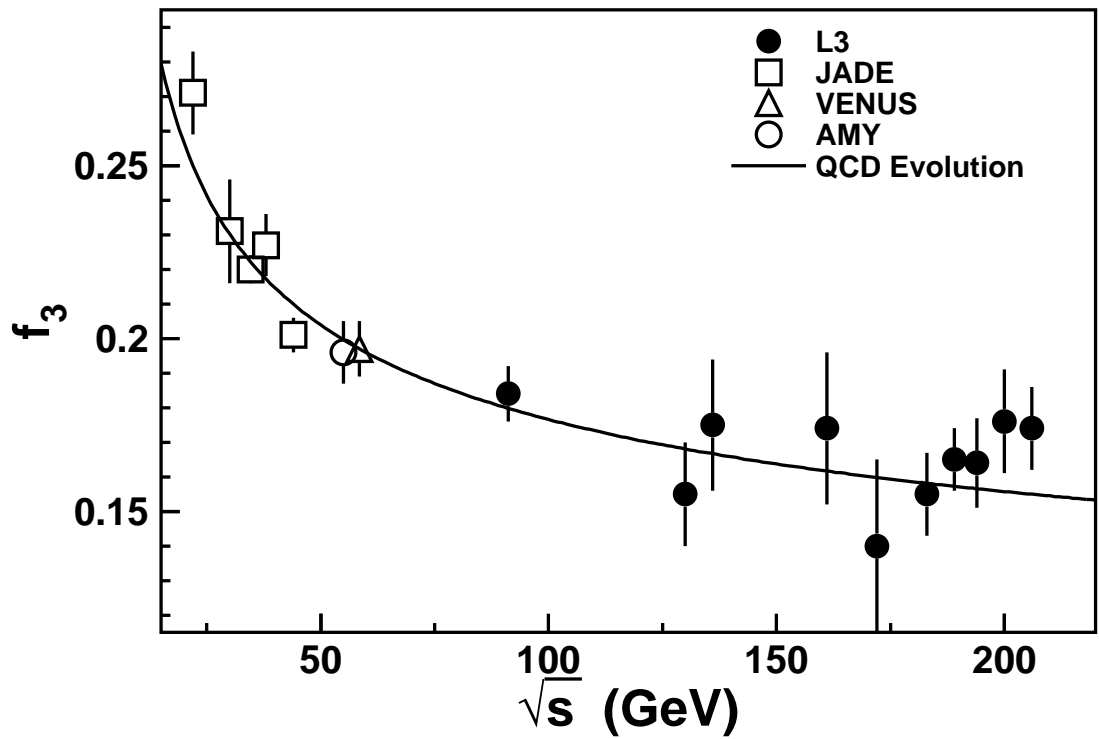}
\end{center}
\caption{Energy evolution of the 3-jet fraction at $\ycJ = 0.08$ with the
           \textsc{Jade} algorithm.
           }
\label{fig:3jetr}
\end{figure}

Figure \ref{fig:3jetr} shows the energy evolution of the 3-jet fraction using the \textsc{Jade}
algorithm at a fixed \ycJ\ of 0.08. The plot shows measurements from the \Lthree\
experiment together with similar measurements done at lower energies
\cite{jade,jade1,mark2,r3-tasso,r3-venus,amy}.
The data clearly demonstrate a decrease of 3-jet fraction with increasing centre-of-mass energy.
This result is in agreement with the running of \as\ with the energy scale as
expected in \QCD, which is also shown.
The curve  corresponds to ${\cal O}(\as^{2})$  \QCD\ calculations with $\as(\MZ) = 0.120$.

\subsection{Comparison of Event Shapes with Monte Carlo Models}\label{sec:mcmodel}
 
The corrected distributions of the shape variables $1-T$, \rhoh, $\bt$, and $\bw$
at the different centre-of-mass energies
below \cite{l3qqg} and above \MZ\ are presented in Tables \ref{tab:thrustqqg1}--\ref{tab:thrust3},
\ref{tab:rhoqqg1}--\ref{tab:rho3}, \ref{tab:btqqg1}--\ref{tab:bt3}, and \ref{tab:bwqqg1}--\ref{tab:bw3},
respectively.  Those of $C$ and $D$ at centre-of-mass energies  above \MZ\ are shown
in Tables \ref{tab:cpar1}--\ref{tab:cpar3} and \ref{tab:dpar1}--\ref{tab:dpar3}, respectively.
Tables for the distributions at the Z pole can be found in Reference~\citen{l3qcd91z}.

\begin{figure}[htbp]
\begin{center}
  \includegraphics*[width=.5\figwidth,bbllx=5,bblly=30,bburx=525,bbury=560]{\mydirfig 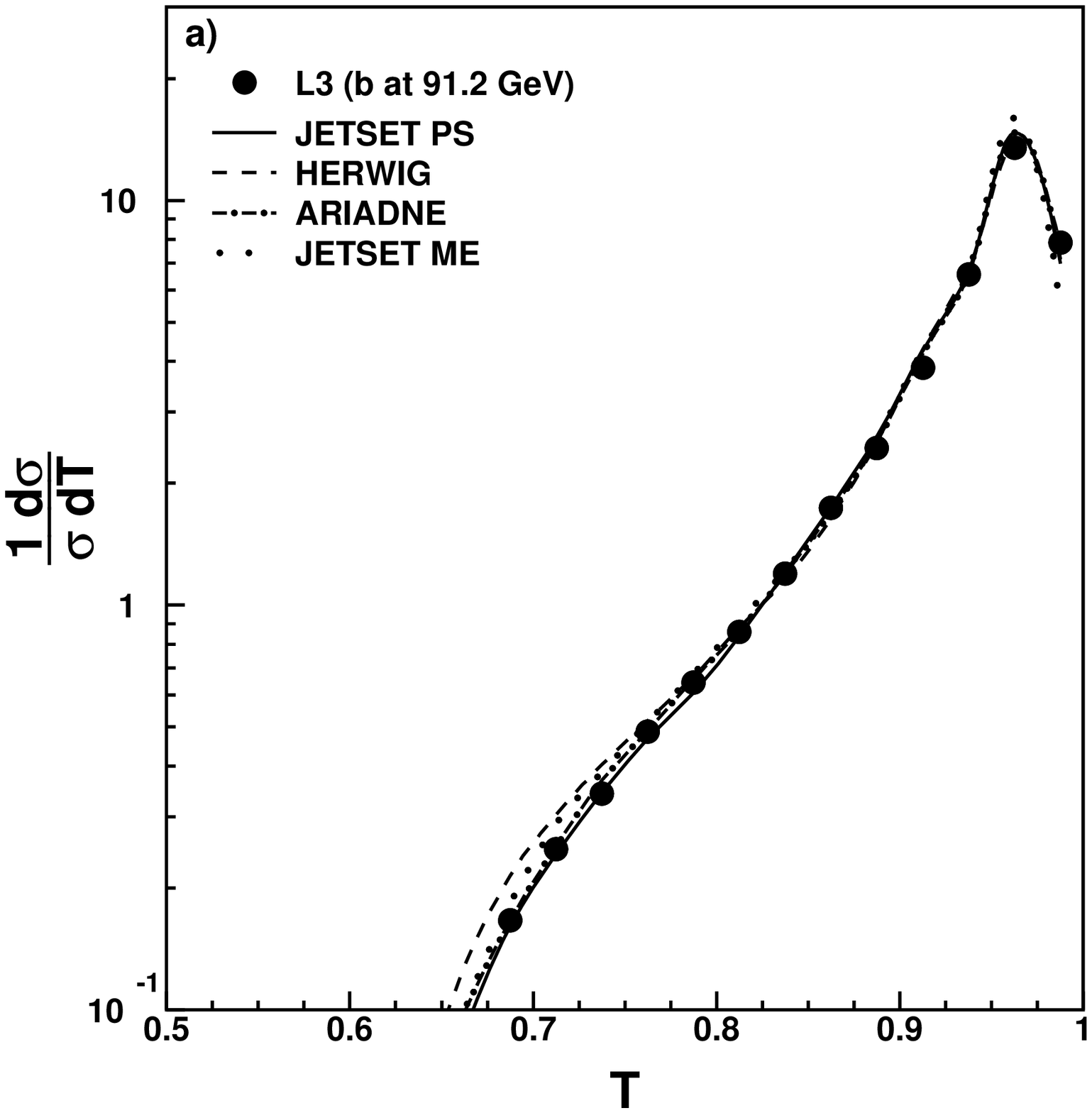}
  \includegraphics*[width=.5\figwidth,bbllx=5,bblly=30,bburx=525,bbury=560]{\mydirfig 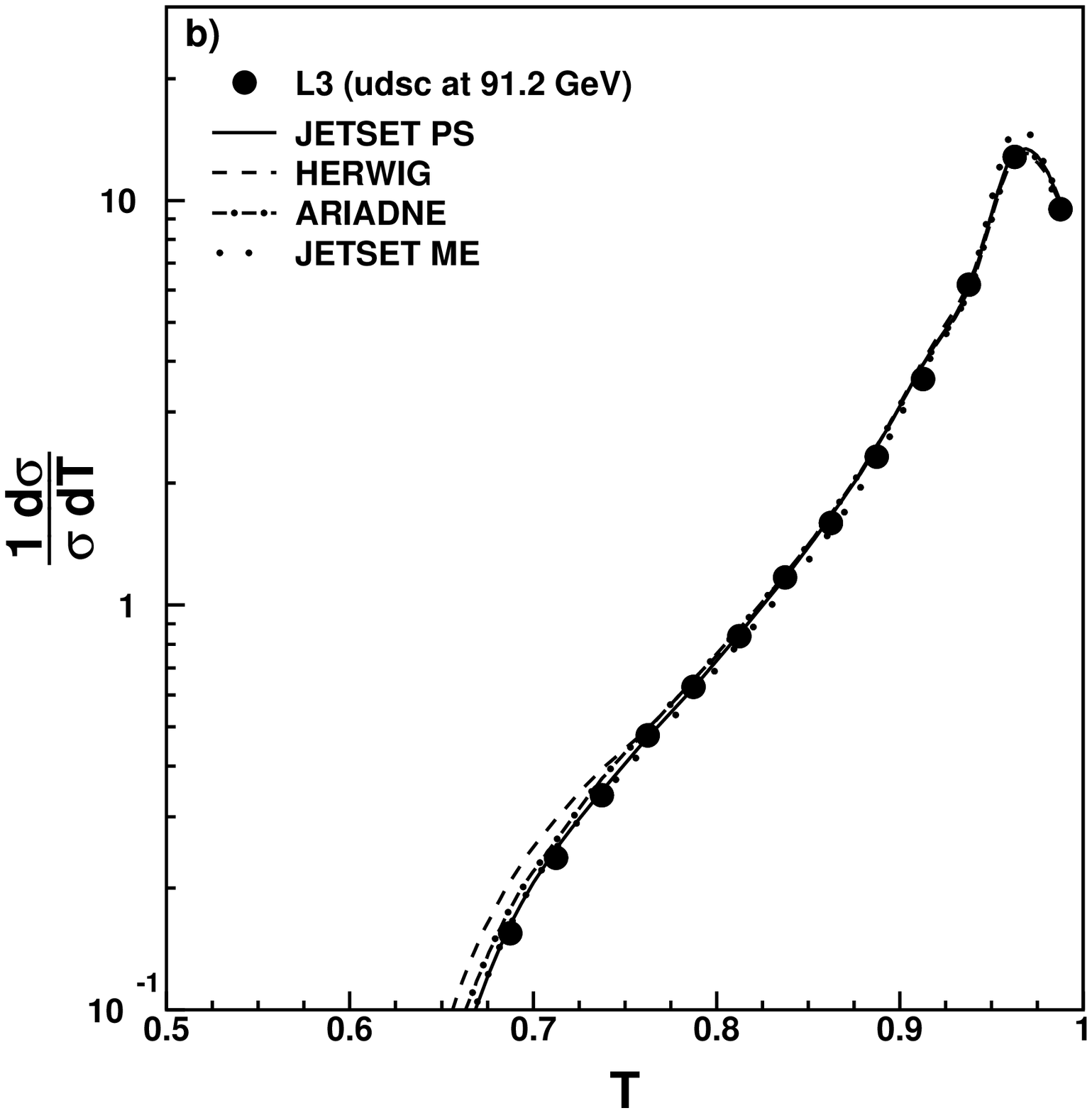}
  \includegraphics*[width=.5\figwidth,bbllx=5,bblly=30,bburx=525,bbury=560]{\mydirfig 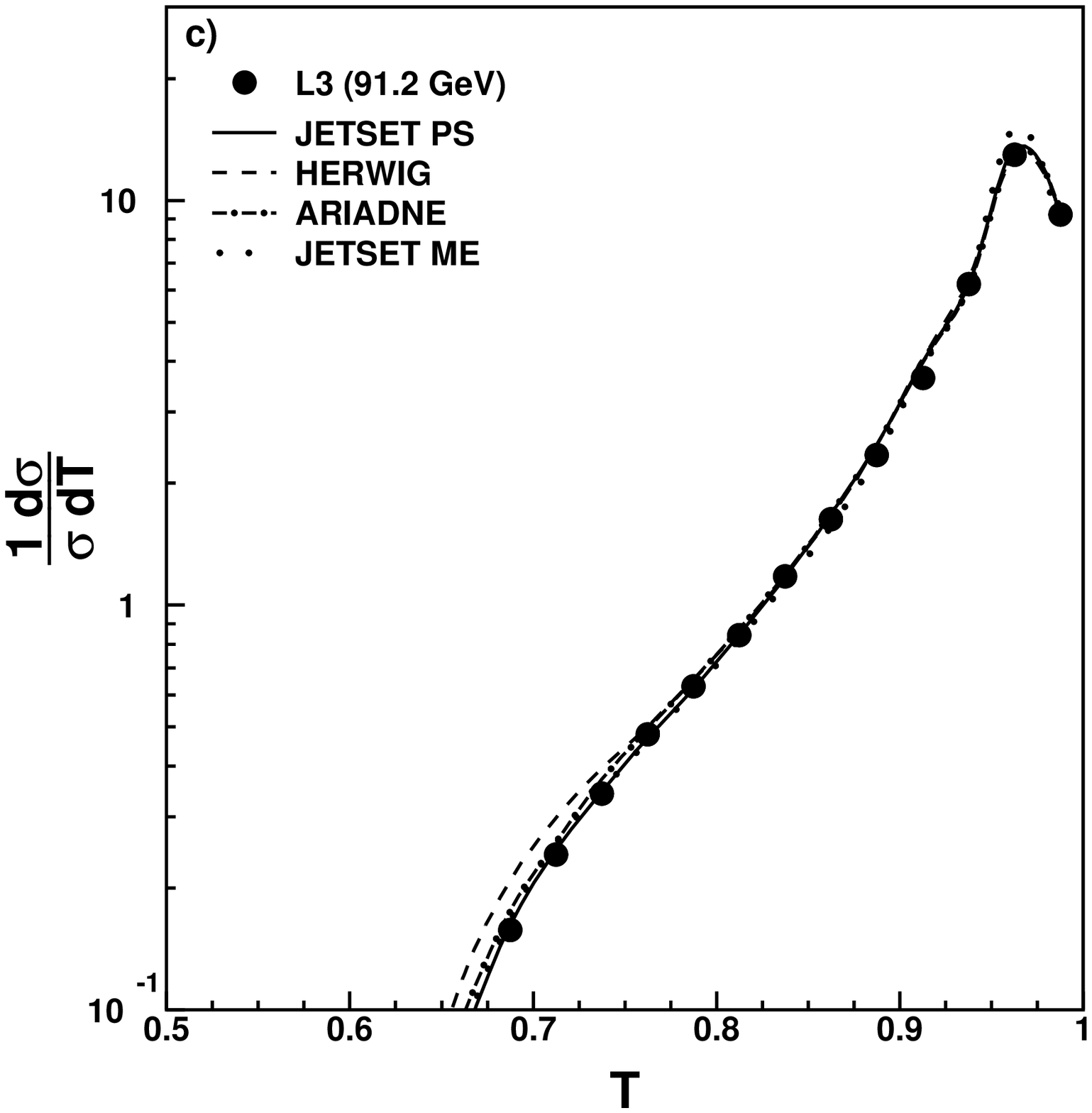}
  \includegraphics*[width=.5\figwidth,bbllx=5,bblly=30,bburx=525,bbury=560]{\mydirfig 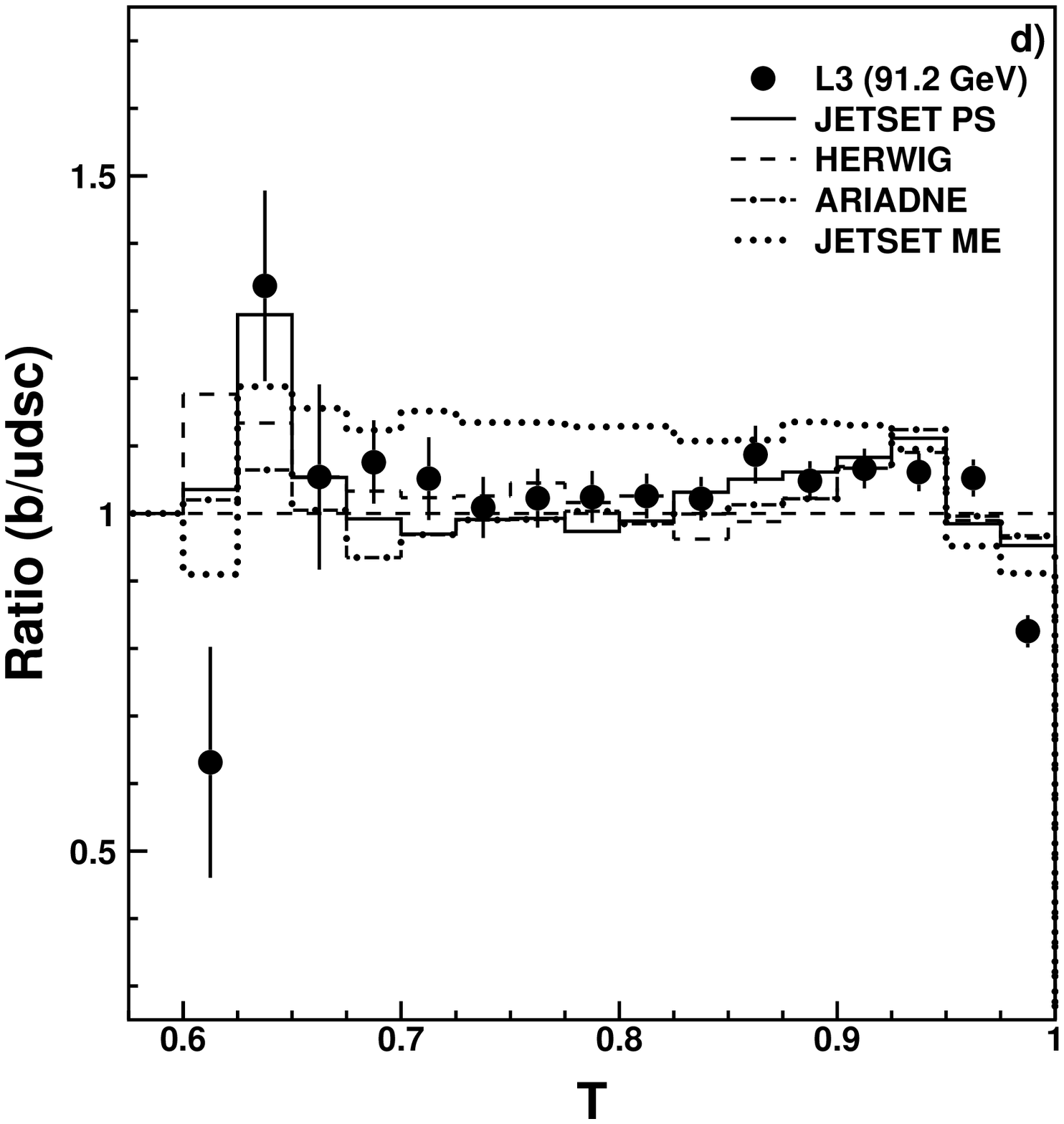}
\end{center}
\caption{Thrust distributions at
         $\rs=91.2\,\GeV$
         for b, udsc, and all quark flavours
         and the ratio b/udsc
         compared to several \QCD\ models.
         }
\label{fig:part-thr}
\end{figure}

\begin{figure}[htbp]
\begin{center}
  \includegraphics*[width=.5\figwidth,bbllx=5,bblly=30,bburx=525,bbury=560]{\mydirfig 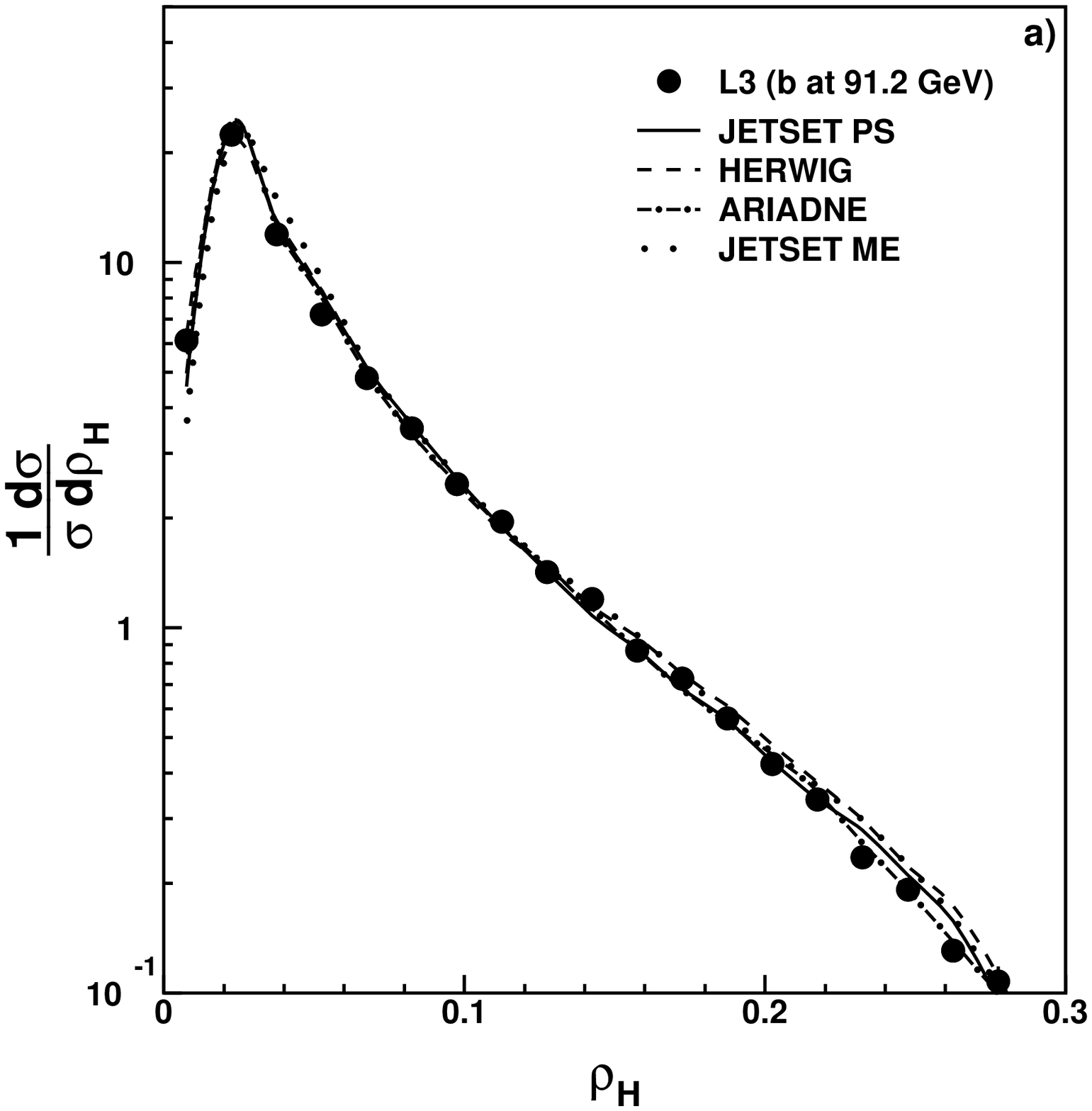}
  \includegraphics*[width=.5\figwidth,bbllx=5,bblly=30,bburx=525,bbury=560]{\mydirfig 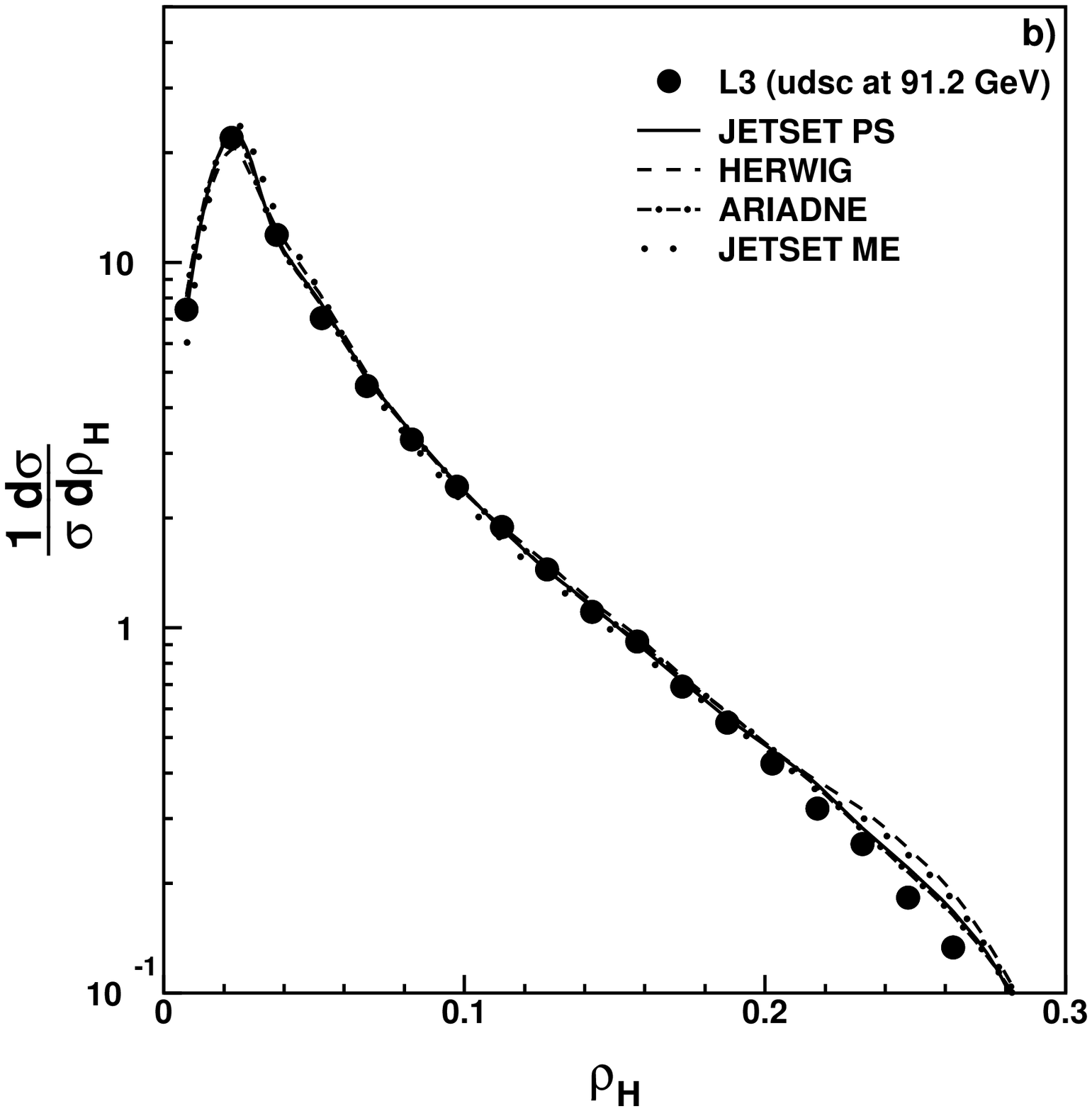}
  \includegraphics*[width=.5\figwidth,bbllx=5,bblly=30,bburx=525,bbury=560]{\mydirfig 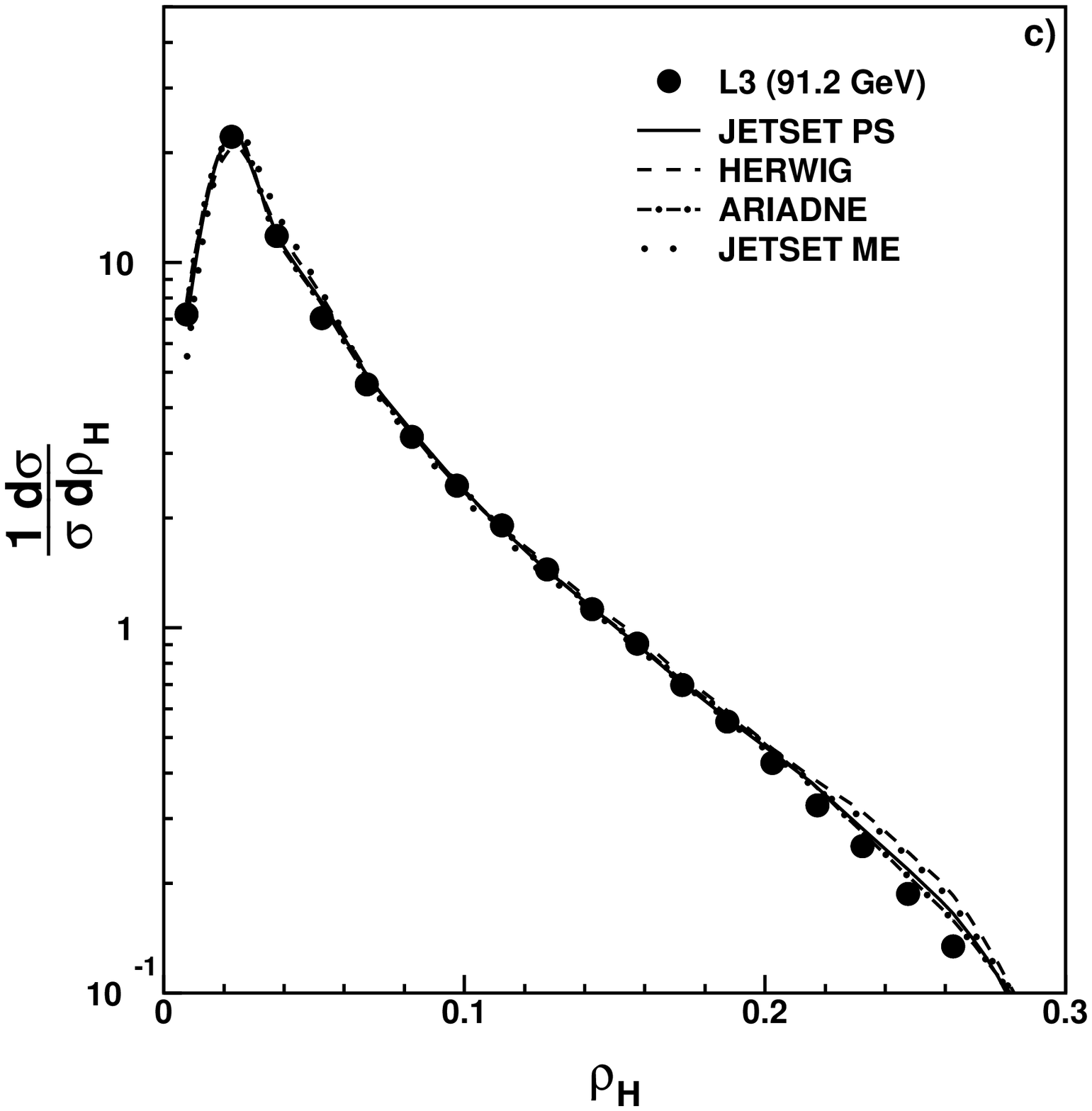}
  \includegraphics*[width=.5\figwidth,bbllx=5,bblly=30,bburx=525,bbury=560]{\mydirfig 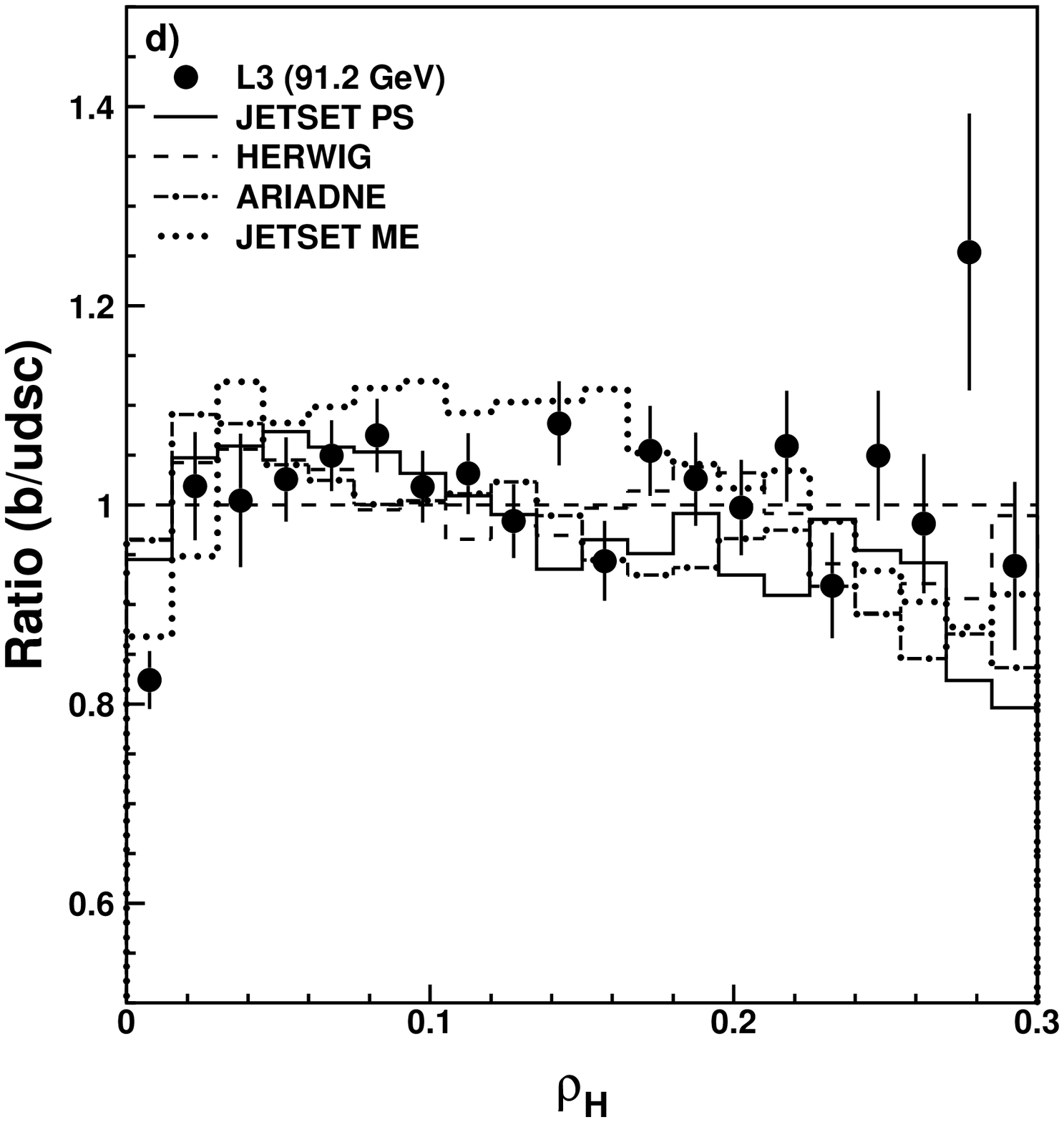}
\end{center}
\caption{Scaled heavy jet mass distributions at
         $\rs= 91.2\,\GeV$
         for b, udsc, and all quark flavours
         and the ratio b/udsc
         compared to several \QCD\ models.
         }
\label{fig:part-rho}
\end{figure}

\begin{figure}[htbp]
\begin{center}
  \includegraphics*[width=.5\figwidth,bbllx=5,bblly=30,bburx=525,bbury=560]{\mydirfig 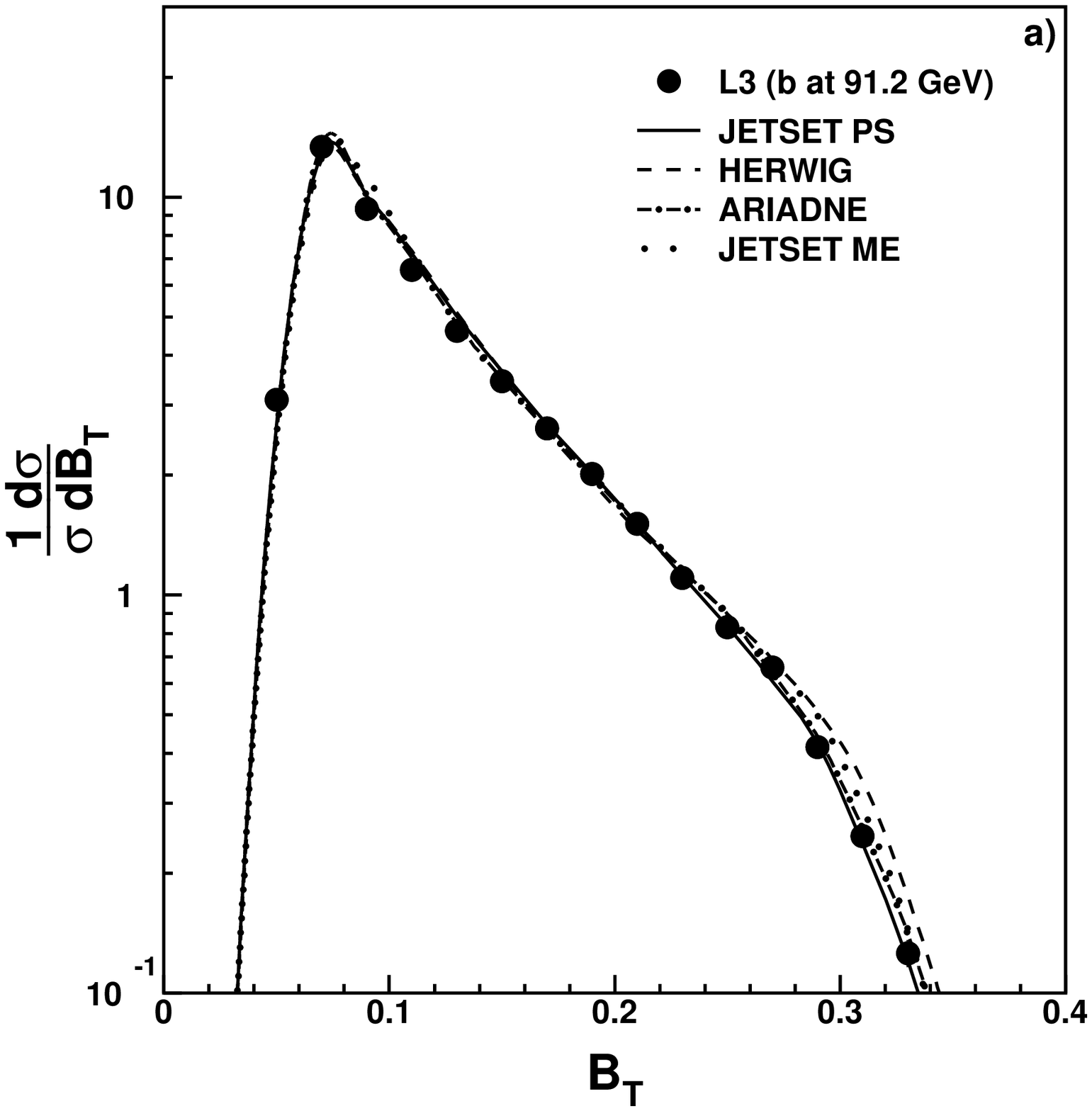}
  \includegraphics*[width=.5\figwidth,bbllx=5,bblly=30,bburx=525,bbury=560]{\mydirfig 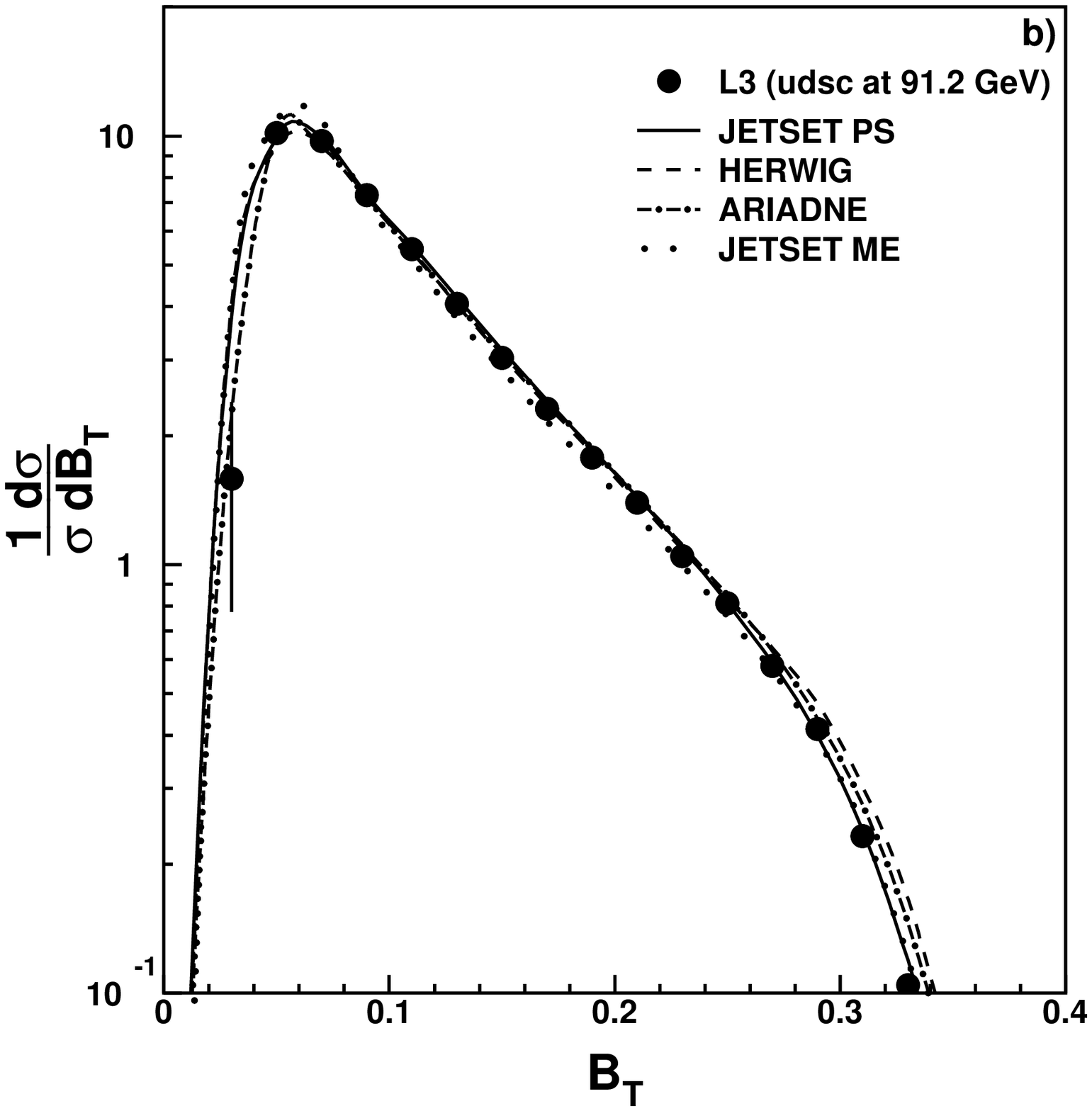}
  \includegraphics*[width=.5\figwidth,bbllx=5,bblly=30,bburx=525,bbury=560]{\mydirfig 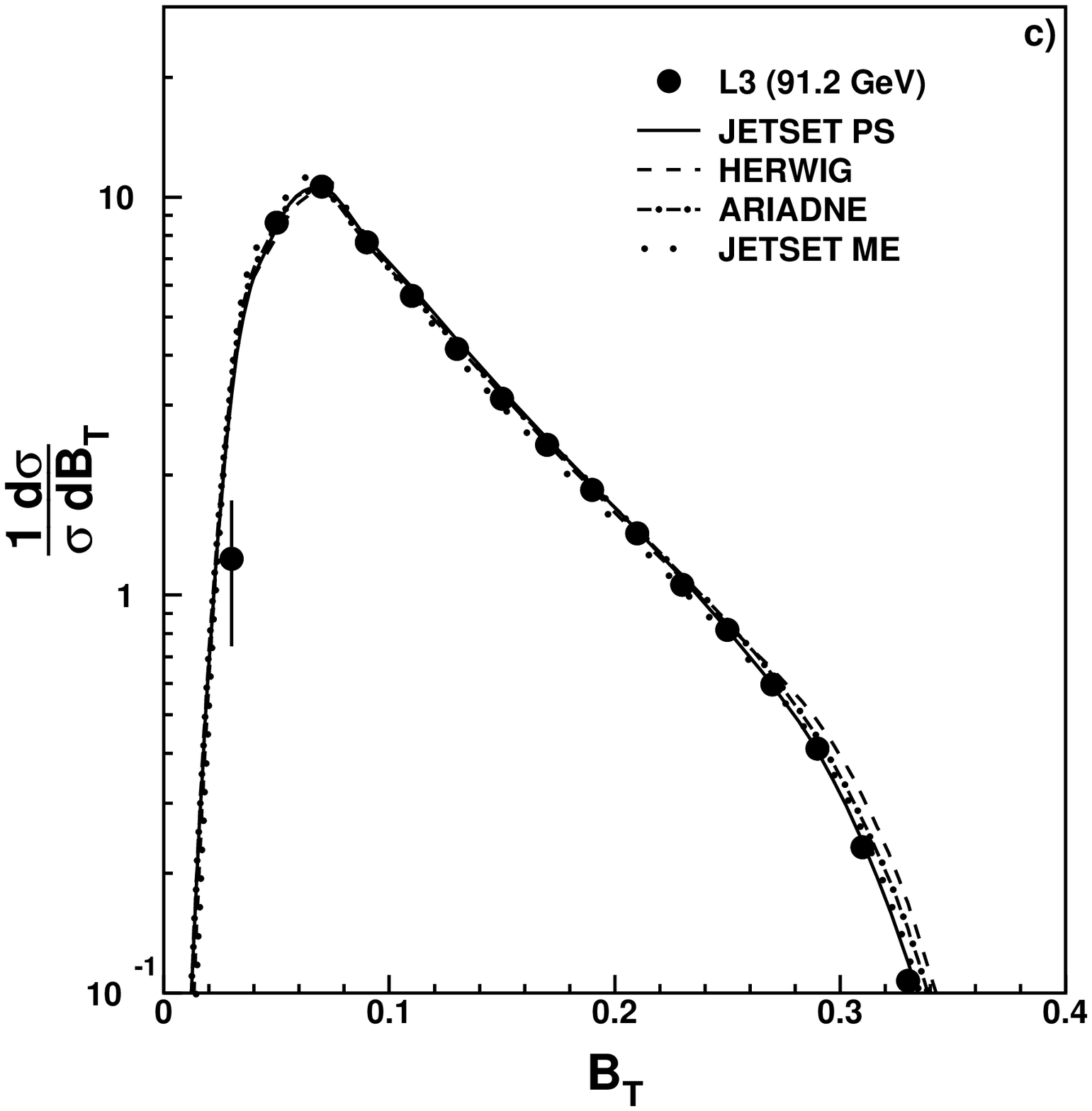}
  \includegraphics*[width=.5\figwidth,bbllx=5,bblly=30,bburx=525,bbury=560]{\mydirfig 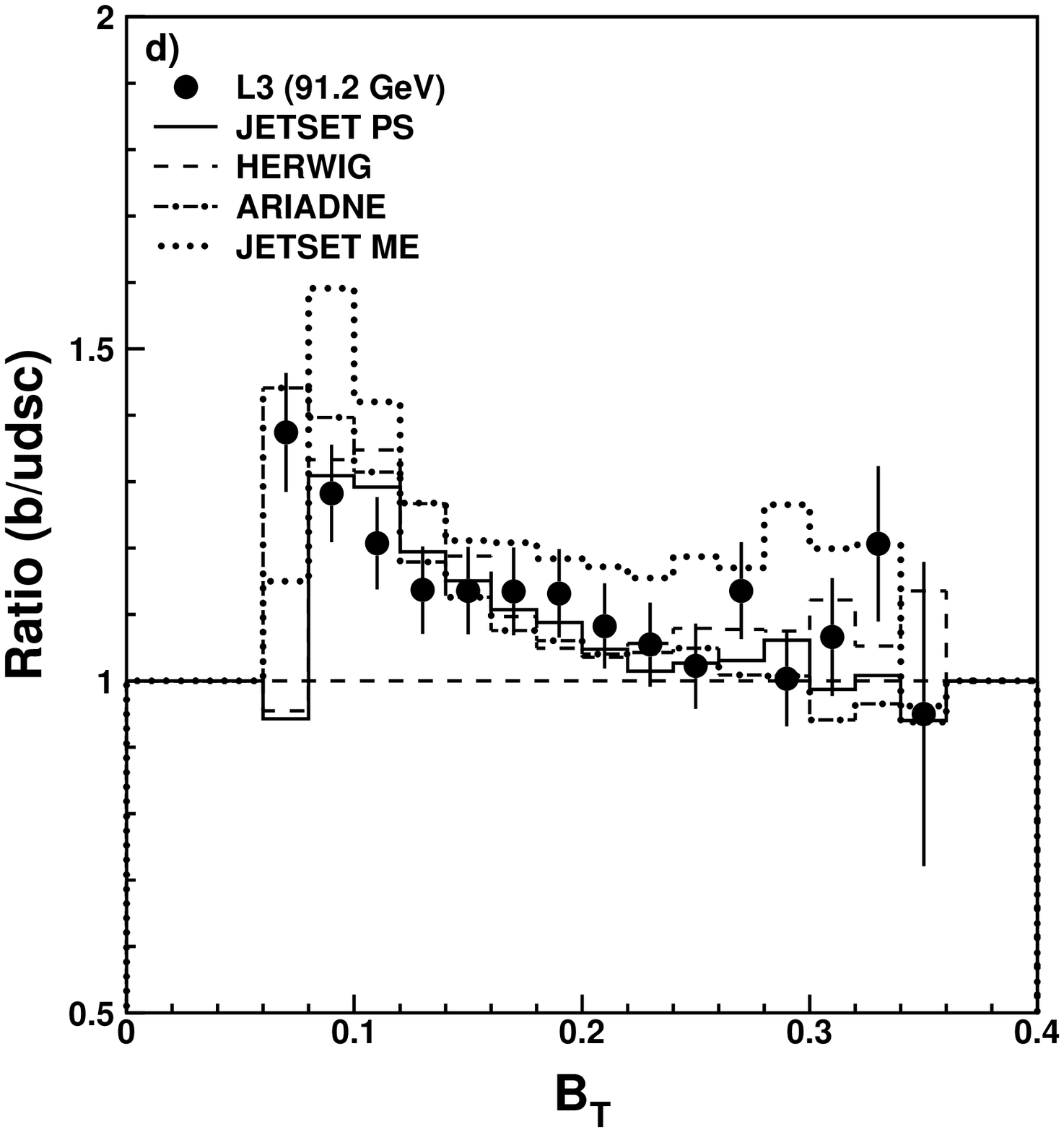}
\end{center}
\caption{Total jet broadening distributions at
         $\rs=91.2\,\GeV$
         for b, udsc, and all quark flavours
         and the ratio b/udsc
         compared to several \QCD\ models.
         }
\label{fig:part-bt}
\end{figure}

\begin{figure}[htbp]
\begin{center}
  \includegraphics*[width=.5\figwidth,bbllx=5,bblly=30,bburx=525,bbury=560]{\mydirfig 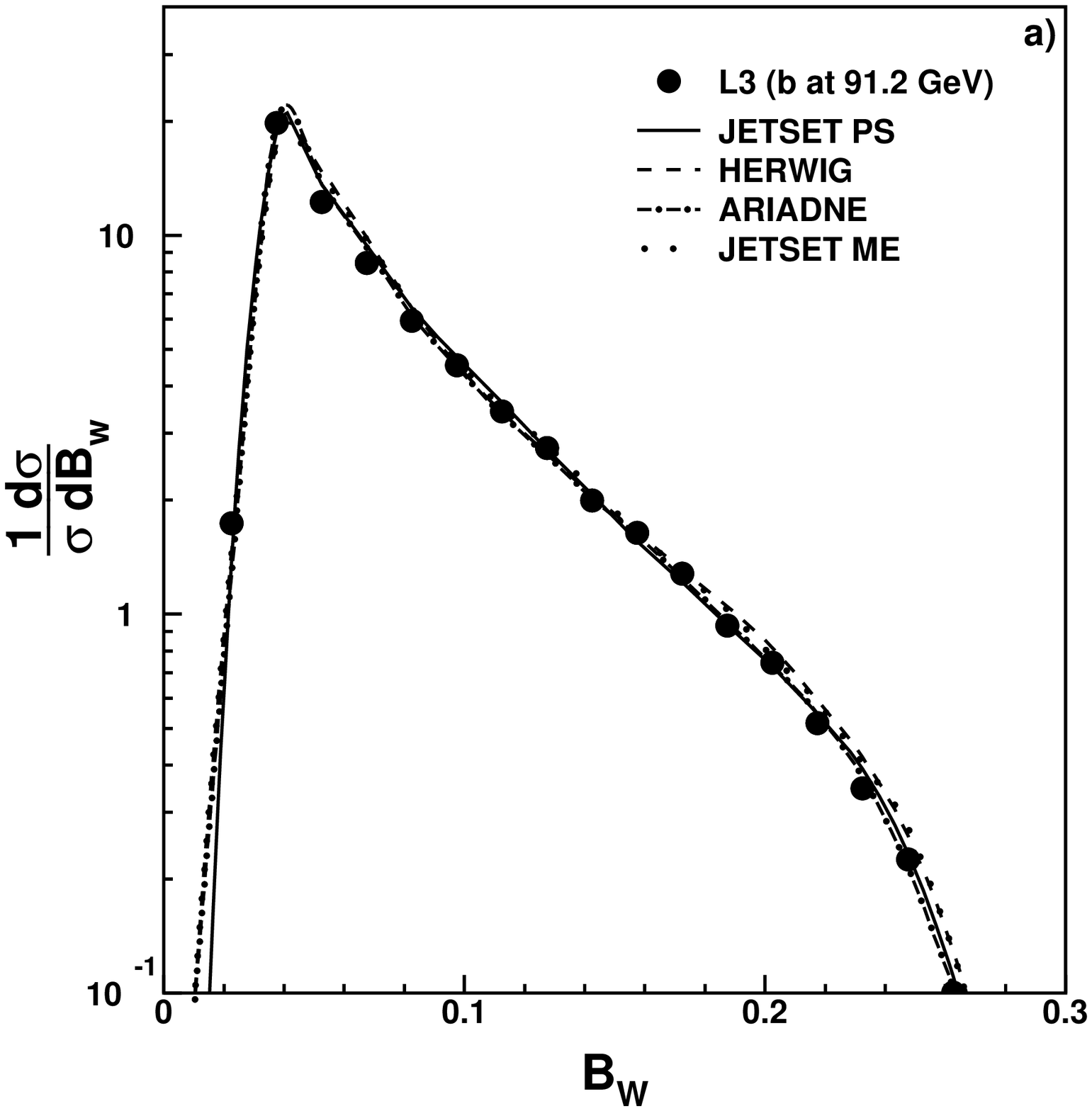}
  \includegraphics*[width=.5\figwidth,bbllx=5,bblly=30,bburx=525,bbury=560]{\mydirfig 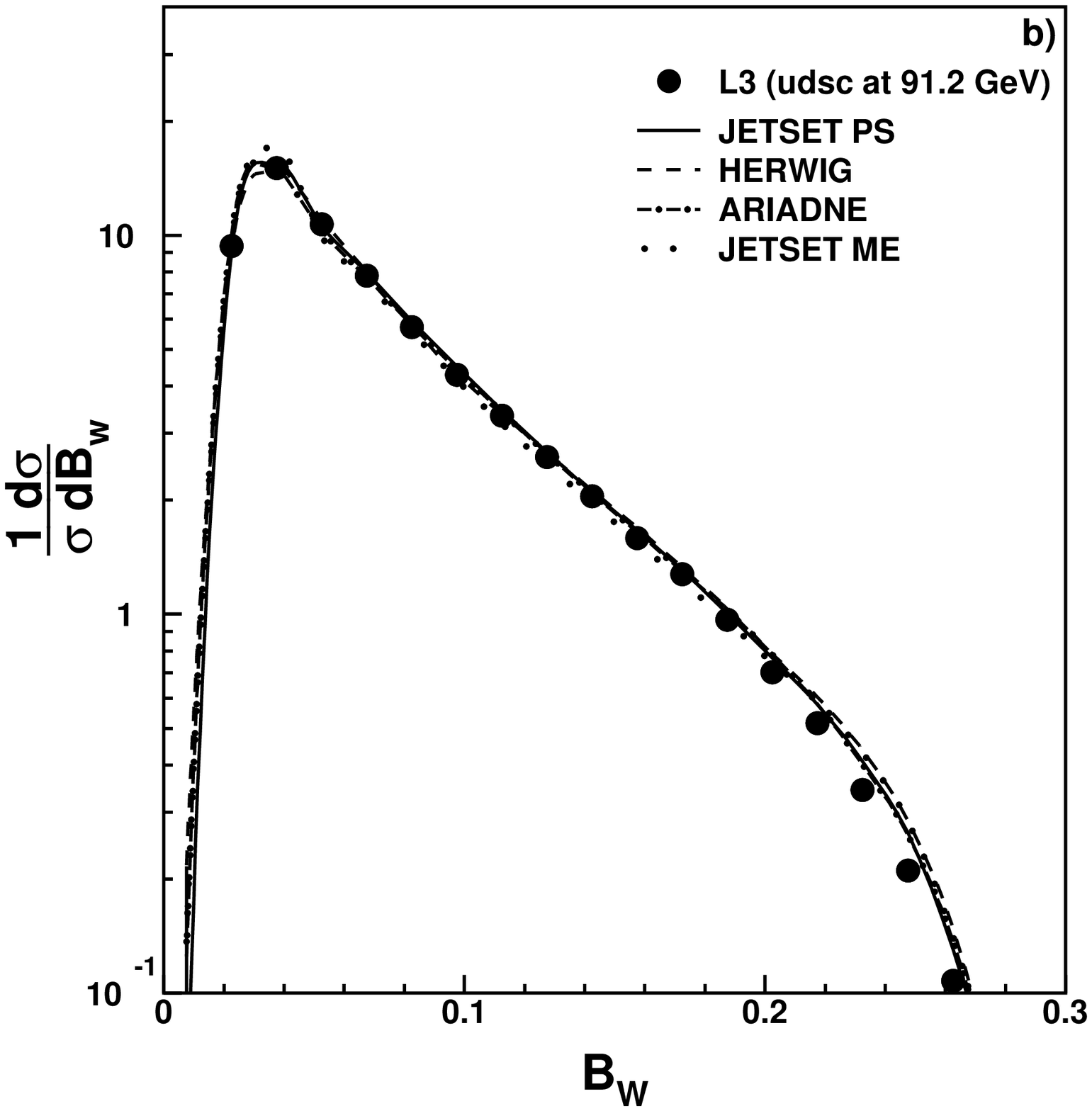}
  \includegraphics*[width=.5\figwidth,bbllx=5,bblly=30,bburx=525,bbury=560]{\mydirfig 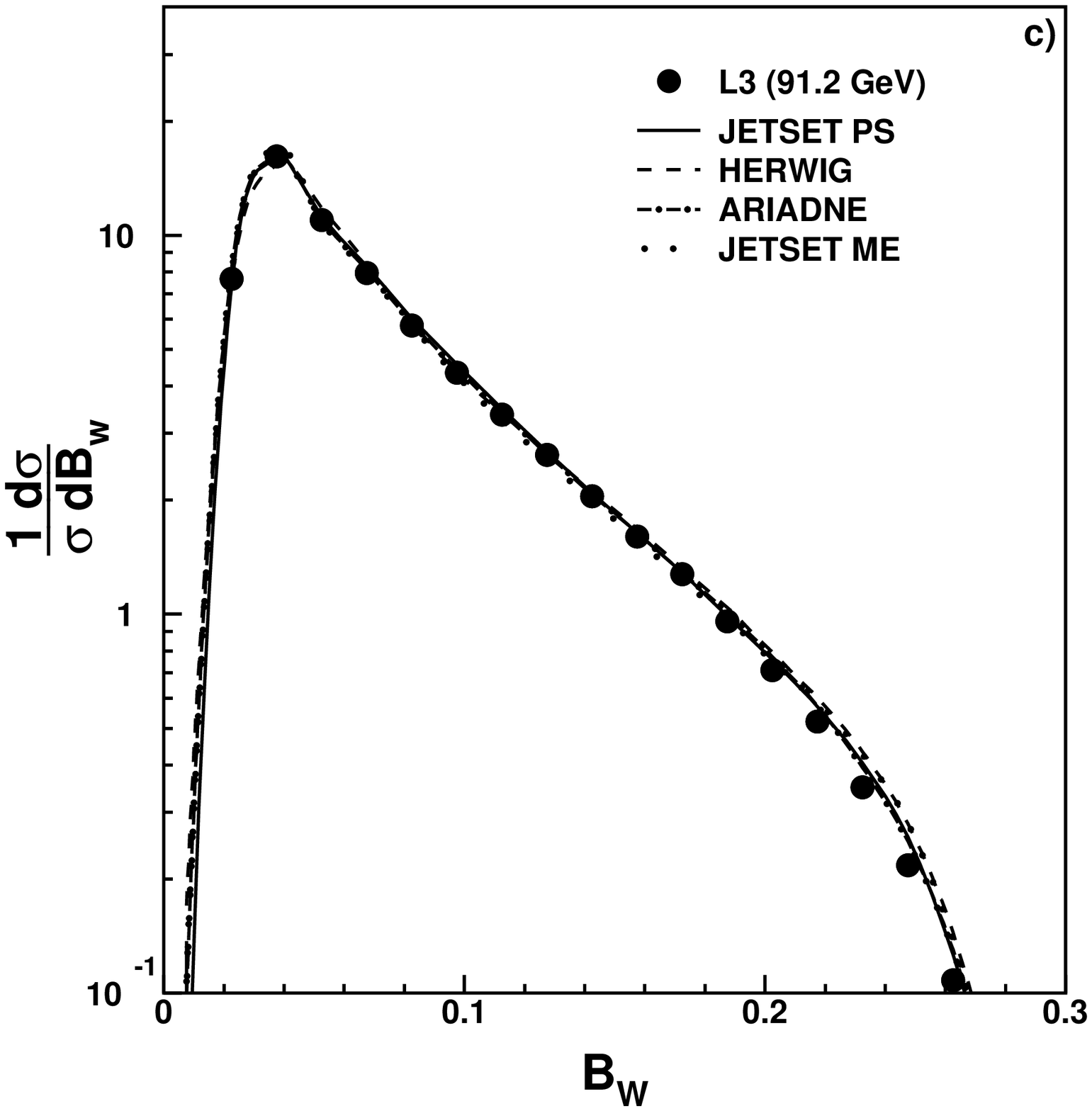}
  \includegraphics*[width=.5\figwidth,bbllx=5,bblly=30,bburx=525,bbury=560]{\mydirfig 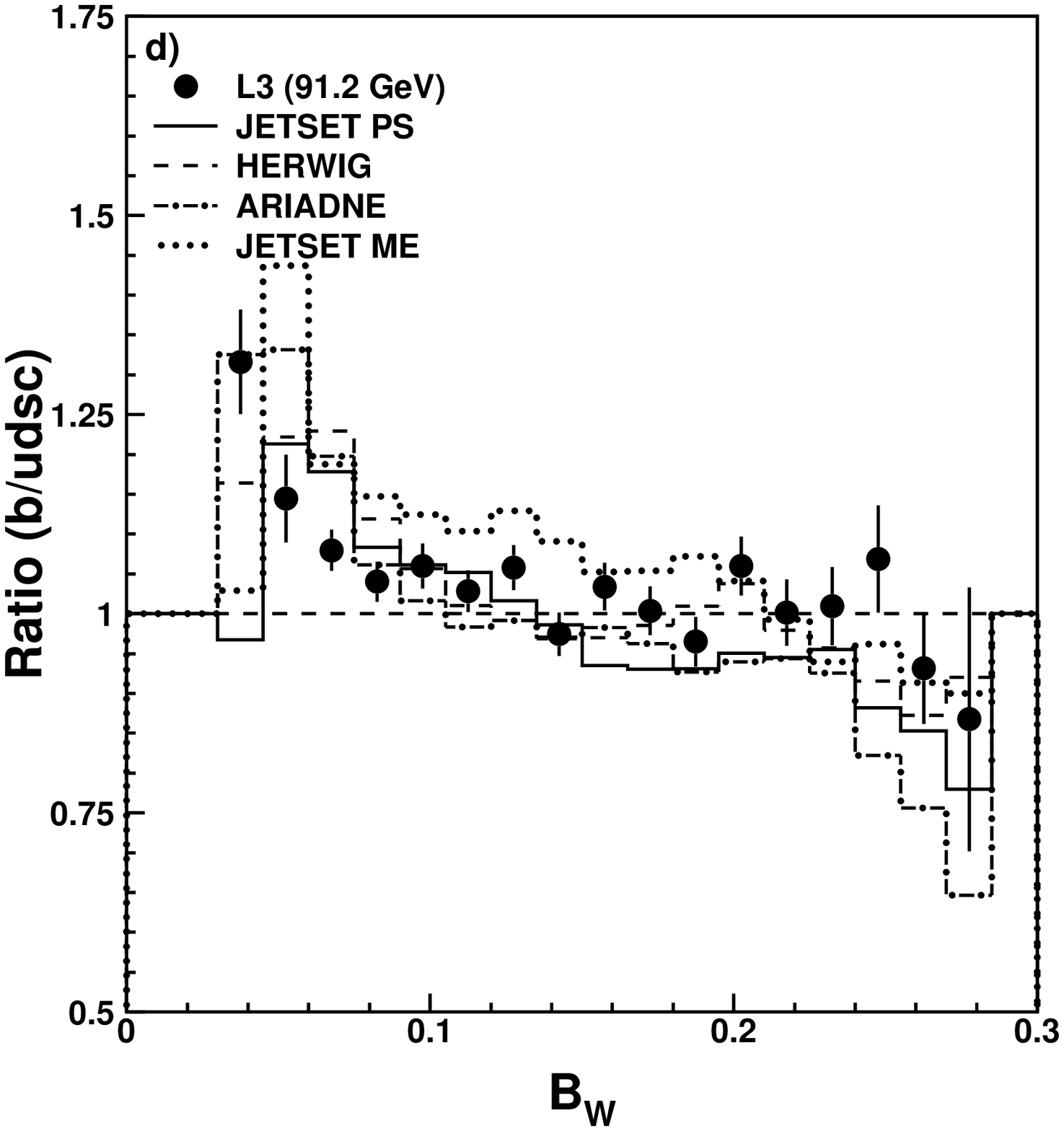}
\end{center}
\caption{Wide jet broadening distributions at
         $\rs=91.2\,\GeV$
         for b, udsc, and all quark flavours
         and the ratio b/udsc
         compared to several \QCD\ models.
         }
\label{fig:part-bw}
\end{figure}

\begin{figure}[htbp]
\begin{center}
  \includegraphics*[width=.5\figwidth,bbllx=5,bblly=30,bburx=525,bbury=560]{\mydirfig 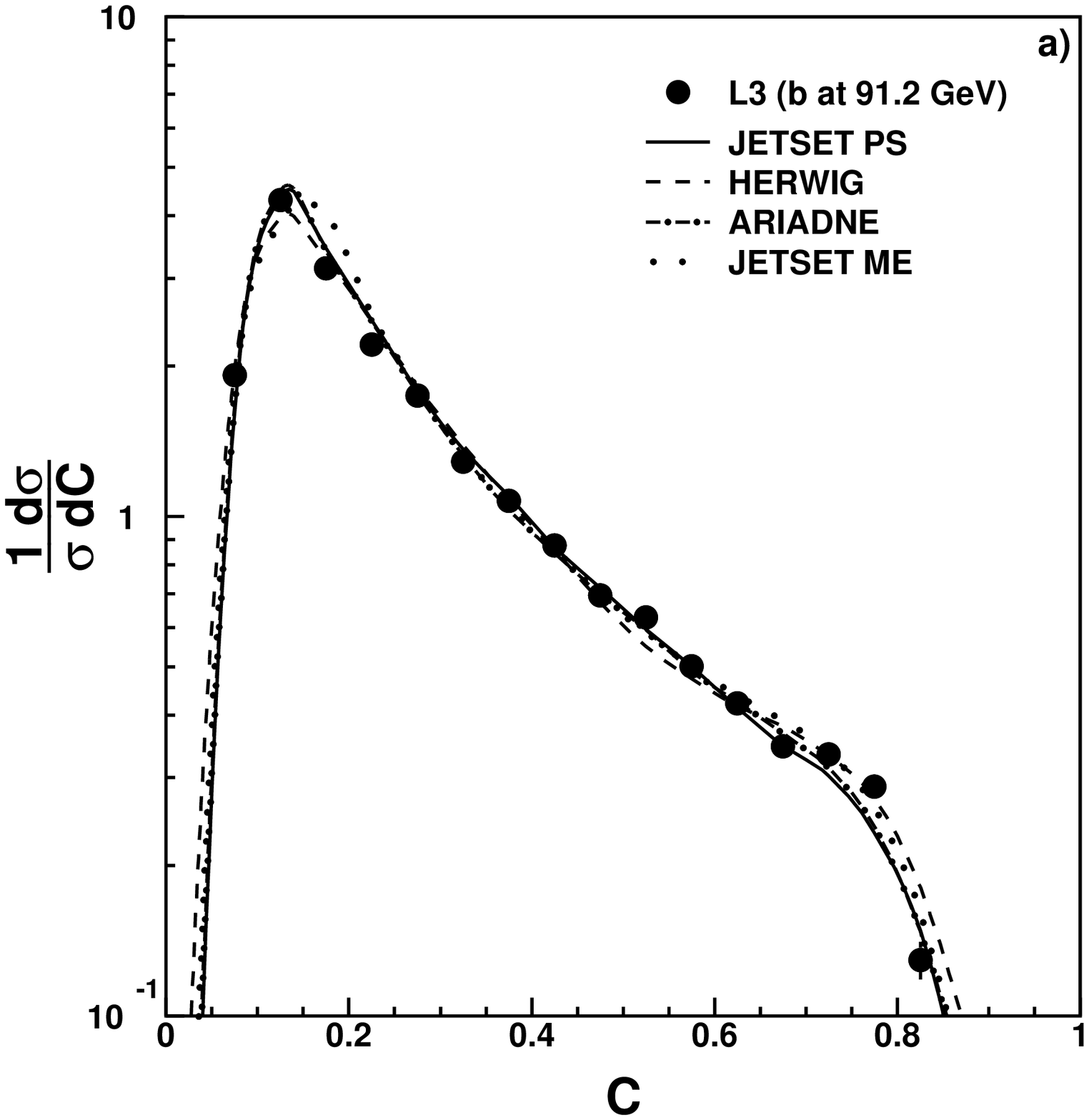}
  \includegraphics*[width=.5\figwidth,bbllx=5,bblly=30,bburx=525,bbury=560]{\mydirfig 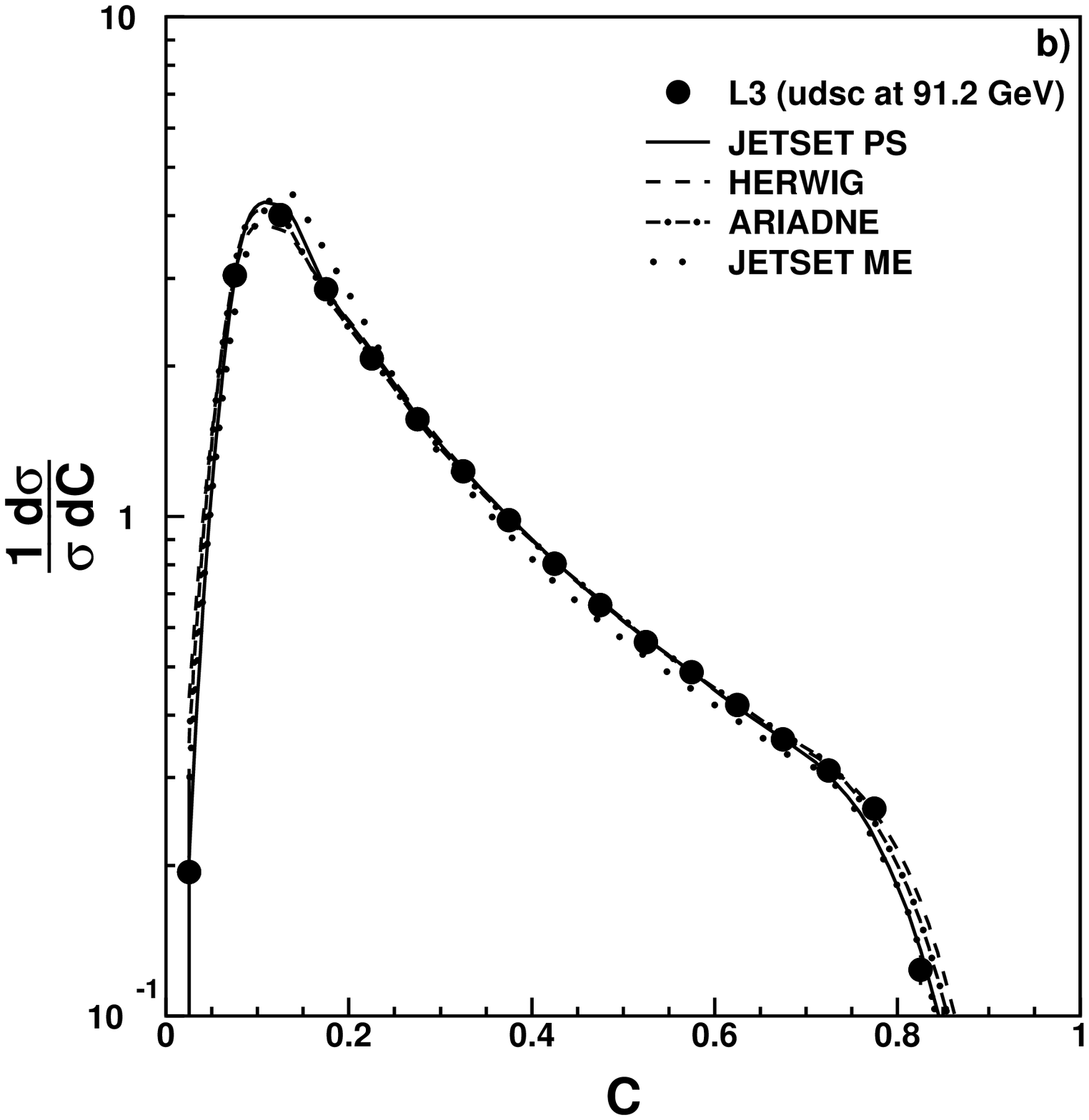}
  \includegraphics*[width=.5\figwidth,bbllx=5,bblly=30,bburx=525,bbury=560]{\mydirfig 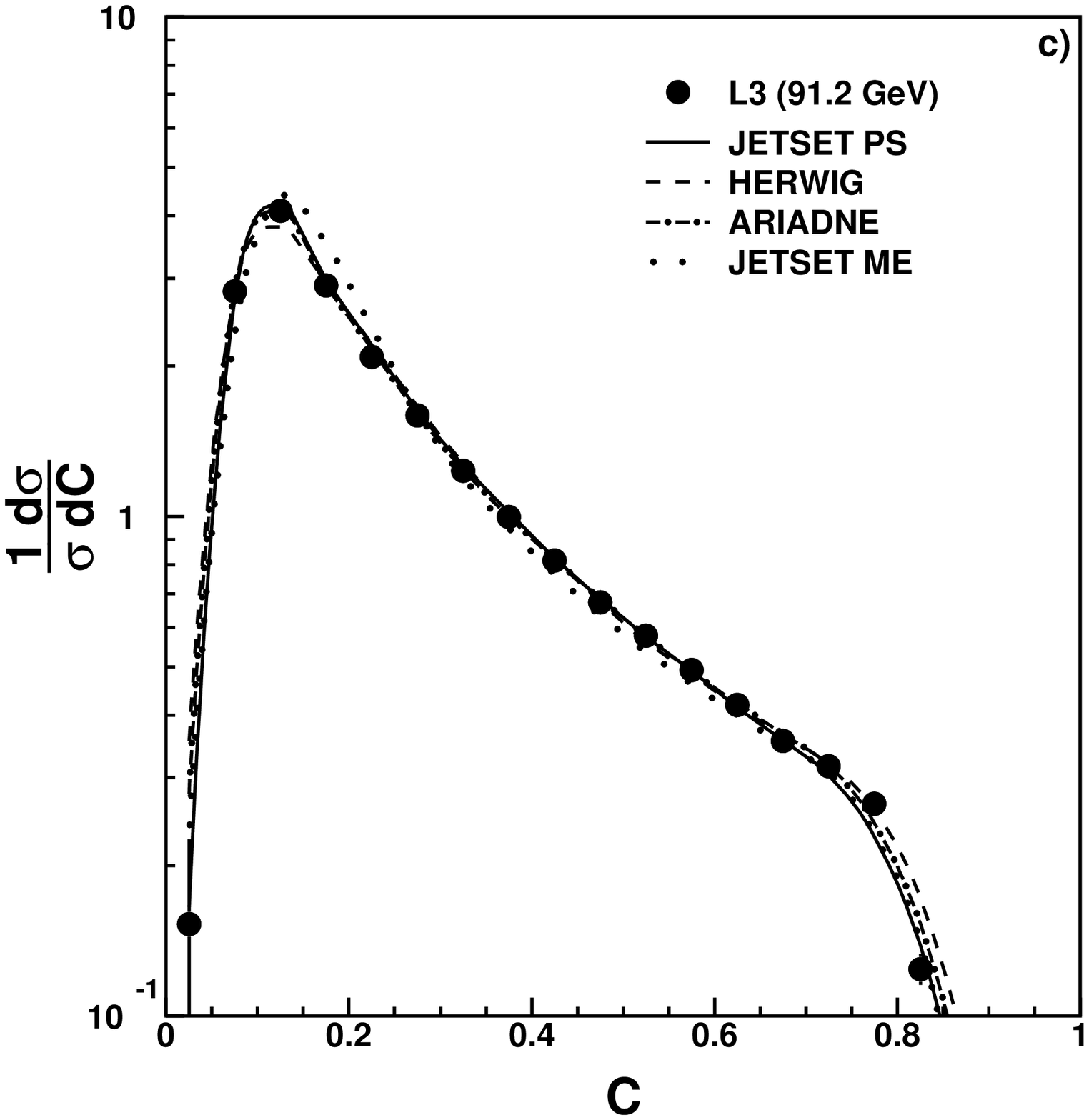}
  \includegraphics*[width=.5\figwidth,bbllx=5,bblly=30,bburx=525,bbury=560]{\mydirfig 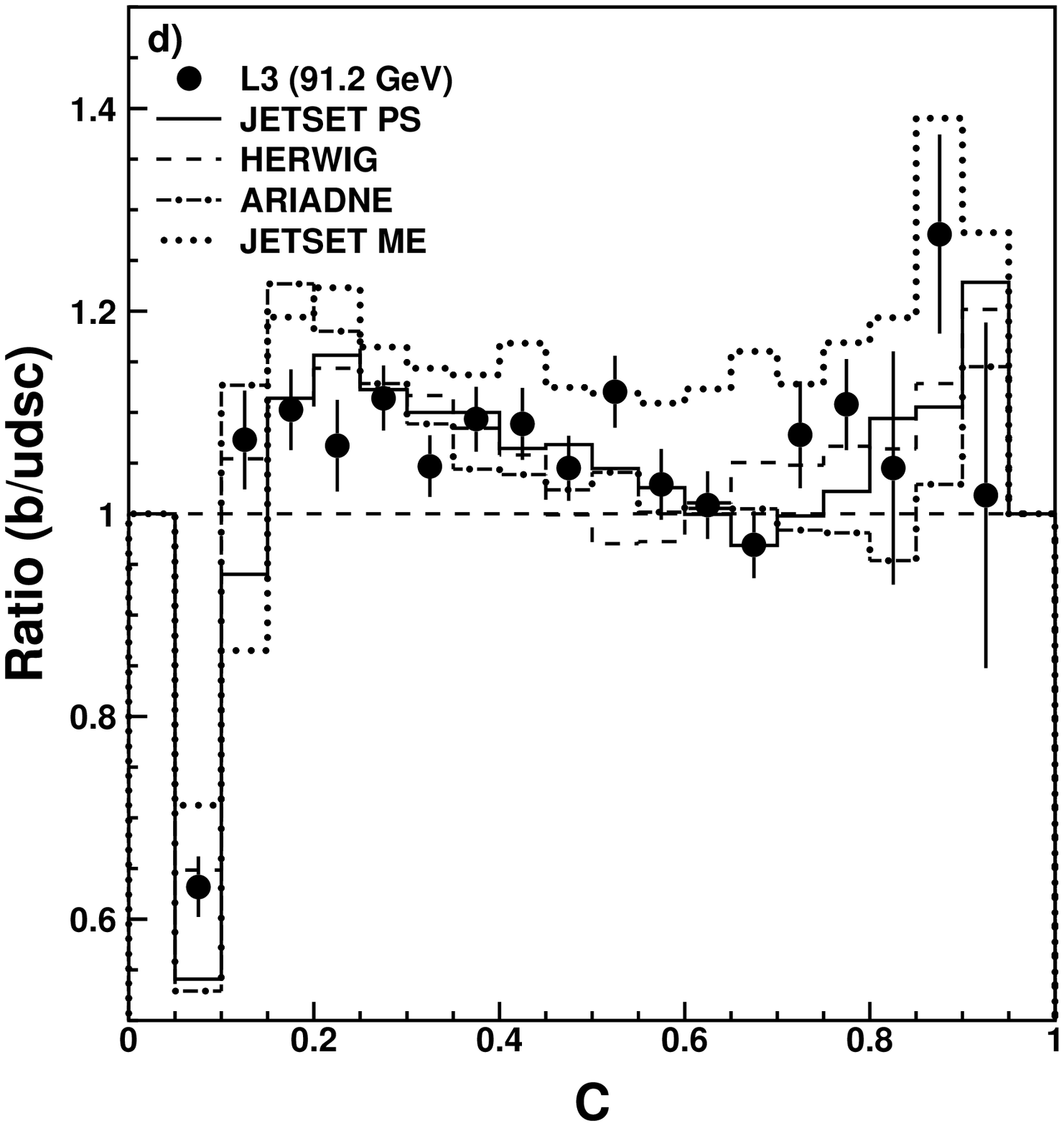}
\end{center}
\caption{$C$-parameter distributions at
         $\rs=91.2\,\GeV$
         for b, udsc, and all quark flavours
         and the ratio b/udsc
         compared to several \QCD\ models.
         }
\label{fig:part-c}
\end{figure}

 
At the Z pole the distributions of the six event-shape variables
$T$, \rhoh, \bt, \bw, $C$ and $D$
are also measured for b and udsc flavours separately.
These distributions, corrected for purity by Monte Carlo,
are summarised in Tables \ref{tab:thrust91}--\ref{tab:dpar91}
and compared with the \textsc{Jetset} PS, \textsc{Herwig} and
\textsc{Ariadne} \QCD\ models in
Figures \ref{fig:part-thr}--\ref{fig:part-d}.
The figures also contain the distributions for all flavours.
The Monte Carlo models provide a reasonable description of the data.
Significant flavour-dependent differences exist,
particularly for the jet broadenings and the $C$- and $D$-parameters.
These differences are reasonably described by the models, with the exception of the \textsc{Jetset} ME model.
 
\begin{figure}[htbp]
\begin{center}
  \includegraphics*[width=.5\figwidth,bbllx=5,bblly=30,bburx=525,bbury=560]{\mydirfig 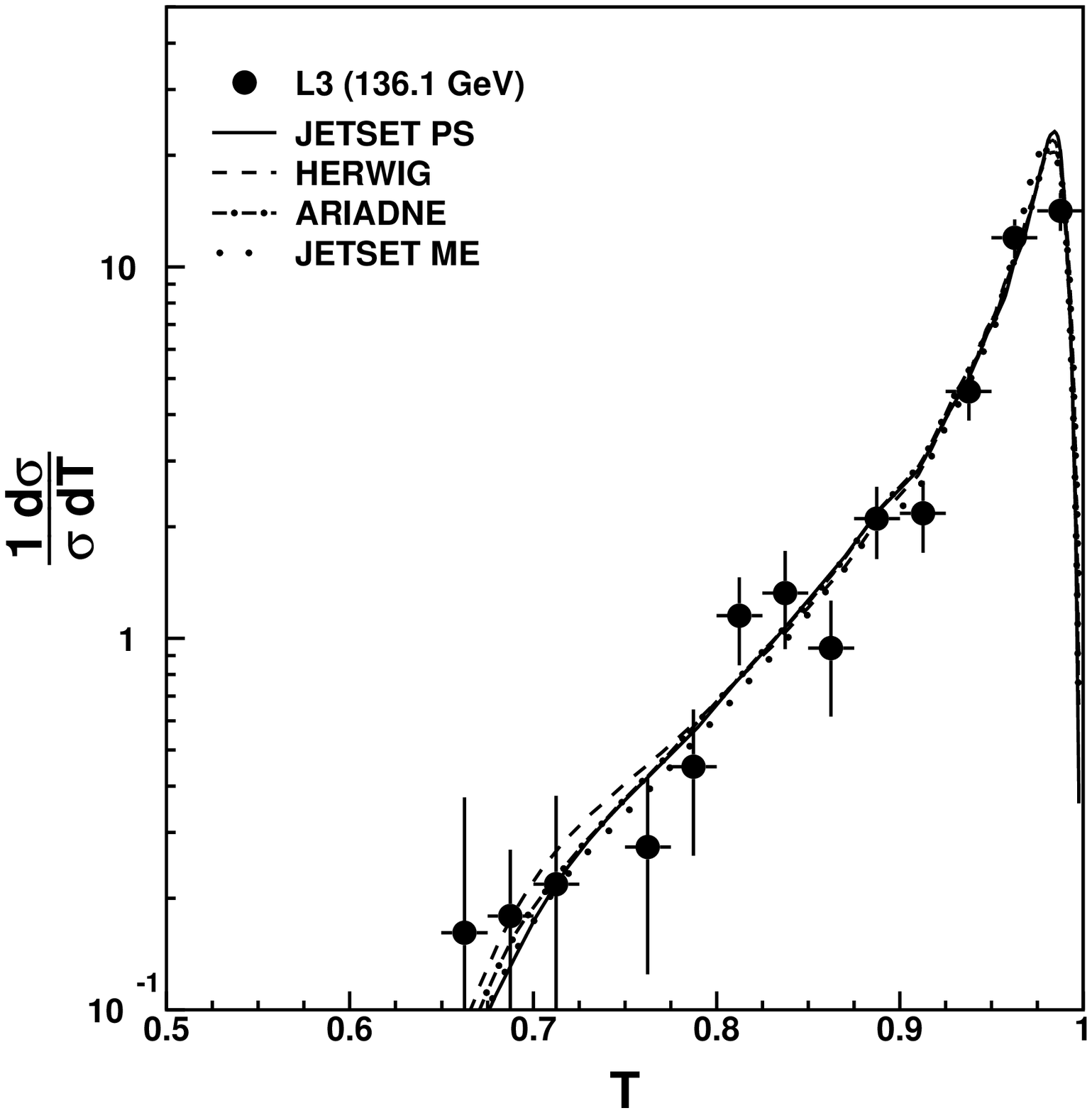}
  \includegraphics*[width=.5\figwidth,bbllx=5,bblly=30,bburx=525,bbury=560]{\mydirfig 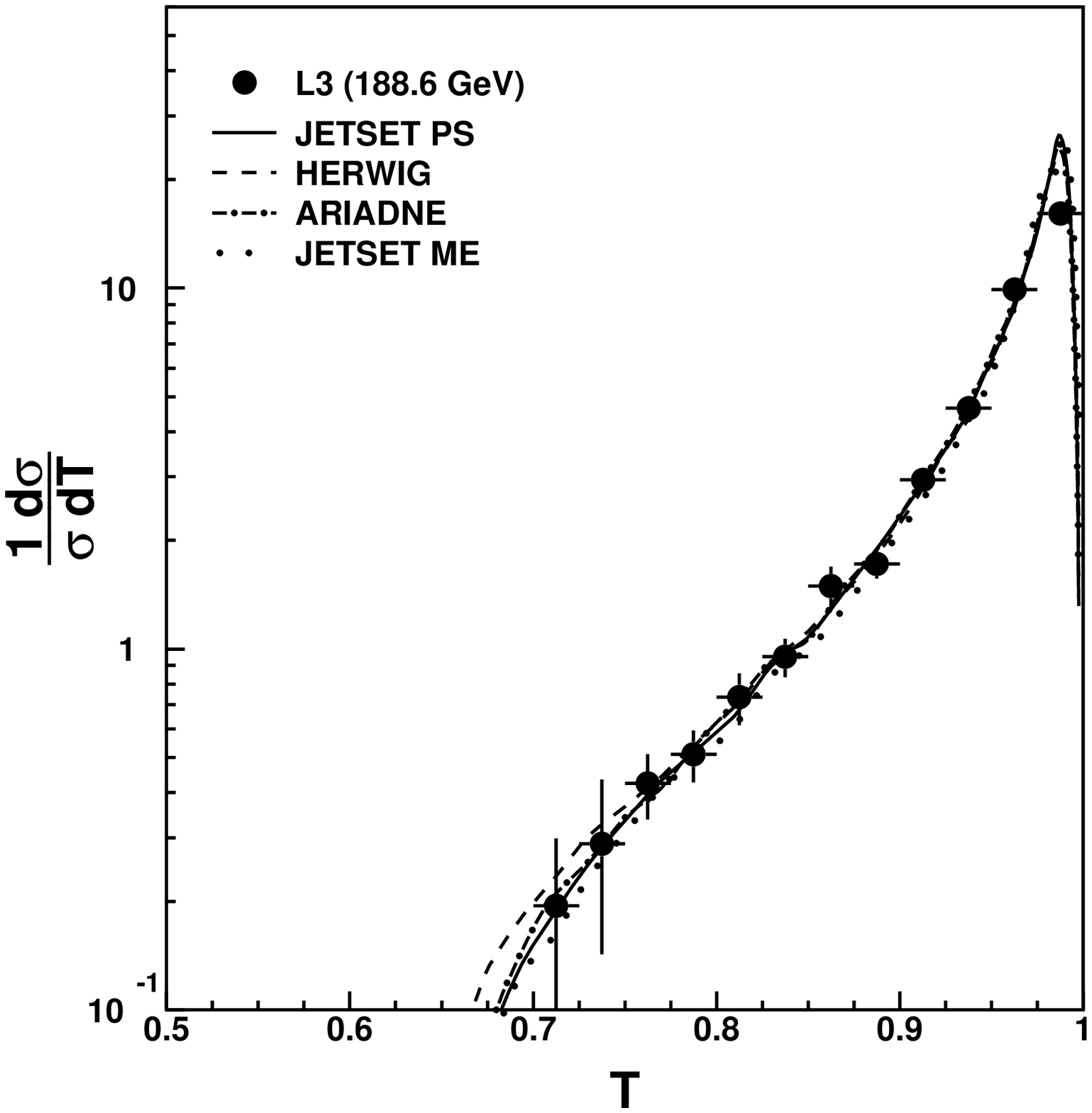}
  \includegraphics*[width=.5\figwidth,bbllx=5,bblly=30,bburx=525,bbury=560]{\mydirfig 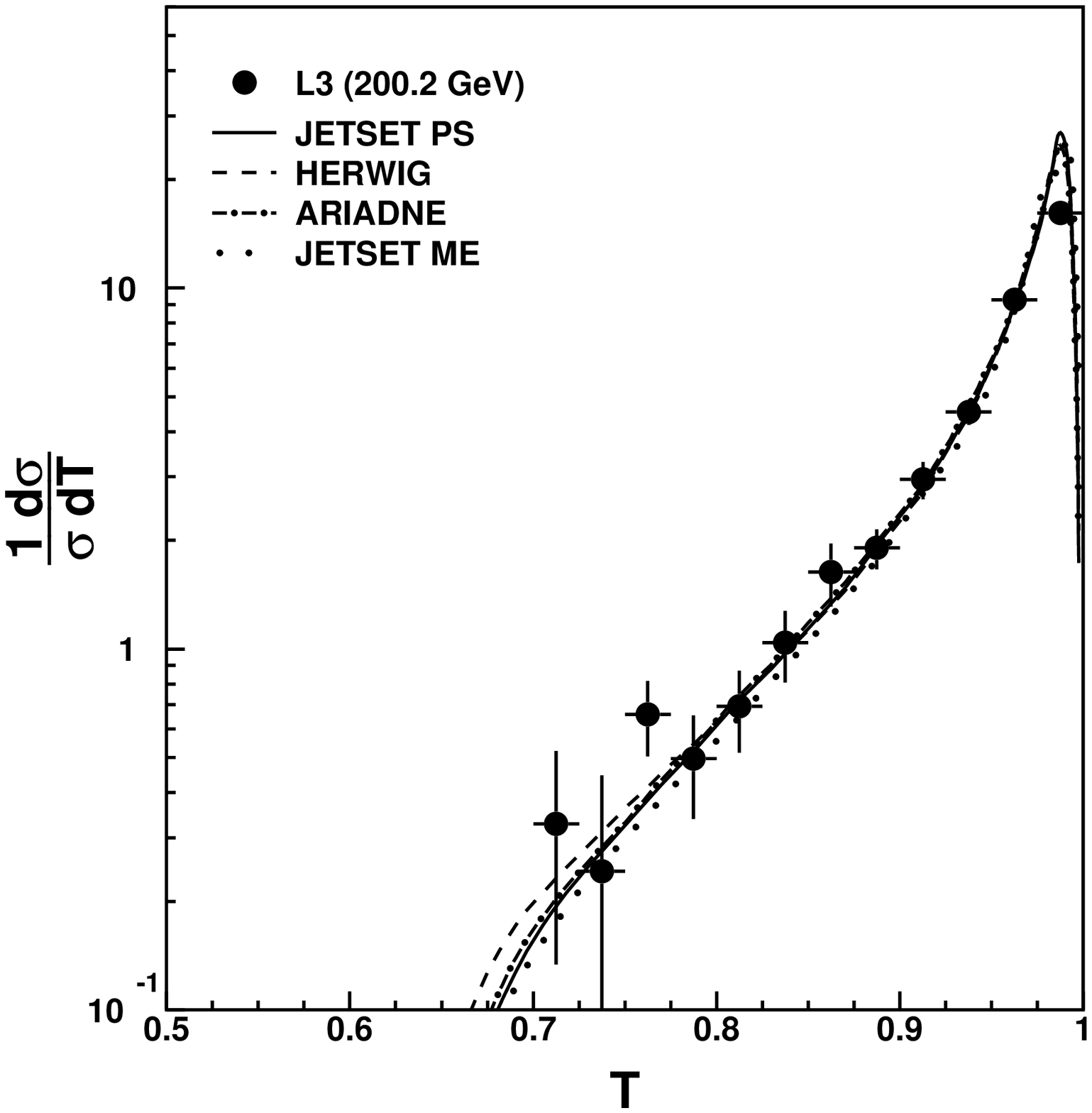}
  \includegraphics*[width=.5\figwidth,bbllx=5,bblly=30,bburx=525,bbury=560]{\mydirfig 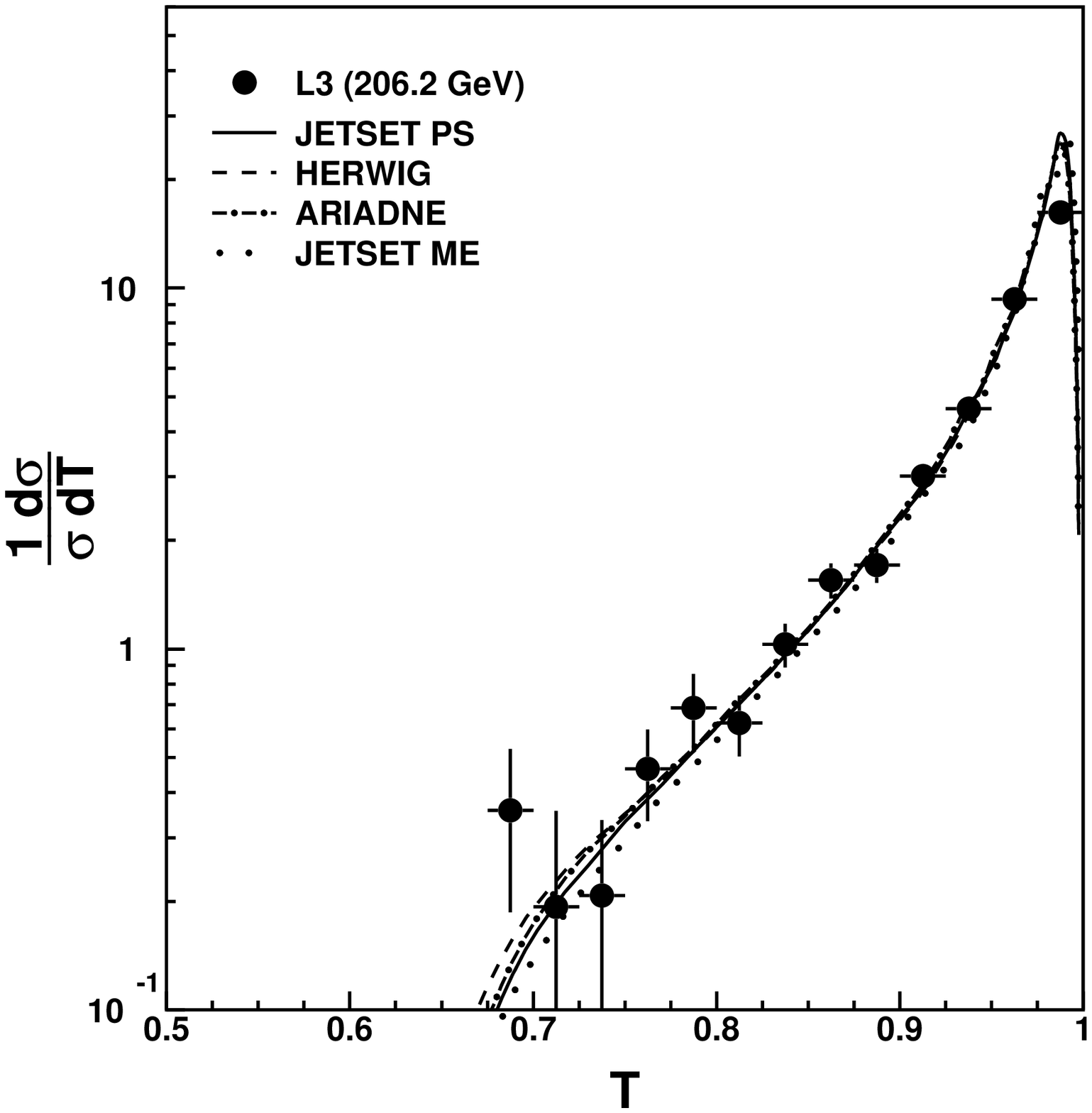}
\end{center}
\caption{Thrust distributions at
         $\langle\rs\,\rangle=
          136.1, 188.6, 200.2 \text{ and } 206.2\,\GeV$
         compared to several \QCD\ models.}
\label{fig:thr}
\end{figure}

\begin{figure}[htbp]
\begin{center}
  \includegraphics*[width=.5\figwidth,bbllx=5,bblly=30,bburx=525,bbury=560]{\mydirfig 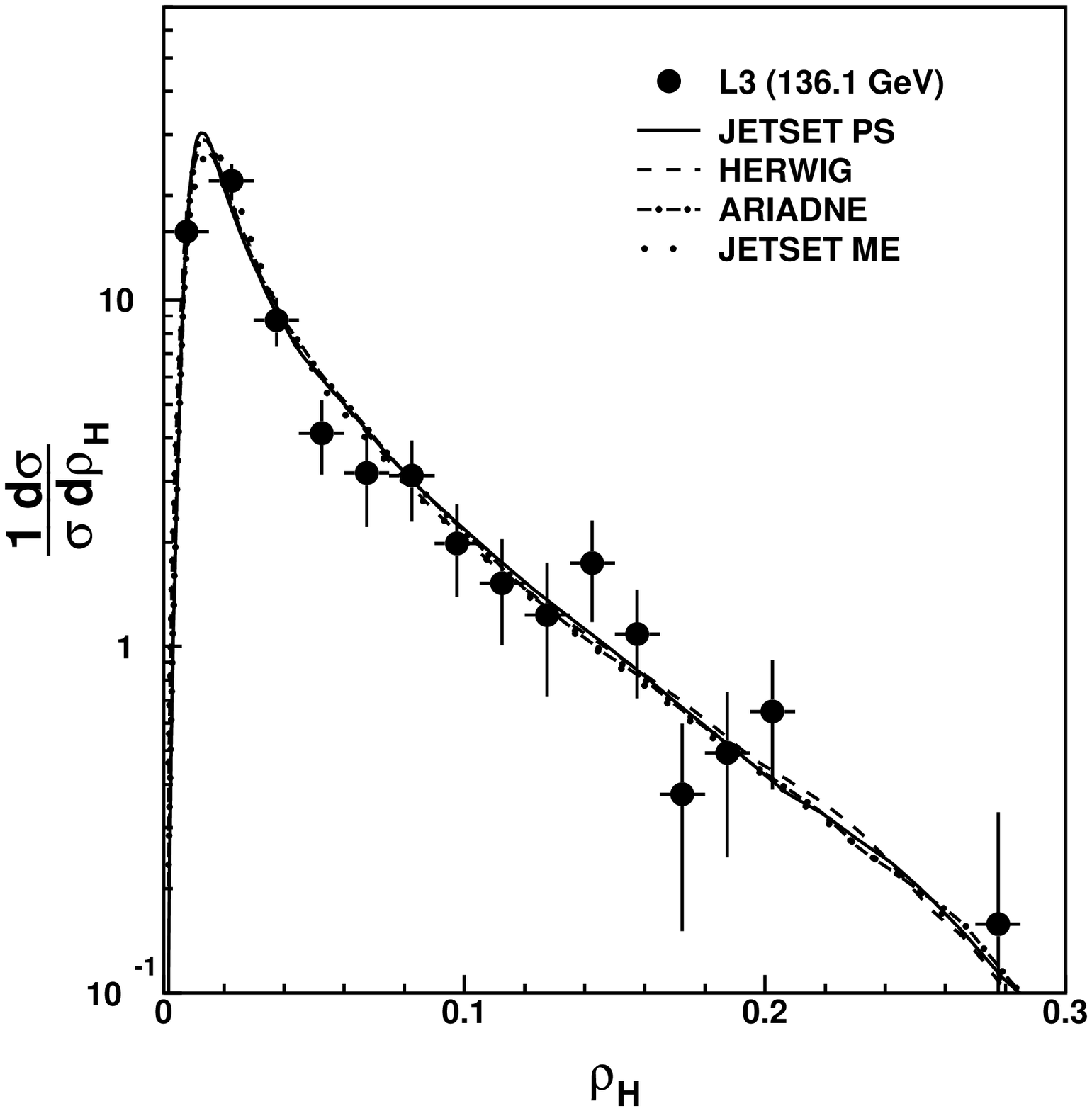}
  \includegraphics*[width=.5\figwidth,bbllx=5,bblly=30,bburx=525,bbury=560]{\mydirfig 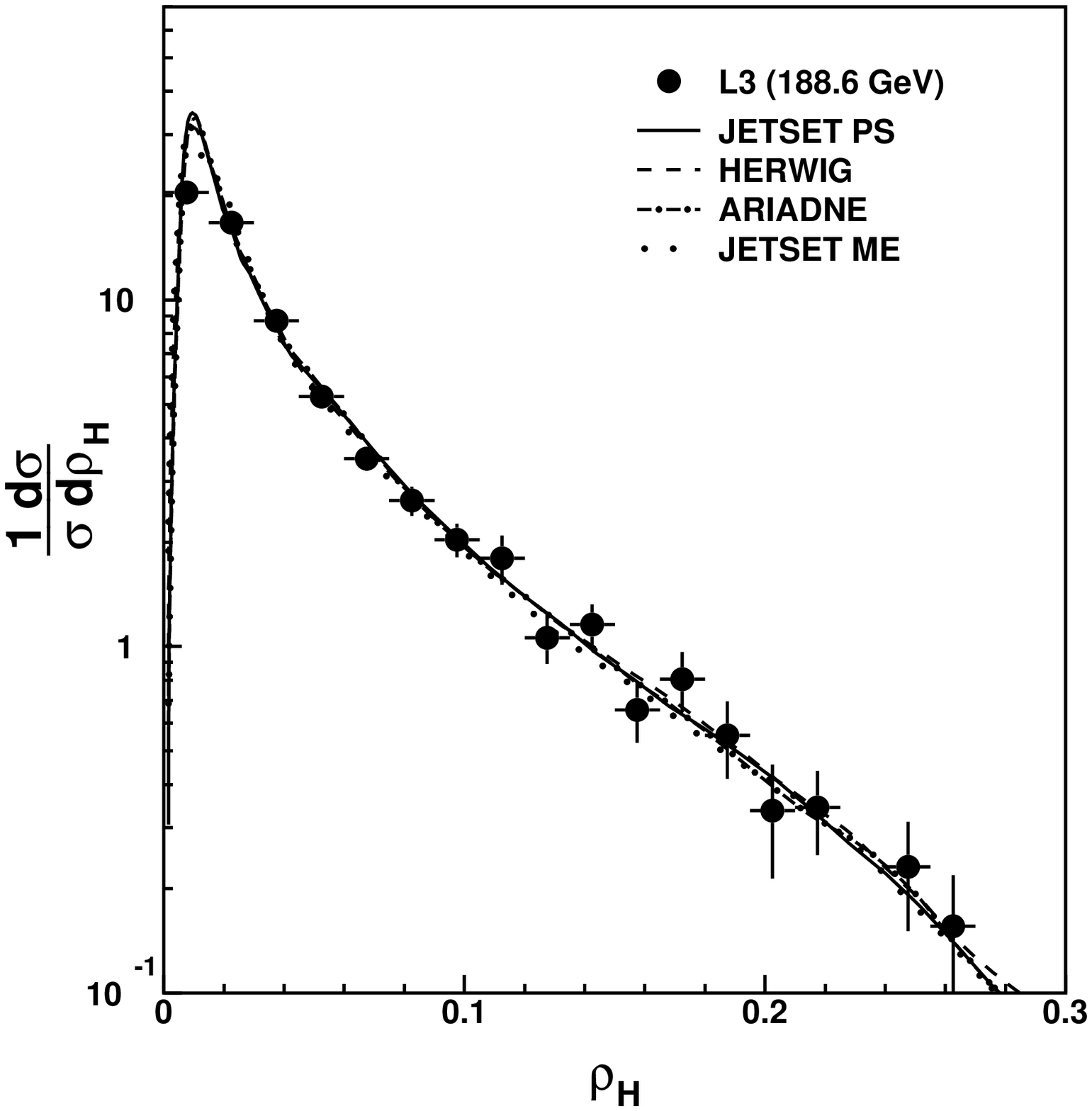}
  \includegraphics*[width=.5\figwidth,bbllx=5,bblly=30,bburx=525,bbury=560]{\mydirfig 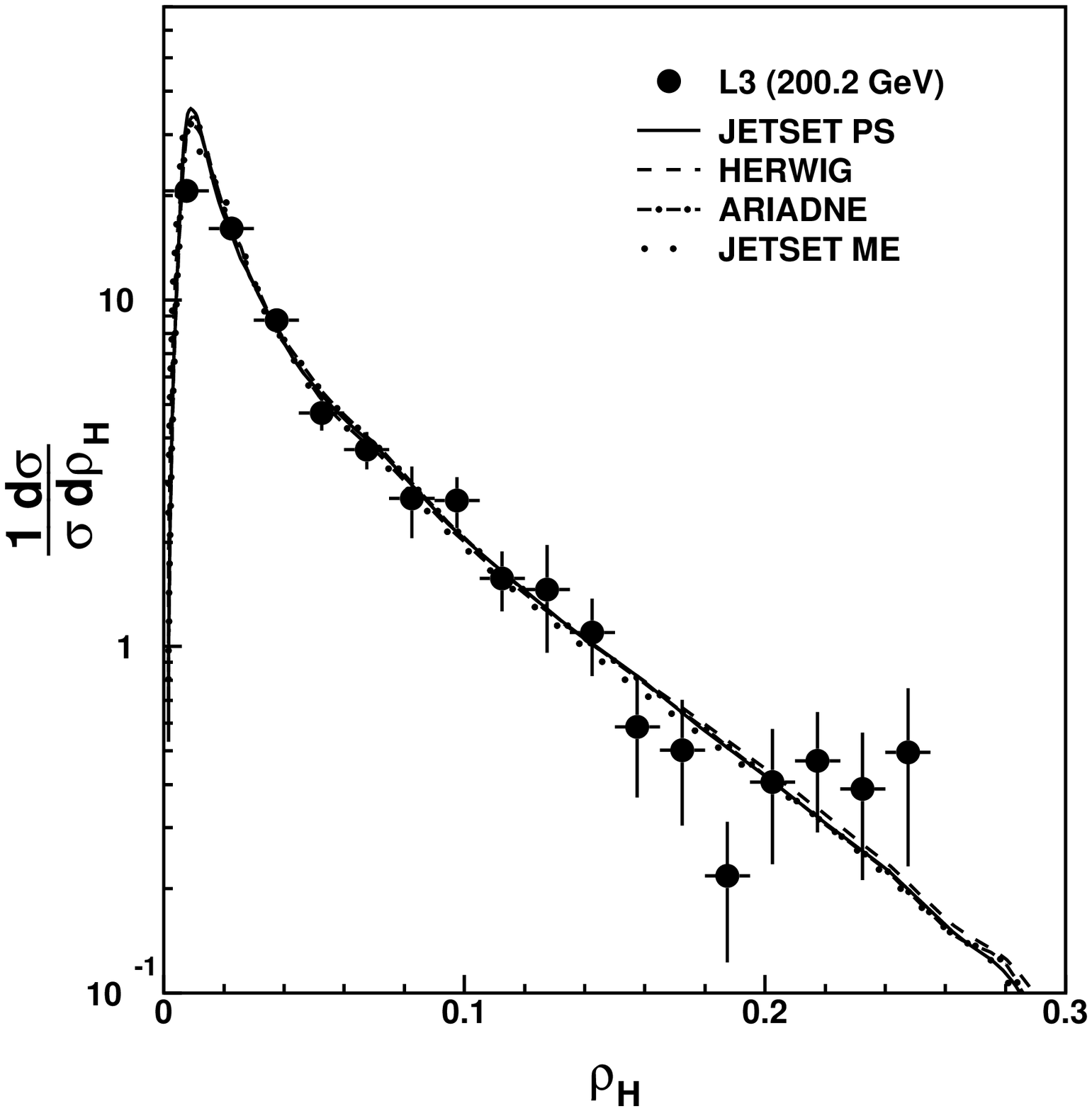}
  \includegraphics*[width=.5\figwidth,bbllx=5,bblly=30,bburx=525,bbury=560]{\mydirfig 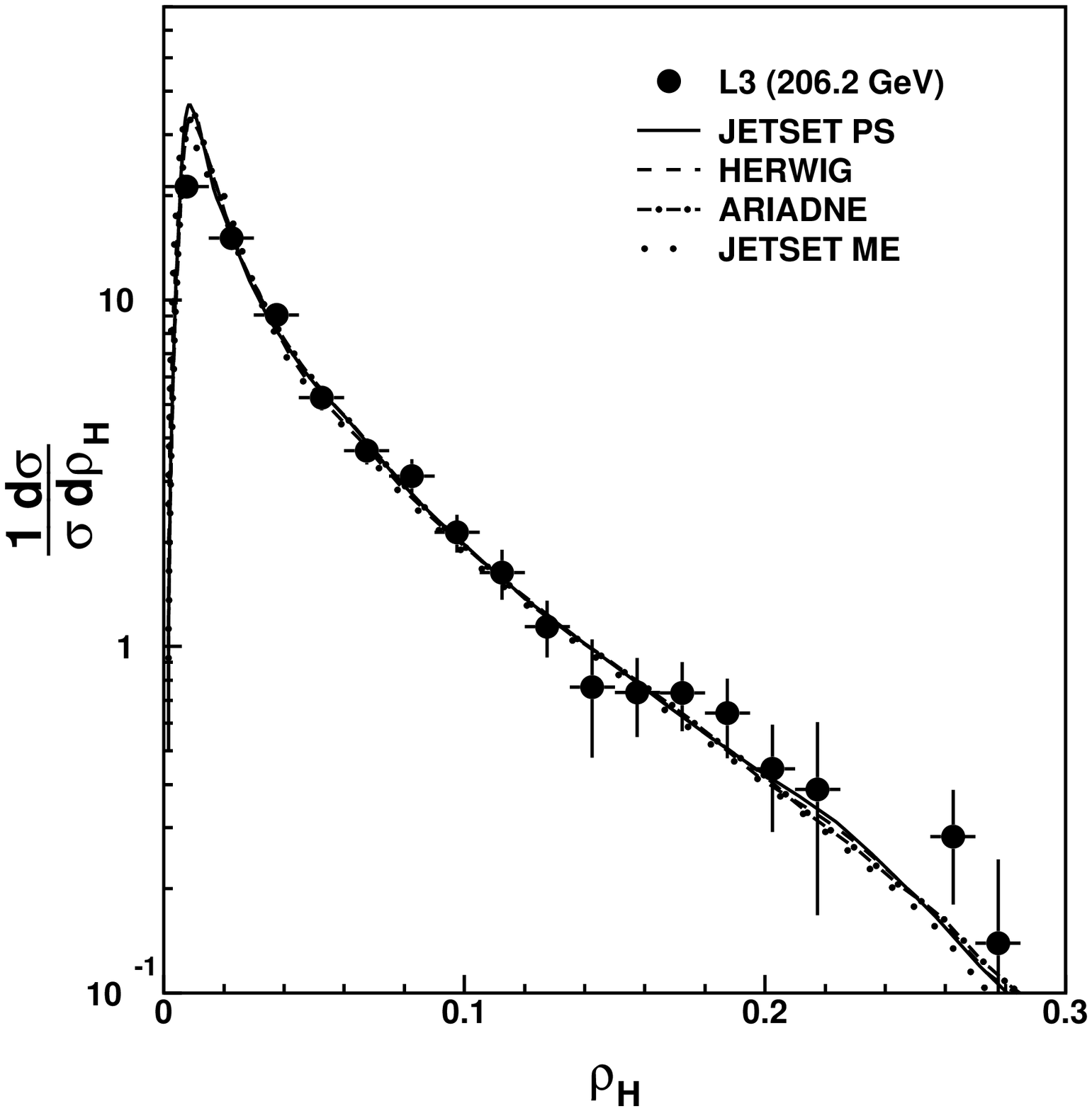}
\end{center}
\caption{Scaled heavy jet mass distributions at
         $\langle\rs\,\rangle=
          136.1, 188.6, 200.2 \text{ and } 206.2\,\GeV$
         compared to several \QCD\ models.}
\label{fig:rho}
\end{figure}

\begin{figure}[htbp]
\begin{center}
  \includegraphics*[width=.5\figwidth,bbllx=5,bblly=30,bburx=525,bbury=560]{\mydirfig 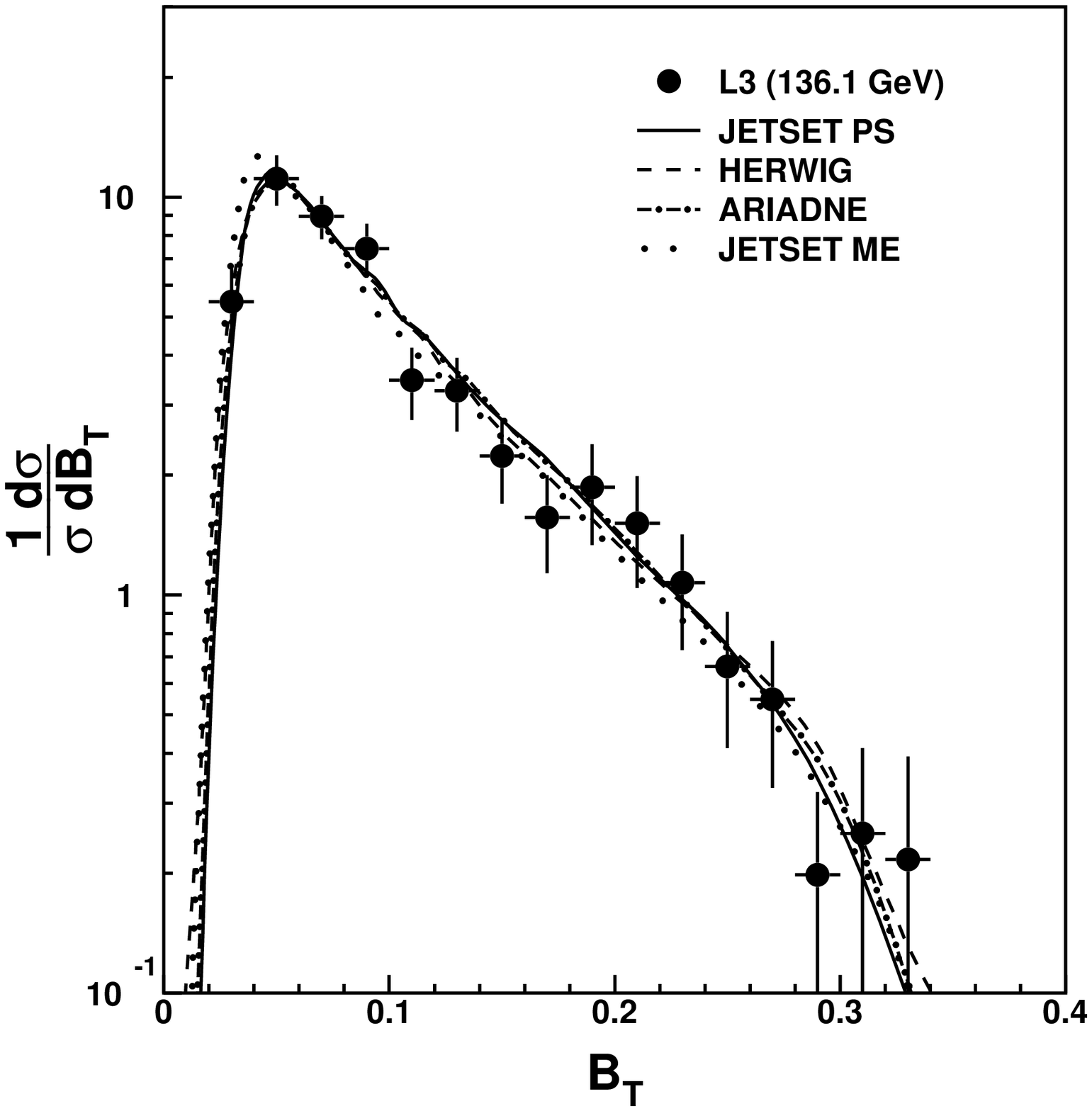}
  \includegraphics*[width=.5\figwidth,bbllx=5,bblly=30,bburx=525,bbury=560]{\mydirfig 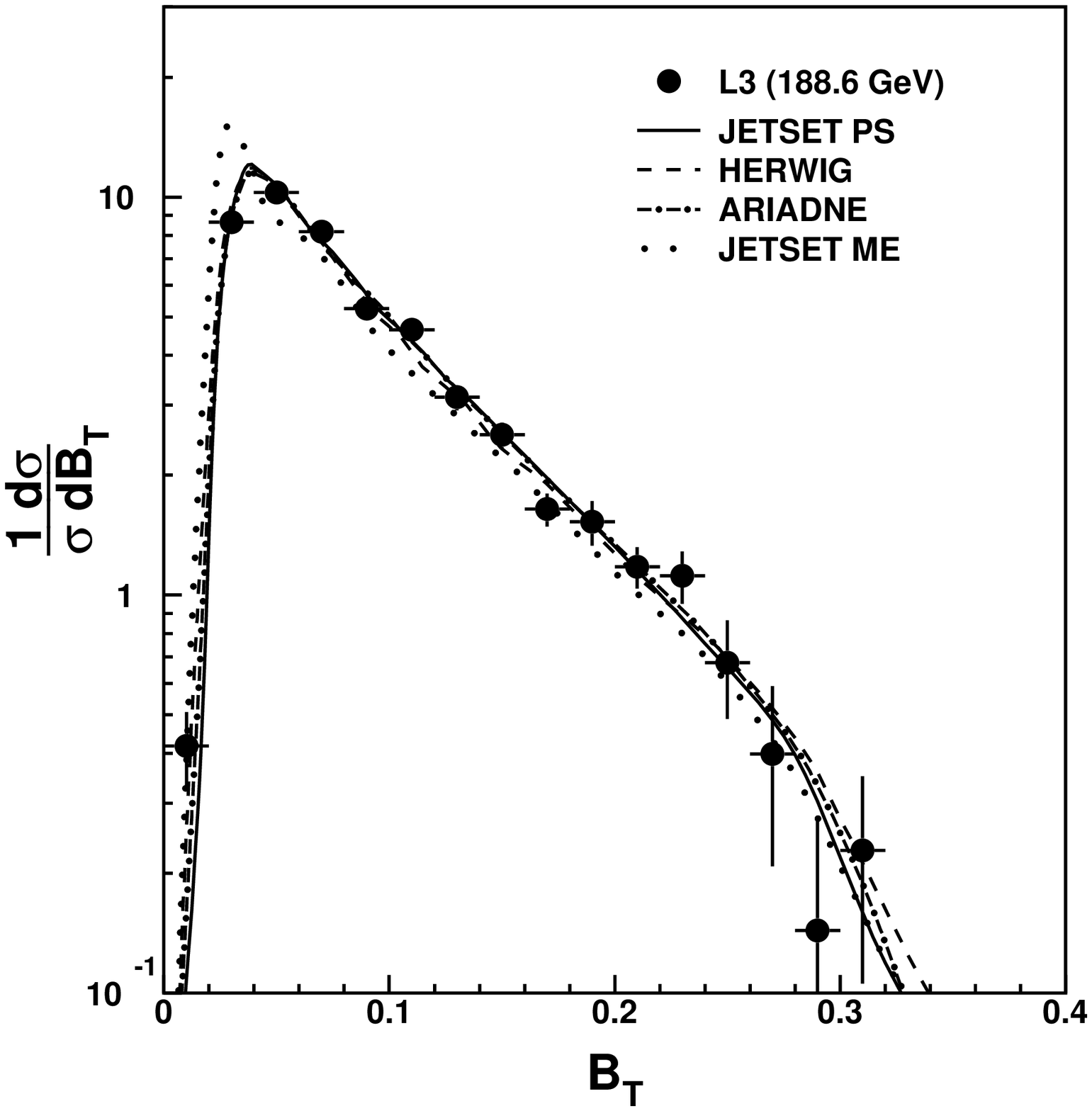}
  \includegraphics*[width=.5\figwidth,bbllx=5,bblly=30,bburx=525,bbury=560]{\mydirfig 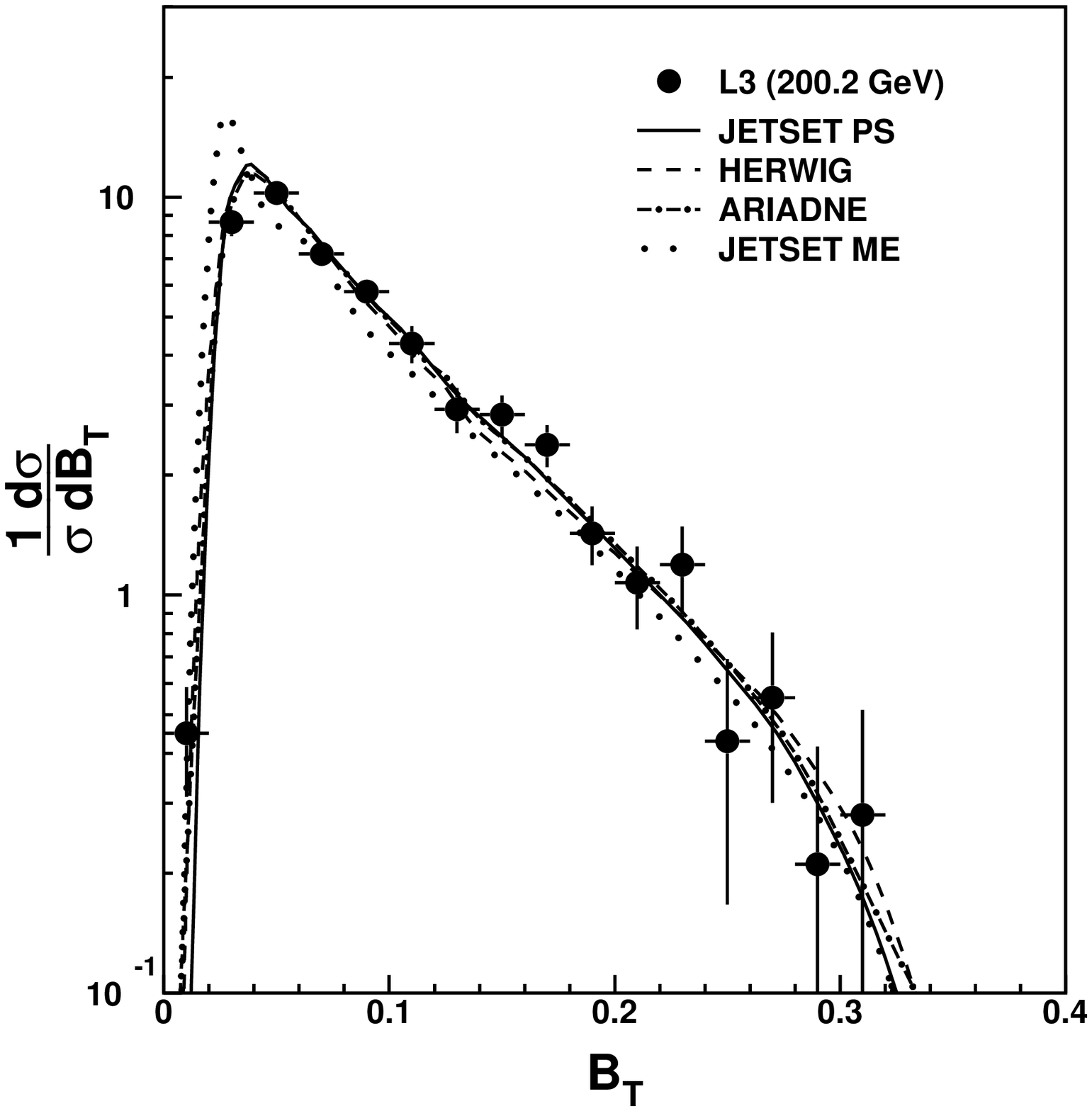}
  \includegraphics*[width=.5\figwidth,bbllx=5,bblly=30,bburx=525,bbury=560]{\mydirfig 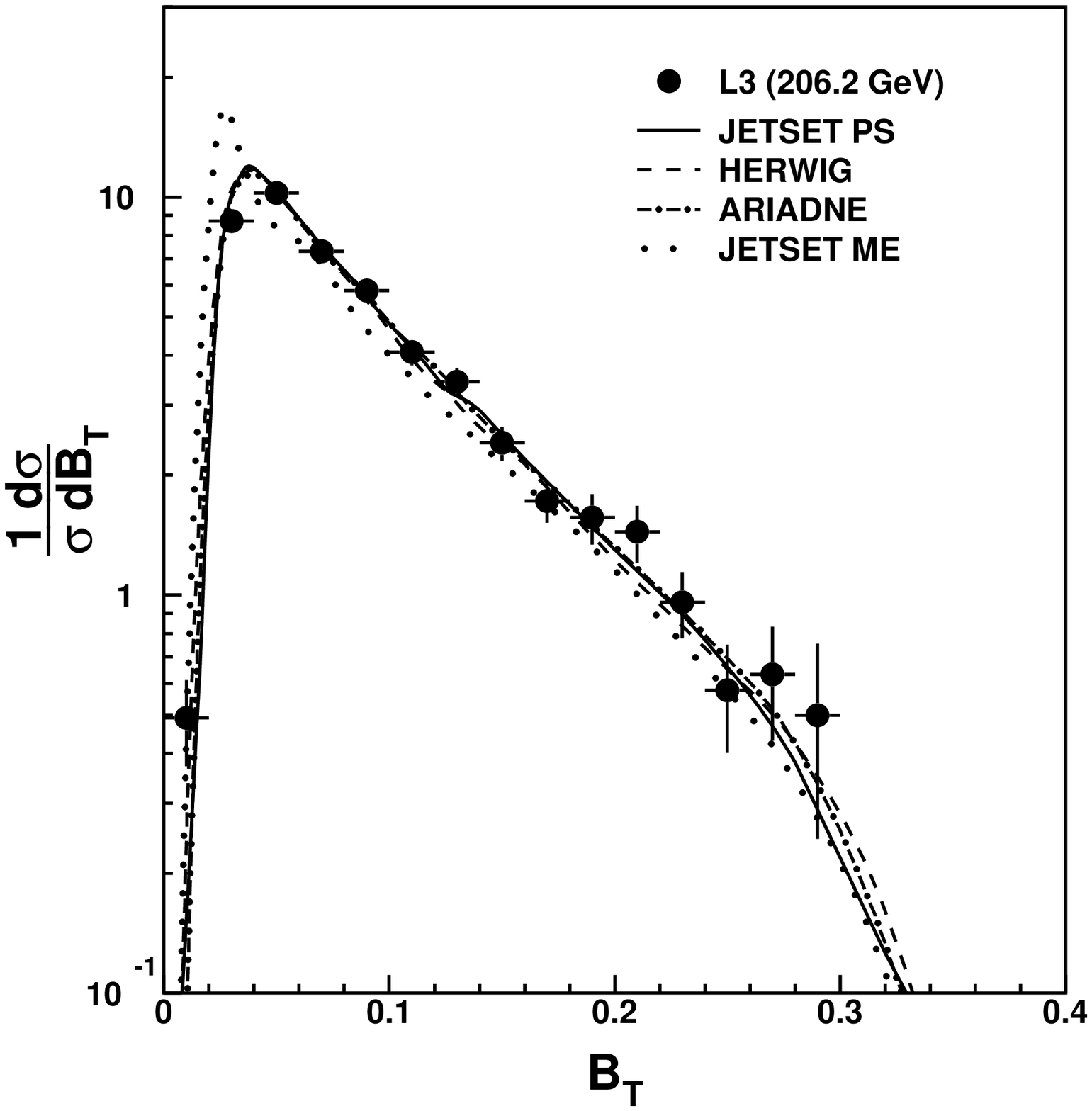}
\end{center}
\caption{Total jet broadening distributions at
         $\langle\rs\,\rangle=
         136.1, 188.6, 200.2 \text{ and } 206.2\,\GeV$
           compared to several \QCD\ models.}
\label{fig:bt}
\end{figure}

\begin{figure}[htbp]
\begin{center}
  \includegraphics*[width=.5\figwidth,bbllx=5,bblly=30,bburx=525,bbury=560]{\mydirfig 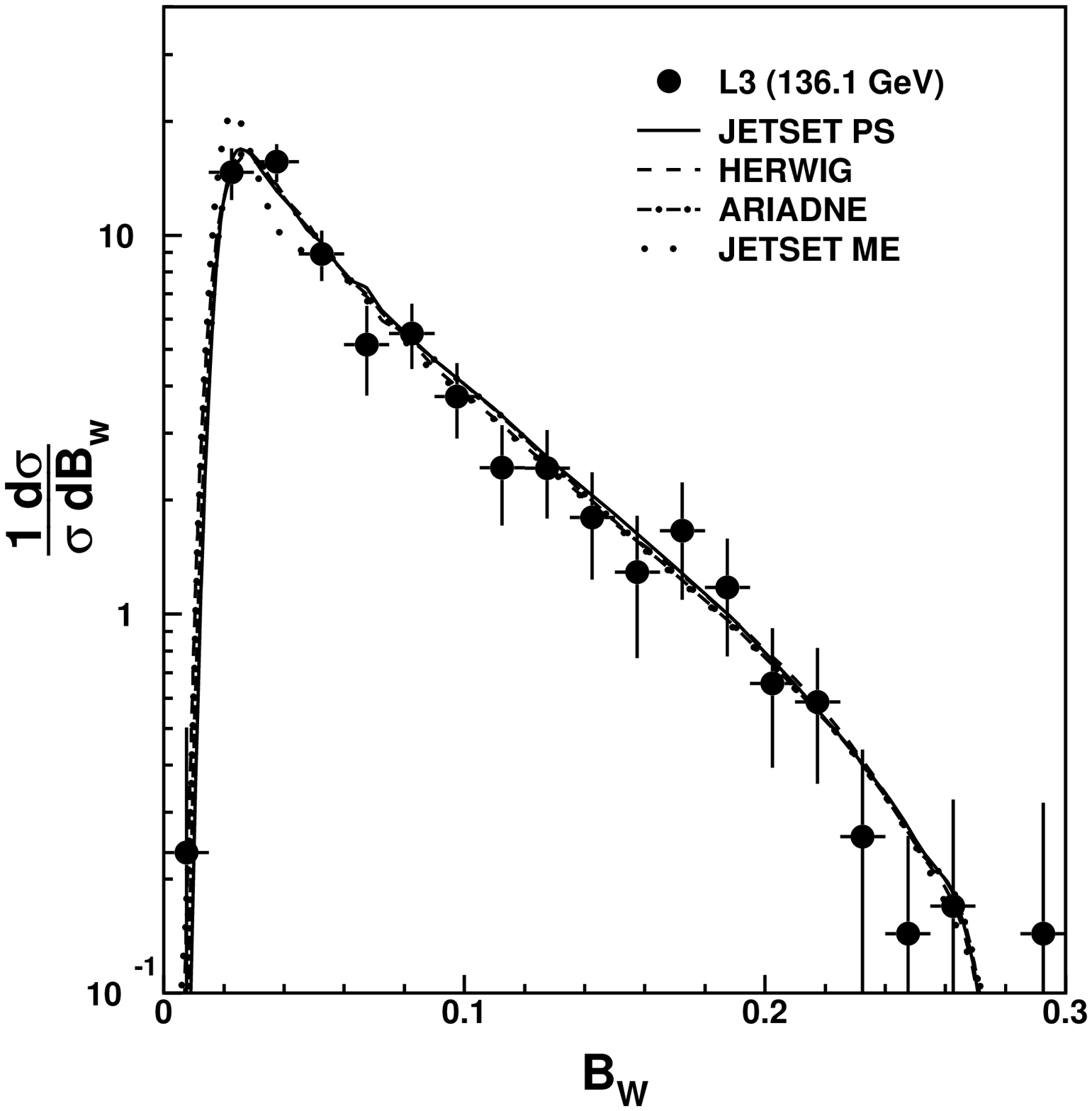}
  \includegraphics*[width=.5\figwidth,bbllx=5,bblly=30,bburx=525,bbury=560]{\mydirfig 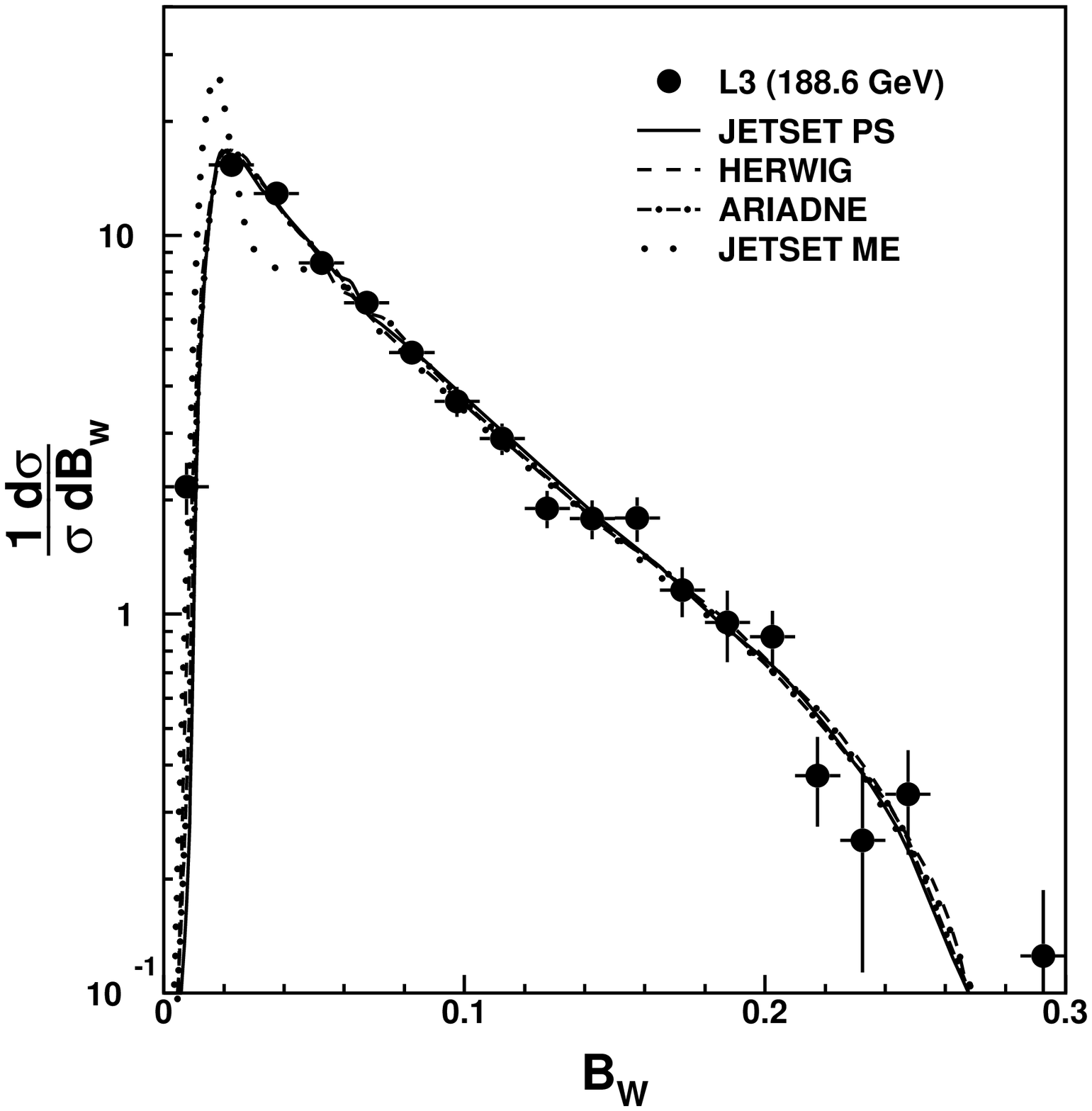}
  \includegraphics*[width=.5\figwidth,bbllx=5,bblly=30,bburx=525,bbury=560]{\mydirfig 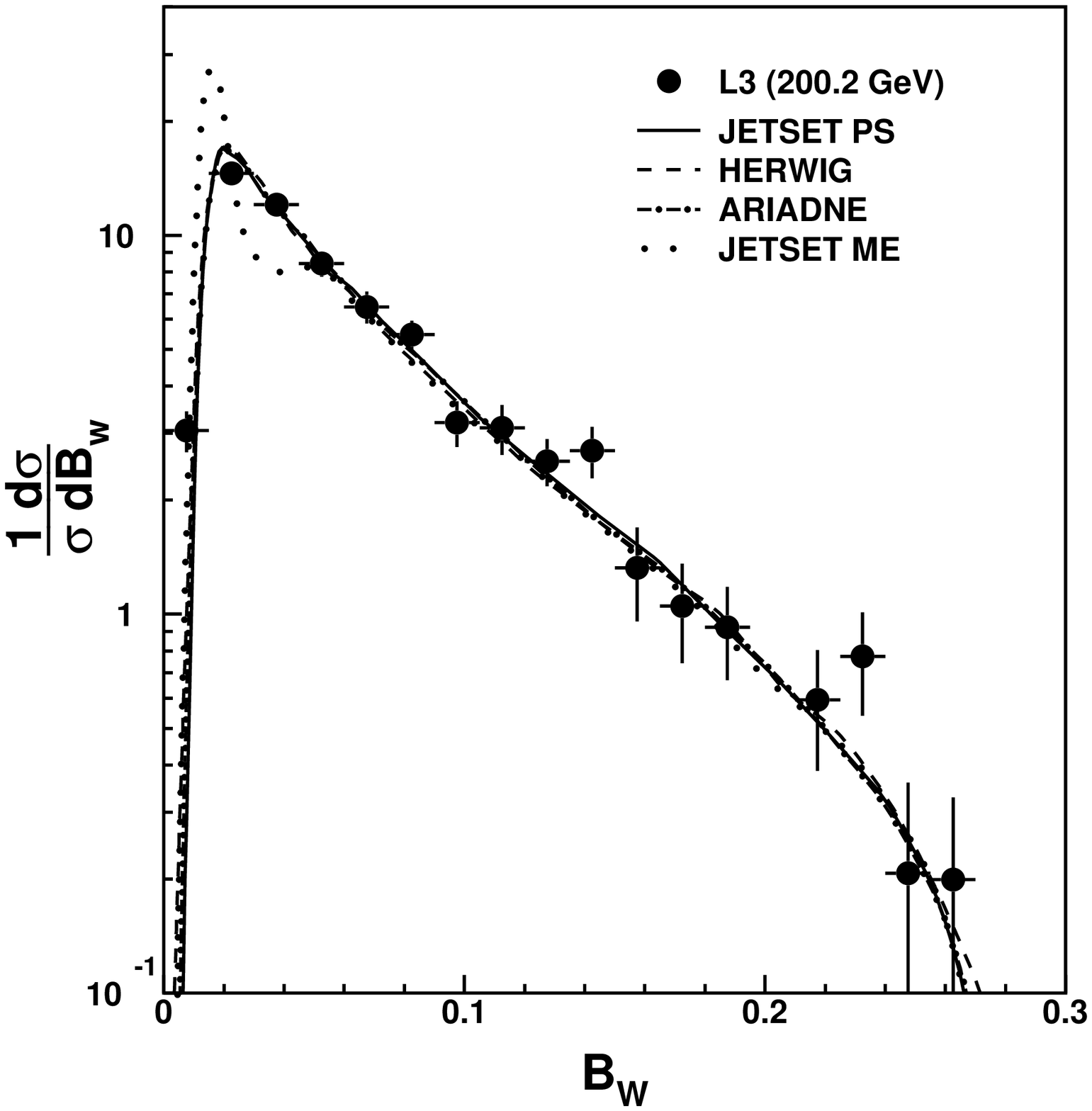}
  \includegraphics*[width=.5\figwidth,bbllx=5,bblly=30,bburx=525,bbury=560]{\mydirfig 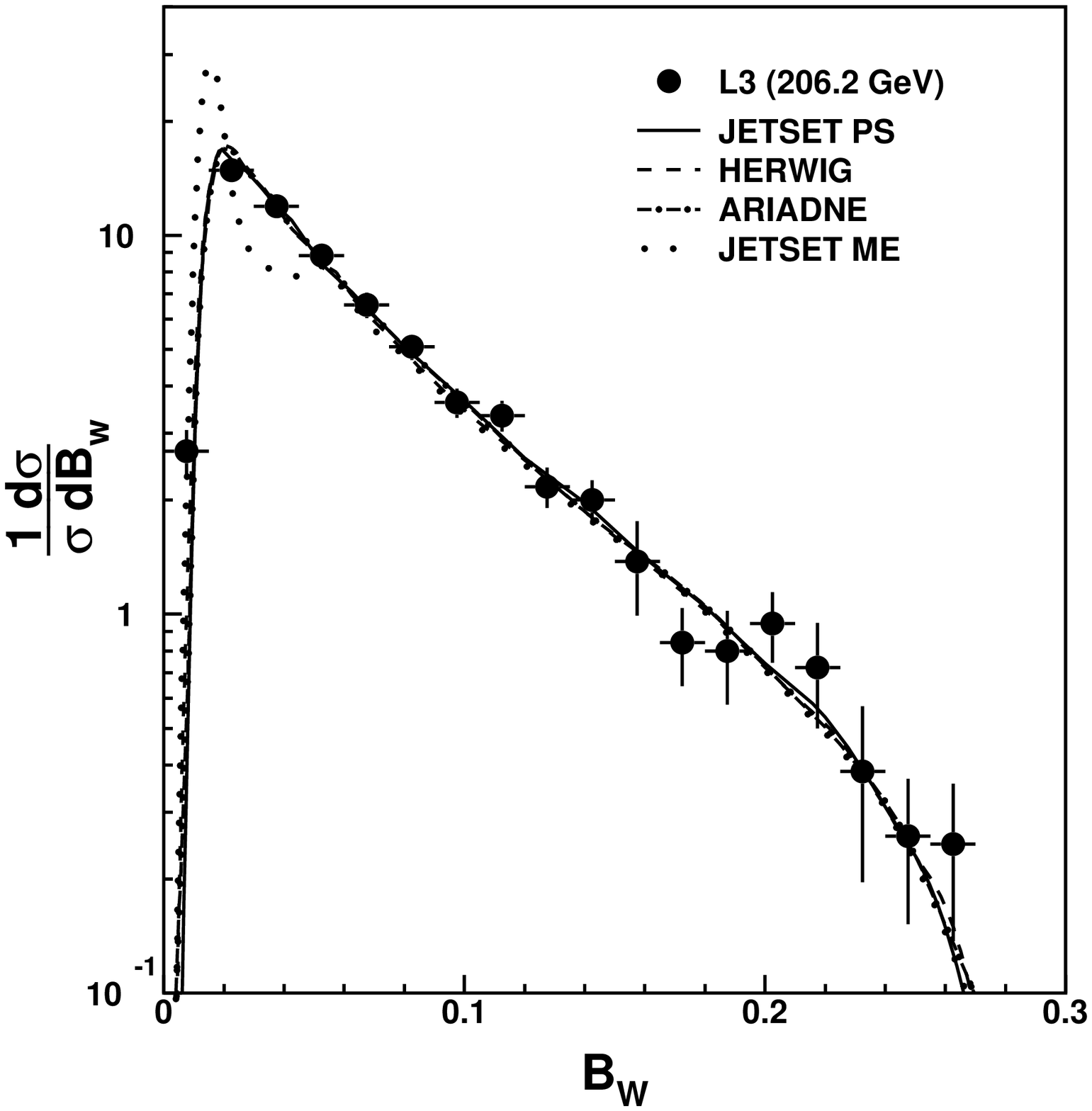}
\end{center}
\caption{Wide jet broadening distributions at
         $\langle\rs\,\rangle=
         136.1, 188.6, 200.2 \text{ and } 206.2\,\GeV$
           compared to several \QCD\ models.}
\label{fig:bw}
\end{figure}

\begin{figure}[htbp]
\begin{center}
  \includegraphics*[width=.5\figwidth,bbllx=5,bblly=30,bburx=525,bbury=560]{\mydirfig 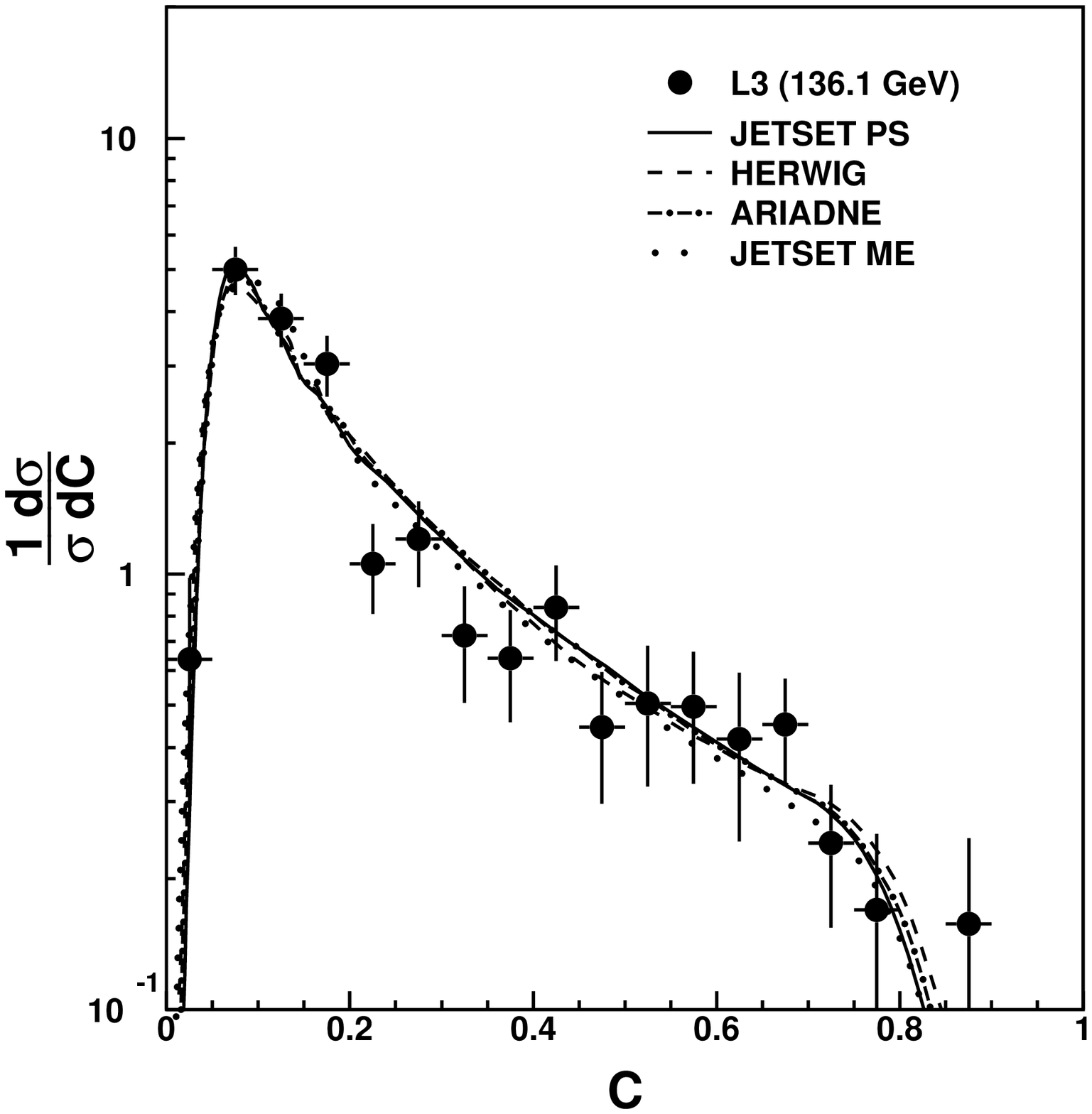}
  \includegraphics*[width=.5\figwidth,bbllx=5,bblly=30,bburx=525,bbury=560]{\mydirfig 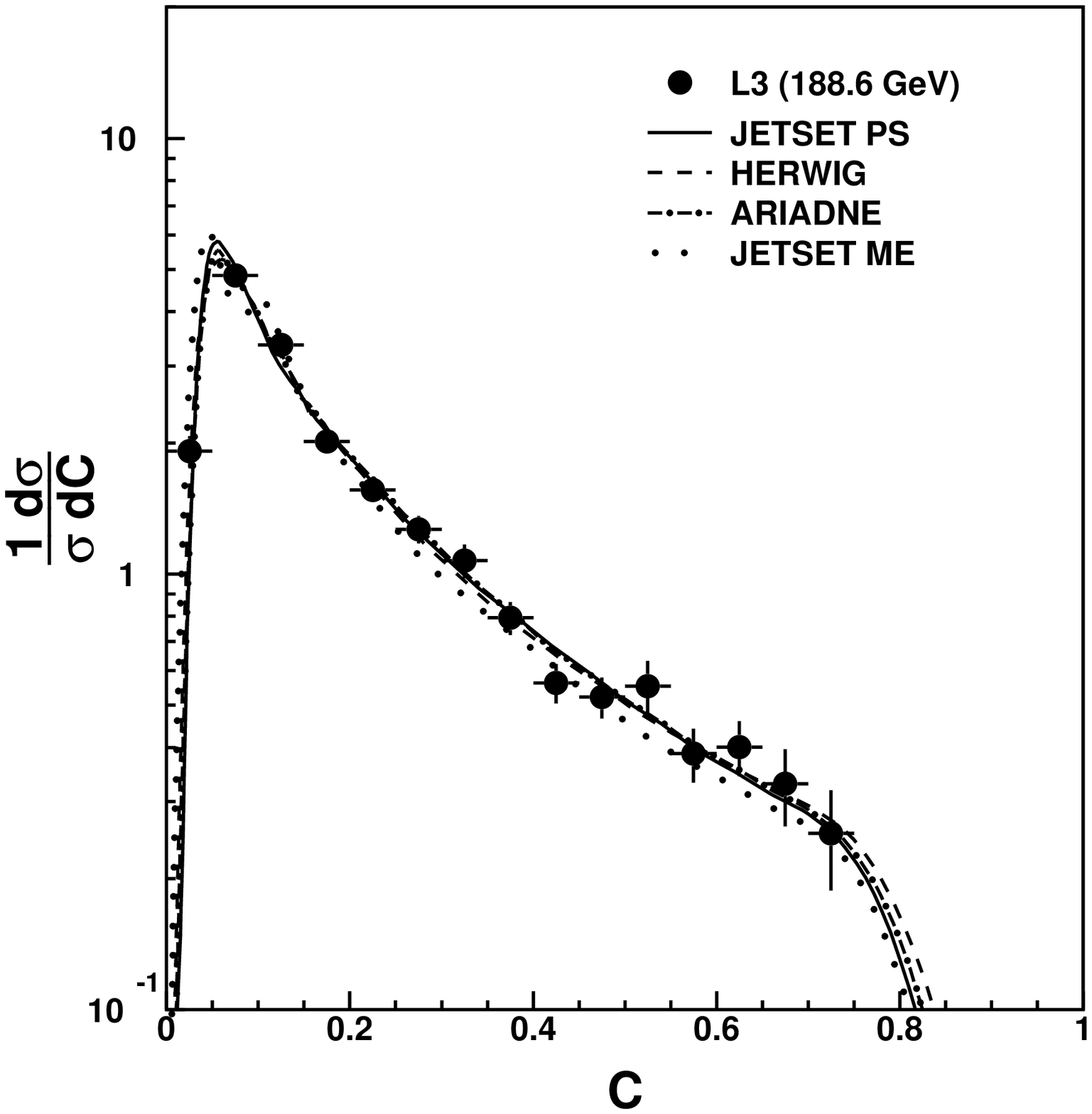}
  \includegraphics*[width=.5\figwidth,bbllx=5,bblly=30,bburx=525,bbury=560]{\mydirfig 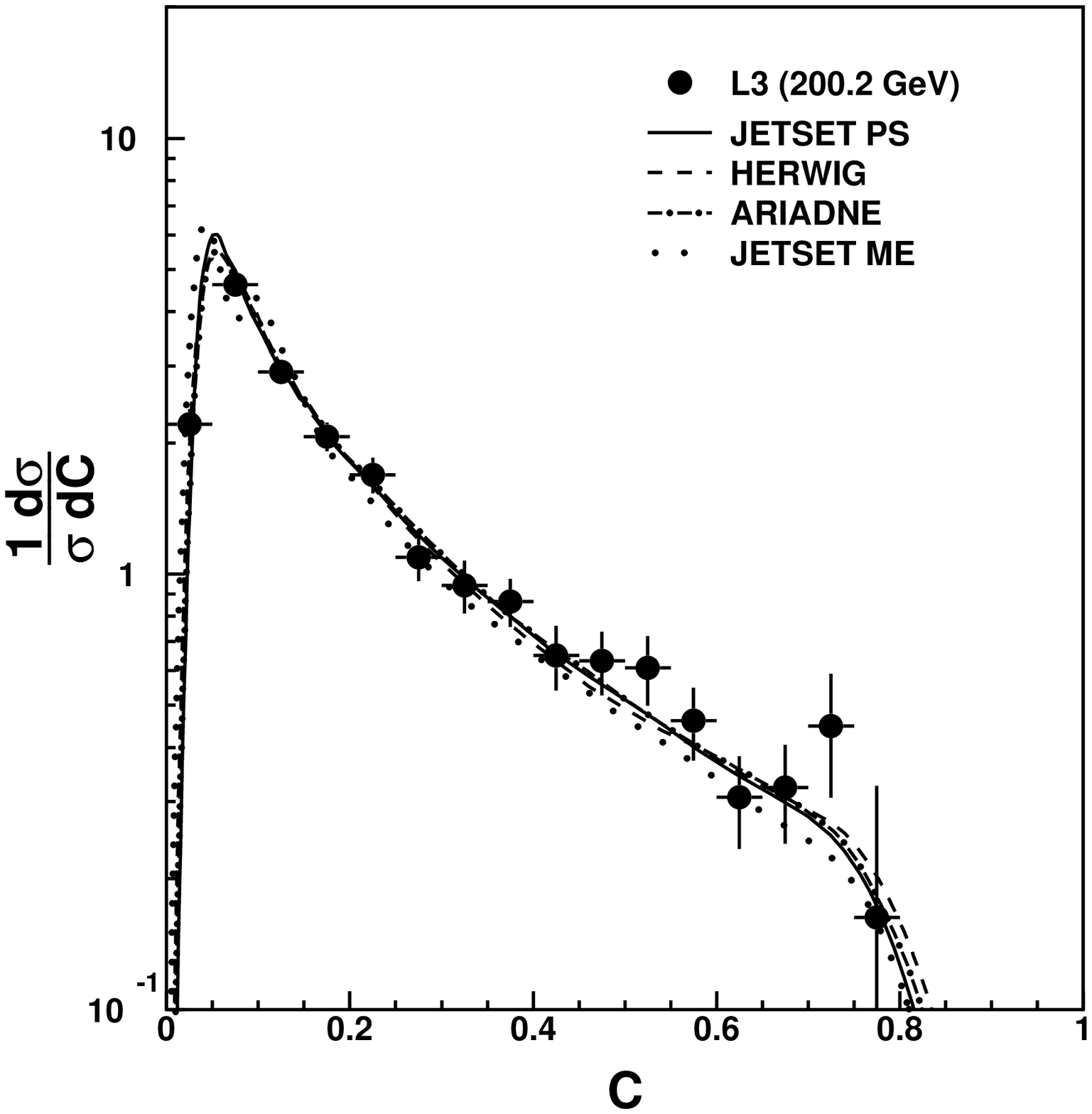}
  \includegraphics*[width=.5\figwidth,bbllx=5,bblly=30,bburx=525,bbury=560]{\mydirfig 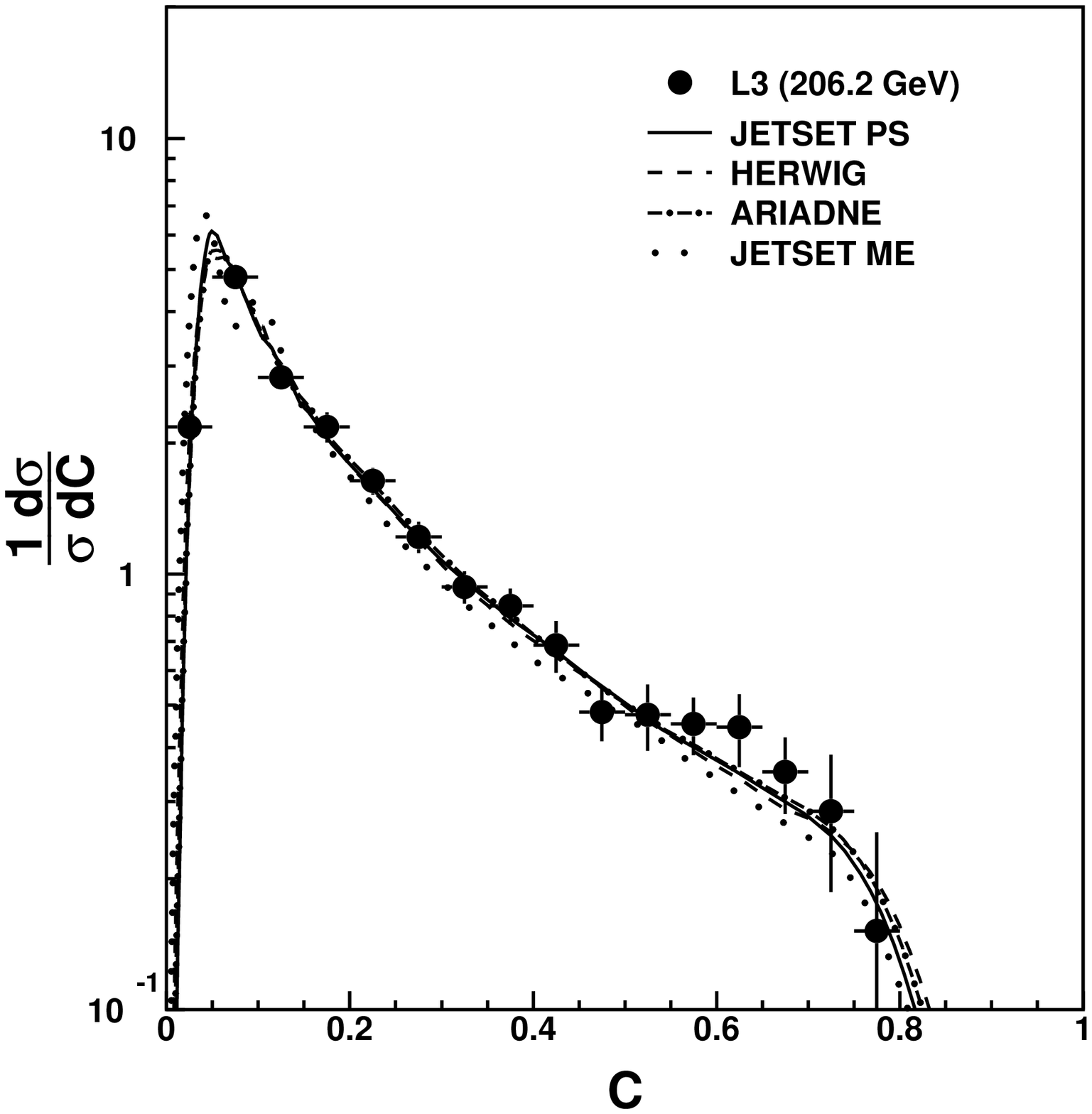}
\end{center}
\caption{$C$-parameter distributions at
         $\langle\rs\,\rangle=
         136.1, 188.6, 200.2 \text{ and } 206.2\,\GeV$
          compared to several \QCD\ models.}
\label{fig:cpar}
\end{figure}

\begin{figure}[htbp]
\begin{center}
  \includegraphics*[width=.5\figwidth,bbllx=5,bblly=30,bburx=525,bbury=560]{\mydirfig 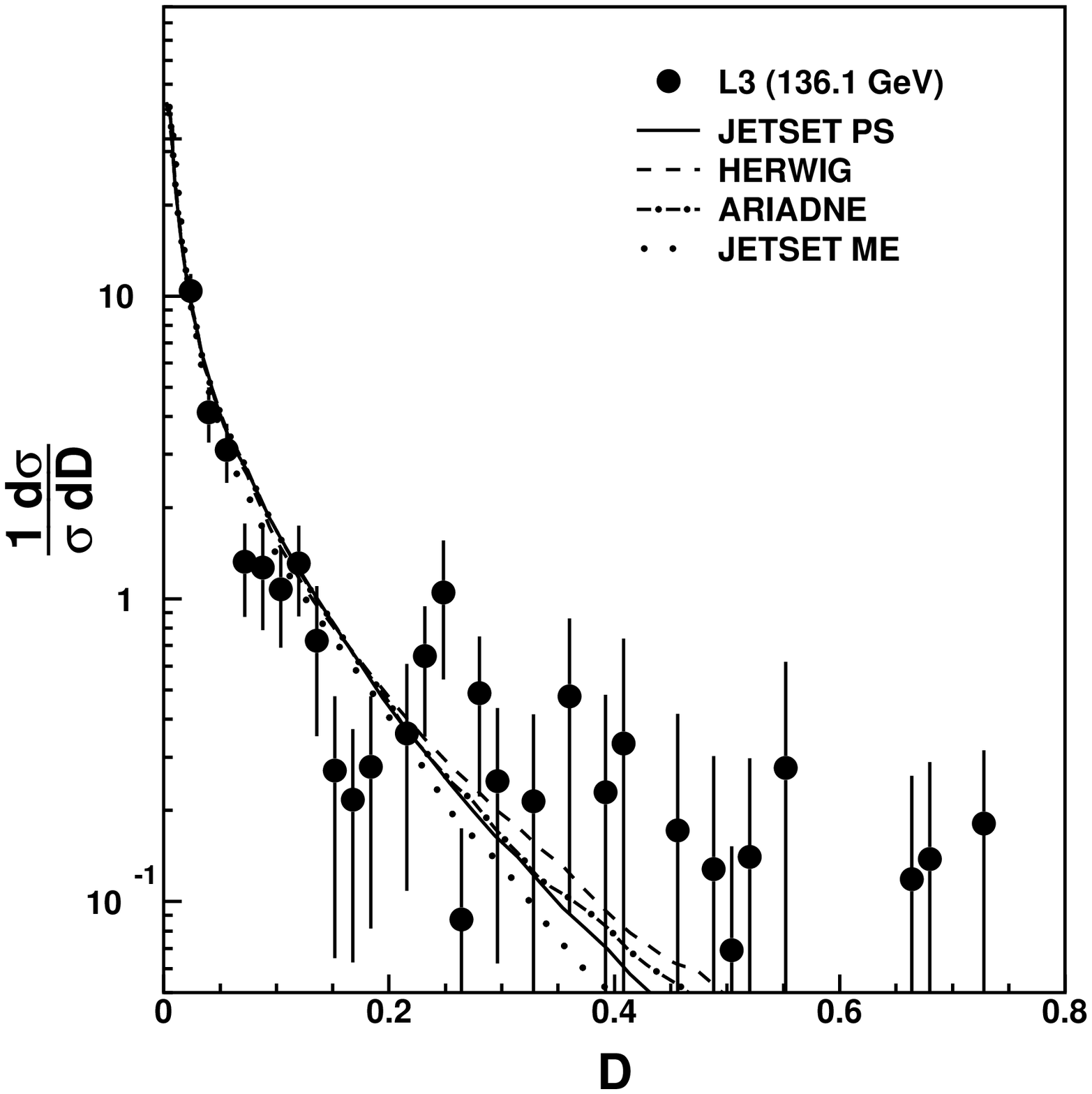}
  \includegraphics*[width=.5\figwidth,bbllx=5,bblly=30,bburx=525,bbury=560]{\mydirfig 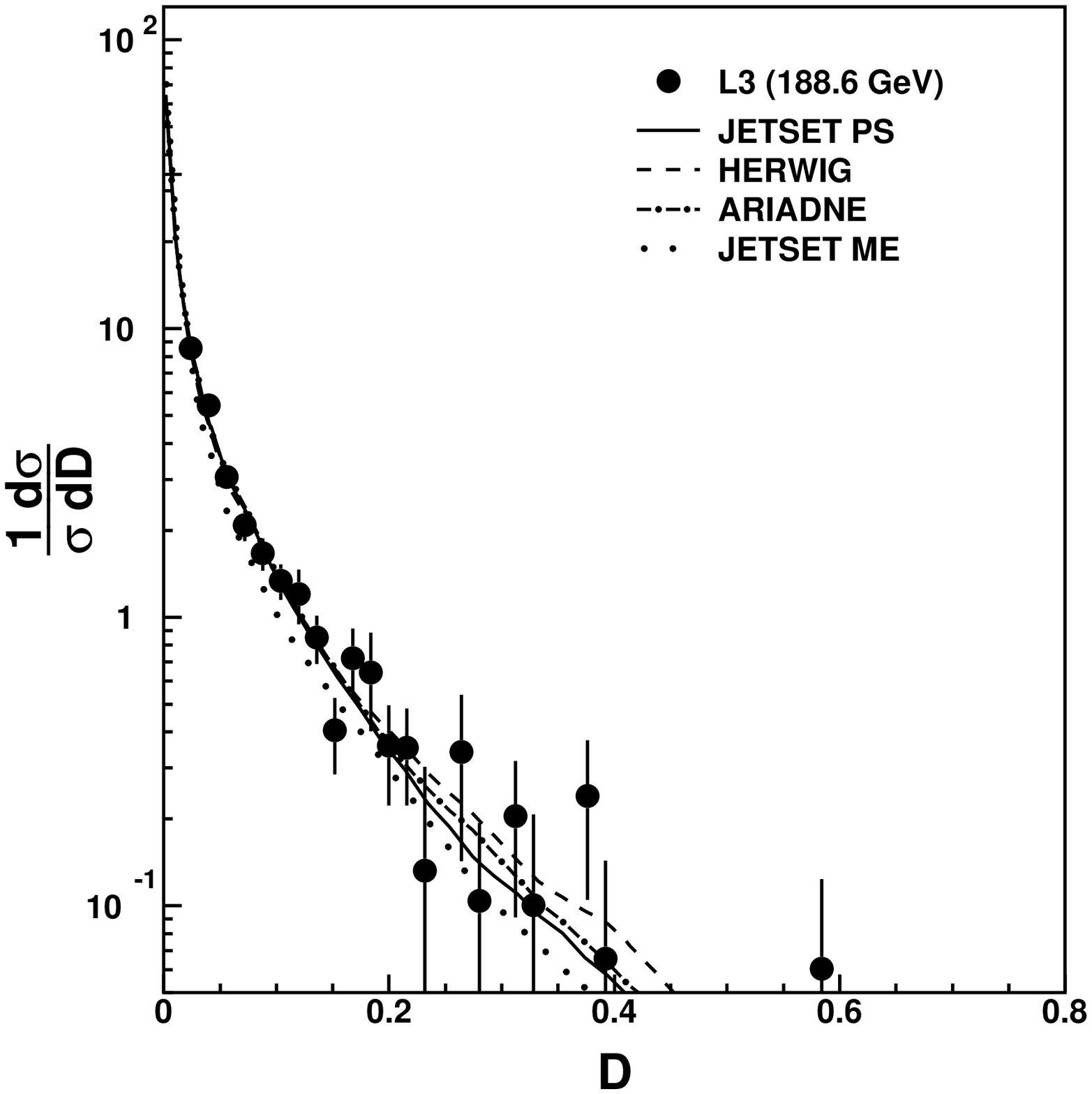}
  \includegraphics*[width=.5\figwidth,bbllx=5,bblly=30,bburx=525,bbury=560]{\mydirfig 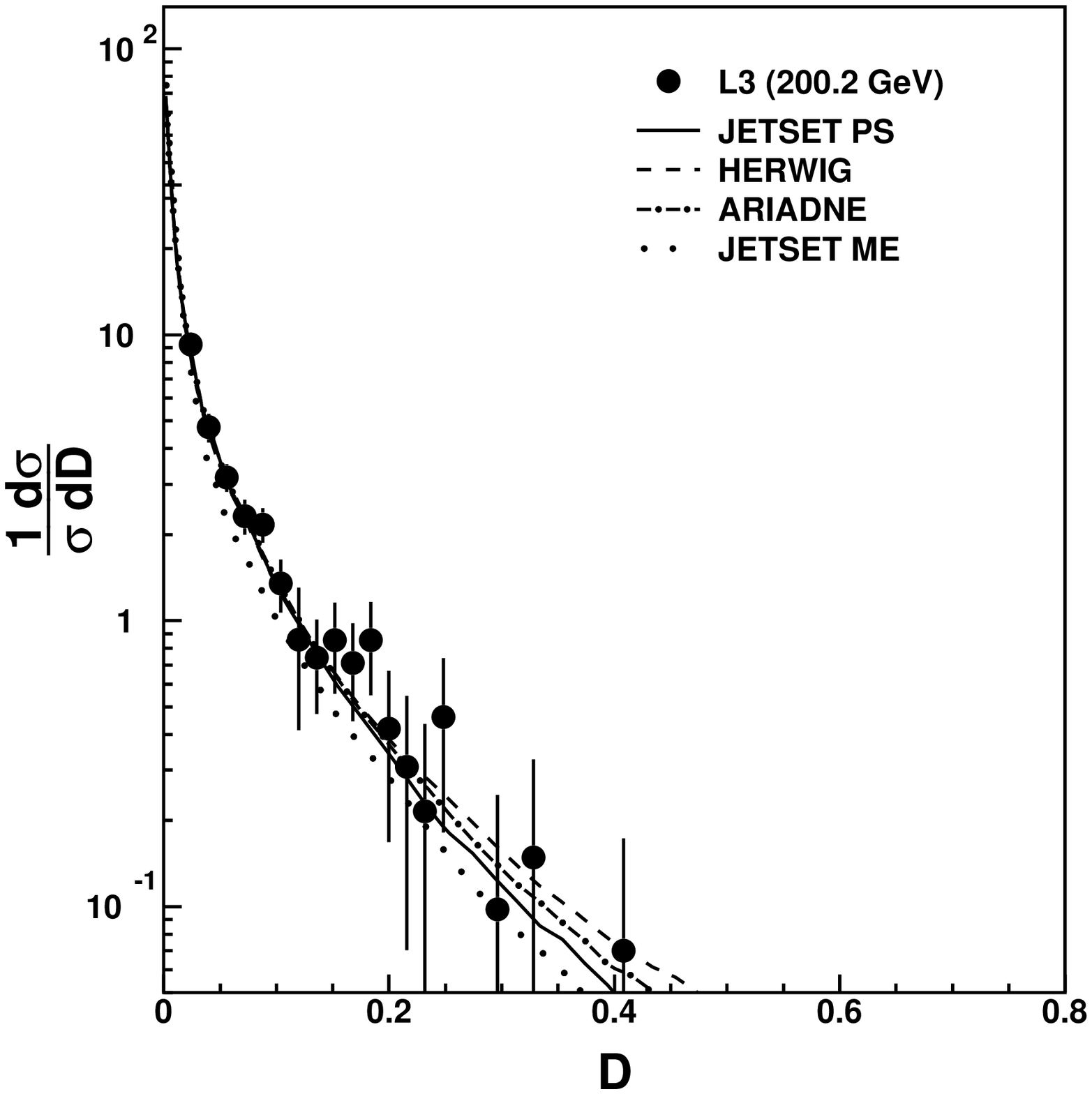}
  \includegraphics*[width=.5\figwidth,bbllx=5,bblly=30,bburx=525,bbury=560]{\mydirfig 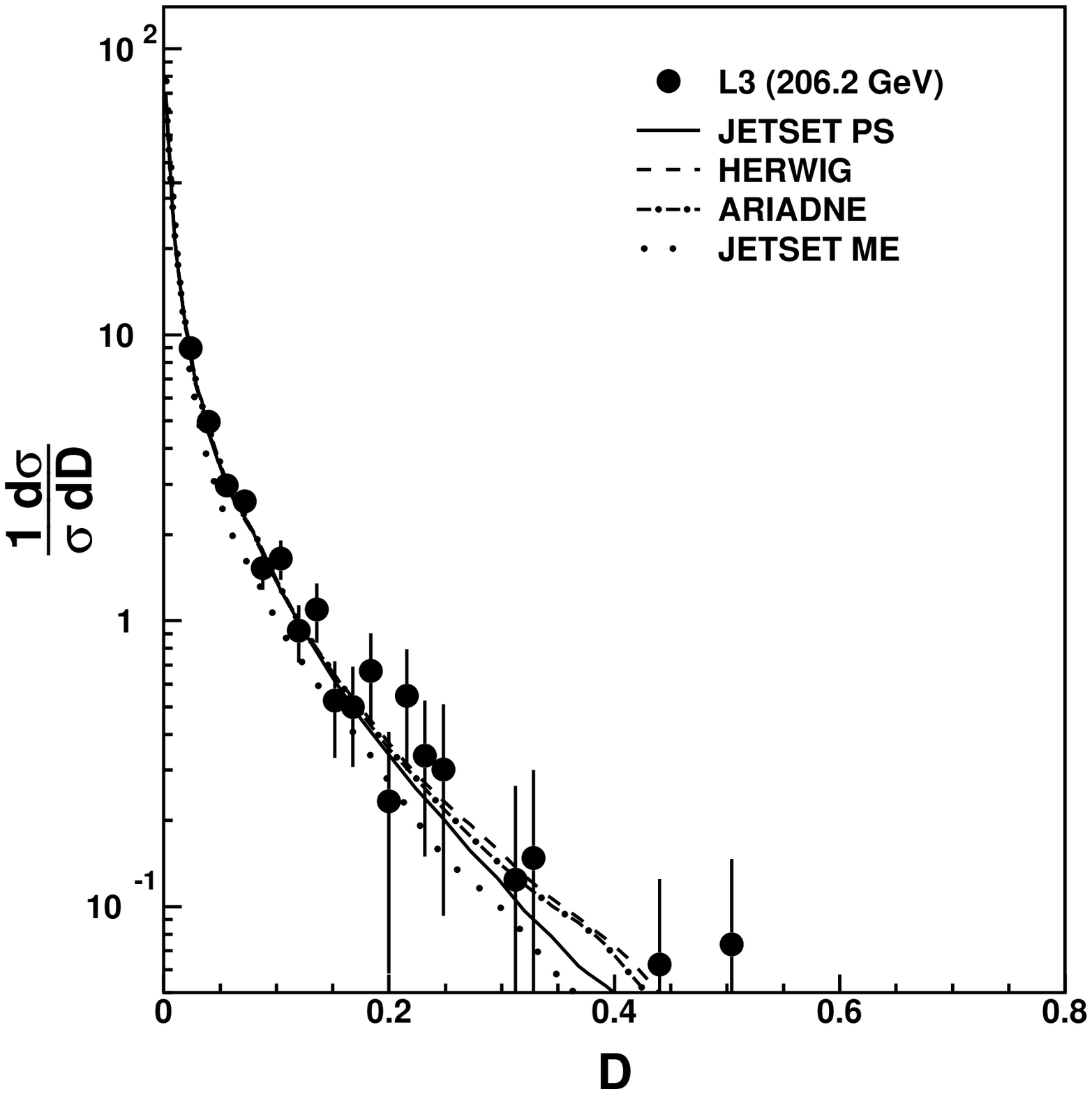}
\end{center}
\caption{$D$-parameter distributions at
         $\langle\rs\,\rangle=
         136.1, 188.6, 200.2 \text{ and } 206.2\,\GeV$
          compared to several \QCD\ models.}
\label{fig:dpar}
\end{figure}

The distributions for high energy are shown in
Figures \ref{fig:thr}--\ref{fig:dpar}.
The agreement is satisfactory, with the exception of the \textsc{Jetset} ME
comparisons for the jet broadenings and the $C$- and $D$-parameters at high energy.
 
\begin{figure}[htbp]
\begin{center}
    \includegraphics*[width=\figwidth]{\mydirfig 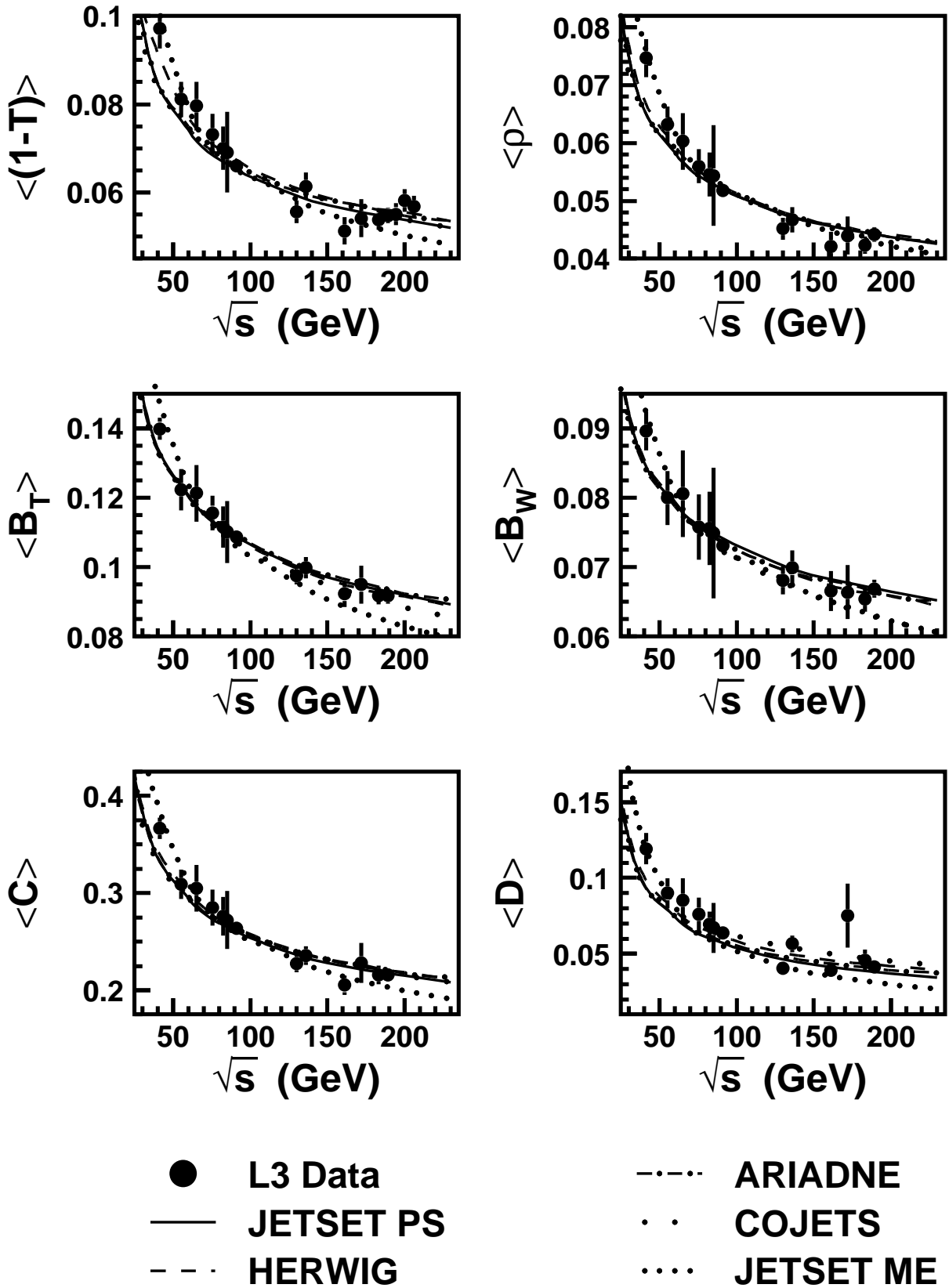}
\end{center}
\caption{The first moments of the six event-shape variables, \onet,
           \rhoh, \bt, \bw, $C$ and $D$, as a
           function of the centre-of-mass energy, compared with several \QCD\ models.}
\label{fig:evol}
\end{figure}

An important test of \QCD\ models is a comparison of the energy
evolution of the event-shape variables.
The energy dependence  of the mean event-shape variables arises mainly
from two sources: the logarithmic energy scale dependence of $\as$ and the
power law behaviour of non-perturbative effects.
The first moments of the six event-shape variables are shown in
Figure~\ref{fig:evol} 
and are also given in Tables \ref{tab:thrustqqg1}--\ref{tab:dpar91}
along with the differential distributions.
Also shown are the energy dependences of these quantities as predicted by
\textsc{Jetset} PS, \textsc{Herwig}, \textsc{Ariadne},
\textsc{Cojets} and \textsc{Jetset} ME.
All models give a good description of the data
with the exception of \textsc{Jetset} ME,
which decreases too rapidly with \rs\ for the jet broadenings and the $C$- and $D$-parameters.
 

\subsection{Power Law Correction Analysis}    \label{sec:power}
 
Rather than the phenomenological fragmentation models of the Monte Carlo programs,
the non-perturbative contribution to event-shape distributions can be described
using a so-called power correction ansatz.  In this approach,
the energy dependence of moments of the event-shape variables are
described \cite{qcdpower,qcdpower1,qcdpower2,qcdpower3,milan} as a sum of the perturbative contribution and
a
power law dependence due to non-perturbative contributions.
The first moment of an event-shape variable, $y$, is written as
\begin{equation} \label{eq:f}
 \langle y\rangle  =  \langle y_{\mathrm{pert}} \rangle \; +\;
                        \langle y_{\mathrm{pow}}\rangle \; ,
\end{equation}
where the perturbative contribution $\langle y_{\mathrm{pert}}\rangle$ has
been calculated \cite{kunznas} to ${\cal{O}}(\as^{2})$:  
\begin{equation} \label{eq:fpert}
   \langle y_{\mathrm{pert}} \rangle = A_y \frac{\as(\mu)}{2\pi}
      + \left(A_y\frac{\beta_0}{2}\log\frac{\mu^2}{s} + B_y\right) \left(\frac{\as(\mu)}{2\pi}\right)^2 \; ,
\end{equation}
where $A_y$ and $B_y$ are coefficients
depending on the event-shape variable, $y$,
which are obtained by integrating \cite{qcd-event2}
the  ${\cal{O}}(\as^{2})$ matrix elements \cite{ert-me},
$\mu$ is the renormalisation scale (taken equal to \rs), and
$\beta_{0} = (11\Nc - 2\Nf)/3$, with $\Nc=3$ the number of
colours and $\Nf$ the number of active flavours.
The power correction term, for $1-T$, $\rhoh$, and $C$,
is given by
\begin{equation} \label{eq:fpow}
  \langle y_{\mathrm{pow}}\rangle  =  c_y F_y {\cal{P}} \; ,
\end{equation}
where the factors $c_y$ and $F_y$ depend on the shape variable $y$,
and ${\cal P}$ is supposed to have the universal form\cite{qcdpower,qcdpower1,qcdpower2,qcdpower3,milan}:
\begin{equation} \label{eq:P}
  {\cal{P}}  =  \frac{4C_{\text{F}}}{\pi^{2}} {\cal{M}} \frac{\mu_{\text{I}}}{\rs}
    \left[ \alpha_0(\mu_\mathrm{I}) - \as(\mu) - \beta_0\frac{\as^{2}(\mu)}{2\pi}
    \left( \ln\frac{\mu}{\mu_\mathrm{I}} + \frac{K}{\beta_0} + 1\right)\right]
\end{equation}
The parameter $\alpha_{0}$ is the average value of $\as$ in the
non-perturbative region below an infrared matching scale $\mu_\mathrm{I}$
($= 2\,\GeV$);
$K = (67/18 - \pi^{2}/6)C_\mathrm{A} - 5\Nf/9$;
and $C_{\text{F}} $, $C_{\text{A}} $ are the SU(3) colour factors.
The so-called Milan factor, ${\cal{M}}$, is 1.49 for $\Nf=3$ \cite{milan}.
The shape-variable dependent coefficients, $A_y$, $B_y$ and $c_y$ are given in Table~\ref{tab:coef}.
For \onet, $\rhoh$, $C$ and $D$, $F_y=1$, while for
the jet broadening variables it is \cite{qcdpower,qcdpower1,qcdpower2,qcdpower3,milan}
\begin{equation}  \label{eq:FB}
   F_y  =  \frac{\pi\sqrt{c_y}}{2\sqrt{C_{\text{F}}\as\left(1+K\frac{\as}{2\pi}\right)}}
           + \frac{3}{4}
           - \frac{\beta_0 c_y}{6 C_{\text{F}}} - 0.61371
\end{equation}
 
Recently, the power law correction term has been calculated for the $D$-parameter \cite{dpar}.
Since $A_D=0$, the leading-order term is second order.
To obtain \NLO\ accuracy, $\langle y_{\mathrm{pert}} \rangle$ must be computed to third order.
This results in an additional term in Equation~(\ref{eq:fpert}): $+2450\left(\as/(2\pi)\right)^3$.
 
\begin{figure}[htbp]
\begin{center}
 \includegraphics*[width=\figwidth]{\mydirfig 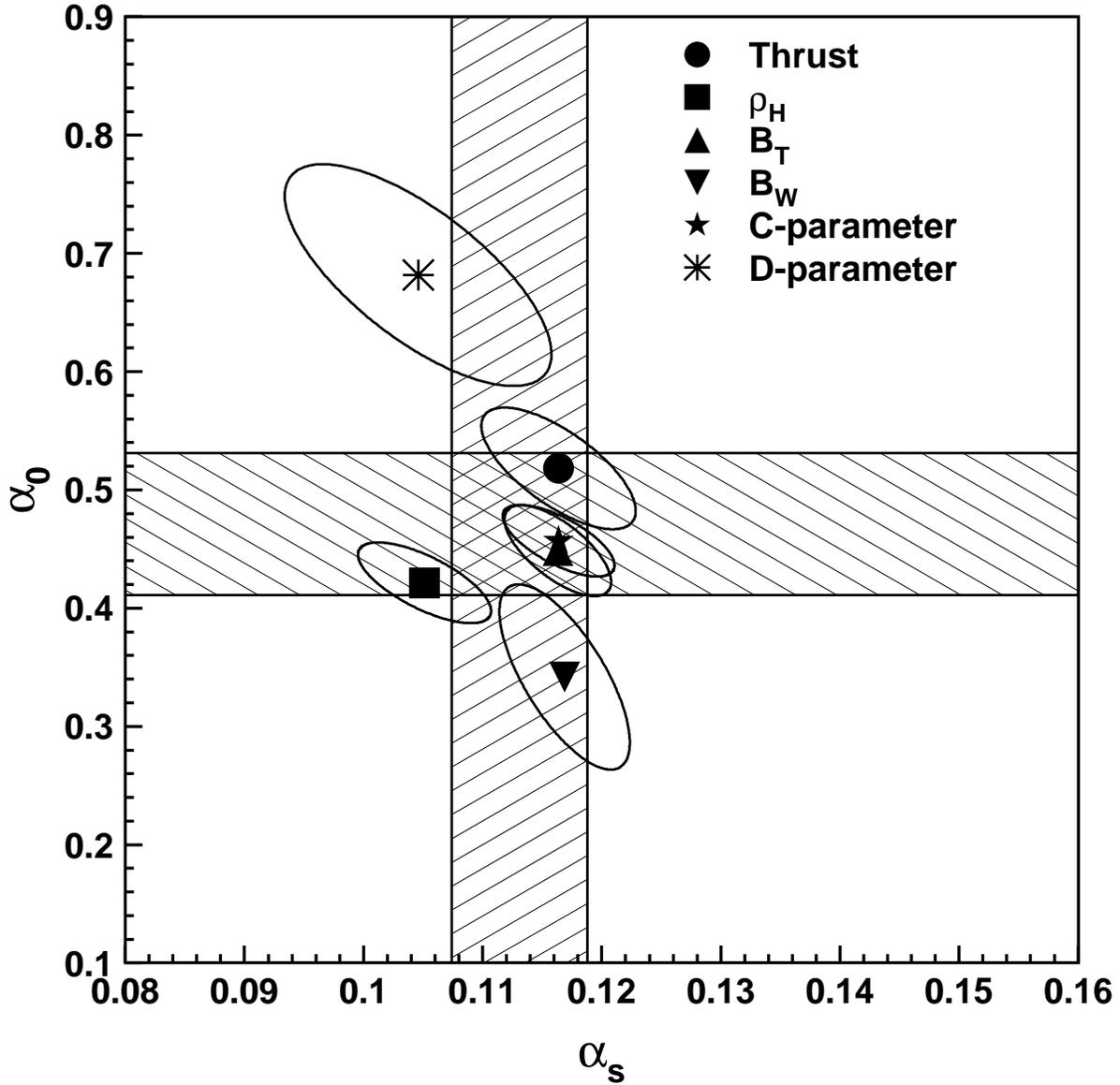}
\end{center}
\caption[The values of \as\ and $\alpha_0$ from fits of the power correction ansatz
          to the first moments of the six event-shape variables,
         \onet, \rhoh, \bt, \bw, $C$ and $D$.]
        {The values of \as\ and $\alpha_0$ from fits of the power correction ansatz
          to the first moments of the six event-shape variables,
         \onet, \rhoh, \bt, \bw, $C$ and $D$.
         The ellipses represent 39\%  two-dimensional confidence intervals including both
         statistical and systematic uncertainties.
         The bands represent unweighted averages of the \as\ and $\alpha_0$ including both
         statistical and systematic uncertainties.
         }
\label{fig:poweralpha}
\end{figure}

\begin{figure}[htbp]
\begin{center}
 \includegraphics*[width=\figwidth]{\mydirfig 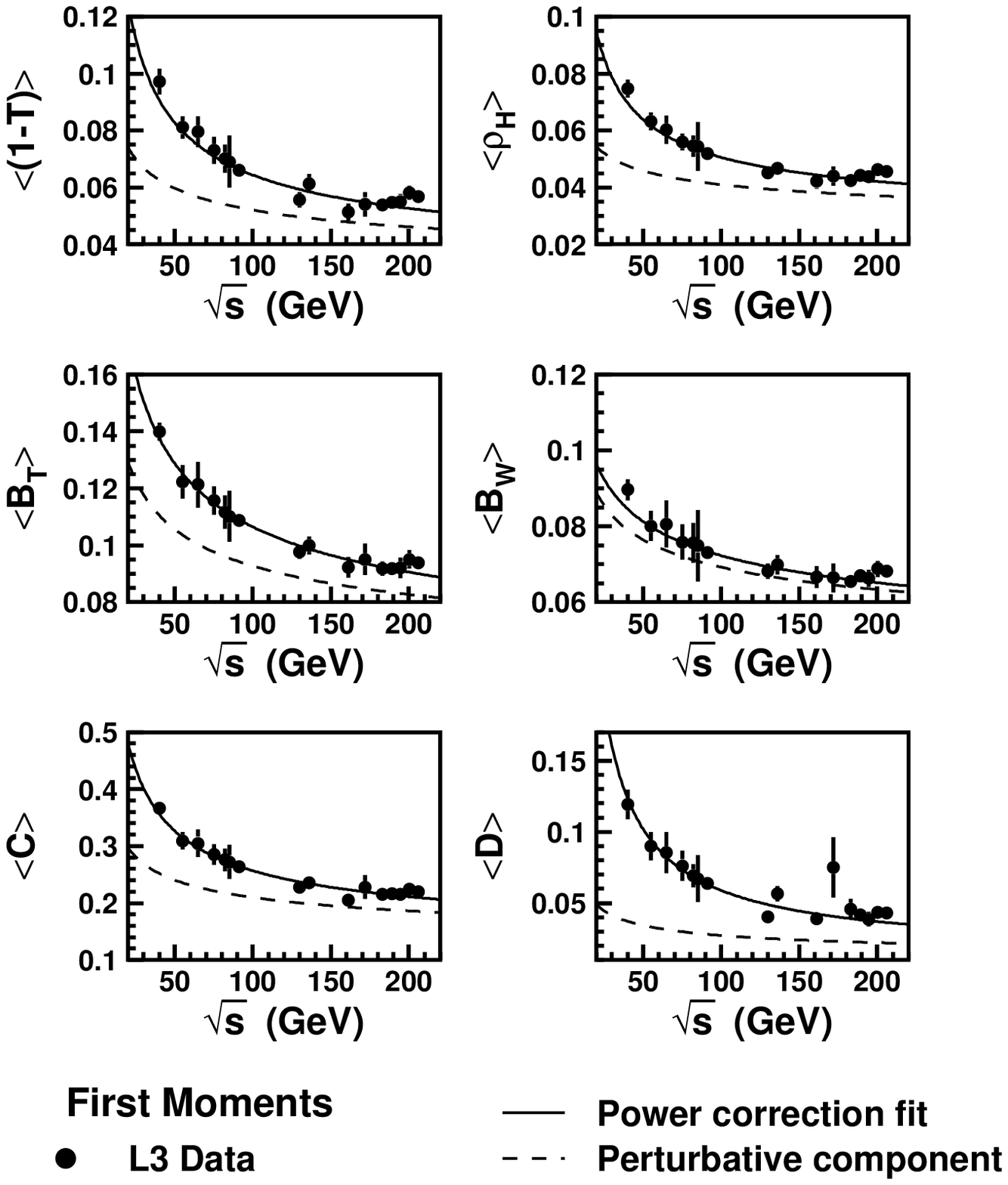}
\end{center}
\caption[The first moments of the six event-shape variables, \onet,
           \rhoh, \bt, \bw, $C$ and $D$ compared to the results
           of a fit including perturbative and power law contributions.]
        {The first moments of the six event-shape variables, \onet,
           \rhoh, \bt, \bw, $C$ and $D$ compared to the results
           of a fit including perturbative and power law contributions,
           Equation~(\ref{eq:f}).
           }
\label{fig:fmom}
\end{figure}

We have carried out fits to the first moments of the six event-shape variables
separately
with $\as(\MZ)$ and $\alpha_{0}$ as free parameters.
The diagonal terms of the covariance matrix between the different energy
points are constructed by summing in quadrature the systematic
and statistical uncertainties.
The off-diagonal terms are obtained from the common systematic uncertainties.
The results of the fits are summarised in Table \ref{tab:moment1} and
shown in Figures \ref{fig:poweralpha} and~\ref{fig:fmom}.
 
The six values of $\alpha_{0}$ obtained from the event-shape variables do not agree well.
The confidence level for the hypothesis that $\alpha_{0}$ is the same for all quantities is only about 3\%
when the systematic uncertainties are treated as uncorrelated, 1\%
if only statistical uncertainties are used.
In particular, the values of $\alpha_{0}$ for $D$ and \bw\ differ by
about 40\% and 30\% in opposite directions from the unweighted average of the six estimates of $\alpha_{0}$:
\begin{equation*}
  \alpha_{0}  =  0.478\  \pm\ 0.054\  \pm\ 0.024 \; .
\end{equation*}
On the other hand, the six estimates of \as\ are consistent with each other,
yielding an unweighted average:
\begin{equation*}
  \as (\MZ )  =  0.1126\ \pm\ 0.0045\ \pm\ 0.0039 \; . 
\end{equation*}
 
The first uncertainty is the average of the statistical uncertainties of the
measurements.
To estimate theoretical uncertainties the renormalisation scale $\mu$ is varied
between $0.5\rs$ and $2.0\rs$ resulting in average variations of
$\pm 0.024$  and $\pm 0.0039$ for
$\alpha_{0}$ and $\as(\MZ )$, respectively.
A variation of $\mu_{I}$ in the range 1--3\,\GeV\ gives an additional
uncertainty on both $\alpha_{0}$ and $\as(\MZ)$ of $\pm 0.0010$.
These two estimates of theoretical uncertainty are combined in
quadrature and quoted as the second uncertainty.
 
\begin{figure}[htbp]
\begin{center}
 \includegraphics*[width=\figwidth]{\mydirfig 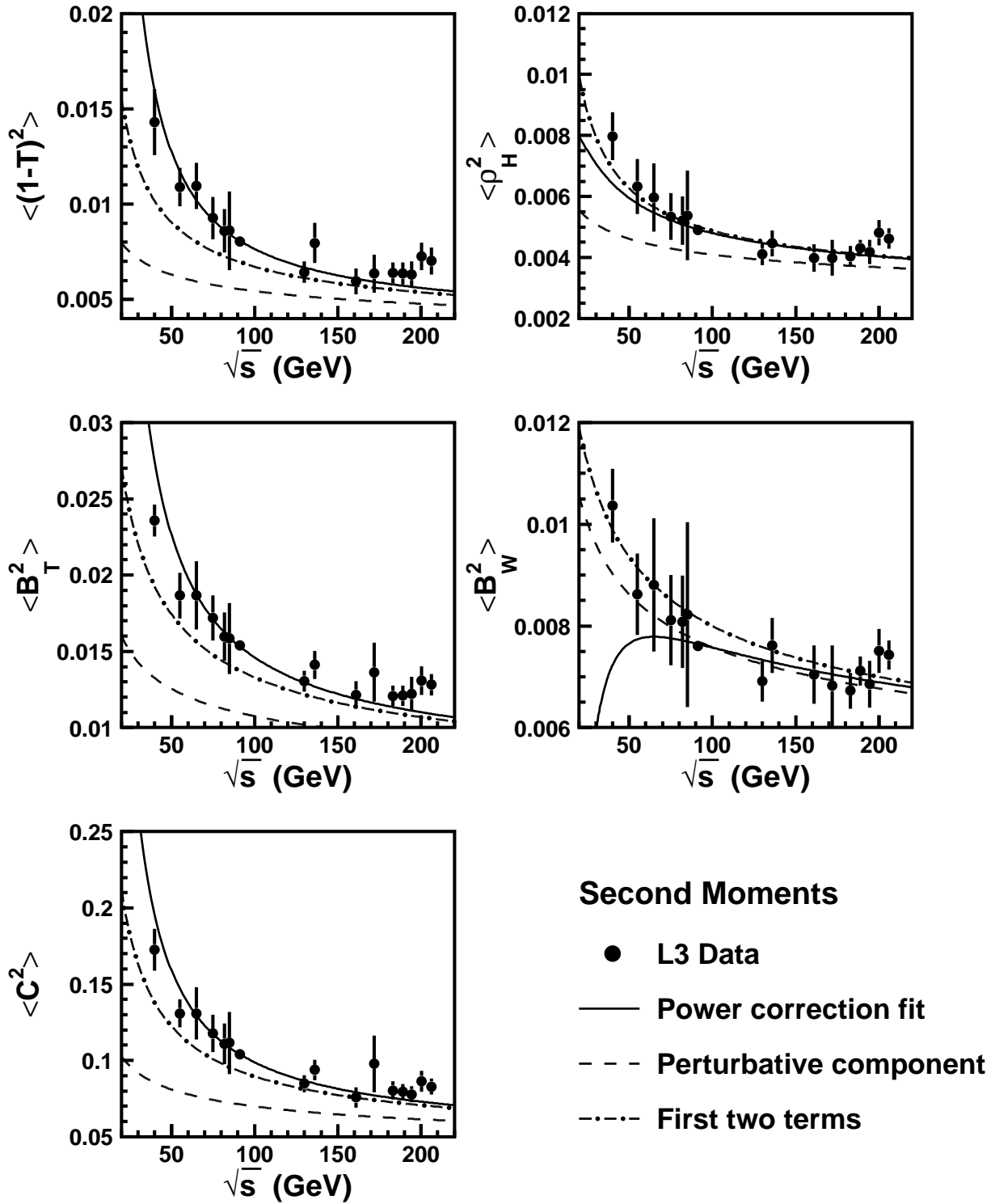}
\end{center}
\caption[The second moments of the five event-shape variables, \onet,
           \rhoh, \bt, \bw\ and $C$ compared to the results
           of a fit including perturbative and power law contributions.]
        {The second moments of the five event-shape variables, \onet,
           \rhoh, \bt, \bw\ and $C$ compared to the results
           of a fit including perturbative and power law contributions,
           Equation (\ref{eq:f2}).
           The parameters  $\alpha_{0}$ and $\as$ are fixed to the values obtained
           by the corresponding fit to the first moment.
           }
\label{fig:smom}
\end{figure}

We have also measured the second moments of these shape variables which
are also given in Tables \ref{tab:thrustqqg1}--\ref{tab:dpar91}.
The energy dependence of these moments has been analysed in terms of power law corrections.
It is expected \cite{salamDW,salamW,salam} that
\begin{equation}  \label{eq:f2}
 \langle y^{2}\rangle  =  \langle y^{2}_{\mathrm{pert}} \rangle
             +  2 \langle y_{\mathrm{pert}}\rangle c_y F_y  {\cal{P}}
             +  {\cal{O}}\left(\frac{1}{s} \right)  \; .
\end{equation}
The ${\cal O}(\frac{1}{s})$ term is expected to be small for \onet, $\rhoh$, $C$ and $D$.
This assumes that the non-perturbative correction to the distributions causes only a shift
in the distributions.
Fits are performed to the second moments.
In the fits, the ${\cal O}(\frac{1}{s})$ term is parametrised as $A_{2}/s$, and both
$\alpha_{0}$ and \as\ are fixed to the
values obtained from the corresponding fits to the first moments.
Figure \ref{fig:smom} shows the second moments compared to these fits.
The contributions of the power term and the ${\cal O}(\frac{1}{s})$ term are shown separately.
The results of the fits are also given in Table~\ref{tab:moment2}.
The contribution of the ${\cal O}(\frac{1}{s})$ term is not negligible
for \onet\ and $C$, contrary to the expectation.
           It                 is negative for $\rhoh$ and $\bw$.
Further, the shape of the fitted curve is unphysical for $\bw$,
and the $\chisq$ of the $\bt$ fit  is unacceptably low.
 
 
Given the mildly discrepant values of $\alpha_0$ and
these problems with the fits to the second moments,
one can conclude that the power correction ansatz gives a good qualitative description, but
that additional terms will be needed to achieve a good quantitative description.

\subsection[Determination of \as\ from event-shape variables]
           {Determination of {\boldmath $\as$} from event-shape variables}\label{sec:alphas}
 
The presently available \QCD\ predictions in fixed-order perturbation theory do not take into account
the effect of emission of more than two gluons.
For variables like \onet, \bt, \bw, \rhoh\ and $C$ this leads to a
poor description of the distributions in kinematic regions where
multi-gluon emission becomes dominant.
It is possible to isolate the leading terms in every order of perturbation
theory and to sum them up in the form of an exponential series.
These calculations have been carried out for the above variables
\cite{qcd-t,qcd-hjm,qcd-jbm,qcd-jbm2,qcd-oth,qcd-c} to next-to-leading log order.
 
For all these variables, denoted by $y$, the cumulative
cross section can be written in the form
\begin{eqnarray}
   R(y,\as) \equiv \int_{0}^{y}\frac{1}{\sigma} \frac{\dd{\sigma}}{\dd{y}}
   & = & C(\as )\Sigma(y,\as) + D(\as,y) \\
   {\mathrm{with}}\;\; C(\as)   & = & 1 + \sum_{n=1}^{\infty} C_{n} \asp^{n} \\
                       D(\as,y) & = & \sum_{n=1}^{\infty} D_{n}(y) \asp^{n} \\
         \Sigma(y,\as) & = & \exp\left[ \sum_{n=1}^{\infty}
                          \sum_{m=1}^{n+1} G_{nm} \asp^{n} L^{m} \right]  \\
                 & \equiv & \exp \left[ L g_{1}(\asp L) + g_{2}(\asp L)
                           + \as\ g_{3}(\asp L) + \cdots \right] \\
            \asp & \equiv & \frac{\as}{2\pi} \\
               L & \equiv & \ln \left( \frac{1}{y} \right) \; .
\end{eqnarray}
In the 2-jet region, $y$ is small.
Therefore, $L$ and the corrections due to large powers of $L$ are large.
 
In the fixed-order calculations \cite{kunznas,ert-me}, one can write
\begin{equation}
   R(y,\as)  =  \as A(y) + \as^{2} B(y) + {\cal{O}}\left(\as^{3}\right) \; .
\end{equation}
Note that  $A_y$ and $B_y$ of Equation (\ref{eq:fpert}) are related to $A(y)$ and $B(y)$.
 
The two approaches are summarised in Table \ref{tab:match}. The first two rows
have been completely computed in the fixed-order calculations and the first
two columns are known to all orders in the recent resummed calculations. In
order to describe the data over a wide kinematic region, it is desirable to
combine the two sets of calculations, avoiding double-counting  of the common
parts. This leads to a number of matching schemes \cite{qcd-oth}.
The simplest one matches the two calculations at a given value
of $y$ and uses a suitable damping function so that the resummed calculations
contribute to the 2-jet region and the fixed-order calculations dominate
in the multi-jet region. A preferable approach would be to combine the
two calculations and subtract the common terms of the two calculations.
This is done by taking the logarithm of the fixed-order calculations and
expanding it as a power series. Then the matching can be done in $\ln R(y)$
(called the `$\ln R$ matching' scheme). Alternatively, a similar
procedure can be performed in the function $R(y)$ rather than in $\ln R(y)$. This
procedure is called the `$R$-matching' scheme. In a variation of this 
scheme, the term $G_{21}\asp^{2}L$ is included in the term of
the exponential and subtracted after exponentiation. This method is called
the `modified $R$-matching' scheme.
 
One has to take care of the additional constraint coming from kinematics,
namely that the cross sections vanish beyond the kinematic limit
\begin{eqnarray}
                        R(y=y_\mathrm{max}) & = & 1 \\
   \frac{\dd{R}}{\dd{y}} (y=y_\mathrm{max}) & = & 0\; .
\end{eqnarray}
These constraints are strictly obeyed in the fixed-order calculations but
they are not valid for the resummed expansion. The first constraint can be
imposed by replacing $R(y)$ by $R(y)-R(y_\mathrm{max})$ for the resummed
calculations. Alternatively, $L$ can be replaced in the resummed term by
$L^{\prime}=\ln{(y^{-1}-y_\mathrm{max}^{-1}+1)}$ in the $\ln R$ matching
scheme to fulfil both conditions.
This possibility is referred to as the `modified $\ln R$' matching scheme.
 
An important improvement of the new \QCD\ calculations with respect
to the second-order formulae is their ability to describe
also the low-$y$ region. One should note that the sub-leading terms
not included beyond next-to-leading logarithmic order
are expected to be relatively small at low $y$.
 
The calculations for the distributions of the five variables are given in the
form of analytical functions
\begin{equation}
    f^\mathrm{pert} (y; s,\as(\mu),\mu)    \;.
\end{equation}
The fixed-order calculations include quark masses, while the resummed calculations
assume massless partons.
To compare the analytical calculations with the experimental distributions,
the effect of hadronisation and decays must be taken into account using Monte Carlo programs.
We use the
parton shower programs \textsc{Jetset}, \textsc{Ariadne} and \textsc{Herwig}
with the tuned parameter values of Section~\ref{sec:tuning}.
The perturbative calculations for a variable $y$ are convoluted
with the probability ${p^{ \text{non-pert}}} (y;y^\prime)$ to obtain the value
$y$, after fragmentation and decays, for a  parton level value $y^\prime$:
\begin{equation}
   f(y) =  \int f^{\text{pert}} (y') \cdot p^{\text{non-pert}} (y;y^\prime) \, \dd{y^\prime} \;.
\end{equation}
The resulting differential cross section $f(y)$ is compared to the
measurements. The correction for hadronisation and decays changes the
perturbative prediction by less than 5\% for the event-shape variables
over a large kinematic range. However, the corrections increase to as much as 20\%
in the extreme 2-jet region.
 
\begin{figure}[htbp]
\begin{center}
   \includegraphics*[width=.95\figwidth]{\mydirfig 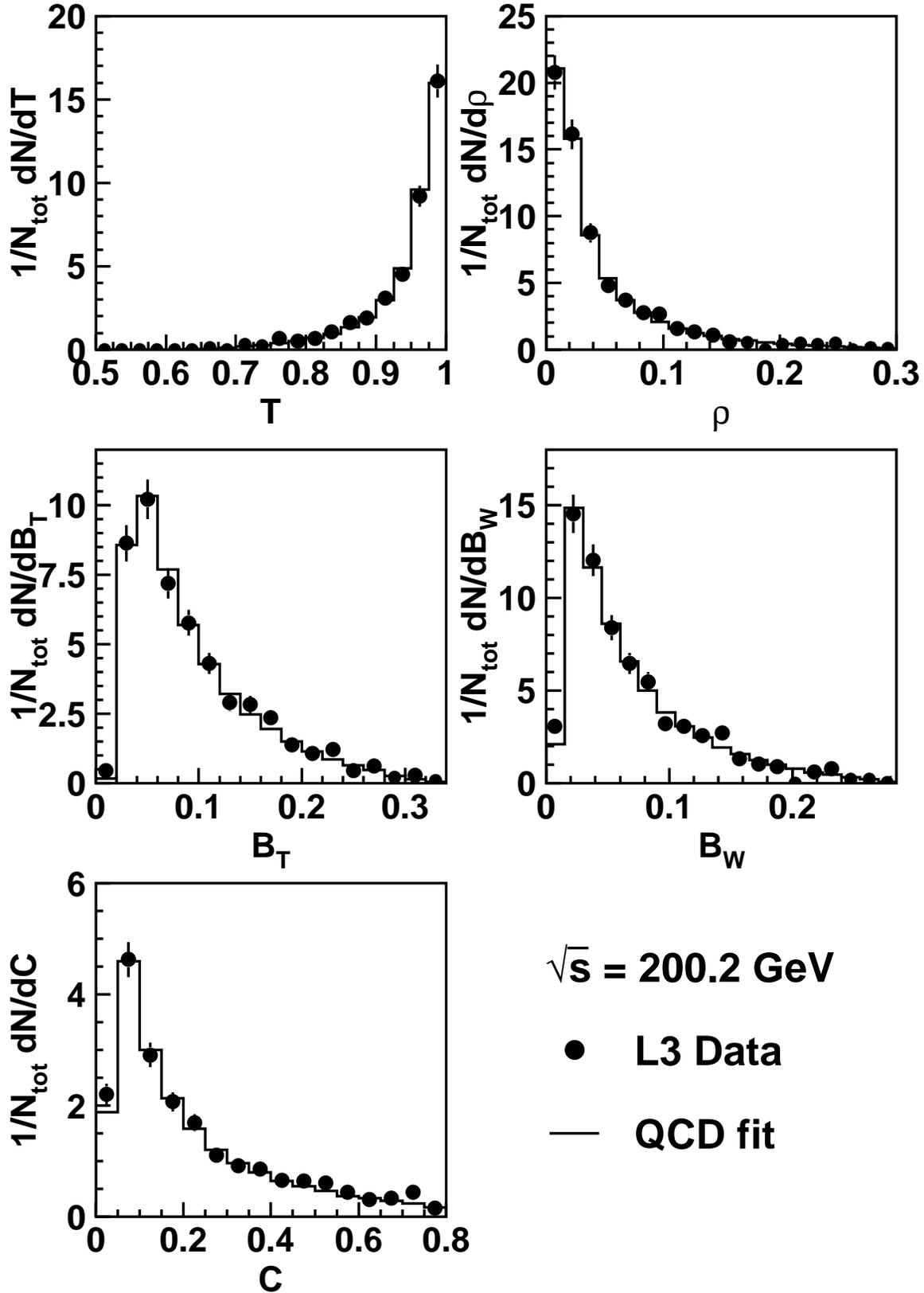}
\end{center}
\caption[Measured distributions,
           at $\langle\rs\,\rangle=200.2\,\GeV$,
           of thrust, $T$, scaled heavy jet mass,
           $\rhoh$, total, $\bt$, and wide, $\bw$, jet broadenings, and $C$-parameter
           compared to fitted \QCD\ predictions.]
        {Measured distributions,
           at $\langle\rs\,\rangle=200.2\,\GeV$,
           of thrust, $T$, scaled heavy jet mass,
           $\rhoh$, total, $\bt$, and wide, $\bw$, jet broadenings, and $C$-parameter
           compared to fitted \QCD\ predictions.
           The error bars  include  systematic as well as statistical uncertainties.}
\label{fig:alsfit}
\end{figure}

To determine $\as$ at each energy point, the measured distributions
are fitted in the ranges  given in Table~\ref{tab:fitrange} to the analytical
predictions, using the modified $\ln R$ matching scheme after corrections for
hadronisation effects. Figure~\ref{fig:alsfit} shows the experimental data
together with the result of the \QCD\ fits for the five variables
at $\langle\rs\,\rangle = 200.2\,\GeV$.
Reasonable fits are obtained at all these energy
points; the \chisq\ per degree of freedom are given in Table~\ref{tab:fitrange}.
 
The $\as$ measurements for the 16 energy points are summarised in
Table~\ref{tab:als} together with their experimental and theoretical
uncertainties. The former includes the statistical and
the experimental systematic uncertainties discussed above.
The latter is obtained from estimates  
of the hadronisation uncertainty and of  the uncalculated
higher orders in the \QCD\ predictions.
 
The hadronisation uncertainty is obtained from the variation in the fitted
value of $\as$ due to hadronisation corrections determined
by comparing  \textsc{Jetset} with \textsc{Herwig} and \textsc{Ariadne}
and by changing the \textsc{Jetset} fragmentation parameters, $b$, $\sigma_{q}$
and \LambdaLLA\ within their uncertainties, listed in Table~\ref{tab:tunmodel},
as well as by turning off Bose-Einstein correlations.
By far the largest uncertainty is that of the fragmentation model, which is
therefore taken as the estimate of the overall hadronisation uncertainty.
It is evaluated as half of the largest difference in \as\ obtained with different models.
 
The uncertainty  coming from uncalculated higher orders in the \QCD\ predictions
is estimated in two independent ways: by varying the renormalisation scale,
$\mu$, and by changing the matching scheme. The scale uncertainty is obtained
by repeating the fit for different values of the renormalisation scale
in the interval $0.5 \rs \leq \mu \leq  2 \rs$. The matching scheme
uncertainty is obtained from half of the maximum spread given by the different
matching schemes.
The largest of these uncertainties  is assigned as the theoretical
uncertainty due to uncalculated higher orders.
 
To obtain a combined value for the strong coupling constant, we take the
unweighted average of the five $\as$ values. The overall theoretical
uncertainty is obtained from the average  hadronisation uncertainty added in
quadrature to the average higher-order uncertainty. A cross-check
of this theoretical uncertainty is obtained from a comparison of $\as$
measurements from the various event-shape variables which are
expected to be differently affected by higher order corrections and
hadronisation effects. Half of the maximum spread in the five $\as$ values is
found to be consistent with the estimated theoretical uncertainty.
 
\begin{figure}[htbp]
\begin{center}
    \includegraphics[width=.8\textwidth]{\mydirfig 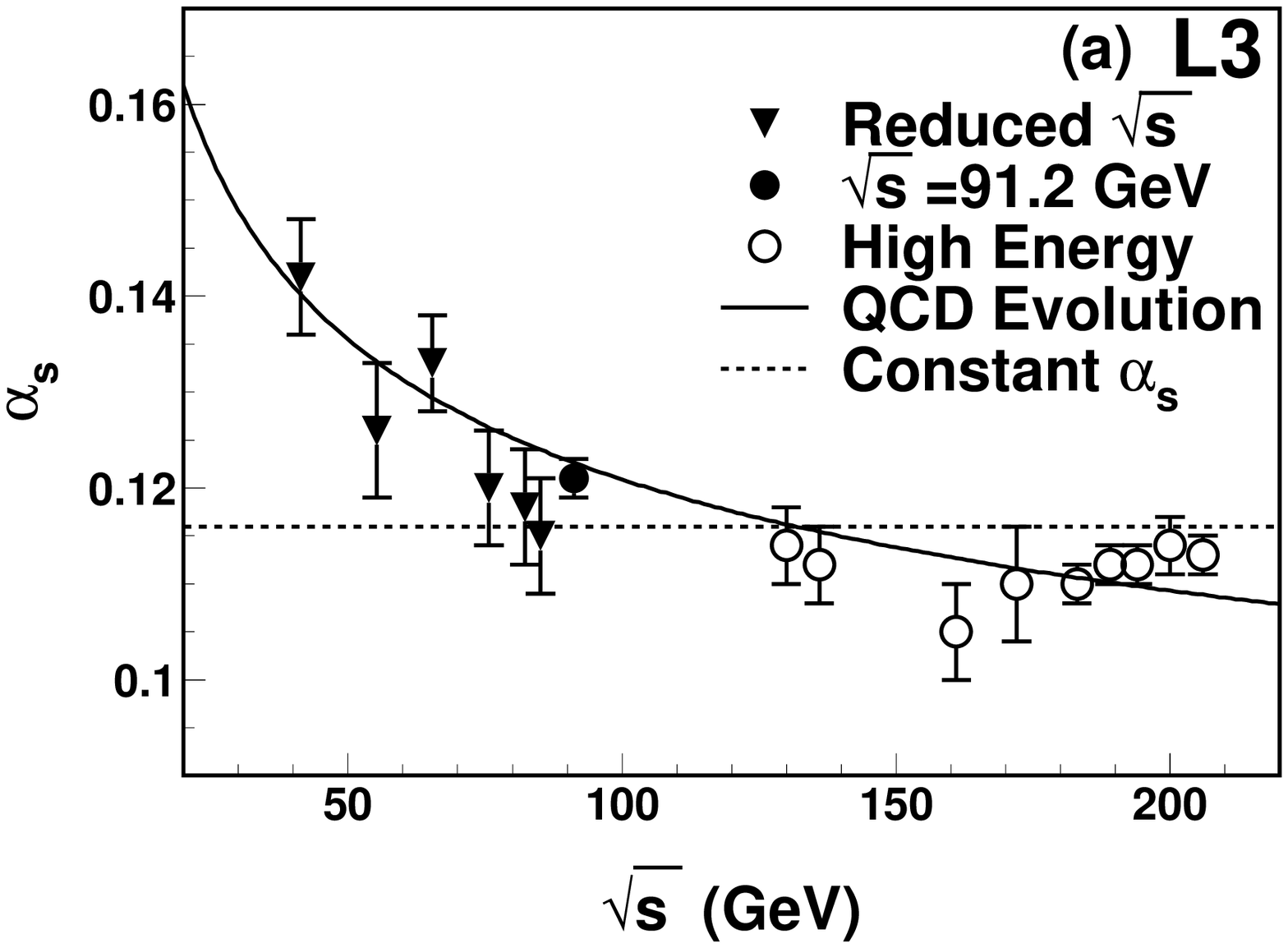}
    \includegraphics[width=.8\textwidth]{\mydirfig 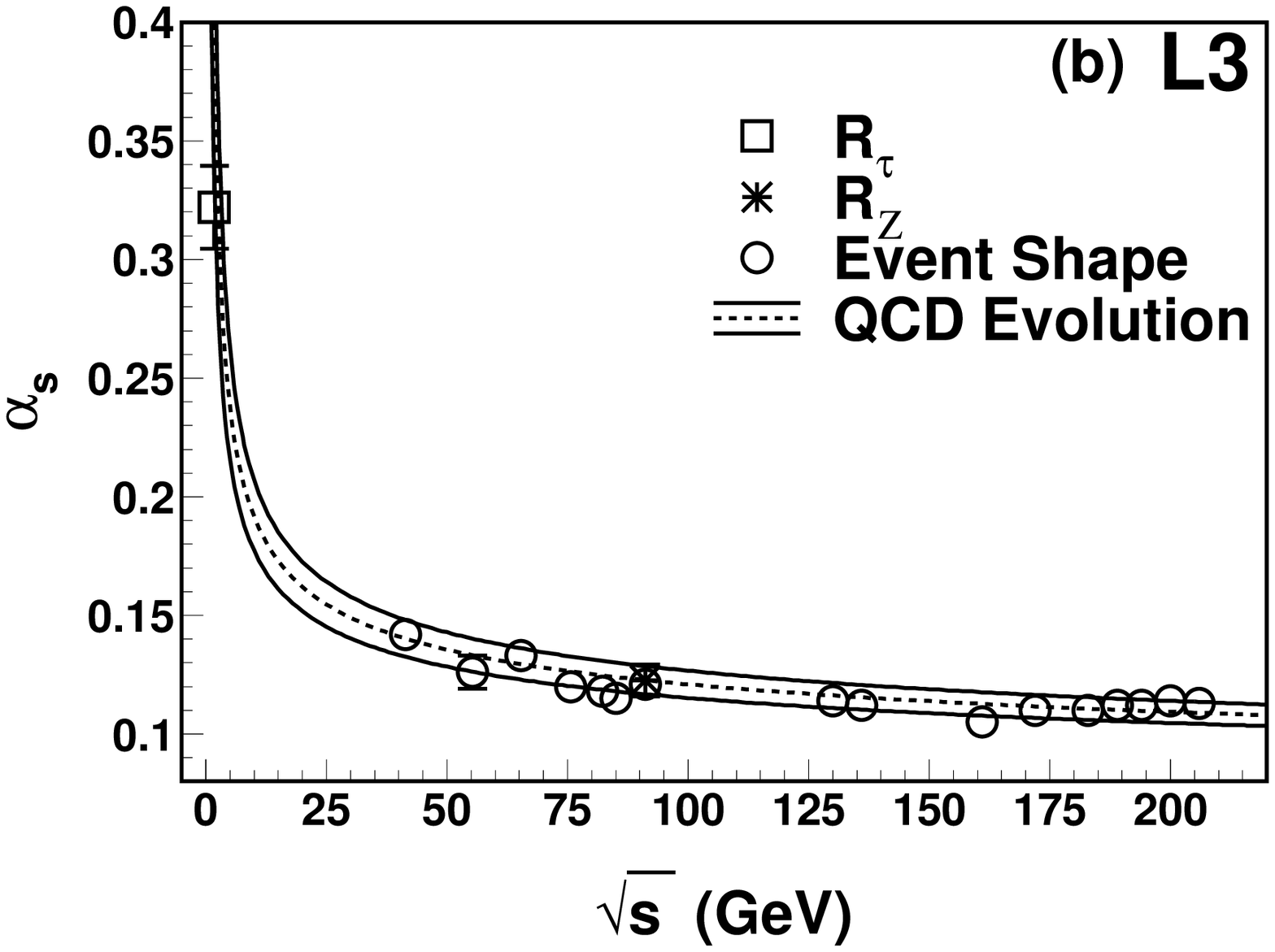}
\end{center}
\caption[Values of $\as$ determined as a function of $\rs$:
           a) from event-shape distributions with  experimental uncertainties only.
           b) from the measurement of the $\tau$ branching fractions into
           leptons.         the Z line shape,                    and event-shape distributions.
           ]
        {Values of $\as$ determined as a function of $\rs$:
           a) from event-shape distributions with  experimental uncertainties only.
           The solid and dashed lines are fits with the energy dependence
           of $\as$ as expected from \QCD\ and with constant $\as$, respectively.
           b) from the measurement of the $\tau$ branching fractions into
           leptons~\cite{l3tau}, the Z line shape~\cite{l3lineshape}, and event-shape distributions.
           The dashed line is a fit of the \QCD\ evolution
           function to the measurements made from event-shape variables.
           The width of the band corresponds to the evolved uncertainty on $\as(\MZ)$.
           }
\label{fig:alsevol}
\end{figure}

The  mean $\as$ values from  the five event-shape distributions are given
in Table~\ref{tab:alscomb2} together with the experimental and theoretical
uncertainties. Figure \ref{fig:alsevol}a compares the energy dependence of
the measured $\as$ values with the prediction from \QCD. The theoretical
uncertainties are strongly correlated between these measurements.
Hence, the energy dependence of $\as$ is investigated using only experimental
uncertainties. The experimental systematic uncertainties on $\as$ are
partially correlated. The background uncertainties are correlated between data
points in the same energy range but not between the low-energy, Z pole and
high-energy data sets. The 16 measurements in Figure~\ref{fig:alsevol}a
are shown with experimental uncertainties only, together with a fit
to the \QCD\ evolution equation \cite{pdg} with $\as(\MZ)$ as a free parameter,
assuming 5 active flavours.
The covariance matrix used in the fit is  obtained 
assuming that
       the experimental systematic uncertainties are
       uncorrelated between the three data sets and that they have a minimum overlap
       correlation between different energies within the same data set. This
       definition consists of assigning to the covariance matrix element
       the smallest of the two squared uncertainties,
       \ie, $\sigma^2_{ij}=\min(\sigma^2_{ii},\sigma^2_{jj})$

The fit, having a \chisq\ of 17.9 for 15 degrees of
freedom corresponding to a confidence level of 27\%,
yields a value of $\as$:
\begin{equation*}
    \as(\MZ)  =  0.1227~\pm~0.0012~\pm~0.0058 \; .
\end{equation*}
The first uncertainty is experimental and the second  theoretical.
The latter is obtained from the result of a fit which includes the theoretical
uncertainties and their correlations.
There are two types of theoretical uncertainties: those associated with the hadronisation corrections,
and those due to uncalculated higher order terms.
For each type, the correlations are determined assuming minimum overlap.
The hadronisation uncertainty is estimated by using different Monte Carlo programs.
Its contribution to the total theoretical uncertainty is $\pm0.0026$.
The uncertainty due to uncalculated higher order terms is estimated by varying the renormalisation scale by
a factor 2 and by using different matching schemes.
This is the largest uncertainty, $\pm0.0052$.
This value of \as\ is consistent with the values measured by other experiments at \MZ\ using event shapes
\cite{alephas,alephas1,alephas2,delphias,delphias1,delphias2,opalas,opalas1,opalas2,opalas3,sldas}.

A fit with constant \as\ gives a \chisq\  of 51.7 for 15 degrees of freedom, corresponding to a confidence
level of $6\times10^{-6}$.
These measurements support the energy evolution of the strong coupling constant predicted by \QCD.

The 
energy evolution of $\as$ depends on the number of active flavours.
A fit with \Nf, as well as \as, as free parameters yields:
\begin{eqnarray*}
  \Nf      & = & 6.9 ~\pm~ 1.3     \\     
  \as(\MZ) & = & 0.1219 \pm 0.0013 \; ,   
\end{eqnarray*}
where the uncertainty is only experimental. 
This result agrees with the expected $\Nf = 5$, and
the \as\ value 
is compatible with that from the fit with \Nf\ fixed to 5.
 
Figure~\ref{fig:alsevol}b summarises the $\as$ values determined by \Lthree\
from the measurement of the $\tau$ branching fractions into leptons
\cite{l3tau}, the Z line shape~\cite{l3lineshape} and event-shape distributions
at various energies, together with the \QCD\ prediction obtained from the fit
to the event-shape measurements only. The width of the band corresponds to the
evolved uncertainty on $\as(\MZ)$. All the measurements are consistent with
the energy evolution of the strong coupling constant predicted by \QCD.
The uncertainties on these measurements are dominated by the theoretical
uncertainty coming from the unknown higher order contributions in the
calculations.  An improved determination of $\as$ from these measurements
thus awaits improved theoretical calculations of these corrections.

\section{Soft Gluon Coherence}\label{sec:softgluon}
The phenomenon of colour coherence in \QCD\ implies destructive interference
in soft gluon emission.
With the assumption of Local Parton Hadron Duality (\LPHD) \cite{lphd,lphd2},
colour coherence can be studied
in charged particle distributions, in particular in the multiplicity distribution and in
the charged particle momentum spectrum of
the variable $\xi = \ln (1/x)$, where $x$ is the momentum scaled by the beam energy.
 
To study these distributions, events are selected using cuts very similar to those of Section~\ref{sec:sel}.
Well measured charged tracks are selected and the event is required to be in the barrel region of
the detector by demanding that the thrust axis calculated from calorimeter clusters be more than
$42.3^\circ$ from the beam axis 
and that calculated from charged tracks more than
$45.6^\circ$.
%
 
\subsection{Charged Particle Multiplicity}\label{sec:chmul}
 
The dynamics of hadron production can be probed using the charged
particle multiplicity distribution which is found to be very sensitive
to the parameters of the \QCD\ models.
 
The measured distributions are corrected for the remaining estimated
background using Monte Carlo on a bin-by-bin basis.
The distributions are then corrected for resolution
and acceptance, using a matrix unfolding method.
At the Z pole, the high statistics warrant a more refined method, and
the matrix unfolding is iterated in a Bayesian procedure\cite{l3mult, dagost}.
In this correction procedure, we assume all
particles with mean lifetime greater than $3.3\times 10^{-10}$\,s to be stable.

The systematic uncertainties are determined as
for the global event-shape variables with one additional
contribution corresponding to a variation of
the quality criteria for track selection.
 
\begin{figure}[htbp]
\begin{center}
  \includegraphics*[width=0.5\figwidth]{\mydirfig 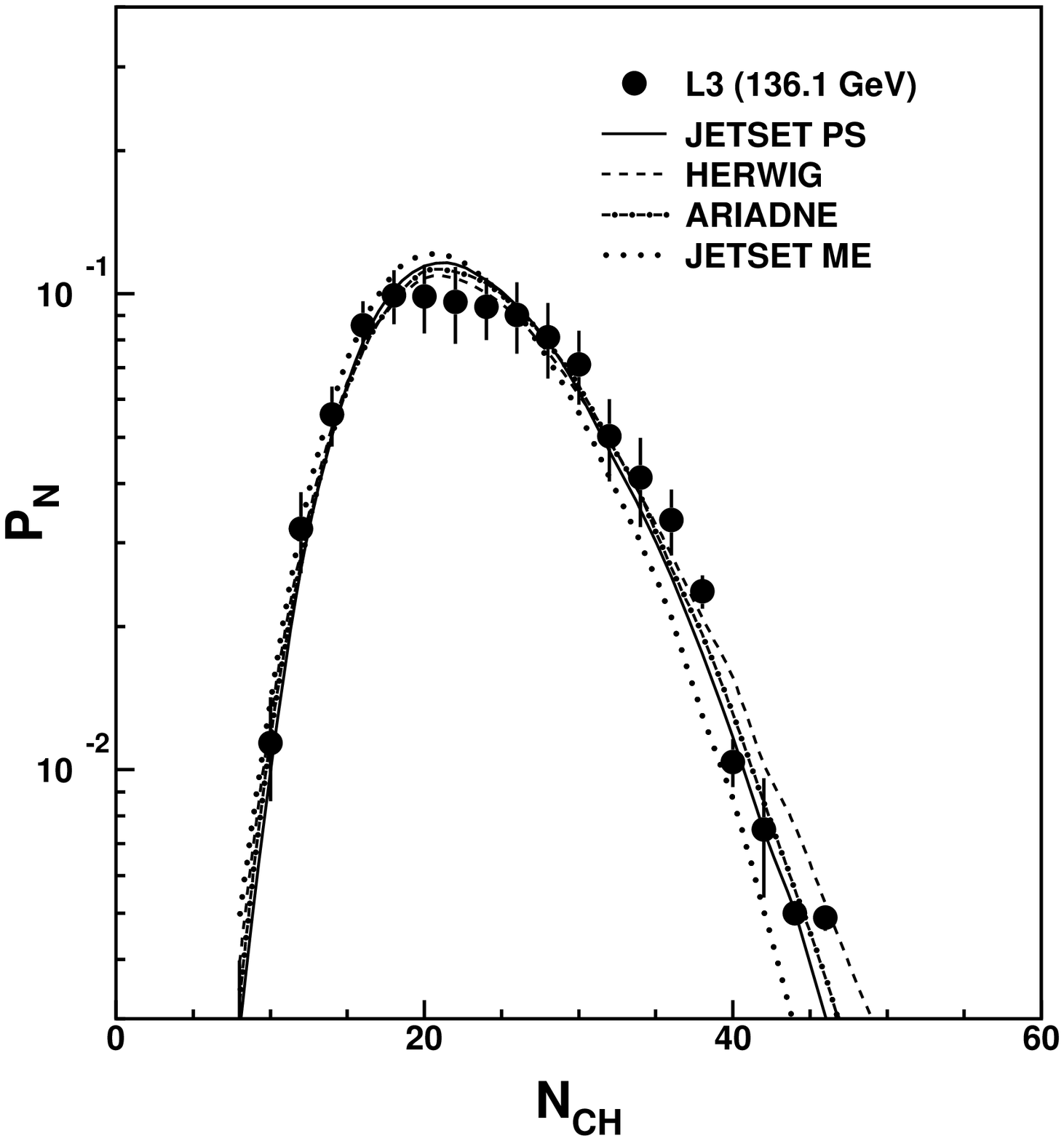}
  \includegraphics*[width=0.5\figwidth]{\mydirfig 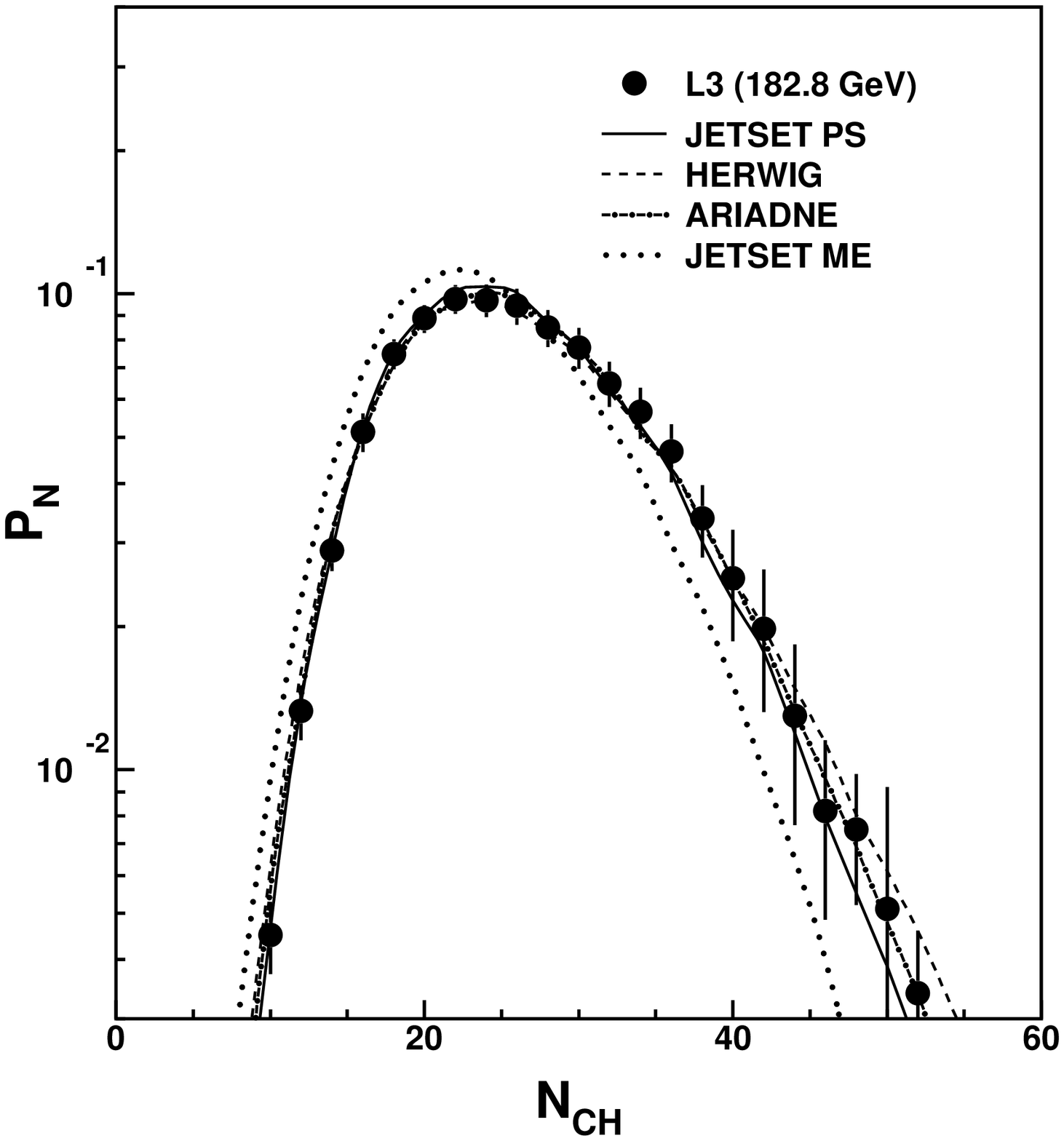}
  \includegraphics*[width=0.5\figwidth]{\mydirfig 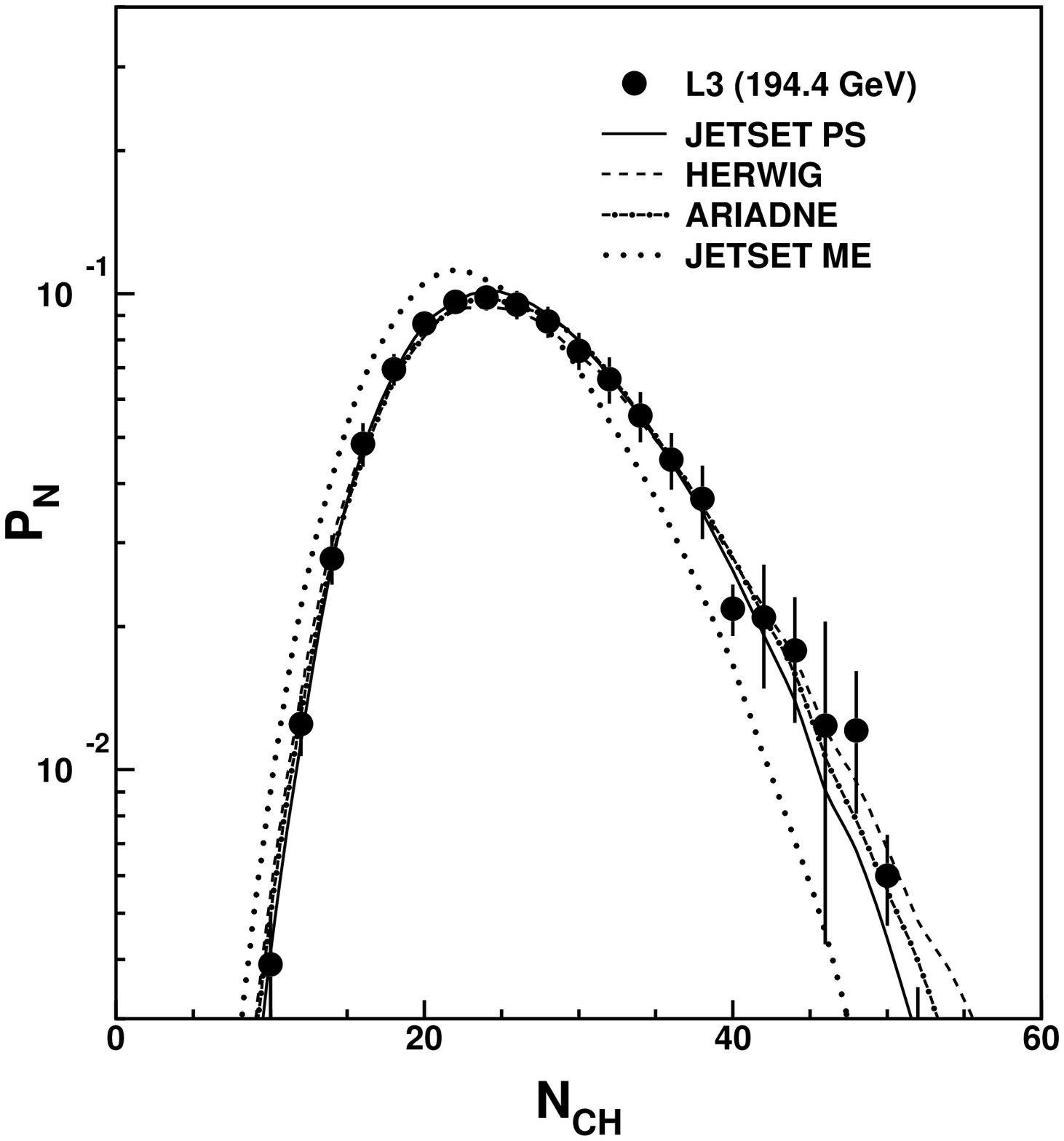}
  \includegraphics*[width=0.5\figwidth]{\mydirfig 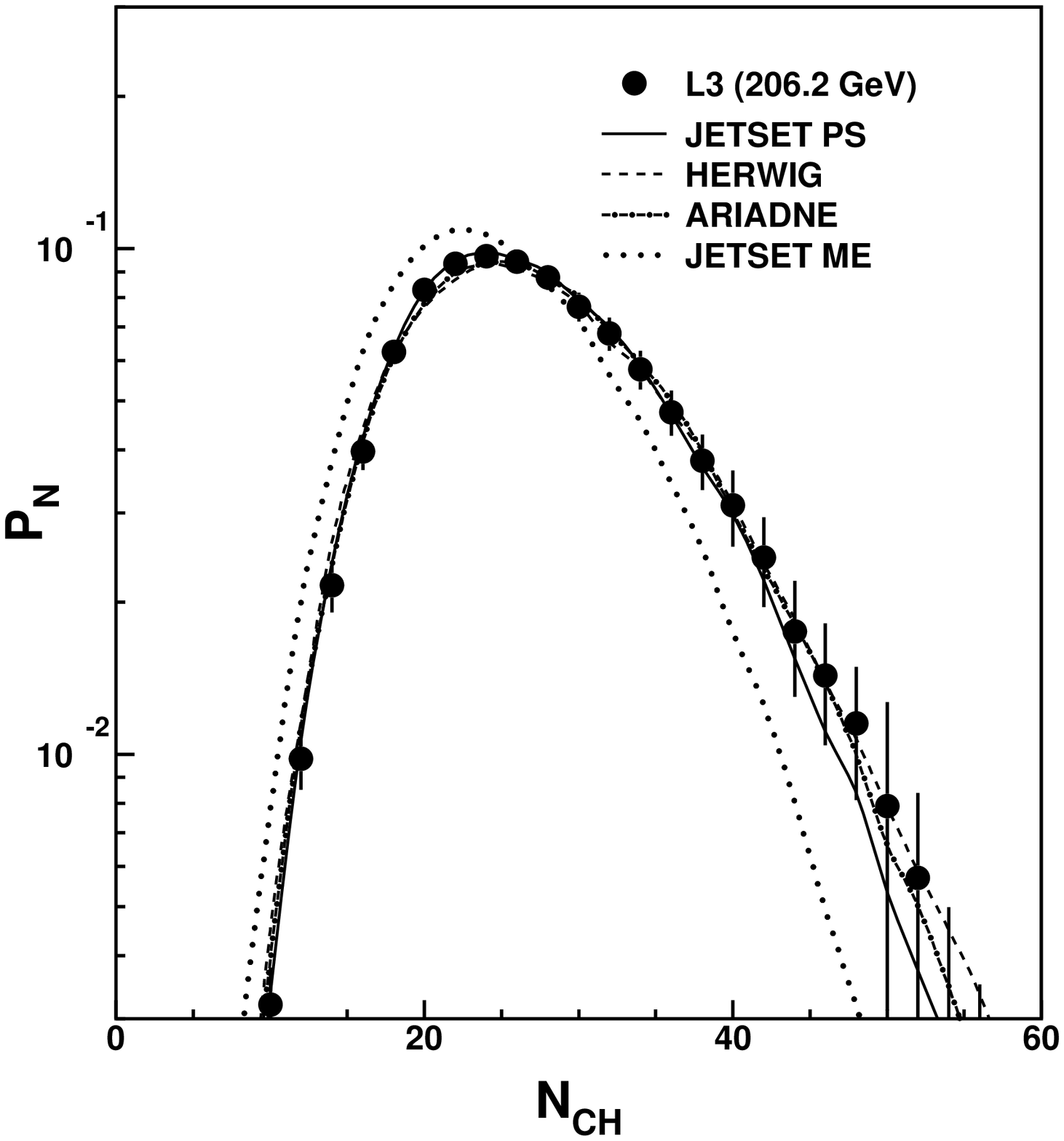}
\end{center}
\caption{Charged particle multiplicity distributions, normalised to unity, at
         $\rs=136.1$, $182.8$, $194.4$ and $206.2\,\GeV$
         compared to several \QCD\ models.
         }
\label{fig:chmul}
\end{figure}

The corrected distributions together with the mean charged particle
multiplicities, $\langle\nch\rangle$, at the Z pole and above are
summarised in Tables
\ref{tab:nch0}--\ref{tab:nch3}.
Figures \ref{fig:chmulZ} and \ref{fig:chmul} show the
measured charged particle multiplicity distributions at centre-of-mass
energies of 91.2, 136.1, 182.8, 194.4 and 206.2\,\GeV\ compared to the different
Monte Carlo models tuned to the Z-pole data (\cf\ Section \ref{sec:tuning}).
At 91.2\,\GeV, \textsc{Jetset PS} agrees well with the data for all, b  and udsc quarks.
This is not the case for \textsc{Herwig}, whose distributions are too broad.
At higher energies, this feature of \textsc{Herwig} remains, while \textsc{Jetset PS} continues to provide a
good description. The matrix element version of \textsc{Jetset} produces too few particles at high energy.

\begin{figure}[htbp]
\begin{center}
 \includegraphics*[width=0.5\figwidth]{\mydirfig 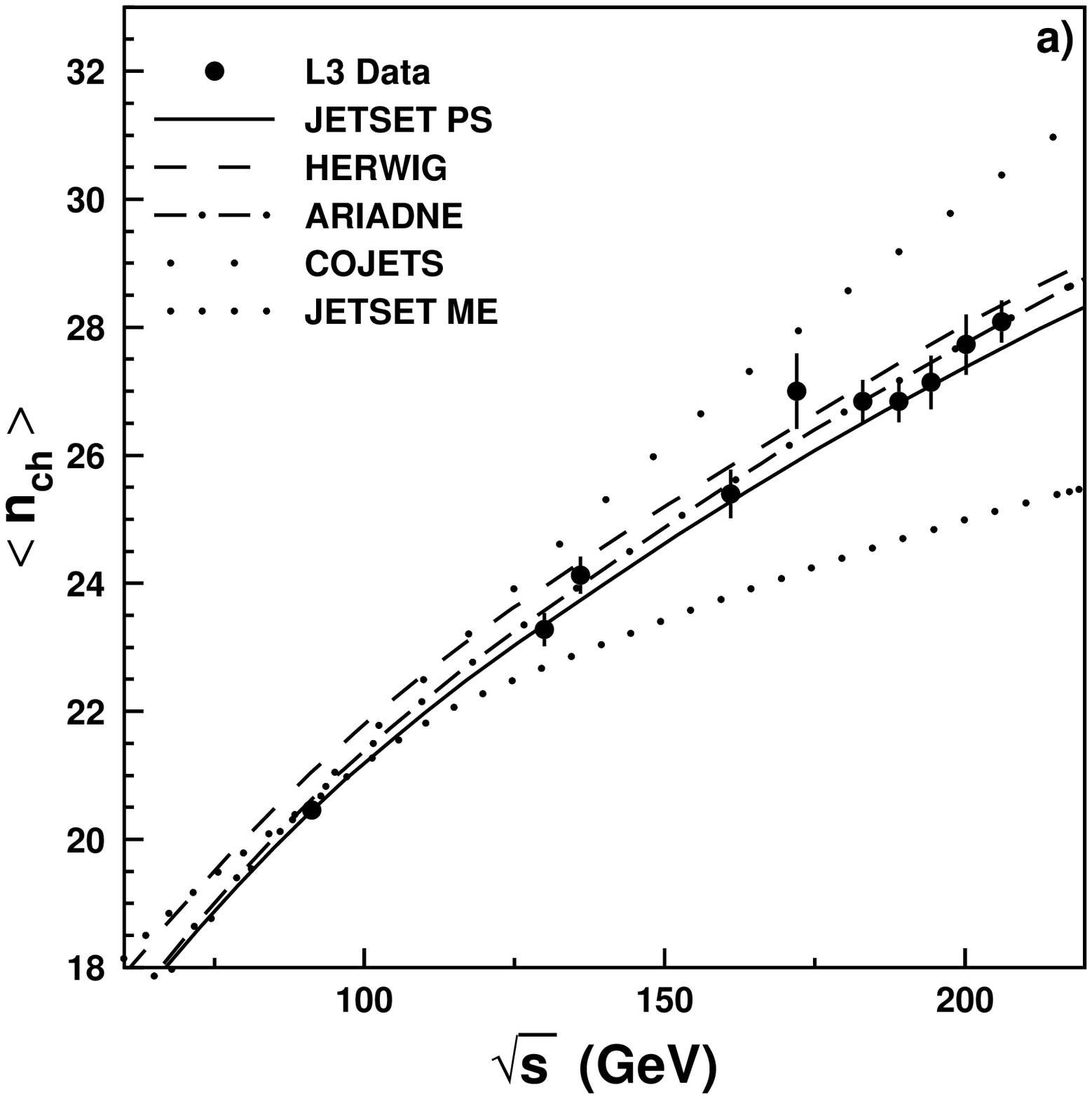}
 \includegraphics*[width=0.5\figwidth]{\mydirfig 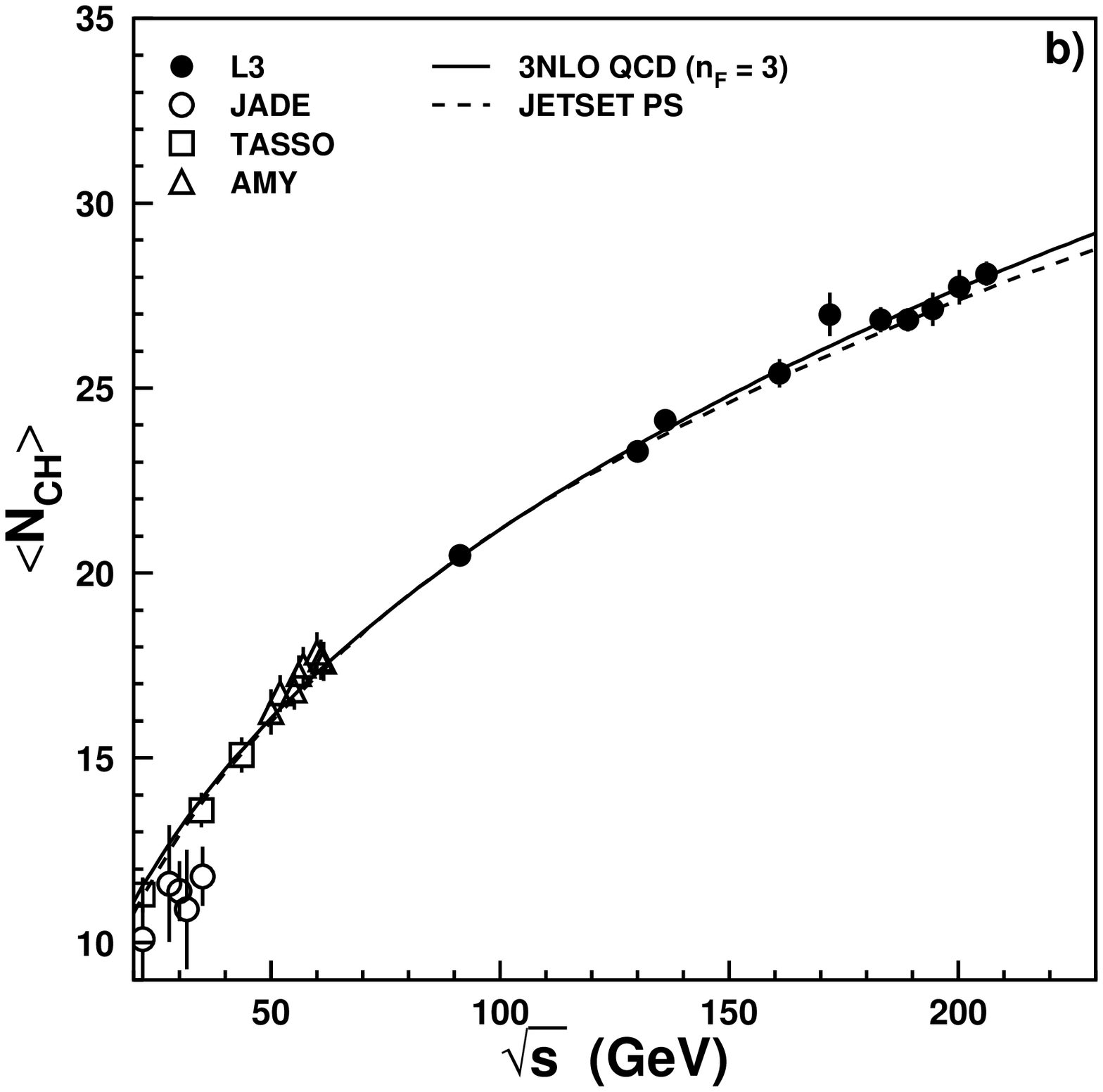}
\end{center}
\caption{The mean charged particle multiplicity,
           $\langle\nch\rangle$, as a function of the centre-of-mass energy,
           (a) compared to several \QCD\ models,
           (b) fitted to the \NNNLO\ prediction of \QCD\ with local parton hadron duality,
           assuming 3 active flavours.
           }
\label{fig:meanch}
\end{figure}

Figure~\ref{fig:meanch}a shows the evolution of mean charged particle
multiplicity with centre-of-mass energy compared to several \QCD\ models.
The parameters of the models are the same at all energies.
We find that the energy dependence predicted by the parton shower models
\textsc{Jetset}, \textsc{Herwig} and \textsc{Ariadne}, which include \QCD\ coherence effects,
are in agreement with the measured mean multiplicities.
However, parton shower models with no \QCD\ coherence effects, such as
\textsc{Cojets} and \textsc{Jetset} ME, do not explain the observed energy dependence.
\textsc{Cojets} predicts a faster energy evolution, while \textsc{Jetset} ME,
which has low parton multiplicity before fragmentation
due to the  ${\cal{O}}(\as^{2})$ calculation, would need retuning at each
centre-of-mass energy.
 
The mean charged particle multiplicity
in gluon jets has been calculated,
in the framework of \LPHD, 
to next-to-next-to-next-to leading order (\NNNLO) 
\cite{3nlo1}:
\begin{equation} \label{eq:3nlog}
   \langle\nch(y)\rangle_\mathrm{g} = {\cal N} y^{-a_1 C^2} \exp\left[2C\sqrt{y} + \delta_\mathrm{g}(y)\right]
\end{equation}
where ${\cal N}$ is an overall normalization constant, $y=\ln(\mu/\Lambda)$,
$C=\sqrt{4N_\mathrm{C}\;/\beta_0}$ and
\begin{equation}  \label{eq:3nlo23}
   \delta_\mathrm{g}(y) = \frac{C}{\sqrt{y}}\left\{2a_2 C^2 + \frac{\beta_1}{\beta_0}\left[2+\ln(2y)\right]\right\}
                        + \frac{C^2}{y}\left\{a_3 C^2 - \frac{a_1\beta_1}{\beta_0^2}\left[1+\ln(2y)\right]\right\}
\end{equation}
The leading order (LO) prediction is given by  $\exp\left[2C\sqrt{y}\right]$. The factor in front of this
exponential arises in \NLO.  The first term in $\delta_\mathrm{g}(y)$ is the \NNLO\ contribution, and the
second term that of \NNNLO. 
The prediction for quark jets or \epem\ events, is then
\begin{equation}   \label{eq:3nloq}
      \langle\nch(y)\rangle_\mathrm{F} = \frac{\langle\nch(y)\rangle_\mathrm{g}}{r(y)}
\end{equation}
where $r(y)$, the ratio of multiplicities of gluon and quark jets, is given by
\begin{equation}   \label{eq:3nlor}
   r(y) = r_0\left(1-r_1\gamma_0-r_2\gamma_0^2-r_3\gamma_0^3\right) \; .
\end{equation}
Here $\gamma_0$ is an anomalous dimension, which is related to \as\ by
\begin{equation}     \label{eq:gamma0}
  \gamma_0 = \sqrt{\frac{2\as\Nc}{\pi}}
\end{equation}
and $r_0=\Nc/C_\mathrm{F}=4$.
The coefficients $a_i$ and $r_i$, $i=1, 2, 3$ have been calculated in Reference \citen{3nlo2}
and are given in Table~\ref{tab:3nlo2}.

Figure~\ref{fig:meanch}b shows the mean charged particle multiplicity as measured
by this experiment together with measurements of other \ee\ experiments at
lower \cite{mean:JADE,mean:TASSO1,mean:TASSO2,mean:AMY1,mean:AMY2} centre-of-mass energies.
The prediction of \JETSET\ PS is also shown.
Fits of \LO\ through \NNNLO\ with $\Nf=3$ or 5
are performed
to the data from \TASSO, \AMY\ and this experiment using $\mu=\sqrt{s}$.
The result for \NNNLO\ with $\Nf=3$ is shown in Figure~\ref{fig:meanch}b.
The results of all of the fits are given in Table~\ref{tab:nchfit} in terms of the value of
$\as(\MZ)$ calculated from
\begin{equation}  \label{eq:alphas}
  \as(\mu) = \frac{2\pi}{\beta_0 y}
              \left[1-\frac{\beta_1\ln(2\ln(\mu/\Lambda))}{\beta_0^2 \ln(\mu/\Lambda)}\right] \;.
\end{equation}
 
The description of \nch\ \vs\ \rs\ is good in all cases, the resulting curves being
nearly indistinguishable. 
Also the choice of number of active flavours makes little difference.
However the value of \as\ obtained in the fit increases steadily from \LO\ to \NNNLO, and
the values of \as\ from the \NNNLO\ fits are consistent with that obtained from event shapes.

\subsection{Inclusive Particle Spectrum}\label{sec:ksi}
 
The suppression of low momentum hadron production as a consequence of colour coherence
is studied in terms of
the variable $\xi$. 
The observed distribution is corrected for the effect of background,
detector resolution and acceptance.
At $\sqrt{s}=91.2\,\GeV$ this is done using a matrix unfolding method as
for the charged particle multiplicity distribution.
For the other energies it is done on a bin-by-bin basis using Monte Carlo events.
 
\begin{figure}[htbp]
\begin{center}
   \includegraphics*[width=0.5\figwidth]{\mydirfig 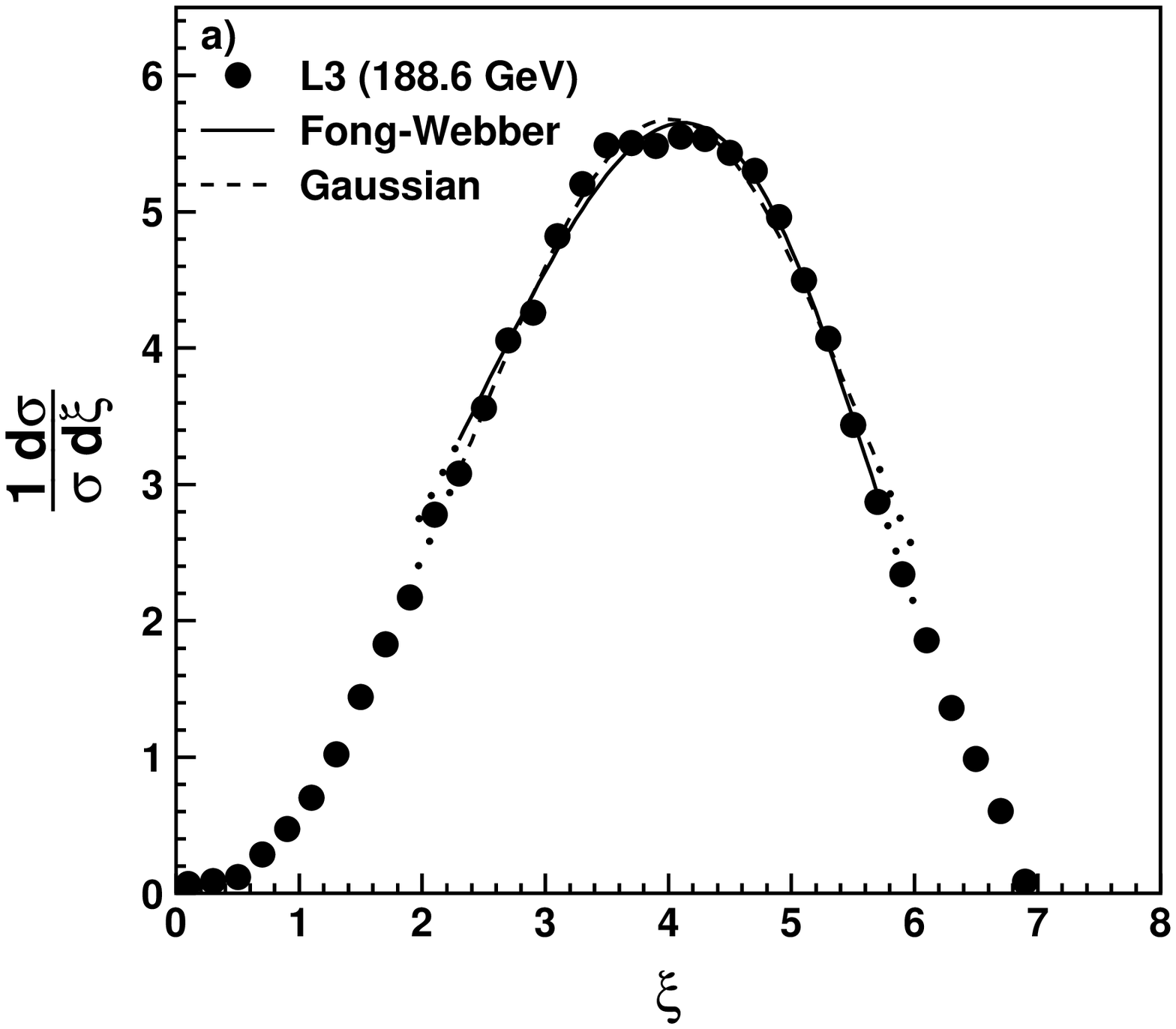}
   \includegraphics*[width=0.5\figwidth]{\mydirfig 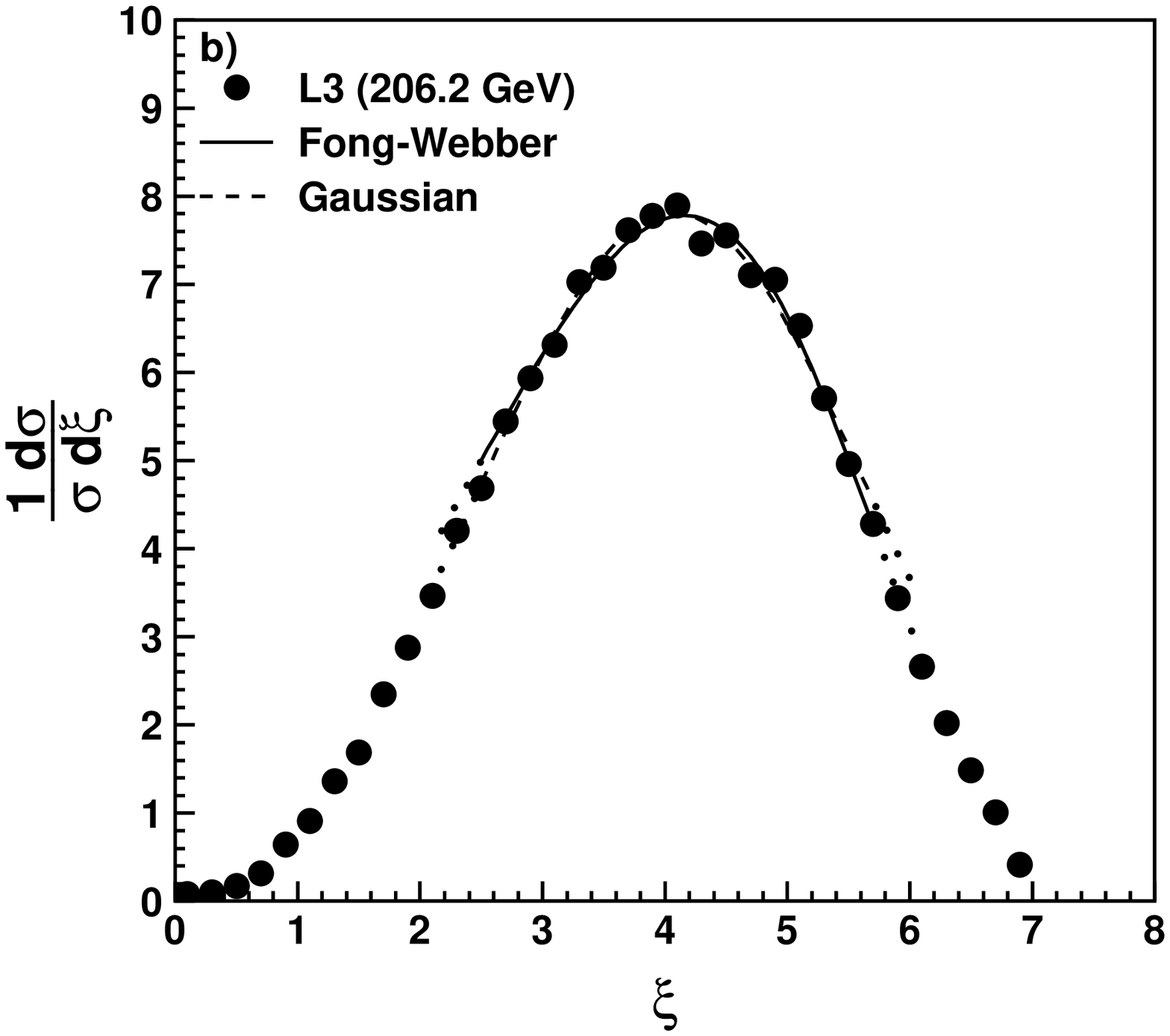}
\end{center}
\caption{Corrected $\xi$-spectra at $\rs= 188.6\,\GeV$  and $\rs=206.2\,\GeV$
           together with the results of fits to Gaussian and Fong-Webber parametrisations.}
\label{fig:ksi}
\end{figure}

The corrected spectra for all flavours as well as for non-b and b quarks
at $\rs = 91.2\,\GeV$ are shown in Figure \ref{fig:ksiZ}
and summarised in Table \ref{tab:ksi0}.
{\scshape Jetset} overestimates the central region.
This may be due to its tuning,
which only uses the charged-particle multiplicity distribution and global event-shape data.
The description provided by \textsc{Herwig} is in general poorer, particularly for the b-flavour events.
The corrected distributions at $\rs = 188.6$ and 206.2\,\GeV\ are shown
in Figure \ref{fig:ksi}.
The corrected distributions at $\rs > 130\,\GeV$ are
summarised in Tables
\ref{tab:ksi1}--\ref{tab:ksi3}.
The asymptotic shape of the $\xi$ spectrum is predicted to be
Gaussian~\cite{mlla1,gaus1,gaus1a,gaus2}.
However, at finite energies the shape is affected by
destructive interference in soft gluon emission.
With next-to-leading order corrections~\cite{sgaus}, one expects
a skewed platykurtic shape (often called a skewed Gaussian) for the $\xi$ distribution.
This implies a narrower $\xi$-peak shifted towards higher $\xi$-values,
skewed and flattened towards lower $\xi$-values,
with the high-$\xi$ tail falling off faster than Gaussian.
The smooth lines in Figure~\ref{fig:ksi} are the results of fits
to the corrected distributions of both a Gaussian and the
Fong-Webber parametrisation of the skewed Gaussian\cite{sgaus},
which reproduces the expected \MLLA\ shape around the peak value,
The fit range is restricted  to values of
$\xi$ where the distribution is within 60\% of its maximum value.
In the fit,
the systematic uncertainties are taken to be fully correlated.
Both parametrisations give a reasonable description of the data.
The Fong-Webber curve also provides a good description for large $\xi$ where
the Gaussian is systematically too high. However, at small $\xi$ the Gaussian fits better.

The results of the fits at $\sqrt{s}=91.2\,\GeV$ are shown in Table \ref{tab:xiZ}.
The systematic uncertainty is estimated by
repeating the fits changing
 (a) the quality cuts on track selection;
 (b) the hadronic selection criteria to vary the backgrounds
      within one standard deviation;
 (c) the model, using \textsc{Herwig} for
      detector corrections instead of \textsc{Pythia}.
Half of the maximum spread is assigned as the systematic uncertainty.
As expected,
the values obtained
from the  Fong-Webber fits are systematically higher than those obtained using the Gaussian parametrisation.
The difference is about 0.03, independent of flavour.
Thus the                      flavour dependence of \xistar\ is independent of the choice of the fit function.
 
We observe a flavour dependence of the peak position, \xistar,  more clearly shown by the ratios,
\begin{eqnarray*}
     \xistar_\mathrm{udsc}/ \xistar_\mathrm{all}  &=&   1.008 \pm0.003\pm0.002    \\
     \xistar_\mathrm{b}/ \xistar_\mathrm{all}     &=&   0.975 \pm0.003\pm0.008\quad,
\end{eqnarray*}
where the first uncertainty is statistical and the second systematic.
The small size of the resulting systematic uncertainty is due to the fact that
most of the systematic uncertainty cancels when    forming these ratios.
Moreover, these ratios are insensitive to the fit parametrisation,
the small difference being assigned as an additional systematic uncertainty.

The peak positions \xistar\ of the $\xi$ distribution as well as the
\chisq/d.o.f and the confidence level of the fits obtained with the
skewed Gaussian for all the energy points are summarised in Table~\ref{tab:ksi}.
The systematic uncertainty includes, in addition to those mentioned above, a contribution
of half the difference between the result using the Gaussian and the Fong-Webber parametrisations.
 
\begin{figure}[htbp]
\begin{center}
   \includegraphics*[width=\figwidth]{\mydirfig 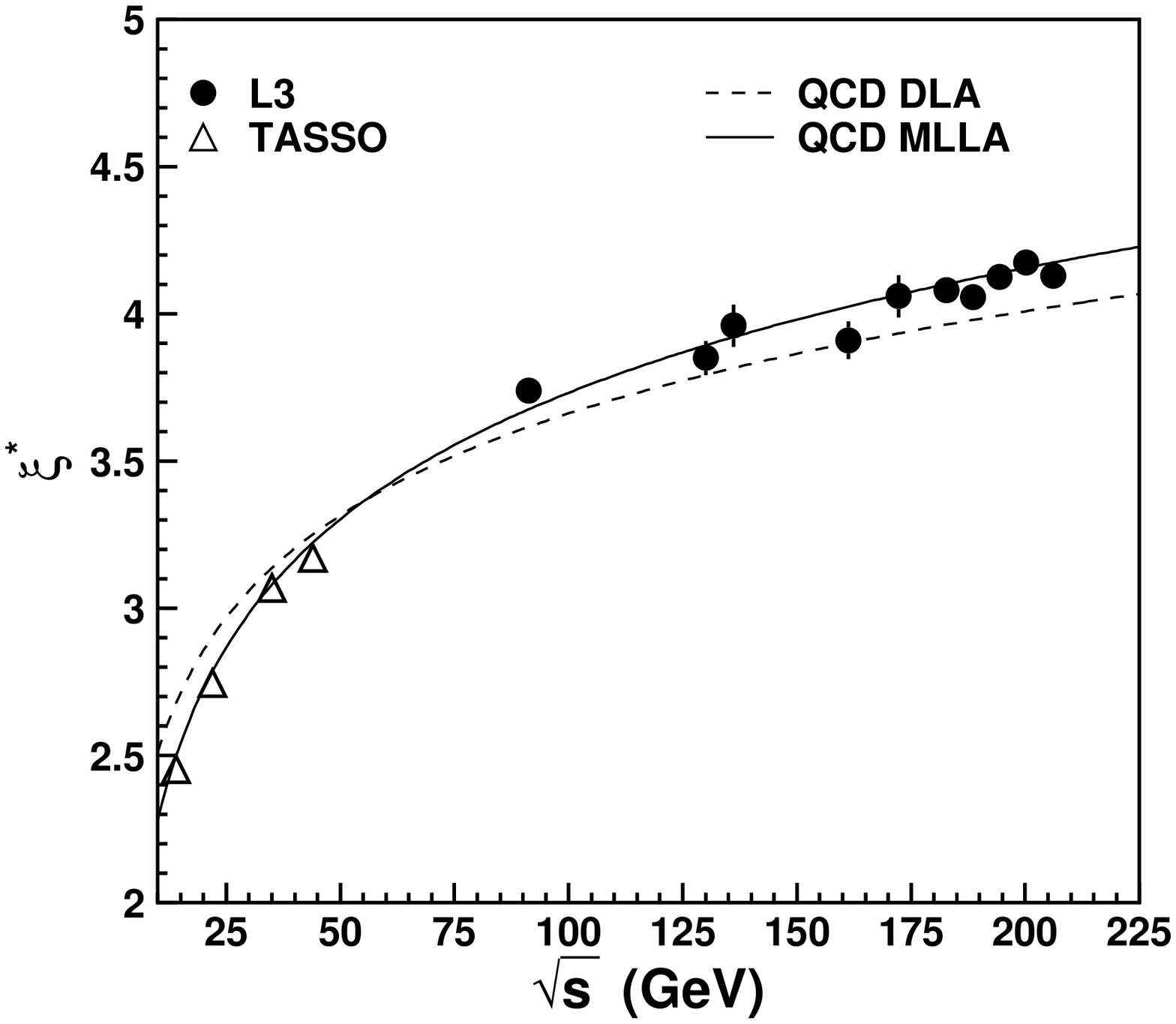}
\end{center}
\caption{Energy evolution of $\xistar$.
         The solid and dashed lines are fits to the \Lthree\ and \TASSO\ data
         with Modified Leading Log Approximation (\MLLA)
         and Double Log Approximation (\DLA) \QCD.}
\label{fig:ksistar}
\end{figure}

Figure~\ref{fig:ksistar} shows the measured values of \xistar\
together with earlier measurements
 \cite{ksi-l3,ksi-l3a,ksi-tasso} 
as a function of centre-of-mass energy.
The energy evolution of \xistar\ has been fitted using the \QCD\ prediction
\begin{equation}   \label{eq:xistar}
  \xistar(s)  =  y \left( \frac{1}{2} + \sqrt{\frac{C}{y}}
                                       - \frac{C}{y} \right),
\end{equation}
where $y =  \ln(\mu/\Lambda)$
and
$C=a^2/(16\Nc\beta_0)$ with
$a = \left[11\Nc/3\vphantom{^2)}\right] + \left[(2\Nf)/(3\Nc^2)\right]$.
We choose $\mu=\rs/2$.
The first term is given by the double logarithm approximation (\DLA),
and the correction terms arise in the next-to-leading order~\cite{mlla1,mlla2,mlla3} (\MLLA) \QCD\ predictions.
In the fits, the systematic uncertainties among the \TASSO\ points are treated as fully correlated.
The same is true of the \Lthree\ points with $\rs>130\,\GeV.$
We find that the data are in better agreement with \QCD\ predictions
computed to next-to leading orders.
The fit of the \Lthree\  and \TASSO\ data to the \DLA\ parametrisation gives
a \chisq\ of 110 for 13 degrees of freedom 
whereas the \MLLA\ predictions give a fit with \chisq\ of 26 for 13
degrees of freedom, corresponding to a confidence level of 2\%.  
 
It should be recalled that the suppression of hadron production at very small
momenta resulting in a bell shape of the $\xi$ distribution
is expected on purely kinematical grounds due to finite hadron masses.
Soft gluon coherence, however, increases this suppression and is manifested
in the energy dependence of $\xistar$. The change with
energy would be approximately two times larger
without any destructive interference.

\section{Summary}\label{sec:summ}
 
Distributions of event-shape variables in hadronic events from \ee\ annihilation at centre-of-mass
energies from 30\,\GeV\ to 209\,\GeV\ have been measured.
These distributions as well as the energy dependence of their first
moments are well described by parton shower models.
 
Jet fractions have been measured using the \textsc{Jade}, Durham and Cambridge
algorithms as a function of the jet resolution parameters. The parton
shower models are in good agreement with the measured jet fractions. The energy
evolution of the 3-jet fraction at a fixed jet resolution parameter is in
agreement with ${\cal O}(\as^2)$ \QCD\ calculations.
 
The energy dependence of the first two moments has been compared to
second order perturbative \QCD\ with power law corrections for the
non-perturbative effects.
The  fits to the six event-shape variables give consistent values of \as,
which are somewhat lower than that obtained by the event-shape analysis.
However, the values of $\alpha_0$ are not consistent, differing by as much as 40\% from their average.
Further, the contribution from a
${\cal O}(\frac{1}{s})$ term  in describing the second moments
of \onet\ and $C$  is not small in contradiction to expectations.
This implies that the power law correction can at best be described as semi-quantitative.
 
The event-shape distributions are compared to second order \QCD\ calculations
combined with resummed leading and next-to-leading logarithmic terms.  The data are
well described by these calculations at all energies. The measurements
demonstrate the running of $\as$ as expected in \QCD\ with a value of
$\as(\MZ) = 0.1227\pm0.0012 \text{(exp)} \pm0.0058 \text{(th)}$.
The uncertainties on these measurements are dominated by the theoretical uncertainty coming
from unknown higher order contributions in the calculations.
An improved determination of \as\ from these measurements thus awaits improved theoretical
calculations.
 
The energy evolution of the charged particle multiplicity as well as the
inclusive charged particle momentum spectrum show evidence of soft gluon
suppression. Energy evolution of the peak position $\xistar$ of the inclusive
$\xi$ spectrum is described adequately by the next-to-leading order \QCD\
calculation including gluon interference effects.
 

\clearpage
\typeout{   }     
\typeout{Using author list for paper 287 -  }
\typeout{$Modified: Jul 15 2001 by smele $}
\typeout{!!!!  This should only be used with document option a4p!!!!}
\typeout{   }
%
%
%
%
%
%

\newcount\tutecount  \tutecount=0
\def\tutenum#1{\global\advance\tutecount by 1 \xdef#1{\the\tutecount}}
\def\tute#1{$^{#1}$}
\tutenum\aachen            
\tutenum\nikhef            
\tutenum\mich              
\tutenum\lapp              
\tutenum\basel             
\tutenum\lsu               
\tutenum\beijing           
\tutenum\bologna           
\tutenum\tata              
\tutenum\ne                
\tutenum\bucharest         
\tutenum\budapest          
\tutenum\mit               
\tutenum\panjab            
\tutenum\debrecen          
\tutenum\dublin            
\tutenum\florence          
\tutenum\cern              
\tutenum\wl                
\tutenum\geneva            
\tutenum\hamburg           
\tutenum\hefei             
\tutenum\lausanne          
\tutenum\lyon              
\tutenum\madrid            
\tutenum\florida           
\tutenum\milan             
\tutenum\moscow            
\tutenum\naples            
\tutenum\cyprus            
\tutenum\nymegen           
\tutenum\caltech           
\tutenum\perugia           
\tutenum\peters            
\tutenum\cmu               
\tutenum\potenza           
\tutenum\prince            
\tutenum\riverside         
\tutenum\rome              
\tutenum\salerno           
\tutenum\ucsd              
\tutenum\sofia             
\tutenum\korea             
\tutenum\taiwan            
\tutenum\tsinghua          
\tutenum\purdue            
\tutenum\psinst            
\tutenum\zeuthen           
\tutenum\eth               

{
\parskip=0pt
\noindent
{\bf The L3 Collaboration:}
\ifx\selectfont\undefined
 \baselineskip=10.8pt
 \baselineskip\baselinestretch\baselineskip
 \normalbaselineskip\baselineskip
 \ixpt
\else
 \fontsize{9}{10.8pt}\selectfont
\fi
\medskip
\tolerance=10000
\hbadness=5000
\raggedright
\hsize=162truemm\hoffset=0mm
\def\r{\rlap,}
\noindent

P.Achard\r\tute\geneva\ 
O.Adriani\r\tute{\florence}\ 
M.Aguilar-Benitez\r\tute\madrid\ 
J.Alcaraz\r\tute{\madrid}\ 
G.Alemanni\r\tute\lausanne\
J.Allaby\r\tute\cern\
A.Aloisio\r\tute\naples\ 
M.G.Alviggi\r\tute\naples\
H.Anderhub\r\tute\eth\ 
V.P.Andreev\r\tute{\lsu,\peters}\
% ...
\newpage
\begin{list}{A}{\itemsep=0pt plus 0pt minus 0pt\parsep=0pt plus 0pt minus 0pt
                \topsep=0pt plus 0pt minus 0pt}
\item[\aachen]
 III. Physikalisches Institut, RWTH, D-52056 Aachen, Germany$^{\S}$
\item[\nikhef] National Institute for High Energy Physics, NIKHEF, 
     and University of Amsterdam, NL-1009 DB Amsterdam, The Netherlands
\item[\mich] University of Michigan, Ann Arbor, MI 48109, USA
\item[\lapp] Laboratoire d'Annecy-le-Vieux de Physique des Particules, 
     LAPP,IN2P3-CNRS, BP 110, F-74941 Annecy-le-Vieux CEDEX, France
\item[\basel] Institute of Physics, University of Basel, CH-4056 Basel,
     Switzerland
\item[\lsu] Louisiana State University, Baton Rouge, LA 70803, USA
\item[\beijing] Institute of High Energy Physics, IHEP, 
  100039 Beijing, China$^{\triangle}$ 
\item[\bologna] University of Bologna and INFN-Sezione di Bologna, 
     I-40126 Bologna, Italy
\item[\tata] Tata Institute of Fundamental Research, Mumbai (Bombay) 400 005, India
\item[\ne] Northeastern University, Boston, MA 02115, USA
\item[\bucharest] Institute of Atomic Physics and University of Bucharest,
     R-76900 Bucharest, Romania
\item[\budapest] Central Research Institute for Physics of the 
     Hungarian Academy of Sciences, H-1525 Budapest 114, Hungary$^{\ddag}$
\item[\mit] Massachusetts Institute of Technology, Cambridge, MA 02139, USA
\item[\panjab] Panjab University, Chandigarh 160 014, India
\item[\debrecen] KLTE-ATOMKI, H-4010 Debrecen, Hungary$^\P$
\item[\dublin] Department of Experimental Physics,
  University College Dublin, Belfield, Dublin 4, Ireland
\item[\florence] INFN Sezione di Firenze and University of Florence, 
     I-50125 Florence, Italy
\item[\cern] European Laboratory for Particle Physics, CERN, 
     CH-1211 Geneva 23, Switzerland
\item[\wl] World Laboratory, FBLJA  Project, CH-1211 Geneva 23, Switzerland
\item[\geneva] University of Geneva, CH-1211 Geneva 4, Switzerland
\item[\hamburg] University of Hamburg, D-22761 Hamburg, Germany
\item[\hefei] Chinese University of Science and Technology, USTC,
      Hefei, Anhui 230 029, China$^{\triangle}$
\item[\lausanne] University of Lausanne, CH-1015 Lausanne, Switzerland
\item[\lyon] Institut de Physique Nucl\'eaire de Lyon, 
     IN2P3-CNRS,Universit\'e Claude Bernard, 
     F-69622 Villeurbanne, France
\item[\madrid] Centro de Investigaciones Energ{\'e}ticas, 
     Medioambientales y Tecnol\'ogicas, CIEMAT, E-28040 Madrid,
     Spain${\flat}$ 
\item[\florida] Florida Institute of Technology, Melbourne, FL 32901, USA
\item[\milan] INFN-Sezione di Milano, I-20133 Milan, Italy
\item[\moscow] Institute of Theoretical and Experimental Physics, ITEP, 
     Moscow, Russia
\item[\naples] INFN-Sezione di Napoli and University of Naples, 
     I-80125 Naples, Italy
\item[\cyprus] Department of Physics, University of Cyprus,
     Nicosia, Cyprus
\item[\nymegen] University of Nijmegen and NIKHEF, 
     NL-6525 ED Nijmegen, The Netherlands
\item[\caltech] California Institute of Technology, Pasadena, CA 91125, USA
\item[\perugia] INFN-Sezione di Perugia and Universit\`a Degli 
     Studi di Perugia, I-06100 Perugia, Italy   
\item[\peters] Nuclear Physics Institute, St. Petersburg, Russia
\item[\cmu] Carnegie Mellon University, Pittsburgh, PA 15213, USA
\item[\potenza] INFN-Sezione di Napoli and University of Potenza, 
     I-85100 Potenza, Italy
\item[\prince] Princeton University, Princeton, NJ 08544, USA
\item[\riverside] University of Californa, Riverside, CA 92521, USA
\item[\rome] INFN-Sezione di Roma and University of Rome, ``La Sapienza",
     I-00185 Rome, Italy
\item[\salerno] University and INFN, Salerno, I-84100 Salerno, Italy
\item[\ucsd] University of California, San Diego, CA 92093, USA
\item[\sofia] Bulgarian Academy of Sciences, Central Lab.~of 
     Mechatronics and Instrumentation, BU-1113 Sofia, Bulgaria
\item[\korea]  The Center for High Energy Physics, 
     Kyungpook National University, 702-701 Taegu, Republic of Korea
\item[\taiwan] National Central University, Chung-Li, Taiwan, China
\item[\tsinghua] Department of Physics, National Tsing Hua University,
      Taiwan, China
\item[\purdue] Purdue University, West Lafayette, IN 47907, USA
\item[\psinst] Paul Scherrer Institut, PSI, CH-5232 Villigen, Switzerland
\item[\zeuthen] DESY, D-15738 Zeuthen, Germany
\item[\eth] Eidgen\"ossische Technische Hochschule, ETH Z\"urich,
     CH-8093 Z\"urich, Switzerland
\item[\S]  Supported by the German Bundesministerium 
        f\"ur Bildung, Wissenschaft, Forschung und Technologie.
\item[\ddag] Supported by the Hungarian OTKA fund under contract
numbers T019181, F023259 and T037350.
\item[\P] Also supported by the Hungarian OTKA fund under contract
  number T026178.
\item[$\flat$] Supported also by the Comisi\'on Interministerial de Ciencia y 
        Tecnolog{\'\i}a.
\item[$\sharp$] Also supported by CONICET and Universidad Nacional de La Plata,
        CC 67, 1900 La Plata, Argentina.
\item[$\triangle$] Supported by the National Natural Science
  Foundation of China.
\end{list}
}
\vfill


\clearpage
 
\begin{table}[htbp]
\begin{center}\begin{tabular}{|c|c|r@{.}l|r@{.}l|c|c|r|}\hline
\rule{0pt}{11pt}
Type &$\rs$    &\multicolumn{2}{c|}{$\langle\rs\rangle$}&\multicolumn{2}{c|}{Integrated}& Selection & Sample &\multicolumn{1}{c|}{Selected}
\\
 of  &         &\multicolumn{2}{c|}{}                   &\multicolumn{2}{c|}{Luminosity}& Efficiency& Purity &\multicolumn{1}{c|}{events}
\\
Event&(\GeV)   &\multicolumn{2}{c|}{(\GeV)}             &\multicolumn{2}{c|}{(\pb)}     &   (\%)    &  (\%)  &   \\ \hline
Reduced& 30--50     &    41&4 &\phantom{00} 142&4 &  48.3     & 68.4   &    1247\phantom{.}     \\
Centre-& 50--60     &    55&3           & 142&4      &  41.0     & 78.0   &    1047\phantom{.}  \\
of-    & 60--70     &    65&4           & 142&4      &  35.2     & 86.0   &    1575\phantom{.}  \\
Mass   & 70--80     &    75&7           & 142&4      &  29.9     & 89.0   &    2938\phantom{.}  \\
Energy & 80--84     &    82&3           & 142&4      &  27.4     & 90.5   &    2091\phantom{.}  \\
       & 84--86     &    85&1           & 142&4      &  27.5     & 87.0   &    1607\phantom{.}  \\  \hline
Z pole &\phantom{1}91.2
                    &    91&2           &   8&3      &  98.5     & 99.8   &  248100\phantom{.}  \\  \hline
       &129.9--130.4&   130&1           &   6&1      &  90.0     & 80.6   &     556\phantom{.}  \\
       &135.9--140.1&   136&1           &   5&9      &  89.0     & 81.5   &     414\phantom{.}  \\
High   &161.2--164.7&   161&3           &  10&8      &  89.0     & 81.2   &     424\phantom{.}  \\
Energy &170.3--172.5&   172&3           &  10&2      &  84.8     & 82.6   &     325\phantom{.}  \\
       &180.8--184.2&   182&8           &  55&3      &  84.2     & 82.4   &    1500\phantom{.}  \\
       &188.4--189.9&   188&6           & 176&8      &  87.8     & 81.1   &    4479\phantom{.}  \\
       &191.4--196.0&   194&4           & 112&2      &  82.8     & 81.4   &    2403\phantom{.}  \\
       &199.2--203.8&   200&2           & 117&0      &  85.7     & 80.6   &    2456\phantom{.}  \\
       &201.5--209.1&   206&2           & 207&6      &  86.0     & 78.8   &    4146\phantom{.}  \\
\hline\end{tabular}\end{center}
\caption[Summary of integrated luminosity, selection efficiency, sample
           purity and number of selected hadronic events at the different
           energies used in this analysis.]
        {Summary of integrated luminosity, selection efficiency, sample
           purity and number of selected hadronic events at the different
           energies used in this analysis.
           The energies below $\rs=91.2\,\GeV$
           are obtained from the full data sample at the Z pole, by
           selecting events with an isolated high energy photon.
           }
\label{tab:events}
\end{table}

\begin{table}[htbp]
\begin{center}
\begin{tabular}{|c|c|c|c|c|c|c|c|c|c|}\hline
$\rs$ &  & \multicolumn{4}{c|}{\SRC} & \multicolumn{4}{c|}{\ECLU} \\  \cline{3-10}
(\GeV)&  & RMS & $\sigma_{1}$ & $\sigma_{2}$ & $f_1$
         & RMS & $\sigma_{1}$ & $\sigma_{2}$ & $f_1$                                                 \\
\hline
               &${\evis}/{\rs}$       & 0.135 & 0.125 & 0.225 & 0.86 & 0.099 & 0.059 & 0.118 & 0.63  \\
\phantom{1}91.2&$\Delta\theta$ (mrad) &  44.3 &  34.9 &  60.0 & 0.71 &  45.1 &  33.1 &  59.4 & 0.64  \\
               &$\Delta\phi$ (mrad)   &  57.5 &  36.8 &  88.3 & 0.70 &  51.4 &  31.8 &  85.6 & 0.75  \\
\hline
               &${\evis}/{\rs}$       & 0.120 & 0.107 & 0.173 & 0.85 & 0.095 & 0.040 & 0.121 & 0.49  \\
188.6          &$\Delta\theta$ (mrad) &  57.4 &  33.5 & 103.8 & 0.79 &  60.7 &  33.4 & 108.3 & 0.75  \\
               &$\Delta\phi$ (mrad)   &  47.9 &  32.0 &  85.6 & 0.79 &  43.3 &  25.8 &  69.7 & 0.78  \\
\hline\end{tabular}
\end{center}
\caption[Resolution of total energy measurement and jet angles as obtained
           in the hadronic data at $\rs= 91.2\,\GeV$ and $\rs= 188.6\,\GeV$]
        {Resolution of total energy measurement and jet angles as obtained
           in the hadronic data at $\rs= 91.2\,\GeV$ and $\rs= 188.6\,\GeV$.
         The RMS is of the data.
         the $\sigma$ are the standard deviations from a fit to a sum of two Gaussian functions.
         The fraction of events in the narrower Gaussian, $f_1$, is also given.
         }
\label{tab:angr}
\end{table}

 
 
\begin{table}[htbp]\begin{center}
\begin{tabular}{|l|l|r@{$\;\pm\;$}l|}\hline
\multicolumn{1}{|c|}{Model} & \multicolumn{1}{c|}{Parameter} &
                              \multicolumn{2}{c|}{Fit Value}   \\ \hline
                 &  \LambdaLLA\ (\GeV)        &  0.311 & 0.034 \\
{\textsc{Jetset} 7.4 PS} & $\sQ$ (\GeV)       &  0.411 & 0.034 \\
                 & $b$ (\GeV$^{-2}$)          &  0.886 & 0.120 \\ \hline
                 &  \LambdaAR\    (\GeV)      &  0.254 & 0.024 \\
{\textsc{Ariadne} 4.06} & $\sQ$ (\GeV)        &  0.384 & 0.025 \\
                 & $b$ (\GeV$^{-2}$)          &  0.772 & 0.075 \\ \hline
                 & $\Lambda_{\text{\ME}}$ (\GeV) &  0.152 & 0.007 \\
{\textsc{Jetset} 7.4 ME} & $\sQ$ (\GeV)       &  0.430 & 0.026 \\
                 & $b$ (\GeV$^{-2}$)          &  0.310 & 0.016 \\ \hline
                 &  \LambdaMLLA\  (\GeV)      &  0.184 & 0.015 \\
{\textsc{Herwig} 5.9} &     \CLMAX\ (\GeV)    &  3.911 & 0.196 \\
                 &    \CLPOW                  &  2.000 & 0.482 \\ \hline
\end{tabular}\end{center}
\caption[Tuned parameters for the Monte Carlo models used in this study.]
        {Tuned parameters for the Monte Carlo models
        \cite{jetset-pythia,herwig59,ariadne}
        used in this study.
         }
\label{tab:tunmodel}
\end{table}
 
\begin{table}[htbp]\begin{center}
\begin{tabular}{|l|l|r@{$\;$}l|}\hline
\multicolumn{1}{|c|}{Model} & \multicolumn{1}{c|}{Parameter} &
                              \multicolumn{2}{c|}{Fit value}      \\ \hline
No BE     & \LambdaLLA\ (\GeV)           &  0.266 & $\pm\; 0.008$ \\
All       & $\sigma_\mathrm{Q}$ (\GeV)   &  0.393 & $\pm\; 0.004$ \\
flavours  & $b$                          &  0.874 & $\pm\; 0.014$ \\ \hline
No BE     & \LambdaLLA\ (\GeV)  &           0.258 & $\pm\; 0.002$ \\
udsc      & $\sigma_\mathrm{Q}$ (\GeV)   &  0.390 & $\pm\; 0.015$ \\
flavours  & $b$                          &  0.776 & $\pm\; 0.006$ \\ \hline
          & \LambdaLLA\ (\GeV)  &           0.270 & $\;^{+~0.002}_{-~0.004}$
                                                           \rule{0pt}{12pt} \\
\BEtt     & $\sigma_\mathrm{Q}$ (\GeV)   &  0.420 & $\pm\; 0.008$ \\
All       & $b$                          &  0.750 & $\pm\; 0.031$ \\
flavours  & $\lambda_\mathrm{BE}$        &  1.100 & $\;^{+~0.1}_{-~0.5}$\\
          & $r_\mathrm{BE}^{-1}$ (\GeV)  &  0.400 & $\pm\; 0.051$ \\ \hline
          & \LambdaLLA\ (\GeV)           &  0.268 & $\pm\; 0.003$ \\
\BEtt     & $\sigma_\mathrm{Q}$ (\GeV)   &  0.421 & $\pm\; 0.004$ \\
udsc      & $b$                          &  0.741 & $\pm\; 0.010$  \\
flavours  & $\lambda_\mathrm{BE}$        &  0.900 & $\;^{+~0.2}_{-~0.1}$ \\
          & $r_\mathrm{BE}^{-1}$ (\GeV)  &  0.425 & $\pm\; 0.041$ \\ \hline
\end{tabular}\end{center}
\caption[Tuned parameters for the {\textsc{Pythia}} 6.2 parton shower program
         for udsc-quarks and for all flavours, without Bose-Einstein correlations and with these
         correlations using the \BEtt\ Gaussian model.]
        {Tuned parameters for the {\textsc{Pythia}} 6.2 parton shower program \cite{pythiasix}
         for udsc-quarks and for all flavours, without Bose-Einstein correlations and with these
         correlations using the \BEtt\ Gaussian model \cite{lundbe1}.
         The cut-off parameter and the Lund
         fragmentaion parameter were kept fixed at {$Q_0 = 1.0\,\GeV$} and $a=0.5$
         and the Peterson fragmentation parameters for heavy quarks at
         $\epsilon_\mathrm{c}=0.03$ and $\epsilon_\mathrm{b}=0.002$.
         }
\label{tab:tunpythia}
\end{table}

\begin{table}[htbp]\begin{center}
\begin{tabular}{|c|l|r@{\;}l|r@{\;}l|}\cline{3-6}
\multicolumn{2}{c|}{\ }  & \multicolumn{4}{c|}{Fit Value}   \\  \cline{2-6}
\multicolumn{1}{c|}{\ }  & \multicolumn{1}{c|}{Parameter} &
                           \multicolumn{2}{c|}{All flavours} &
                           \multicolumn{2}{c|}{udsc  flavours} \\ \hline
               & $\Lambda_\mathrm{AR}$ (\GeV) & 0.223 & $\pm\; 0.002$
                                              & 0.225 & \multicolumn{1}{l|}{$^{+~0.002}_{-~0.003}$
                                                           \rule{0pt}{12pt} }    \\
               & $p_\mathrm{t}$ cut-off (\GeV)&  0.65 & \multicolumn{1}{l|}{$^{+~0.05}_{-~0.02}$}
                                              & 0.60  & $\pm\; 0.03$             \\
   No BE       & $\sigma_\mathrm{Q}$ (\GeV)   & 0.424 & $\pm\;  0.002$
                                              & 0.436 & $\pm\;  0.004$           \\
               & $a$                          & 0.106 & $\pm\;  0.006$
                                              & 0.112 & \multicolumn{1}{l|}{$^{+~0.006}_{-~0.012}$} \\
               & $b$                          & 0.62  & \multicolumn{1}{l|}{$^{+~0.02}_{-~0.04}$}
                                              & 0.62  & $\pm\; 0.02$                                \\
\hline
               & $\Lambda_\mathrm{AR}$ (\GeV) &  0.227 & $\pm\;  0.002$
                                              &  0.227 & $\pm\;  0.003$ \\
\hline
\end{tabular}\end{center}
\caption[Tuned parameters for \textsc{Ariadne} 4.12
        with hadronisation by \textsc{Pythia} 6.2
        for udsc-quarks and for all flavours,
        without Bose-Einstein correlations and using the \BEtt\ Gaussian  model.]
        {Tuned parameters for \textsc{Ariadne} 4.12 \cite{ariadne}
        with hadronisation by \textsc{Pythia} 6.2   \cite{pythiasix}
        for udsc-quarks and for all flavours,
        without Bose-Einstein correlations and using the \BEtt\ Gaussian model \cite{lundbe1}.}
\label{tab:tunariadne}
\end{table}

\begin{table}[htbp]\begin{center}
\end{center}
\caption[Tuned parameters for {\textsc{Herwig}} 6.2                  using the
           {\textsc{Pythia}} 6.2               particle decay routines
           without Bose-Einstein correlations and using the \BEtt\ Gaussian model.]
        {Tuned parameters for {\textsc{Herwig}} 6.2 \cite{herwigsix} using the
           {\textsc{Pythia}} 6.2 \cite{pythiasix} particle decay routines
           without Bose-Einstein correlations and using the \BEtt\ Gaussian model \cite{lundbe1}.
           In both cases the parameter \DECWT\ is fixed at the value 0.70.
           }
\label{tab:tunherwig}
\end{table}
\clearpage

\begin{table}{\small\begin{center}
\begin{tabular}{|c|r@{$\;\pm\;$}c@{$\;\pm\;$}l|r@{$\;\pm\;$}c@{$\;\pm\;$}l|r@{$\;\pm\;$}c@{$\;\pm\;$}l|r@{$\;\pm\;$}c@{$\;\pm\;$}l|}
\hline
 \ycJ\rule[-4pt]{0pt}{16pt}
& \multicolumn{3}{c|}{2-Jet Fraction} & \multicolumn{3}{c|}{3-Jet Fraction}
& \multicolumn{3}{c|}{4-Jet Fraction} & \multicolumn{3}{c|}{5-Jet Fraction} \\ \hline
0.001&0.013&0.006&0.002&0.161&0.019&0.002&0.316&0.024&0.002&0.271&0.022&0.004 \\
0.002&0.065&0.013&0.002&0.341&0.026&0.003&0.355&0.025&0.004&0.187&0.018&0.004 \\
0.004&0.148&0.018&0.002&0.478&0.031&0.003&0.293&0.022&0.002&0.070&0.010&0.002 \\
0.006&0.222&0.022&0.002&0.535&0.032&0.005&0.204&0.018&0.004&0.035&0.007&0.002 \\
0.008&0.287&0.024&0.002&0.529&0.031&0.003&0.155&0.016&0.005&0.029&0.007&0.001 \\
0.010&0.323&0.026&0.003&0.525&0.031&0.002&0.133&0.014&0.003&0.019&0.005&0.001 \\
0.020&0.491&0.032&0.007&0.445&0.028&0.005&0.061&0.009&0.003&0.003&0.002&0.001 \\
\hline\end{tabular}\end{center}}
\caption[Jet fraction using the \textsc{Jade} algorithm as a function of the jet resolution parameter \ycJ\ at $\rs= 130.1\,\GeV$.]
        {Jet fraction using the \textsc{Jade} algorithm as a function of the jet resolution parameter \ycJ\ at $\rs= 130.1\,\GeV$.
The first uncertainty is statistical, the second systematic.}
\label{tab:jade130}
\end{table}
 
\begin{table}{\small\begin{center}
\begin{tabular}{|c|r@{$\;\pm\;$}c@{$\;\pm\;$}l|r@{$\;\pm\;$}c@{$\;\pm\;$}l|r@{$\;\pm\;$}c@{$\;\pm\;$}l|r@{$\;\pm\;$}c@{$\;\pm\;$}l|}
\hline
 \ycJ\rule[-4pt]{0pt}{16pt}
 & \multicolumn{3}{c|}{2-Jet Fraction} & \multicolumn{3}{c|}{3-Jet Fraction}
 & \multicolumn{3}{c|}{4-Jet Fraction} & \multicolumn{3}{c|}{5-Jet Fraction} \\ \hline
0.001&0.019&0.007&0.001&0.175&0.022&0.003&0.305&0.028&0.004&0.236&0.024&0.001 \\
0.002&0.061&0.013&0.003&0.344&0.031&0.006&0.326&0.028&0.009&0.157&0.019&0.005 \\
0.004&0.149&0.021&0.002&0.464&0.034&0.002&0.272&0.025&0.005&0.092&0.014&0.002 \\
0.006&0.206&0.024&0.003&0.512&0.036&0.004&0.224&0.023&0.004&0.047&0.010&0.004 \\
0.008&0.266&0.027&0.003&0.523&0.036&0.001&0.175&0.020&0.005&0.034&0.009&0.001 \\
0.010&0.319&0.030&0.005&0.504&0.035&0.004&0.148&0.018&0.002&0.030&0.008&0.002 \\
0.020&0.507&0.037&0.004&0.406&0.031&0.005&0.080&0.012&0.003&0.008&0.004&0.001 \\
\hline\end{tabular}\end{center}}
\caption[Jet fraction using  the \textsc{Jade} algorithm as a function of the jet resolution parameter \ycJ\ at $\rs=136.1\,\GeV$.]
        {Jet fraction using  the \textsc{Jade} algorithm as a function of the jet resolution parameter \ycJ\ at $\rs=136.1\,\GeV$.
The first uncertainty is statistical, the second systematic.}
\label{tab:jade136}
\end{table}
 
\begin{table}{\small\begin{center}
\begin{tabular}{|c|r@{$\;\pm\;$}c@{$\;\pm\;$}l|r@{$\;\pm\;$}c@{$\;\pm\;$}l|r@{$\;\pm\;$}c@{$\;\pm\;$}l|r@{$\;\pm\;$}c@{$\;\pm\;$}l|}
\hline
 \ycJ\rule[-4pt]{0pt}{16pt}
& \multicolumn{3}{c|}{2-Jet Fraction} & \multicolumn{3}{c|}{3-Jet Fraction}
& \multicolumn{3}{c|}{4-Jet Fraction} & \multicolumn{3}{c|}{5-Jet Fraction} \\ \hline
0.001&0.062&0.016&0.006&0.228&0.029&0.009&0.313&0.030&0.005&0.262&0.026&0.001 \\
0.002&0.108&0.021&0.007&0.369&0.035&0.011&0.366&0.031&0.006&0.132&0.018&0.010 \\
0.004&0.196&0.027&0.006&0.493&0.039&0.010&0.257&0.025&0.005&0.054&0.013&0.008 \\
0.006&0.278&0.032&0.004&0.504&0.038&0.007&0.199&0.022&0.004&0.019&0.009&0.004 \\
0.008&0.337&0.035&0.007&0.503&0.037&0.006&0.153&0.020&0.008&0.007&0.007&0.002 \\
0.010&0.387&0.037&0.007&0.489&0.036&0.003&0.120&0.018&0.006&0.004&0.006&0.001 \\
0.020&0.529&0.042&0.010&0.420&0.033&0.009&0.051&0.013&0.002&\multicolumn{3}{c|}{} \\
\hline\end{tabular}\end{center}}
\caption[Jet fraction using  the \textsc{Jade} algorithm as a function of the jet resolution parameter \ycJ\ at $\rs=161.3\,\GeV$.]
        {Jet fraction using  the \textsc{Jade} algorithm as a function of the jet resolution parameter \ycJ\ at $\rs=161.3\,\GeV$.
The first uncertainty is statistical, the second systematic.}
\label{tab:jade161}
\end{table}

\begin{table}{\small\begin{center}
\begin{tabular}{|c|r@{$\;\pm\;$}c@{$\;\pm\;$}l|r@{$\;\pm\;$}c@{$\;\pm\;$}l|r@{$\;\pm\;$}c@{$\;\pm\;$}l|r@{$\;\pm\;$}c@{$\;\pm\;$}l|}
\hline
 \ycJ\rule[-4pt]{0pt}{16pt}
& \multicolumn{3}{c|}{2-Jet Fraction} & \multicolumn{3}{c|}{3-Jet Fraction}
& \multicolumn{3}{c|}{4-Jet Fraction} & \multicolumn{3}{c|}{5-Jet Fraction} \\  \hline
0.001&0.055&0.020&0.003&0.186&0.030&0.004&0.340&0.036&0.007&0.253&0.029&0.021 \\
0.002&0.115&0.026&0.009&0.342&0.037&0.012&0.290&0.031&0.012&0.183&0.025&0.009 \\
0.004&0.170&0.029&0.010&0.455&0.041&0.028&0.255&0.029&0.009&0.083&0.020&0.006 \\
0.006&0.245&0.034&0.006&0.466&0.040&0.019&0.200&0.026&0.009&0.049&0.017&0.006 \\
0.008&0.287&0.036&0.008&0.463&0.040&0.011&0.177&0.026&0.007&0.037&0.016&0.007 \\
0.010&0.323&0.038&0.012&0.448&0.039&0.009&0.157&0.025&0.007&0.038&0.017&0.009 \\
0.020&0.491&0.045&0.013&0.399&0.036&0.010&0.068&0.021&0.014&0.040&0.020&0.011 \\
\hline\end{tabular}\end{center}}
\caption[Jet fraction using  the \textsc{Jade} algorithm as a function of the jet resolution parameter \ycJ\ at $\rs=172.3\,\GeV$.]
        {Jet fraction using  the \textsc{Jade} algorithm as a function of the jet resolution parameter \ycJ\ at $\rs=172.3\,\GeV$.
The first uncertainty is statistical, the second systematic.}
\label{tab:jade172}
\end{table}
 
\begin{table}{\small\begin{center}
\begin{tabular}{|c|r@{$\;\pm\;$}c@{$\;\pm\;$}l|r@{$\;\pm\;$}c@{$\;\pm\;$}l|r@{$\;\pm\;$}c@{$\;\pm\;$}l|r@{$\;\pm\;$}c@{$\;\pm\;$}l|}
\hline
 \ycJ\rule[-4pt]{0pt}{16pt}
& \multicolumn{3}{c|}{2-Jet Fraction} & \multicolumn{3}{c|}{3-Jet Fraction}
& \multicolumn{3}{c|}{4-Jet Fraction} & \multicolumn{3}{c|}{5-Jet Fraction}  \\ \hline
0.001&0.028&0.004&0.002&0.209&0.012&0.002&0.345&0.016&0.004&0.230&0.014&0.004 \\
0.002&0.069&0.007&0.002&0.363&0.016&0.004&0.345&0.016&0.005&0.169&0.014&0.002 \\
0.004&0.154&0.011&0.003&0.477&0.019&0.002&0.284&0.016&0.007&0.077&0.011&0.007 \\
0.006&0.231&0.013&0.005&0.498&0.019&0.002&0.216&0.015&0.004&0.045&0.010&0.004 \\
0.008&0.293&0.015&0.005&0.493&0.019&0.002&0.177&0.014&0.004&0.026&0.009&0.006 \\
0.010&0.365&0.017&0.006&0.472&0.019&0.004&0.144&0.013&0.004&0.018&0.009&0.002 \\
0.020&0.514&0.020&0.006&0.413&0.018&0.004&0.067&0.011&0.005&0.006&0.005&0.001 \\
\hline\end{tabular}\end{center}}
\caption[Jet fraction using  the \textsc{Jade} algorithm as a function of the jet resolution parameter \ycJ\ at $\rs=182.8\,\GeV$.]
        {Jet fraction using  the \textsc{Jade} algorithm as a function of the jet resolution parameter \ycJ\ at $\rs=182.8\,\GeV$.
The first uncertainty is statistical, the second systematic.}
\label{tab:jade183}
\end{table}
 
\begin{table}{\small\begin{center}
\begin{tabular}{|c|r@{$\;\pm\;$}c@{$\;\pm\;$}l|r@{$\;\pm\;$}c@{$\;\pm\;$}l|r@{$\;\pm\;$}c@{$\;\pm\;$}l|r@{$\;\pm\;$}c@{$\;\pm\;$}l|}
\hline
 \ycJ\rule[-4pt]{0pt}{16pt}
& \multicolumn{3}{c|}{2-Jet Fraction} & \multicolumn{3}{c|}{3-Jet Fraction}
& \multicolumn{3}{c|}{4-Jet Fraction} & \multicolumn{3}{c|}{5-Jet Fraction}  \\ \hline
0.001&0.038&0.003&0.001&0.215&0.008&0.005&0.318&0.009&0.002&0.252&0.009&0.004 \\
0.002&0.082&0.005&0.003&0.356&0.010&0.006&0.345&0.010&0.009&0.166&0.008&0.004 \\
0.004&0.162&0.007&0.004&0.472&0.012&0.006&0.286&0.009&0.010&0.070&0.007&0.005 \\
0.006&0.235&0.009&0.007&0.503&0.012&0.004&0.220&0.009&0.006&0.042&0.007&0.002 \\
0.008&0.289&0.010&0.008&0.510&0.012&0.004&0.174&0.008&0.007&0.026&0.006&0.004 \\
0.010&0.334&0.010&0.009&0.511&0.012&0.007&0.142&0.008&0.005&0.012&0.005&0.003 \\
0.020&0.507&0.013&0.008&0.439&0.011&0.008&0.054&0.006&0.002&\multicolumn{3}{c|}{} \\
\hline\end{tabular}\end{center}}
\caption[Jet fraction using  the \textsc{Jade} algorithm as a function of the jet resolution parameter \ycJ\ at $\rs=188.6\,\GeV$.]
        {Jet fraction using  the \textsc{Jade} algorithm as a function of the jet resolution parameter \ycJ\ at $\rs=188.6\,\GeV$.
The first uncertainty is statistical, the second systematic.}
\label{tab:jade189}
\end{table}

\begin{table}{\small\begin{center}
\begin{tabular}{|c|r@{$\;\pm\;$}c@{$\;\pm\;$}l|r@{$\;\pm\;$}c@{$\;\pm\;$}l|r@{$\;\pm\;$}c@{$\;\pm\;$}l|r@{$\;\pm\;$}c@{$\;\pm\;$}l|}
\hline
 \ycJ\rule[-4pt]{0pt}{16pt}
& \multicolumn{3}{c|}{2-Jet Fraction} & \multicolumn{3}{c|}{3-Jet Fraction}
& \multicolumn{3}{c|}{4-Jet Fraction} & \multicolumn{3}{c|}{5-Jet Fraction}  \\ \hline
0.001&0.031&0.004&0.002&0.213&0.010&0.004&0.332&0.012&0.007&0.244&0.012&0.011 \\
0.002&0.078&0.006&0.004&0.351&0.012&0.004&0.375&0.014&0.009&0.142&0.012&0.008 \\
0.004&0.154&0.008&0.008&0.483&0.015&0.004&0.275&0.014&0.010&0.078&0.012&0.009 \\
0.006&0.218&0.010&0.010&0.531&0.016&0.007&0.217&0.013&0.006&0.034&0.012&0.004 \\
0.008&0.273&0.011&0.011&0.547&0.016&0.012&0.173&0.013&0.008&0.007&0.009&0.003 \\
0.010&0.321&0.012&0.010&0.527&0.016&0.011&0.151&0.013&0.005&0.001&0.007&0.001 \\
0.020&0.499&0.015&0.008&0.454&0.016&0.009&0.044&0.012&0.005&0.003&0.008&0.003 \\
\hline\end{tabular}\end{center}}
\caption[Jet fraction using  the \textsc{Jade} algorithm as a function of the jet resolution parameter \ycJ\ at $\rs=194.4\,\GeV$.]
        {Jet fraction using  the \textsc{Jade} algorithm as a function of the jet resolution parameter \ycJ\ at $\rs=194.4\,\GeV$.
The first uncertainty is statistical, the second systematic.}
\label{tab:jade194}
\end{table}

\begin{table}{\small\begin{center}
\begin{tabular}{|c|r@{$\;\pm\;$}c@{$\;\pm\;$}l|r@{$\;\pm\;$}c@{$\;\pm\;$}l|r@{$\;\pm\;$}c@{$\;\pm\;$}l|r@{$\;\pm\;$}c@{$\;\pm\;$}l|}
\hline
 \ycJ\rule[-4pt]{0pt}{16pt}
& \multicolumn{3}{c|}{2-Jet Fraction} & \multicolumn{3}{c|}{3-Jet Fraction}
& \multicolumn{3}{c|}{4-Jet Fraction} & \multicolumn{3}{c|}{5-Jet Fraction}  \\ \hline
0.001&0.036&0.004&0.002&0.207&0.009&0.007&0.324&0.012&0.006&0.248&0.012&0.006 \\
0.002&0.072&0.006&0.003&0.351&0.012&0.005&0.338&0.013&0.006&0.171&0.012&0.004 \\
0.004&0.155&0.009&0.007&0.473&0.015&0.004&0.282&0.013&0.006&0.073&0.010&0.007 \\
0.006&0.226&0.010&0.007&0.507&0.015&0.012&0.229&0.013&0.012&0.023&0.009&0.008 \\
0.008&0.285&0.012&0.010&0.522&0.016&0.012&0.182&0.013&0.012&0.010&0.008&0.009 \\
0.010&0.329&0.013&0.011&0.519&0.016&0.009&0.148&0.012&0.006&0.004&0.008&0.004 \\
0.020&0.493&0.015&0.011&0.435&0.015&0.009&0.072&0.012&0.006&\multicolumn{3}{c|}{} \\
\hline\end{tabular}\end{center}}
\caption[Jet fraction using  the \textsc{Jade} algorithm as a function of the jet resolution parameter \ycJ\ at $\rs=200.2\,\GeV$.]
        {Jet fraction using  the \textsc{Jade} algorithm as a function of the jet resolution parameter \ycJ\ at $\rs=200.2\,\GeV$.
The first uncertainty is statistical, the second systematic.}
\label{tab:jade200}
\end{table}
 
\begin{table}{\small\begin{center}
\begin{tabular}{|c|r@{$\;\pm\;$}c@{$\;\pm\;$}l|r@{$\;\pm\;$}c@{$\;\pm\;$}l|r@{$\;\pm\;$}c@{$\;\pm\;$}l|r@{$\;\pm\;$}c@{$\;\pm\;$}l|}
\hline
 \ycJ\rule[-4pt]{0pt}{16pt}
& \multicolumn{3}{c|}{2-Jet Fraction} & \multicolumn{3}{c|}{3-Jet Fraction}
& \multicolumn{3}{c|}{4-Jet Fraction} & \multicolumn{3}{c|}{5-Jet Fraction}  \\ \hline
0.001&0.030&0.003&0.002&0.216&0.008&0.004&0.337&0.010&0.004&0.240&0.009&0.005 \\
0.002&0.075&0.004&0.003&0.367&0.010&0.006&0.341&0.010&0.008&0.157&0.009&0.008 \\
0.004&0.149&0.007&0.005&0.478&0.011&0.002&0.279&0.010&0.009&0.080&0.009&0.003 \\
0.006&0.213&0.008&0.008&0.510&0.012&0.004&0.222&0.010&0.008&0.052&0.008&0.005 \\
0.008&0.272&0.009&0.010&0.518&0.012&0.007&0.180&0.010&0.007&0.025&0.007&0.008 \\
0.010&0.323&0.010&0.010&0.513&0.012&0.008&0.149&0.010&0.006&0.014&0.007&0.005 \\
0.020&0.496&0.012&0.009&0.451&0.012&0.007&0.054&0.009&0.007&\multicolumn{3}{c|}{} \\
\hline\end{tabular}\end{center}}
\caption[Jet fraction using  the \textsc{Jade} algorithm as a function of the jet resolution parameter \ycJ\ at $\rs=206.2\,\GeV$.]
        {Jet fraction using  the \textsc{Jade} algorithm as a function of the jet resolution parameter \ycJ\ at $\rs=206.2\,\GeV$.
The first uncertainty is statistical, the second systematic.}
\label{tab:jade206}
\end{table}
 
\clearpage
 
\begin{table}{\small\begin{center}
\begin{tabular}{|c|r@{$\;\pm\;$}c@{$\;\pm\;$}l|r@{$\;\pm\;$}c@{$\;\pm\;$}l|r@{$\;\pm\;$}c@{$\;\pm\;$}l|r@{$\;\pm\;$}c@{$\;\pm\;$}l|}
\hline
 \ycD\rule[-4pt]{0pt}{16pt}
& \multicolumn{3}{c|}{2-Jet Fraction} & \multicolumn{3}{c|}{3-Jet Fraction}
& \multicolumn{3}{c|}{4-Jet Fraction} & \multicolumn{3}{c|}{5-Jet Fraction}  \\ \hline
0.001&0.234&0.022&0.003&0.367&0.026&0.006&0.271&0.022&0.005&0.089&0.012&0.003 \\
0.002&0.395&0.028&0.005&0.369&0.026&0.003&0.195&0.018&0.007&0.031&0.007&0.002 \\
0.004&0.502&0.031&0.008&0.374&0.026&0.005&0.116&0.014&0.003&0.008&0.004&0.001 \\
0.006&0.580&0.034&0.006&0.349&0.025&0.004&0.068&0.010&0.003&0.003&0.002&0.001 \\
0.008&0.624&0.035&0.005&0.322&0.024&0.005&0.054&0.009&0.002&\multicolumn{3}{c|}{} \\
0.010&0.654&0.036&0.006&0.305&0.023&0.006&0.041&0.008&0.002&\multicolumn{3}{c|}{} \\
0.020&0.770&0.039&0.007&0.218&0.018&0.007&0.012&0.004&0.001&\multicolumn{3}{c|}{} \\
\hline\end{tabular}\end{center}}
\caption[Jet fraction using  the $\kt$ algorithm as a function of the jet resolution parameter \ycD\ at $\rs=130.1\,\GeV$.]
        {Jet fraction using  the $\kt$ algorithm as a function of the jet resolution parameter \ycD\ at $\rs=130.1\,\GeV$.
The first uncertainty is statistical, the second systematic.}
\label{tab:kt130}
\end{table}
 
\begin{table}{\small\begin{center}
\begin{tabular}{|c|r@{$\;\pm\;$}c@{$\;\pm\;$}l|r@{$\;\pm\;$}c@{$\;\pm\;$}l|r@{$\;\pm\;$}c@{$\;\pm\;$}l|r@{$\;\pm\;$}c@{$\;\pm\;$}l|}
\hline
 \ycD\rule[-4pt]{0pt}{16pt}
& \multicolumn{3}{c|}{2-Jet Fraction} & \multicolumn{3}{c|}{3-Jet Fraction}
& \multicolumn{3}{c|}{4-Jet Fraction} & \multicolumn{3}{c|}{5-Jet Fraction}  \\ \hline
0.001&0.224&0.025&0.008&0.370&0.030&0.006&0.219&0.023&0.004&0.115&0.017&0.002 \\
0.002&0.374&0.031&0.009&0.369&0.030&0.010&0.185&0.021&0.006&0.053&0.011&0.001 \\
0.004&0.515&0.036&0.007&0.352&0.029&0.009&0.101&0.015&0.007&0.032&0.009&0.003 \\
0.006&0.586&0.039&0.004&0.316&0.027&0.003&0.079&0.013&0.004&0.020&0.007&0.002 \\
0.008&0.622&0.040&0.003&0.303&0.027&0.002&0.062&0.012&0.004&0.013&0.006&0.002 \\
0.010&0.649&0.041&0.003&0.290&0.026&0.003&0.051&0.010&0.004&0.011&0.006&0.002 \\
0.020&0.763&0.045&0.003&0.209&0.021&0.003&0.026&0.008&0.003&0.002&0.002&0.001 \\
\hline\end{tabular}\end{center}}
\caption[Jet fraction using  the $\kt$ algorithm as a function of the jet resolution parameter \ycD\ at $\rs=136.1\,\GeV$.]
        {Jet fraction using  the $\kt$ algorithm as a function of the jet resolution parameter \ycD\ at $\rs=136.1\,\GeV$.
The first uncertainty is statistical, the second systematic.}
\label{tab:kt136}
\end{table}
 
\begin{table}{\small\begin{center}
\begin{tabular}{|c|r@{$\;\pm\;$}c@{$\;\pm\;$}l|r@{$\;\pm\;$}c@{$\;\pm\;$}l|r@{$\;\pm\;$}c@{$\;\pm\;$}l|r@{$\;\pm\;$}c@{$\;\pm\;$}l|}
\hline
 \ycD\rule[-4pt]{0pt}{16pt}
& \multicolumn{3}{c|}{2-Jet Fraction} & \multicolumn{3}{c|}{3-Jet Fraction}
& \multicolumn{3}{c|}{4-Jet Fraction} & \multicolumn{3}{c|}{5-Jet Fraction}  \\ \hline
0.001&0.252&0.030&0.008&0.382&0.033&0.001&0.253&0.026&0.006&0.107&0.017&0.002 \\
0.002&0.343&0.034&0.012&0.470&0.035&0.010&0.160&0.022&0.006&0.027&0.010&0.004 \\
0.004&0.491&0.039&0.013&0.403&0.033&0.010&0.106&0.019&0.004&\multicolumn{3}{c|}{} \\
0.006&0.577&0.042&0.013&0.370&0.031&0.014&0.052&0.014&0.006&\multicolumn{3}{c|}{} \\
0.008&0.619&0.044&0.014&0.349&0.030&0.014&0.032&0.012&0.002&\multicolumn{3}{c|}{} \\
0.010&0.656&0.045&0.013&0.323&0.029&0.015&0.021&0.011&0.003&\multicolumn{3}{c|}{} \\
0.020&0.763&0.048&0.004&0.228&0.024&0.005&0.009&0.008&0.003&\multicolumn{3}{c|}{} \\
\hline\end{tabular}\end{center}}
\caption[Jet fraction using  the $\kt$ algorithm as a function of the jet resolution parameter \ycD\ at $\rs = 161.3\,\GeV$.]
        {Jet fraction using  the $\kt$ algorithm as a function of the jet resolution parameter \ycD\ at $\rs = 161.3\,\GeV$.
The first uncertainty is statistical, the second systematic.}
\label{tab:kt161}
\end{table}
 
\begin{table}[htbp]{\small\begin{center}
\begin{tabular}{|c|r@{$\;\pm\;$}c@{$\;\pm\;$}l|r@{$\;\pm\;$}c@{$\;\pm\;$}l|r@{$\;\pm\;$}c@{$\;\pm\;$}l|r@{$\;\pm\;$}c@{$\;\pm\;$}l|}
\hline
 \ycD\rule[-4pt]{0pt}{16pt}
& \multicolumn{3}{c|}{2-Jet Fraction} & \multicolumn{3}{c|}{3-Jet Fraction}
& \multicolumn{3}{c|}{4-Jet Fraction} & \multicolumn{3}{c|}{5-Jet Fraction}  \\ \hline
0.001&0.252&0.033&0.009&0.363&0.036&0.013&0.255&0.030&0.014&0.095&0.021&0.005 \\
0.002&0.365&0.038&0.005&0.361&0.035&0.010&0.193&0.028&0.007&0.036&0.017&0.007 \\
0.004&0.458&0.041&0.010&0.335&0.033&0.013&0.135&0.027&0.007&0.011&0.014&0.004 \\
0.006&0.454&0.038&0.016&0.273&0.028&0.029&0.047&0.018&0.014&0.014&0.014&0.005 \\
0.008&0.619&0.048&0.021&0.293&0.032&0.005&0.033&0.020&0.010&0.048&0.025&0.014 \\
0.010&0.657&0.049&0.019&0.252&0.030&0.004&0.032&0.021&0.006&0.052&0.025&0.014 \\
0.020&0.707&0.050&0.025&0.185&0.026&0.006&0.093&0.042&0.021&0.015&0.009&0.007 \\
\hline\end{tabular}\end{center}}
\caption[Jet fraction using  the $\kt$ algorithm as a function of the jet resolution parameter \ycD\ at $\rs = 172.3\,\GeV$.]
        {Jet fraction using  the $\kt$ algorithm as a function of the jet resolution parameter \ycD\ at $\rs = 172.3\,\GeV$.
The first uncertainty is statistical, the second systematic.}
\label{tab:kt172}
\end{table}

\begin{table}[htbp]{\small\begin{center}
\begin{tabular}{|c|r@{$\;\pm\;$}c@{$\;\pm\;$}l|r@{$\;\pm\;$}c@{$\;\pm\;$}l|r@{$\;\pm\;$}c@{$\;\pm\;$}l|r@{$\;\pm\;$}c@{$\;\pm\;$}l|}
\hline
 \ycD\rule[-4pt]{0pt}{16pt}
& \multicolumn{3}{c|}{2-Jet Fraction} & \multicolumn{3}{c|}{3-Jet Fraction}
& \multicolumn{3}{c|}{4-Jet Fraction} & \multicolumn{3}{c|}{5-Jet Fraction} \\ \hline
0.001&0.265&0.014&0.005&0.385&0.017&0.003&0.241&0.015&0.008&0.082&0.011&0.003 \\
0.002&0.392&0.017&0.007&0.396&0.017&0.004&0.159&0.014&0.011&0.040&0.010&0.004 \\
0.004&0.524&0.020&0.009&0.347&0.016&0.003&0.112&0.013&0.009&0.014&0.009&0.002 \\
0.006&0.594&0.021&0.007&0.320&0.016&0.003&0.080&0.013&0.007&0.001&0.007&0.001 \\
0.008&0.640&0.022&0.006&0.287&0.015&0.004&0.071&0.013&0.003&0.002&0.007&0.002 \\
0.010&0.668&0.023&0.003&0.271&0.015&0.004&0.061&0.013&0.005&\multicolumn{3}{c|}{} \\
0.020&0.785&0.025&0.004&0.201&0.013&0.002&0.014&0.010&0.004&0.001&0.003&0.001 \\
\hline\end{tabular}\end{center}}
\caption[Jet fraction using  the $\kt$ algorithm as a function of the jet resolution parameter \ycD\ at $\rs= 182.8\,\GeV$.]
        {Jet fraction using  the $\kt$ algorithm as a function of the jet resolution parameter \ycD\ at $\rs= 182.8\,\GeV$.
The first uncertainty is statistical, the second systematic.}
\label{tab:kt183}
\end{table}
 
\begin{table}[htbp]{\small\begin{center}
\begin{tabular}{|c|r@{$\;\pm\;$}c@{$\;\pm\;$}l|r@{$\;\pm\;$}c@{$\;\pm\;$}l|r@{$\;\pm\;$}c@{$\;\pm\;$}l|r@{$\;\pm\;$}c@{$\;\pm\;$}l|}
\hline
 \ycD\rule[-4pt]{0pt}{16pt}
& \multicolumn{3}{c|}{2-Jet Fraction} & \multicolumn{3}{c|}{3-Jet Fraction}
& \multicolumn{3}{c|}{4-Jet Fraction} & \multicolumn{3}{c|}{5-Jet Fraction}  \\ \hline
0.001&0.267&0.009&0.005&0.386&0.010&0.003&0.231&0.009&0.007&0.095&0.007&0.005 \\
0.002&0.395&0.011&0.007&0.399&0.010&0.005&0.161&0.008&0.006&0.043&0.007&0.004 \\
0.004&0.520&0.013&0.009&0.373&0.010&0.007&0.099&0.008&0.005&0.007&0.005&0.003 \\
0.006&0.582&0.013&0.010&0.346&0.010&0.007&0.068&0.007&0.006&\multicolumn{3}{c|}{} \\
0.008&0.634&0.014&0.010&0.325&0.010&0.006&0.040&0.007&0.005&0.001&0.004&0.002 \\
0.010&0.670&0.014&0.008&0.298&0.009&0.005&0.031&0.006&0.004&\multicolumn{3}{c|}{} \\
0.020&0.786&0.015&0.009&0.206&0.008&0.008&0.007&0.006&0.005&\multicolumn{3}{c|}{} \\
\hline\end{tabular}\end{center}}
\caption[Jet fraction using  the $\kt$ algorithm as a function of the jet resolution parameter \ycD\ at $\rs=188.6\,\GeV$.]
        {Jet fraction using  the $\kt$ algorithm as a function of the jet resolution parameter \ycD\ at $\rs=188.6\,\GeV$.
The first uncertainty is statistical, the second systematic.}
\label{tab:kt189}
\end{table}
 
\begin{table}[htbp]{\small\begin{center}
\begin{tabular}{|c|r@{$\;\pm\;$}c@{$\;\pm\;$}l|r@{$\;\pm\;$}c@{$\;\pm\;$}l|r@{$\;\pm\;$}c@{$\;\pm\;$}l|r@{$\;\pm\;$}c@{$\;\pm\;$}l|}
\hline
 \ycD\rule[-4pt]{0pt}{16pt}
& \multicolumn{3}{c|}{2-Jet Fraction} & \multicolumn{3}{c|}{3-Jet Fraction}
& \multicolumn{3}{c|}{4-Jet Fraction} & \multicolumn{3}{c|}{5-Jet Fraction}  \\ \hline
0.001&0.259&0.011&0.011&0.397&0.013&0.007&0.228&0.012&0.017&0.080&0.011&0.010 \\
0.002&0.389&0.013&0.013&0.416&0.014&0.009&0.161&0.014&0.016&0.028&0.011&0.005 \\
0.004&0.511&0.015&0.015&0.404&0.014&0.011&0.079&0.015&0.021&0.002&0.006&0.005 \\
0.006&0.581&0.017&0.014&0.368&0.014&0.009&0.051&0.016&0.011&\multicolumn{3}{c|}{} \\
0.008&0.630&0.017&0.010&0.326&0.014&0.012&0.045&0.017&0.018&\multicolumn{3}{c|}{} \\
0.010&0.679&0.018&0.013&0.303&0.014&0.006&0.018&0.016&0.013&\multicolumn{3}{c|}{} \\
0.020&0.766&0.019&0.013&0.215&0.012&0.011&0.019&0.018&0.018&\multicolumn{3}{c|}{} \\
\hline\end{tabular}\end{center}}
\caption[Jet fraction using  the $\kt$ algorithm as a function of the jet resolution parameter \ycD\ at $\rs = 194.4\,\GeV$.]
        {Jet fraction using  the $\kt$ algorithm as a function of the jet resolution parameter \ycD\ at $\rs = 194.4\,\GeV$.
The first uncertainty is statistical, the second systematic.}
\label{tab:kt194}
\end{table}
 
\begin{table}[htbp]{\small\begin{center}
\begin{tabular}{|c|r@{$\;\pm\;$}c@{$\;\pm\;$}l|r@{$\;\pm\;$}c@{$\;\pm\;$}l|r@{$\;\pm\;$}c@{$\;\pm\;$}l|r@{$\;\pm\;$}c@{$\;\pm\;$}l|}
\hline
 \ycD\rule[-4pt]{0pt}{16pt}
& \multicolumn{3}{c|}{2-Jet Fraction} & \multicolumn{3}{c|}{3-Jet Fraction}
& \multicolumn{3}{c|}{4-Jet Fraction} & \multicolumn{3}{c|}{5-Jet Fraction}  \\ \hline
0.001&0.267&0.011&0.007&0.369&0.013&0.009&0.245&0.013&0.008&0.104&0.011&0.004 \\
0.002&0.372&0.013&0.010&0.398&0.014&0.011&0.179&0.013&0.010&0.048&0.010&0.007 \\
0.004&0.492&0.015&0.013&0.396&0.014&0.007&0.104&0.013&0.007&0.006&0.008&0.007 \\
0.006&0.573&0.016&0.015&0.368&0.014&0.006&0.054&0.012&0.011&0.005&0.007&0.004 \\
0.008&0.619&0.017&0.015&0.340&0.014&0.008&0.041&0.012&0.009&\multicolumn{3}{c|}{} \\
0.010&0.661&0.018&0.016&0.313&0.014&0.011&0.026&0.012&0.011&\multicolumn{3}{c|}{} \\
0.020&0.764&0.019&0.013&0.226&0.012&0.010&0.010&0.014&0.006&\multicolumn{3}{c|}{} \\
\hline\end{tabular}\end{center}}
\caption[Jet fraction using  the $\kt$ algorithm as a function of the jet resolution parameter \ycD\ at $\rs = 200.2\,\GeV$.]
        {Jet fraction using  the $\kt$ algorithm as a function of the jet resolution parameter \ycD\ at $\rs = 200.2\,\GeV$.
The first uncertainty is statistical, the second systematic.}
\label{tab:kt200}
\end{table}
 
\begin{table}[htbp]{\small\begin{center}
\begin{tabular}{|c|r@{$\;\pm\;$}c@{$\;\pm\;$}l|r@{$\;\pm\;$}c@{$\;\pm\;$}l|r@{$\;\pm\;$}c@{$\;\pm\;$}l|r@{$\;\pm\;$}c@{$\;\pm\;$}l|}
\hline
 \ycD\rule[-4pt]{0pt}{16pt}
& \multicolumn{3}{c|}{2-Jet Fraction} & \multicolumn{3}{c|}{3-Jet Fraction}
& \multicolumn{3}{c|}{4-Jet Fraction} & \multicolumn{3}{c|}{5-Jet Fraction}  \\ \hline
0.001&0.251&0.008&0.005&0.394&0.010&0.005&0.237&0.010&0.011&0.093&0.008&0.004 \\
0.002&0.379&0.010&0.008&0.412&0.011&0.006&0.169&0.010&0.009&0.034&0.008&0.004 \\
0.004&0.509&0.012&0.011&0.390&0.011&0.007&0.090&0.010&0.013&0.011&0.007&0.002 \\
0.006&0.583&0.013&0.013&0.367&0.011&0.006&0.048&0.009&0.014&0.002&0.005&0.002 \\
0.008&0.627&0.014&0.013&0.343&0.011&0.008&0.030&0.009&0.013&\multicolumn{3}{c|}{} \\
0.010&0.666&0.014&0.012&0.316&0.011&0.009&0.018&0.009&0.010&\multicolumn{3}{c|}{} \\
0.020&0.765&0.015&0.009&0.234&0.010&0.004&0.001&0.009&0.001&\multicolumn{3}{c|}{} \\
\hline\end{tabular}\end{center}}
\caption[Jet fraction using  the $\kt$ algorithm as a function of the jet resolution parameter \ycD\ at $\rs = 206.2\,\GeV$.]
        {Jet fraction using  the $\kt$ algorithm as a function of the jet resolution parameter \ycD\ at $\rs = 206.2\,\GeV$.
The first uncertainty is statistical, the second systematic.}
\label{tab:kt206}
\end{table}
 
\begin{table}[htbp]{\small\begin{center}
\begin{tabular}{|c|r@{$\;\pm\;$}c@{$\;\pm\;$}l|r@{$\;\pm\;$}c@{$\;\pm\;$}l|r@{$\;\pm\;$}c@{$\;\pm\;$}l|r@{$\;\pm\;$}c@{$\;\pm\;$}l|}
\hline
 \ycD\rule[-4pt]{0pt}{16pt}
& \multicolumn{3}{c|}{2-Jet Fraction} & \multicolumn{3}{c|}{3-Jet Fraction}
& \multicolumn{3}{c|}{4-Jet Fraction} & \multicolumn{3}{c|}{5-Jet Fraction}  \\ \hline
0.001&0.336&0.013&0.008&0.380&0.013&0.007&0.214&0.013&0.010&0.064&0.010&0.004 \\
0.002&0.435&0.014&0.010&0.379&0.014&0.011&0.157&0.013&0.013&0.024&0.009&0.006 \\
0.004&0.538&0.016&0.013&0.367&0.014&0.011&0.088&0.012&0.010&0.005&0.008&0.004 \\
0.006&0.605&0.017&0.015&0.339&0.014&0.008&0.050&0.011&0.010&0.006&0.007&0.005 \\
0.008&0.647&0.018&0.014&0.317&0.014&0.008&0.035&0.011&0.007&\multicolumn{3}{c|}{} \\
0.010&0.680&0.018&0.015&0.294&0.013&0.011&0.026&0.011&0.010&\multicolumn{3}{c|}{} \\
0.020&0.781&0.020&0.013&0.212&0.012&0.009&0.005&0.007&0.004&0.001&0.002&0.001 \\
\hline\end{tabular}\end{center}}
\caption[Jet fraction using  the Cambridge algorithm as a function of the jet resolution parameter \ycD\ at $\rs= 200.2\,\GeV$.]
        {Jet fraction using  the Cambridge algorithm as a function of the jet resolution parameter \ycD\ at $\rs= 200.2\,\GeV$.
The first uncertainty is statistical, the second systematic.}
\label{tab:camb200}
\end{table}
 
\begin{table}[htbp]{\small\begin{center}
\begin{tabular}{|c|r@{$\;\pm\;$}c@{$\;\pm\;$}l|r@{$\;\pm\;$}c@{$\;\pm\;$}l|r@{$\;\pm\;$}c@{$\;\pm\;$}l|r@{$\;\pm\;$}c@{$\;\pm\;$}l|}
\hline
 \ycD\rule[-4pt]{0pt}{16pt}
& \multicolumn{3}{c|}{2-Jet Fraction} & \multicolumn{3}{c|}{3-Jet Fraction}
& \multicolumn{3}{c|}{4-Jet Fraction} & \multicolumn{3}{c|}{5-Jet Fraction} \\ \hline
0.001&0.339&0.010&0.007&0.401&0.011&0.008&0.183&0.009&0.012&0.073&0.008&0.007 \\
0.002&0.434&0.011&0.009&0.408&0.011&0.007&0.132&0.009&0.011&0.024&0.007&0.003 \\
0.004&0.551&0.013&0.011&0.374&0.011&0.006&0.070&0.009&0.011&0.005&0.006&0.003 \\
0.006&0.608&0.013&0.012&0.344&0.011&0.007&0.045&0.009&0.013&0.002&0.005&0.002 \\
0.008&0.651&0.014&0.012&0.320&0.011&0.006&0.029&0.009&0.012&\multicolumn{3}{c|}{} \\
0.010&0.688&0.014&0.012&0.292&0.010&0.006&0.020&0.009&0.010&\multicolumn{3}{c|}{} \\
0.020&0.779&0.015&0.008&0.213&0.009&0.004&0.008&0.009&0.006&\multicolumn{3}{c|}{} \\
\hline\end{tabular}\end{center}}
\caption[Jet fraction using  the Cambridge algorithm as a function of the jet resolution parameter \ycD\ at $\rs= 206.2\,\GeV$.]
        {Jet fraction using  the Cambridge algorithm as a function of the jet resolution parameter \ycD\ at $\rs= 206.2\,\GeV$.
The first uncertainty is statistical, the second systematic.}
\label{tab:camb206}
\end{table}
 
\clearpage

\begin{table}{\small\begin{center}
\begin{tabular}{|@{\hspace{1.5em}}r@{--}l|r@{$\;\pm$}c@{$\pm\;$}l|r@{$\;\pm$}c@{$\pm\;$}l|r@{$\;\pm$}c@{$\pm\;$}l|}
\cline{3-11}
\multicolumn{2}{c|}{} &
\multicolumn{9}{c|}{\rule[-13pt]{0pt}{32pt}${\displaystyle\frac{1}{\sigma}\cdot\frac{\ds}{\dd{(\onet)}}}$}
\\ \hline
\multicolumn{2}{|c|}{\onet} & \multicolumn{3}{c|}{\rule{0pt}{11pt}at $\rs = 41.4\,\GeV$} &
\multicolumn{3}{c|}{at $\rs = 55.3\,\GeV$} &
\multicolumn{3}{c|}{at $\rs = 65.4\,\GeV$} \\  \hline
0.000&0.025&  1.636& 0.322& 0.591&  3.604& 0.491& 1.252&  4.737& 0.392& 1.292 \\
0.025&0.045&  9.120& 0.914& 1.122& 14.245& 1.066& 2.225& 12.412& 0.716& 1.566 \\
0.045&0.065&  9.768& 1.084& 1.612&  8.897& 0.789& 0.920&  9.618& 0.606& 1.444 \\
0.065&0.085&  7.872& 0.919& 1.014&  6.593& 0.714& 1.244&  7.229& 0.600& 0.534 \\
0.085&0.115&  4.156& 0.478& 0.272&  2.989& 0.439& 0.588&  3.604& 0.317& 0.768 \\
0.115&0.145&  3.547& 0.452& 0.699&  2.237& 0.310& 1.434&  1.722& 0.217& 0.507 \\
0.145&0.175&  2.240& 0.393& 0.376&  2.229& 0.371& 0.539&  1.172& 0.180& 0.240 \\
0.175&0.210&  1.244& 0.292& 0.329&  1.112& 0.254& 0.231&  0.902& 0.161& 0.222 \\
0.210&0.250&  0.882& 0.189& 0.182&  0.531& 0.166& 0.160&  0.809& 0.149& 0.144 \\
0.250&0.290&  0.598& 0.185& 0.133&  0.420& 0.146& 0.121&  0.417& 0.087& 0.207 \\
0.290&0.330&  0.283& 0.137& 0.147&  0.140& 0.084& 0.072&  0.269& 0.089& 0.057 \\
0.330&0.370&  0.218& 0.065& 0.086&  0.168& 0.073& 0.064&  0.253& 0.080& 0.089 \\
0.370&0.420&  0.056& 0.032& 0.055&  0.046& 0.033& 0.023&  0.000& 0.000& 0.000 \\
\hline\multicolumn{2}{|l|}{First Moment} &
 0.0971&0.0030&0.0034& 0.0811&0.0027&0.0029& 0.0796&0.0021&0.0051 \\
\multicolumn{2}{|l|}{Second Moment} &
 0.0143&0.0009&0.0015& 0.0109&0.0008&0.0006& 0.0109&0.0006&0.0010 \\
\hline\end{tabular}\end{center}}
\caption[Differential distribution for event thrust at $\rs = 41.4$, 55.3 and 65.4\,\GeV.]
        {Differential distribution for event thrust at $\rs = 41.4$, 55.3 and 65.4\,\GeV.
The first uncertainty is statistical, the second systematic.}
\label{tab:thrustqqg1}
\end{table}
 

\begin{table}{\small\begin{center}
\begin{tabular}{|@{\hspace{1.5em}}r@{--}l|r@{$\;\pm$}c@{$\pm\;$}l|r@{$\;\pm$}c@{$\pm\;$}l|r@{$\;\pm$}c@{$\pm\;$}l|}
\cline{3-11}
\multicolumn{2}{c|}{} &
\multicolumn{9}{c|}{\rule[-13pt]{0pt}{32pt}${\displaystyle\frac{1}{\sigma}\cdot\frac{\ds}{\dd{(\onet)}}}$}
\\ \hline
\multicolumn{2}{|c|}{\onet} & \multicolumn{3}{c|}{\rule{0pt}{11pt}at $\rs = 75.7\,\GeV$} &
\multicolumn{3}{c|}{at $\rs = 82.3\,\GeV$} &
\multicolumn{3}{c|}{at $\rs = 85.1\,\GeV$}  \\ \hline
0.000&0.025&  5.754& 0.287& 1.262&  6.277& 0.364& 1.402&  7.602& 0.480& 1.411 \\
0.025&0.045& 15.307& 0.564& 1.003& 15.551& 0.655& 0.472& 15.215& 0.753& 0.801 \\
0.045&0.065&  7.994& 0.406& 1.077&  8.884& 0.509& 0.346&  8.031& 0.619& 0.908 \\
0.065&0.085&  5.056& 0.338& 0.325&  4.322& 0.384& 1.376&  3.804& 0.427& 0.533 \\
0.085&0.115&  3.569& 0.239& 0.418&  3.408& 0.286& 0.117&  3.078& 0.342& 0.267 \\
0.115&0.145&  1.932& 0.183& 0.269&  1.485& 0.203& 0.346&  1.641& 0.260& 1.148 \\
0.145&0.175&  1.462& 0.157& 0.162&  1.460& 0.169& 0.314&  1.563& 0.237& 0.398 \\
0.175&0.210&  0.944& 0.121& 0.173&  0.894& 0.143& 0.210&  1.116& 0.190& 0.319 \\
0.210&0.250&  0.491& 0.082& 0.120&  0.617& 0.140& 0.126&  0.502& 0.159& 0.147 \\
0.250&0.290&  0.397& 0.084& 0.094&  0.354& 0.110& 0.113&  0.296& 0.114& 0.042 \\
0.290&0.330&  0.186& 0.058& 0.036&  0.104& 0.055& 0.033&  0.191& 0.073& 0.080 \\
0.330&0.370&  0.092& 0.039& 0.027&  0.078& 0.033& 0.025&  0.047& 0.034& 0.025 \\
0.370&0.420&  0.009& 0.009& 0.005&  0.000& 0.000& 0.000&  0.000& 0.000& 0.000 \\
\hline\multicolumn{2}{|l|}{First Moment} &
 0.0731&0.0015&0.0045& 0.0700&0.0018&0.0046& 0.0691&0.0022&0.0088 \\
\multicolumn{2}{|l|}{Second Moment} &
 0.0093&0.0004&0.0010& 0.0086&0.0005&0.0010& 0.0086&0.0006&0.0020 \\
\hline\end{tabular}\end{center}}
\caption[Differential distribution for event thrust at $\rs = 75.7$, 82.3 and 85.1\,\GeV.]
        {Differential distribution for event thrust at $\rs = 75.7$, 82.3 and 85.1\,\GeV.
The first uncertainty is statistical, the second systematic.}
\label{tab:thrustqqg2}
\end{table}
 
\begin{table}{\small\begin{center}
\begin{tabular}{|@{\hspace{1.5em}}r@{--}l|r@{$\;\pm$}c@{$\pm\;$}l|r@{$\;\pm$}c@{$\pm\;$}l|r@{$\;\pm$}c@{$\pm\;$}l|}
\cline{3-11}
\multicolumn{2}{c|}{} &
\multicolumn{9}{c|}{\rule[-13pt]{0pt}{32pt}${\displaystyle\frac{1}{\sigma}\cdot\frac{\ds}{\dd{(\onet)}}}$}
\\ \hline
\multicolumn{2}{|c|}{\onet} & \multicolumn{3}{c|}{\rule{0pt}{11pt}at $\rs =130.1\,\GeV$} &
\multicolumn{3}{c|}{at $\rs = 136.1\,\GeV$} &
\multicolumn{3}{c|}{at $\rs = 161.3\,\GeV$}  \\ \hline
0.000& 0.025& 14.875& 1.223& 1.053& 14.171& 1.371& 0.929& 19.149& 1.709& 0.557 \\
0.025& 0.050& 11.125& 0.900& 0.613& 11.992& 1.107& 0.920&  8.102& 0.922& 0.564 \\
0.050& 0.075&  5.057& 0.615& 0.376&  4.622& 0.688& 0.331&  3.535& 0.582& 0.253 \\
0.075& 0.100&  2.763& 0.439& 0.144&  2.174& 0.464& 0.104&  3.123& 0.543& 0.202 \\
0.100& 0.125&  2.046& 0.352& 0.189&  2.099& 0.439& 0.144&  1.626& 0.397& 0.159 \\
0.125& 0.150&  1.552& 0.303& 0.063&  0.941& 0.310& 0.100&  1.331& 0.362& 0.087 \\
\hline\end{tabular}\end{center}}
\caption[Differential distribution for event thrust at $\rs = 130.1$, 136.1 and 161.3\,\GeV.]
        {Differential distribution for event thrust at $\rs = 130.1$, 136.1 and 161.3\,\GeV.
The first uncertainty is statistical, the second systematic.}
\label{tab:thrust1}
\end{table}
 
\begin{table}{\small\begin{center}
\begin{tabular}{|@{\hspace{1.5em}}r@{--}l|r@{$\;\pm$}c@{$\pm\;$}l|r@{$\;\pm$}c@{$\pm\;$}l|r@{$\;\pm$}c@{$\pm\;$}l|}
\cline{3-11}
\multicolumn{2}{c|}{} &
\multicolumn{9}{c|}{\rule[-13pt]{0pt}{32pt}${\displaystyle\frac{1}{\sigma}\cdot\frac{\ds}{\dd{(\onet)}}}$}
\\ \hline
\multicolumn{2}{|c|}{\onet} & \multicolumn{3}{c|}{\rule{0pt}{11pt}at $\rs =172.3\,\GeV$} &
\multicolumn{3}{c|}{at $\rs = 182.8\,\GeV$} &
\multicolumn{3}{c|}{at $\rs = 188.6\,\GeV$}  \\ \hline
0.000& 0.025& 17.499& 1.879& 0.867& 16.920& 0.740& 1.403& 16.067& 0.471& 0.436 \\
0.025& 0.050&  8.458& 1.016& 0.874&  9.107& 0.534& 0.851&  9.908& 0.330& 0.175 \\
0.050& 0.075&  4.705& 0.774& 0.336&  5.099& 0.400& 0.494&  4.641& 0.223& 0.132 \\
0.075& 0.100&  2.454& 0.569& 0.327&  2.655& 0.295& 0.065&  2.940& 0.183& 0.079 \\
0.100& 0.125&  2.087& 0.557& 0.192&  1.896& 0.253& 0.066&  1.721& 0.141& 0.053 \\
0.125& 0.150&  1.550& 0.490& 0.277&  1.328& 0.214& 0.196&  1.494& 0.134& 0.140 \\
\hline\end{tabular}\end{center}}
\caption[Differential distribution for event thrust at $\rs= 172.3$, 182.8 and 188.6\,\GeV.]
        {Differential distribution for event thrust at $\rs= 172.3$, 182.8 and 188.6\,\GeV.
The first uncertainty is statistical, the second systematic.}
\label{tab:thrust2}
\end{table}
 
\begin{table}[htbp]{\small\begin{center}
\begin{tabular}{|@{\hspace{1.5em}}r@{--}l|r@{$\;\pm$}c@{$\pm\;$}l|r@{$\;\pm$}c@{$\pm\;$}l|r@{$\;\pm$}c@{$\pm\;$}l|}
\cline{3-11}
\multicolumn{2}{c|}{} &
\multicolumn{9}{c|}{\rule[-13pt]{0pt}{32pt}${\displaystyle\frac{1}{\sigma}\cdot\frac{\ds}{\dd{(\onet)}}}$}
\\ \hline
\multicolumn{2}{|c|}{\onet} & \multicolumn{3}{c|}{\rule{0pt}{11pt}at $\rs =194.4\,\GeV$} &
\multicolumn{3}{c|}{at $\rs = 200.2\,\GeV$} &
\multicolumn{3}{c|}{at $\rs = 206.2\,\GeV$}  \\ \hline
0.000& 0.025& 16.177& 0.571& 0.436& 16.119& 0.578& 0.444& 16.202& 0.466& 0.509 \\
0.025& 0.050&  9.285& 0.416& 0.248&  9.277& 0.420& 0.270&  9.321& 0.326& 0.297 \\
0.050& 0.075&  5.302& 0.340& 0.261&  4.534& 0.305& 0.175&  4.637& 0.240& 0.169 \\
0.075& 0.100&  2.823& 0.253& 0.175&  2.947& 0.280& 0.200&  3.013& 0.198& 0.106 \\
0.100& 0.125&  1.837& 0.224& 0.152&  1.904& 0.204& 0.129&  1.707& 0.155& 0.101 \\
0.125& 0.150&  1.229& 0.188& 0.159&  1.636& 0.195& 0.263&  1.553& 0.151& 0.085 \\
\hline\end{tabular}\end{center}}
\caption[Differential distribution for event thrust at $\rs = 194.4$, 200.2 and 206.2\,\GeV.]
        {Differential distribution for event thrust at $\rs = 194.4$, 200.2 and 206.2\,\GeV.
The first uncertainty is statistical, the second systematic.}
\label{tab:thrust3}
\end{table}

\begin{table}{\small\begin{center}
\begin{tabular}{|@{\hspace{1.5em}}r@{--}l|r@{$\;\pm$}c@{$\pm\;$}l|r@{$\;\pm$}c@{$\pm\;$}l|r@{$\;\pm$}c@{$\pm\;$}l|}
\cline{3-11}
\multicolumn{2}{c|}{} &
\multicolumn{9}{c|}{\rule[-13pt]{0pt}{32pt}${\displaystyle\frac{1}{\sigma}\cdot\frac{\ds}{\dd{\rhoh}}}$}
\\ \hline
\multicolumn{2}{|c|}{\rhoh} & \multicolumn{3}{c|}{\rule{0pt}{11pt}at $\rs = 41.4\,\GeV$} &
\multicolumn{3}{c|}{at $\rs = 55.3\,\GeV$} &
\multicolumn{3}{c|}{at $\rs = 65.4\,\GeV$}  \\ \hline
0.000&0.015&  0.183& 0.038& 0.858&  1.329& 0.195& 0.622&  2.063& 0.190& 0.924 \\
0.015&0.027&  6.119& 0.642& 0.465& 13.192& 1.067& 3.301& 14.082& 0.856& 0.654 \\
0.027&0.039& 12.263& 1.432& 3.194& 19.031& 1.637& 2.785& 20.593& 1.349& 1.384 \\
0.039&0.051& 16.680& 1.925& 2.251& 11.201& 1.317& 3.402& 12.065& 0.940& 1.158 \\
0.051&0.066& 10.037& 1.223& 0.515&  9.616& 1.190& 5.665&  7.040& 0.705& 0.482 \\
0.066&0.084&  5.589& 0.819& 0.901&  4.482& 0.658& 1.981&  5.751& 0.557& 0.262 \\
0.084&0.102&  6.752& 0.967& 1.918&  3.114& 0.597& 0.390&  3.361& 0.450& 0.207 \\
0.102&0.126&  2.550& 0.432& 0.578&  3.591& 0.606& 0.951&  1.803& 0.270& 0.530 \\
0.126&0.153&  2.029& 0.430& 0.510&  1.312& 0.333& 0.463&  0.928& 0.182& 0.217 \\
0.153&0.183&  1.615& 0.328& 0.312&  0.559& 0.215& 0.153&  1.116& 0.243& 0.218 \\
0.183&0.216&  0.683& 0.210& 0.194&  0.691& 0.192& 0.203&  0.517& 0.117& 0.067 \\
0.216&0.252&  0.211& 0.130& 0.204&  0.213& 0.117& 0.062&  0.405& 0.095& 0.090 \\
0.252&0.300&  0.193& 0.099& 0.113&  0.191& 0.079& 0.098&  0.107& 0.047& 0.179 \\
\hline\multicolumn{2}{|l|}{First Moment} &
 0.0747&0.0023&0.0023& 0.0632&0.0021&0.0023& 0.0603&0.0015&0.0047 \\
\multicolumn{2}{|l|}{Second Moment} &
 0.0080&0.0006&0.0005& 0.0063&0.0005&0.0008& 0.0060&0.0003&0.0011 \\
\hline\end{tabular}\end{center}}
\caption[Differential distribution for scaled heavy jet mass at $\rs = 41.4$, 55.3 and 65.4\,\GeV.]
        {Differential distribution for scaled heavy jet mass at $\rs = 41.4$, 55.3 and 65.4\,\GeV.
The first uncertainty is statistical, the second systematic.}
\label{tab:rhoqqg1}
\end{table}
 
\begin{table}{\small\begin{center}
\begin{tabular}{|@{\hspace{1.5em}}r@{--}l|r@{$\;\pm$}c@{$\pm\;$}l|r@{$\;\pm$}c@{$\pm\;$}l|r@{$\;\pm$}c@{$\pm\;$}l|}
\cline{3-11}
\multicolumn{2}{c|}{} &
\multicolumn{9}{c|}{\rule[-13pt]{0pt}{32pt}${\displaystyle\frac{1}{\sigma}\cdot\frac{\ds}{\dd{\rhoh}}}$}
\\ \hline
\multicolumn{2}{|c|}{\rhoh} & \multicolumn{3}{c|}{\rule{0pt}{11pt}at $\rs = 75.7\,\GeV$} &
\multicolumn{3}{c|}{at $\rs = 82.3\,\GeV$} &
\multicolumn{3}{c|}{at $\rs = 85.1\,\GeV$}  \\ \hline
0.000&0.015&  3.092& 0.166& 1.346&  3.493& 0.220& 1.479&  5.058& 0.339& 2.311 \\
0.015&0.027& 22.135& 0.853& 1.200& 21.585& 0.976& 0.641& 24.679& 1.247& 1.747 \\
0.027&0.039& 15.391& 0.773& 0.976& 18.963& 1.044& 1.066& 15.534& 1.135& 1.193 \\
0.039&0.051& 10.937& 0.687& 2.336& 10.236& 0.787& 0.875&  8.703& 0.983& 1.224 \\
0.051&0.066&  6.363& 0.494& 1.029&  5.674& 0.560& 1.575&  4.801& 0.687& 0.316 \\
0.066&0.084&  4.329& 0.357& 1.063&  3.804& 0.407& 0.896&  3.787& 0.531& 0.673 \\
0.084&0.102&  3.554& 0.350& 0.398&  2.835& 0.341& 0.275&  2.764& 0.386& 0.916 \\
0.102&0.126&  2.110& 0.231& 0.270&  1.717& 0.245& 0.305&  1.764& 0.328& 0.235 \\
0.126&0.153&  1.010& 0.139& 0.171&  1.406& 0.207& 0.154&  1.732& 0.300& 0.330 \\
0.153&0.183&  0.915& 0.135& 0.107&  0.435& 0.145& 0.146&  0.978& 0.210& 0.328 \\
0.183&0.216&  0.415& 0.105& 0.188&  1.017& 0.210& 0.279&  0.394& 0.129& 0.064 \\
0.216&0.252&  0.347& 0.081& 0.107&  0.119& 0.049& 0.020&  0.305& 0.104& 0.090 \\
0.252&0.300&  0.066& 0.025& 0.016&  0.074& 0.041& 0.043&  0.099& 0.048& 0.028 \\
\hline\multicolumn{2}{|l|}{First Moment} &
 0.0560&0.0011&0.0027& 0.0546&0.0015&0.0035& 0.0544&0.0017&0.0085 \\
\multicolumn{2}{|l|}{Second Moment} &
 0.0053&0.0002&0.0007& 0.0052&0.0003&0.0007& 0.0054&0.0004&0.0014 \\
\hline\end{tabular}\end{center}}
\caption[Differential distribution for scaled heavy jet mass at $\rs = 75.7$, 82.3 and 85.1\,\GeV.]
        {Differential distribution for scaled heavy jet mass at $\rs = 75.7$, 82.3 and 85.1\,\GeV.
The first uncertainty is statistical, the second systematic.}
\label{tab:rhoqqg2}
\end{table}
 
\begin{table}[htbp]{\small\begin{center}
\begin{tabular}{|@{\hspace{1.5em}}r@{--}l|r@{$\;\pm$}c@{$\pm\;$}l|r@{$\;\pm$}c@{$\pm\;$}l|r@{$\;\pm$}c@{$\pm\;$}l|}
\cline{3-11}
\multicolumn{2}{c|}{} &
\multicolumn{9}{c|}{\rule[-13pt]{0pt}{32pt}${\displaystyle\frac{1}{\sigma}\cdot\frac{\ds}{\dd{\rhoh}}}$}
\\ \hline
\multicolumn{2}{|c|}{\rhoh} & \multicolumn{3}{c|}{\rule{0pt}{11pt}at $\rs =130.1\,\GeV$} &
\multicolumn{3}{c|}{at $\rs= 136.1\,\GeV$} &
\multicolumn{3}{c|}{at $\rs= 161.3\,\GeV$}  \\ \hline
0.000& 0.015& 16.201& 1.462& 1.335& 15.749& 1.715& 1.470& 24.327& 2.252& 0.887 \\
0.015& 0.030& 18.462& 1.621& 0.676& 22.091& 2.114& 1.609& 14.684& 1.771& 1.083 \\
0.030& 0.045& 10.243& 1.201& 0.510&  8.738& 1.266& 0.641&  7.505& 1.167& 0.301 \\
0.045& 0.060&  6.326& 0.939& 0.457&  4.134& 0.888& 0.455&  4.597& 0.944& 0.402 \\
0.060& 0.075&  3.810& 0.687& 0.030&  3.169& 0.740& 0.612&  3.471& 0.835& 0.346 \\
0.075& 0.090&  3.533& 0.649& 0.200&  3.110& 0.755& 0.323&  3.026& 0.759& 0.504 \\
\hline\end{tabular}\end{center}}
\caption[Differential distribution for scaled heavy jet mass at $\rs=130.1$, 136.1 and 161.3\,\GeV.]
        {Differential distribution for scaled heavy jet mass at $\rs=130.1$, 136.1 and 161.3\,\GeV.
The first uncertainty is statistical, the second systematic.}
\label{tab:rho1}
\end{table}
 
\begin{table}[htbp]{\small\begin{center}
\begin{tabular}{|@{\hspace{1.5em}}r@{--}l|r@{$\;\pm$}c@{$\pm\;$}l|r@{$\;\pm$}c@{$\pm\;$}l|r@{$\;\pm$}c@{$\pm\;$}l|}
\cline{3-11}
\multicolumn{2}{c|}{} &
\multicolumn{9}{c|}{\rule[-13pt]{0pt}{32pt}${\displaystyle\frac{1}{\sigma}\cdot\frac{\ds}{\dd{\rhoh}}}$}
\\ \hline
\multicolumn{2}{|c|}{\rhoh} & \multicolumn{3}{c|}{\rule{0pt}{11pt}at $\rs =172.3\,\GeV$} &
\multicolumn{3}{c|}{at $\rs= 182.8\,\GeV$} &
\multicolumn{3}{c|}{at $\rs= 188.6\,\GeV$}  \\ \hline
0.000& 0.015& 20.528& 2.362& 0.824& 21.698& 0.974& 2.301& 20.410& 0.620& 0.687 \\
0.015& 0.030& 14.956& 1.896& 1.607& 15.979& 0.996& 1.428& 16.738& 0.606& 0.365 \\
0.030& 0.045&  9.336& 1.486& 0.294&  9.381& 0.748& 0.703&  8.711& 0.417& 0.250 \\
0.045& 0.060&  4.641& 1.102& 0.946&  5.032& 0.553& 0.471&  5.259& 0.332& 0.243 \\
0.060& 0.075&  4.249& 1.046& 0.485&  3.422& 0.465& 0.490&  3.489& 0.272& 0.065 \\
0.075& 0.090&  3.184& 0.951& 0.621&  2.524& 0.400& 0.043&  2.641& 0.242& 0.081 \\
\hline\end{tabular}\end{center}}
\caption[Differential distribution for scaled heavy jet mass at $\rs= 172.3$, 182.8 and 188.6\,\GeV.]
        {Differential distribution for scaled heavy jet mass at $\rs= 172.3$, 182.8 and 188.6\,\GeV.
The first uncertainty is statistical, the second systematic.}
\label{tab:rho2}
\end{table}
 
\begin{table}[htbp]{\small\begin{center}
\begin{tabular}{|@{\hspace{1.5em}}r@{--}l|r@{$\;\pm$}c@{$\pm\;$}l|r@{$\;\pm$}c@{$\pm\;$}l|r@{$\;\pm$}c@{$\pm\;$}l|}
\cline{3-11}
\multicolumn{2}{c|}{} &
\multicolumn{9}{c|}{\rule[-13pt]{0pt}{32pt}${\displaystyle\frac{1}{\sigma}\cdot\frac{\ds}{\dd{\rhoh}}}$}
\\ \hline
\multicolumn{2}{|c|}{\rhoh} & \multicolumn{3}{c|}{\rule{0pt}{11pt}at $\rs =194.4\,\GeV$} &
\multicolumn{3}{c|}{at $\rs= 200.2\,\GeV$} &
\multicolumn{3}{c|}{at $\rs= 206.2\,\GeV$}  \\ \hline
0.000& 0.015& 20.870& 0.769& 0.662& 20.671& 0.765& 0.753& 21.199& 0.633& 0.788 \\
0.015& 0.030& 16.523& 0.771& 0.644& 16.098& 0.768& 0.439& 15.088& 0.576& 0.357 \\
0.030& 0.045&  8.747& 0.588& 0.549&  8.739& 0.577& 0.393&  9.048& 0.452& 0.384 \\
0.045& 0.060&  5.854& 0.512& 0.461&  4.727& 0.435& 0.302&  5.218& 0.354& 0.208 \\
0.060& 0.075&  3.817& 0.422& 0.256&  3.701& 0.393& 0.220&  3.669& 0.300& 0.082 \\
0.075& 0.090&  2.036& 0.317& 0.209&  2.676& 0.346& 0.519&  3.101& 0.283& 0.233 \\
\hline\end{tabular}\end{center}}
\caption[Differential distribution for scaled heavy jet mass at $\rs =194.4$, 200.2 and 206.2\,\GeV.]
        {Differential distribution for scaled heavy jet mass at $\rs =194.4$, 200.2 and 206.2\,\GeV.
The first uncertainty is statistical, the second systematic.}
\label{tab:rho3}
\end{table}

\begin{table}{\small\begin{center}
\begin{tabular}{|@{\hspace{1.5em}}r@{--}l|r@{$\;\pm$}c@{$\pm\;$}l|r@{$\;\pm$}c@{$\pm\;$}l|r@{$\;\pm$}c@{$\pm\;$}l|}
\cline{3-11}
\multicolumn{2}{c|}{} &
\multicolumn{9}{c|}{\rule[-13pt]{0pt}{32pt}${\displaystyle\frac{1}{\sigma}\cdot\frac{\ds}{\dd{\bt}}}$}
\\ \hline
\multicolumn{2}{|c|}{\bt} & \multicolumn{3}{c|}{\rule{0pt}{11pt}at $\rs = 41.4\,\GeV$} &
\multicolumn{3}{c|}{at $\rs = 55.3\,\GeV$} &
\multicolumn{3}{c|}{at $\rs = 65.4\,\GeV$}  \\ \hline
0.000&0.040&  0.229& 0.098& 0.073&  0.445& 0.131& 0.243&  0.232& 0.054& 0.116 \\
0.040&0.070&  2.056& 0.276& 0.316&  5.108& 0.492& 1.090&  6.125& 0.399& 1.515 \\
0.070&0.100&  7.981& 0.685& 0.974&  9.488& 0.688& 1.665&  9.335& 0.493& 0.537 \\
0.100&0.130&  7.756& 0.749& 0.268&  5.946& 0.558& 0.559&  6.114& 0.421& 0.223 \\
0.130&0.160&  4.514& 0.548& 0.564&  4.592& 0.538& 0.696&  4.384& 0.383& 0.487 \\
0.160&0.190&  3.935& 0.532& 0.921&  2.816& 0.397& 0.845&  2.529& 0.274& 0.157 \\
0.190&0.220&  2.887& 0.485& 0.632&  2.260& 0.406& 0.270&  1.289& 0.210& 0.829 \\
0.220&0.250&  1.817& 0.346& 0.128&  1.136& 0.245& 0.360&  1.452& 0.202& 0.222 \\
0.250&0.280&  0.732& 0.210& 0.217&  0.704& 0.224& 0.146&  0.940& 0.159& 0.293 \\
0.280&0.310&  0.849& 0.274& 0.233&  0.408& 0.173& 0.149&  0.384& 0.118& 0.073 \\
0.310&0.340&  0.372& 0.117& 0.159&  0.168& 0.083& 0.085&  0.402& 0.125& 0.121 \\
0.340&0.400&  0.064& 0.035& 0.021&  0.058& 0.034& 0.010&  0.035& 0.022& 0.019 \\
\hline\multicolumn{2}{|l|}{First Moment} &
 0.1399&0.0027&0.0016& 0.1223&0.0025&0.0054& 0.1213&0.0019&0.0079 \\
\multicolumn{2}{|l|}{Second Moment} &
 0.0236&0.0009&0.0005& 0.0187&0.0008&0.0012& 0.0187&0.0006&0.0022 \\
\hline\end{tabular}\end{center}}
\caption[Differential distribution for total jet broadening at $\rs = 41.4$, 55.3 and 65.4\,\GeV.]
        {Differential distribution for total jet broadening at $\rs = 41.4$, 55.3 and 65.4\,\GeV.
The first uncertainty is statistical, the second systematic.}
\label{tab:btqqg1}
\end{table}
 
\begin{table}{\small\begin{center}
\begin{tabular}{|@{\hspace{1.5em}}r@{--}l|r@{$\;\pm$}c@{$\pm\;$}l|r@{$\;\pm$}c@{$\pm\;$}l|r@{$\;\pm$}c@{$\pm\;$}l|}
\cline{3-11}
\multicolumn{2}{c|}{} &
\multicolumn{9}{c|}{\rule[-13pt]{0pt}{32pt}${\displaystyle\frac{1}{\sigma}\cdot\frac{\ds}{\dd{\bt}}}$}
\\ \hline
\multicolumn{2}{|c|}{\bt} & \multicolumn{3}{c|}{\rule{0pt}{11pt}at $\rs = 75.7\,\GeV$} &
\multicolumn{3}{c|}{at $\rs = 82.3\,\GeV$} &
\multicolumn{3}{c|}{at $\rs = 85.1\,\GeV$}  \\ \hline
0.000&0.040&  0.243& 0.030& 0.121&  0.324& 0.049& 0.178&  0.637& 0.086& 0.315 \\
0.040&0.070&  8.069& 0.316& 0.997&  8.448& 0.376& 1.156&  9.054& 0.464& 1.153 \\
0.070&0.100&  8.805& 0.344& 0.227&  9.377& 0.430& 0.158&  8.500& 0.493& 1.133 \\
0.100&0.130&  5.694& 0.300& 0.278&  5.446& 0.339& 0.940&  5.978& 0.439& 0.376 \\
0.130&0.160&  3.663& 0.246& 0.349&  3.437& 0.296& 0.398&  2.353& 0.337& 0.448 \\
0.160&0.190&  2.536& 0.211& 0.113&  2.341& 0.242& 0.596&  2.510& 0.283& 0.650 \\
0.190&0.220&  1.384& 0.154& 0.327&  1.551& 0.186& 0.139&  1.557& 0.245& 0.286 \\
0.220&0.250&  1.420& 0.168& 0.389&  0.999& 0.186& 0.153&  1.465& 0.260& 0.313 \\
0.250&0.280&  0.723& 0.123& 0.119&  0.886& 0.185& 0.189&  0.609& 0.187& 0.177 \\
0.280&0.310&  0.488& 0.113& 0.140&  0.267& 0.115& 0.138&  0.248& 0.115& 0.040 \\
0.310&0.340&  0.184& 0.079& 0.064&  0.115& 0.064& 0.064&  0.185& 0.092& 0.096 \\
0.340&0.400&  0.022& 0.013& 0.009&  0.017& 0.012& 0.010&  0.012& 0.012& 0.015 \\
\hline\multicolumn{2}{|l|}{First Moment} &
 0.1157&0.0015&0.0048& 0.1116&0.0017&0.0057& 0.1102&0.0021&0.0086 \\
\multicolumn{2}{|l|}{Second Moment} &
 0.0172&0.0005&0.0014& 0.0160&0.0006&0.0015& 0.0158&0.0007&0.0022 \\
\hline\end{tabular}\end{center}}
\caption[Differential distribution for total jet broadening at $\rs = 75.7$, 82.3 and 85.1\,\GeV.]
        {Differential distribution for total jet broadening at $\rs = 75.7$, 82.3 and 85.1\,\GeV.
The first uncertainty is statistical, the second systematic.}
\label{tab:btqqg2}
\end{table}
 
\begin{table}[htbp]{\small\begin{center}
\begin{tabular}{|@{\hspace{1.5em}}r@{--}l|r@{$\;\pm$}c@{$\pm\;$}l|r@{$\;\pm$}c@{$\pm\;$}l|r@{$\;\pm$}c@{$\pm\;$}l|}
\cline{3-11}
\multicolumn{2}{c|}{} &
\multicolumn{9}{c|}{\rule[-13pt]{0pt}{32pt}${\displaystyle\frac{1}{\sigma}\cdot\frac{\ds}{\dd{\bt}}}$}
\\ \hline
\multicolumn{2}{|c|}{\bt} & \multicolumn{3}{c|}{\rule{0pt}{11pt}at $\rs =130.1\,\GeV$} &
\multicolumn{3}{c|}{at $\rs= 136.1\,\GeV$} &
\multicolumn{3}{c|}{at $\rs= 161.3\,\GeV$}  \\ \hline
0.000& 0.020&  0.082& 0.086& 0.037&  0.103& 0.077& 0.090&  0.117& 0.090& 0.069 \\
0.020& 0.040&  4.010& 0.684& 0.468&  5.454& 0.965& 0.863&  9.224& 1.407& 0.577 \\
0.040& 0.060& 11.651& 1.173& 0.360& 11.111& 1.340& 0.898&  9.486& 1.353& 0.934 \\
0.060& 0.080&  8.865& 0.932& 0.289&  8.945& 1.077& 0.286&  8.450& 1.071& 0.344 \\
0.080& 0.100&  7.778& 0.884& 0.880&  7.419& 1.042& 0.468&  5.530& 0.841& 0.214 \\
0.100& 0.120&  4.239& 0.638& 0.388&  3.464& 0.669& 0.235&  4.231& 0.740& 0.197 \\
\hline\end{tabular}\end{center}}
\caption[Differential distribution for total jet broadening at $\rs=130.1$, 136.1 and 161.3\,\GeV.]
        {Differential distribution for total jet broadening at $\rs=130.1$, 136.1 and 161.3\,\GeV.
The first uncertainty is statistical, the second systematic.}
\label{tab:bt1}
\end{table}
 
\begin{table}[htbp]{\small\begin{center}
\begin{tabular}{|@{\hspace{1.5em}}r@{--}l|r@{$\;\pm$}c@{$\pm\;$}l|r@{$\;\pm$}c@{$\pm\;$}l|r@{$\;\pm$}c@{$\pm\;$}l|}
\cline{3-11}
\multicolumn{2}{c|}{} &
\multicolumn{9}{c|}{\rule[-13pt]{0pt}{32pt}${\displaystyle\frac{1}{\sigma}\cdot\frac{\ds}{\dd{\bt}}}$}
\\ \hline
\multicolumn{2}{|c|}{\bt} & \multicolumn{3}{c|}{\rule{0pt}{11pt}at $\rs =172.3\,\GeV$} &
\multicolumn{3}{c|}{at $\rs= 182.8\,\GeV$} &
\multicolumn{3}{c|}{at $\rs= 188.6\,\GeV$}  \\ \hline
0.000& 0.020&  0.410& 0.189& 0.117&  0.343& 0.086& 0.471&  0.418& 0.061& 0.065 \\
0.020& 0.040&  8.824& 1.579& 1.483&  8.245& 0.597& 0.670&  8.634& 0.416& 0.383 \\
0.040& 0.060& 10.568& 1.517& 0.462& 10.962& 0.680& 0.389& 10.279& 0.423& 0.276 \\
0.060& 0.080&  6.654& 1.043& 0.747&  7.658& 0.549& 0.191&  8.186& 0.332& 0.100 \\
0.080& 0.100&  6.733& 1.046& 1.026&  5.831& 0.470& 0.391&  5.241& 0.257& 0.196 \\
0.100& 0.120&  3.987& 0.814& 0.155&  4.179& 0.392& 0.240&  4.633& 0.242& 0.156 \\
\hline\end{tabular}\end{center}}
\caption[Differential distribution for total jet broadening at $\rs = 172.3$, 182.8 and 188.6\,\GeV.]
        {Differential distribution for total jet broadening at $\rs = 172.3$, 182.8 and 188.6\,\GeV.
The first uncertainty is statistical, the second systematic.}
\label{tab:bt2}
\end{table}
 
\begin{table}[htbp]{\small\begin{center}
\begin{tabular}{|@{\hspace{1.5em}}r@{--}l|r@{$\;\pm$}c@{$\pm\;$}l|r@{$\;\pm$}c@{$\pm\;$}l|r@{$\;\pm$}c@{$\pm\;$}l|}
\cline{3-11}
\multicolumn{2}{c|}{} &
\multicolumn{9}{c|}{\rule[-13pt]{0pt}{32pt}${\displaystyle\frac{1}{\sigma}\cdot\frac{\ds}{\dd{\bt}}}$}
\\ \hline
\multicolumn{2}{|c|}{\bt} & \multicolumn{3}{c|}{\rule{0pt}{11pt}at $\rs =194.4\,\GeV$} &
\multicolumn{3}{c|}{at $\rs= 200.2\,\GeV$} &
\multicolumn{3}{c|}{at $\rs= 206.2\,\GeV$}  \\ \hline
0.000& 0.020&  0.376& 0.078& 0.124&  0.449& 0.085& 0.109&  0.491& 0.064& 0.101 \\
0.020& 0.040&  8.334& 0.480& 0.415&  8.633& 0.495& 0.405&  8.694& 0.386& 0.387 \\
0.040& 0.060& 10.931& 0.531& 0.377& 10.228& 0.507& 0.201& 10.221& 0.401& 0.228 \\
0.060& 0.080&  7.336& 0.419& 0.283&  7.197& 0.415& 0.180&  7.299& 0.322& 0.124 \\
0.080& 0.100&  6.187& 0.384& 0.236&  5.780& 0.368& 0.118&  5.821& 0.285& 0.260 \\
0.100& 0.120&  4.157& 0.316& 0.188&  4.282& 0.326& 0.327&  4.079& 0.241& 0.176 \\
\hline\end{tabular}\end{center}}
\caption[Differential distribution for total jet broadening at $\rs =194.4$, 200.2 and 206.2\,\GeV.]
        {Differential distribution for total jet broadening at $\rs =194.4$, 200.2 and 206.2\,\GeV.
The first uncertainty is statistical, the second systematic.}
\label{tab:bt3}
\end{table}
 
\clearpage
 
\begin{table}{\small\begin{center}
\begin{tabular}{|@{\hspace{1.5em}}r@{--}l|r@{$\;\pm$}c@{$\pm\;$}l|r@{$\;\pm$}c@{$\pm\;$}l|r@{$\;\pm$}c@{$\pm\;$}l|}
\cline{3-11}
\multicolumn{2}{c|}{} &
\multicolumn{9}{c|}{\rule[-13pt]{0pt}{32pt}${\displaystyle\frac{1}{\sigma}\cdot\frac{\ds}{\dd{\bw}}}$}
\\ \hline
\multicolumn{2}{|c|}{\bw} & \multicolumn{3}{c|}{\rule{0pt}{11pt}at $\rs = 41.4\,\GeV$} &
\multicolumn{3}{c|}{at $\rs = 55.3\,\GeV$} &
\multicolumn{3}{c|}{at $\rs = 65.4\,\GeV$}  \\ \hline
0.000&0.030&  0.641& 0.167& 0.325&  1.498& 0.246& 0.742&  1.264& 0.164& 0.620 \\
0.030&0.050&  8.586& 0.836& 2.308& 13.144& 1.014& 2.525& 13.873& 0.756& 0.784 \\
0.050&0.075& 12.655& 1.026& 0.338& 11.029& 0.802& 1.831& 10.676& 0.583& 0.421 \\
0.075&0.100&  6.548& 0.676& 0.558&  6.442& 0.669& 0.470&  6.186& 0.473& 0.388 \\
0.100&0.125&  5.168& 0.681& 0.818&  3.544& 0.475& 0.573&  4.246& 0.393& 0.873 \\
0.125&0.150&  2.900& 0.497& 0.436&  3.117& 0.479& 1.599&  2.485& 0.312& 0.325 \\
0.150&0.175&  2.101& 0.402& 0.495&  1.703& 0.403& 0.202&  1.445& 0.224& 0.247 \\
0.175&0.200&  1.856& 0.346& 0.134&  0.692& 0.283& 0.544&  1.175& 0.234& 0.697 \\
0.200&0.225&  0.636& 0.223& 0.242&  0.751& 0.240& 0.233&  0.669& 0.156& 0.159 \\
0.225&0.300&  0.166& 0.080& 0.086&  0.137& 0.057& 0.061&  0.167& 0.040& 0.048 \\
\hline\multicolumn{2}{|l|}{First Moment} &
 0.0896&0.0021&0.0018& 0.0800&0.0020&0.0034& 0.0806&0.0014&0.0060 \\
\multicolumn{2}{|l|}{Second Moment} &
 0.0104&0.0005&0.0005& 0.0086&0.0005&0.0006& 0.0088&0.0003&0.0013 \\
\hline\end{tabular}\end{center}}
\caption[Differential distribution for wide jet broadening at $\rs = 41.4$, 55.3 and 65.4\,\GeV.]
        {Differential distribution for wide jet broadening at $\rs = 41.4$, 55.3 and 65.4\,\GeV.
The first uncertainty is statistical, the second systematic.}
\label{tab:bwqqg1}
\end{table}
 
\begin{table}[htbp]{\small\begin{center}
\begin{tabular}{|@{\hspace{1.5em}}r@{--}l|r@{$\;\pm$}c@{$\pm\;$}l|r@{$\;\pm$}c@{$\pm\;$}l|r@{$\;\pm$}c@{$\pm\;$}l|}
\cline{3-11}
\multicolumn{2}{c|}{} &
\multicolumn{9}{c|}{\rule[-13pt]{0pt}{32pt}${\displaystyle\frac{1}{\sigma}\cdot\frac{\ds}{\dd{\bw}}}$}
\\ \hline
\multicolumn{2}{|c|}{\bw} & \multicolumn{3}{c|}{\rule{0pt}{11pt}at $\rs = 75.7\,\GeV$} &
\multicolumn{3}{c|}{at $\rs = 82.3\,\GeV$} &
\multicolumn{3}{c|}{at $\rs = 85.1\,\GeV$}  \\ \hline
0.000&0.030&  2.299& 0.153& 1.123&  2.628& 0.204& 1.255&  3.260& 0.274& 1.577 \\
0.030&0.050& 16.308& 0.576& 0.614& 15.538& 0.645& 0.746& 15.441& 0.763& 1.344 \\
0.050&0.075&  9.129& 0.384& 0.455&  9.862& 0.472& 0.655& 10.235& 0.602& 0.556 \\
0.075&0.100&  5.427& 0.326& 0.847&  4.651& 0.372& 1.406&  3.917& 0.446& 1.180 \\
0.100&0.125&  3.611& 0.273& 0.275&  3.800& 0.337& 0.769&  3.173& 0.396& 0.476 \\
0.125&0.150&  2.499& 0.222& 0.240&  2.442& 0.288& 0.277&  2.226& 0.303& 0.315 \\
0.150&0.175&  1.297& 0.189& 0.238&  1.500& 0.210& 0.250&  1.775& 0.273& 0.369 \\
0.175&0.200&  1.320& 0.176& 0.248&  0.883& 0.186& 0.078&  1.205& 0.248& 0.234 \\
0.200&0.225&  0.449& 0.114& 0.179&  1.026& 0.237& 0.203&  0.686& 0.187& 0.213 \\
0.225&0.300&  0.154& 0.033& 0.021&  0.084& 0.032& 0.038&  0.173& 0.059& 0.057 \\
\hline\multicolumn{2}{|l|}{First Moment} &
 0.0758&0.0011&0.0046& 0.0756&0.0014&0.0051& 0.0749&0.0017&0.0092 \\
\multicolumn{2}{|l|}{Second Moment} &
 0.0081&0.0003&0.0008& 0.0081&0.0003&0.0008& 0.0082&0.0004&0.0018 \\
\hline\end{tabular}\end{center}}
\caption[Differential distribution for wide jet broadening at $\rs = 75.7$, 82.3 and 85.1\,\GeV.]
        {Differential distribution for wide jet broadening at $\rs = 75.7$, 82.3 and 85.1\,\GeV.
The first uncertainty is statistical, the second systematic.}
\label{tab:bwqqg2}
\end{table}
 
\begin{table}{\small\begin{center}
\begin{tabular}{|@{\hspace{1.5em}}r@{--}l|r@{$\;\pm$}c@{$\pm\;$}l|r@{$\;\pm$}c@{$\pm\;$}l|r@{$\;\pm$}c@{$\pm\;$}l|}
\cline{3-11}
\multicolumn{2}{c|}{} &
\multicolumn{9}{c|}{\rule[-13pt]{0pt}{32pt}${\displaystyle\frac{1}{\sigma}\cdot\frac{\ds}{\dd{\bw}}}$}
\\ \hline
\multicolumn{2}{|c|}{\bw} & \multicolumn{3}{c|}{\rule{0pt}{11pt}at $\rs =130.1\,\GeV$} &
\multicolumn{3}{c|}{at $\rs= 136.1\,\GeV$} &
\multicolumn{3}{c|}{at $\rs= 161.3\,\GeV$}  \\ \hline
0.000& 0.015&  0.157& 0.126& 0.053&  0.235& 0.257& 0.077&  2.601& 0.768& 0.708 \\
0.015& 0.030& 12.915& 1.493& 0.732& 14.660& 1.837& 1.348& 16.194& 2.091& 0.795 \\
0.030& 0.045& 14.268& 1.332& 0.963& 15.618& 1.646& 0.640& 10.741& 1.481& 0.779 \\
0.045& 0.060& 10.861& 1.164& 0.876&  8.923& 1.262& 0.465& 10.016& 1.327& 0.223 \\
0.060& 0.075&  6.320& 0.921& 0.272&  5.144& 0.951& 0.977&  5.587& 0.992& 0.551 \\
0.075& 0.090&  5.646& 0.851& 0.436&  5.510& 0.990& 0.430&  4.394& 0.878& 0.277 \\
\hline\end{tabular}\end{center}}
\caption[Differential distribution for wide jet broadening at $\rs= 130.1$, 136.1 and 161.3\,\GeV.]
        {Differential distribution for wide jet broadening at $\rs= 130.1$, 136.1 and 161.3\,\GeV.
The first uncertainty is statistical, the second systematic.}
\label{tab:bw1}
\end{table}
 
\begin{table}{\small\begin{center}
\begin{tabular}{|@{\hspace{1.5em}}r@{--}l|r@{$\;\pm$}c@{$\pm\;$}l|r@{$\;\pm$}c@{$\pm\;$}l|r@{$\;\pm$}c@{$\pm\;$}l|}
\cline{3-11}
\multicolumn{2}{c|}{} &
\multicolumn{9}{c|}{\rule[-13pt]{0pt}{32pt}${\displaystyle\frac{1}{\sigma}\cdot\frac{\ds}{\dd{\bw}}}$}
\\ \hline
\multicolumn{2}{|c|}{\bw} & \multicolumn{3}{c|}{\rule{0pt}{11pt}at $\rs =172.3\,\GeV$} &
\multicolumn{3}{c|}{at $\rs= 182.8\,\GeV$} &
\multicolumn{3}{c|}{at $\rs= 188.6\,\GeV$}  \\ \hline
0.000& 0.015&  2.337& 0.943& 0.433&  1.727& 0.262& 1.415&  2.170& 0.244& 0.237 \\
0.015& 0.030& 15.587& 2.244& 1.365& 16.496& 0.966& 0.402& 15.348& 0.587& 0.353 \\
0.030& 0.045& 11.428& 1.577& 0.771& 12.573& 0.814& 0.592& 12.889& 0.515& 0.440 \\
0.045& 0.060&  8.594& 1.426& 0.877&  8.528& 0.663& 0.295&  8.460& 0.384& 0.198 \\
0.060& 0.075&  8.232& 1.376& 0.656&  6.961& 0.612& 0.839&  6.632& 0.343& 0.322 \\
0.075& 0.090&  4.833& 1.082& 0.614&  4.354& 0.487& 0.238&  4.898& 0.300& 0.149 \\
\hline\end{tabular}\end{center}}
\caption[Differential distribution for wide jet broadening at $\rs = 172.3$, 182.8 and 188.6\,\GeV.]
        {Differential distribution for wide jet broadening at $\rs = 172.3$, 182.8 and 188.6\,\GeV.
The first uncertainty is statistical, the second systematic.}
\label{tab:bw2}
\end{table}
 
\begin{table}[htbp]{\small\begin{center}
\begin{tabular}{|@{\hspace{1.5em}}r@{--}l|r@{$\;\pm$}c@{$\pm\;$}l|r@{$\;\pm$}c@{$\pm\;$}l|r@{$\;\pm$}c@{$\pm\;$}l|}
\cline{3-11}
\multicolumn{2}{c|}{} &
\multicolumn{9}{c|}{\rule[-13pt]{0pt}{32pt}${\displaystyle\frac{1}{\sigma}\cdot\frac{\ds}{\dd{\bw}}}$}
\\ \hline
\multicolumn{2}{|c|}{\bw} & \multicolumn{3}{c|}{\rule{0pt}{11pt}at $\rs =194.4\,\GeV$} &
\multicolumn{3}{c|}{at $\rs = 200.2\,\GeV$} &
\multicolumn{3}{c|}{at $\rs = 206.2\,\GeV$}  \\ \hline
0.000& 0.015&  2.446& 0.261& 0.243&  3.055& 0.310& 0.216&  2.690& 0.226& 0.302 \\
0.015& 0.030& 15.023& 0.740& 0.609& 14.545& 0.722& 0.506& 14.846& 0.576& 0.679 \\
0.030& 0.045& 12.061& 0.612& 0.394& 12.038& 0.617& 0.392& 11.911& 0.476& 0.200 \\
0.045& 0.060&  9.734& 0.555& 0.363&  8.410& 0.525& 0.354&  8.842& 0.420& 0.256 \\
0.060& 0.075&  6.486& 0.479& 0.350&  6.472& 0.474& 0.398&  6.541& 0.373& 0.340 \\
0.075& 0.090&  4.938& 0.429& 0.287&  5.464& 0.438& 0.217&  5.080& 0.329& 0.215 \\
\hline\end{tabular}\end{center}}
\caption[Differential distribution for wide jet broadening at $\rs= 194.4$, 200.2 and 206.2\,\GeV.]
        {Differential distribution for wide jet broadening at $\rs= 194.4$, 200.2 and 206.2\,\GeV.
The first uncertainty is statistical, the second systematic.}
\label{tab:bw3}
\end{table}

\begin{table}{\small\begin{center}
%
\end{center}}
\caption[Differential distribution for $C$-parameter at $\rs= 130.1$, 136.1 and 161.3\,\GeV.]
        {Differential distribution for $C$-parameter at $\rs= 130.1$, 136.1 and 161.3\,\GeV.
The first uncertainty is statistical, the second systematic.}
\label{tab:cpar1}
\end{table}

\begin{table}{\small\begin{center}
%
\end{center}}
\caption[Differential distribution for $C$-parameter at $\rs = 172.3$, 182.8 and 188.6\,\GeV.]
        {Differential distribution for $C$-parameter at $\rs = 172.3$, 182.8 and 188.6\,\GeV.
The first uncertainty is statistical, the second systematic.}
\label{tab:cpar2}
\end{table}
 
\begin{table}{\small\begin{center}
%
\end{center}}
\caption[Differential distribution for $C$-parameter at $\rs = 194.4$, 200.2 and 206.2\,\GeV.]
        {Differential distribution for $C$-parameter at $\rs = 194.4$, 200.2 and 206.2\,\GeV.
The first uncertainty is statistical, the second systematic.}
\label{tab:cpar3}
\end{table}
 
\begin{table}[htbp]{\small\begin{center}
%
\end{center}}
\caption[Differential distribution for $D$-parameter at $\rs= 130.1$, 136.1 and 161.3\,\GeV.]
        {Differential distribution for $D$-parameter at $\rs= 130.1$, 136.1 and 161.3\,\GeV.
The first uncertainty is statistical, the second systematic.}
\label{tab:dpar1}
\end{table}
 
\begin{table}[htbp]{\small\begin{center}
%
\end{center}}
\caption[Differential distribution for $D$-parameter at $\rs= 172.3$, 182.8  and 188.6\,\GeV.]
        {Differential distribution for $D$-parameter at $\rs= 172.3$, 182.8  and 188.6\,\GeV.
The first uncertainty is statistical, the second systematic.}
\label{tab:dpar2}
\end{table}

\begin{table}[htbp]{\small\begin{center}
%
\end{center}}
\caption[Differential distribution for $D$-parameter at $\rs = 194.4$, 200.2 and 206.2\,\GeV.]
        {Differential distribution for $D$-parameter at $\rs = 194.4$, 200.2 and 206.2\,\GeV.
The first uncertainty is statistical, the second systematic.}
\label{tab:dpar3}
\end{table}
 
\clearpage
 
\begin{table}[htbp]\small\begin{center}
%
\end{center}
\caption[Differential distribution for event thrust at $\rs = 91.2\,\GeV$ for udsc and b events.]
        {Differential distribution for event thrust at $\rs = 91.2\,\GeV$ for udsc and b events.
The first uncertainty is statistical, the second systematic.}
\label{tab:thrust91}
\end{table}

\begin{table}[htbp]\begin{center}
%
\end{center}
\caption[Differential distribution for scaled heavy jet mass at $\rs= 91.2\,\GeV$ for udsc and b events.]
        {Differential distribution for scaled heavy jet mass at $\rs= 91.2\,\GeV$ for udsc and b events.
The first uncertainty is statistical, the second systematic.}
\label{tab:rho91}
\end{table}

\begin{table}[htbp]\begin{center}
%
\end{center}
\caption[Differential distribution for total jet broadening at $\rs= 91.2\,\GeV$ for udsc and b events.]
        {Differential distribution for total jet broadening at $\rs= 91.2\,\GeV$ for udsc and b events.
The first uncertainty is statistical, the second systematic.}
\label{tab:bt91}
\end{table}

\begin{table}[htbp]\begin{center}
%
\end{center}
\caption[Differential distribution for wide jet broadening at $\rs = 91.2\,\GeV$ for udsc and b events.]
        {Differential distribution for wide jet broadening at $\rs = 91.2\,\GeV$ for udsc and b events.
The first uncertainty is statistical, the second systematic.}
\label{tab:bw91}
\end{table}

\begin{table}[htbp]\begin{center}
%
\end{center}
\caption[Differential distribution for $C$-parameter at $\rs = 91.2\,\GeV$ for udsc and b events.]
        {Differential distribution for $C$-parameter at $\rs = 91.2\,\GeV$ for udsc and b events.
The first uncertainty is statistical, the second systematic.}
\label{tab:cpar91}
\end{table}

\begin{table}[htbp]\begin{center}
\begin{tabular}{|@{\hspace{1.5em}}r@{--}l|r@{$\;\pm$}c@{$\pm\;$}l|r@{$\;\pm$}c@{$\pm\;$}l|} \hline
\multicolumn{2}{|c|}{$D$} &
\multicolumn{3}{c|}{\rule[-13pt]{0pt}{32pt}${\displaystyle\frac{1}{\sigma}\cdot\frac{\ds}{\dD}}$\ \ (udsc)} &
\multicolumn{3}{c|}{${\displaystyle\frac{1}{\sigma}\cdot\frac{\ds}{\dD}}$\ \ (b)}  \\ \hline
 0.000& 0.032& 17.574& 0.038& 0.324& 16.133& 0.071& 0.331 \\
 0.032& 0.064&  5.154& 0.009& 0.095&  5.432& 0.020& 0.111 \\
 0.064& 0.096&  2.622& 0.009& 0.048&  2.858& 0.018& 0.059 \\
 0.096& 0.128&  1.620& 0.007& 0.030&  1.816& 0.017& 0.037 \\
 0.128& 0.160&  1.111& 0.006& 0.025&  1.307& 0.015& 0.032 \\
 0.160& 0.192&  0.785& 0.005& 0.019&  0.944& 0.013& 0.025 \\
 0.192& 0.224&  0.563& 0.005& 0.017&  0.698& 0.012& 0.024 \\
\hline\end{tabular}\end{center}
\caption[Differential distribution for $D$-parameter at $\rs= 91.2\,\GeV$ for udsc and b events.]
        {Differential distribution for $D$-parameter at $\rs= 91.2\,\GeV$ for udsc and b events.
The first uncertainty is statistical, the second systematic.}
\label{tab:dpar91}
\end{table}
 
\clearpage

\begin{table}[htbp]
\begin{center}
%
\end{center}
\caption[Shape-variable dependent coefficients appearing in the equations
         for measurement of \as\ using the power correction ansatz.]
        {Shape-variable dependent coefficients appearing in the equations
         for measurement of \as\ using the power correction
         ansatz \cite{qcdpower,qcdpower1,qcdpower2,qcdpower3,power,salamDW,salamW,salam}.
         }
\label{tab:coef}
\end{table}

\begin{table}[htbp]
{\begin{center}
%
\end{center}}
\caption[Determination of $\alpha_0$ and $\as(\MZ)$ from
           fits to the first moments of the event-shape distributions
           together with \chisq/d.o.f.]
        {Determination of $\alpha_0$ and $\as(\MZ)$ from
           fits to the first moments of the event-shape distributions
           together with \chisq/d.o.f. (see text).
           The first uncertainty is statistical, the second systematic.
         The unweighted averages are also shown.
         The first uncertainty is the average of the statistical uncertainties,
         the second the theoretical uncertainty.
           }
\label{tab:moment1}
\end{table}

\begin{table}[htbp]
{\begin{center}

\end{center}}
\caption{Results of fits of the power correction ansatz
           to the second moments of event-shape variables.
           Also shown is the ratio
           at $\rs=\MZ$
           of the ${\cal O}(1/s)$ term
           to the lowest-order power correction term.}
\label{tab:moment2}
\end{table}

\renewcommand{\arraystretch}{1.3}
\begin{table}[htbp]\begin{center}
{%
}\end{center}
\caption{Schematic representation of the fixed-order expansion versus
           the logarithmic expansion of theoretical
           predictions to the event-shape variables.}
\label{tab:match}
\end{table}
\renewcommand{\arraystretch}{1.}
 
\begin{table}[htbp]
\begin{center}
{\small%
}\end{center}
\caption{The fit range used to determine $\as$ from event-shape variables
           at different centre-of-mass energies. The \chisq/d.o.f. of the fit are also given.
           The fit ranges are chosen to exclude regions where the statistics is too small.
         }
\label{tab:fitrange}
\end{table}

\begin{sidewaystable}[htbp]
\begin{center}
{\small
%
}\end{center}
\caption [Values of $\as$ measured at different centre-of-mass energies from fits to the event-shape variables.]
         {Values of $\as$ measured at different centre-of-mass energies from fits to the event-shape variables.
          The first uncertainty is statistical, the second systematic.}
\label{tab:als}
\end{sidewaystable}
 
\begin{table}[htbp]
\begin{center}
%
\end{center}
\caption[Combined $\as$ values from the five event-shape variables.]
        {Combined $\as$ values from the five event-shape variables with their   uncertainties.}
\label{tab:alscomb2}
\end{table}


\begin{table}{\scriptsize\begin{center}
\end{center}}
\caption[Charged particle multiplicity distributions at $\rs = 91.2\,\GeV$]
        {Charged particle multiplicity distributions at $\rs = 91.2\,\GeV$
The first uncertainty is statistical, the second systematic.}
\label{tab:nch0}
\end{table}
 
\begin{table}[htbp]{\small\begin{center}
%
\end{center}}
\caption[Charged particle multiplicity distribution at $\rs= 130.1$, 136.1 and 161.3\,\GeV.]
        {Charged particle multiplicity distribution at $\rs= 130.1$, 136.1 and 161.3\,\GeV.
The first uncertainty is statistical, the second systematic.}
\label{tab:nch1}
\end{table}
 
\begin{table}[htbp]
{\small
\begin{center}
%
\end{center}}
\caption[Charged particle multiplicity distribution at $\rs = 172.3$, 182.8 and 188.6\,\GeV.]
        {Charged particle multiplicity distribution at $\rs = 172.3$, 182.8 and 188.6\,\GeV.
The first uncertainty is statistical, the second systematic.}
\label{tab:nch2}
\end{table}
 
\begin{table}[htbp]
{\small
\begin{center}
%
\end{center}
}
\caption[Charged particle multiplicity distribution at $\rs = 194.4$, 200.2 and 206.2\,\GeV.]
        {Charged particle multiplicity distribution at $\rs = 194.4$, 200.2 and 206.2\,\GeV.
The first uncertainty is statistical, the second systematic.}
\label{tab:nch3}
\end{table}
 
\begin{table}[htbp]
\begin{center}
$
%
$
\end{center}
\caption{The perturbative correction coefficients for various numbers of active flavours from
         Reference~\citen{3nlo2}.}
\label{tab:3nlo2}
\end{table}

\begin{table}[htbp]
\begin{center}
%
\end{center}
\caption{Results of fits to \nch\ \vs\ $\sqrt{s}$.
        }
\label{tab:nchfit}
\end{table}

\begin{table}[htbp]
\begin{center}
%
\end{center}
\caption[$\xi$ distributions at $\rs= 91.2\,\GeV$.]
        {$\xi$ distributions at $\rs= 91.2\,\GeV$.
The first uncertainty is statistical, the second systematic.}
\label{tab:ksi0}
\end{table}
 
\begin{table}[htbp]{\small\begin{center}
%
\end{center}}
\caption[$\xi$ distributions at $\rs= 130.1$, 136.1 and 161.3\,\GeV.]
        {$\xi$ distributions at $\rs= 130.1$, 136.1 and 161.3\,\GeV.
The first uncertainty is statistical, the second systematic.}
\label{tab:ksi1}
\end{table}
 
\begin{table}[htbp]{\small\begin{center}
%
\end{center}}
\caption[$\xi$ distributions at $\rs = 172.3$, 182.8 and 188.6\,\GeV.]
        {$\xi$ distributions at $\rs = 172.3$, 182.8 and 188.6\,\GeV.
The first uncertainty is statistical, the second systematic.}
\label{tab:ksi2}
\end{table}
 
\begin{table}[htbp]{\small\begin{center}
\end{center}}
\caption[$\xi$ distributions at $\rs= 194.4$, 200.2 and 206.2\,\GeV.]
        {$\xi$ distributions at $\rs= 194.4$, 200.2 and 206.2\,\GeV.
The first uncertainty is statistical, the second systematic.}
\label{tab:ksi3}
\end{table}
 
\clearpage
 
\begin{table}
\begin{center}
$
%
$
\end{center}
\caption[The peak position, $\xi^\star$, of the $\xi$ distribution from the Gaussian and Fong-Webber fits
         at $\rs=91.2\,\GeV$.]
        {The peak position, $\xi^\star$, of the $\xi$ distribution from the Gaussian and Fong-Webber fits
         at $\rs=91.2\,\GeV$.
         The first uncertainty is statistical, the second systematic.
\label{tab:xiZ}
}
\end{table}
 
\newcolumntype{d}[1]{D{.}{.}{#1}}
\newcolumntype{/}[1]{D{/}{\;/\;}{#1}}
\begin{table}[htbp]{\begin{center}
%
\end{center}}
\caption[The peak position, $\xi^\star$, of the $\xi$ distribution from the Fong-Webber fits
         at different centre-of-mass energies.]
        {The peak position, $\xi^\star$, of the $\xi$ distribution from the Fong-Webber fits
         at different centre-of-mass energies.
         The \chisq\ and number of degrees of freedom of the fits are also shown.
        }
\label{tab:ksi}
\end{table}

\begin{figure}[htbp]
\begin{center}
  \includegraphics[width=.5\figwidth]{\mydirfig 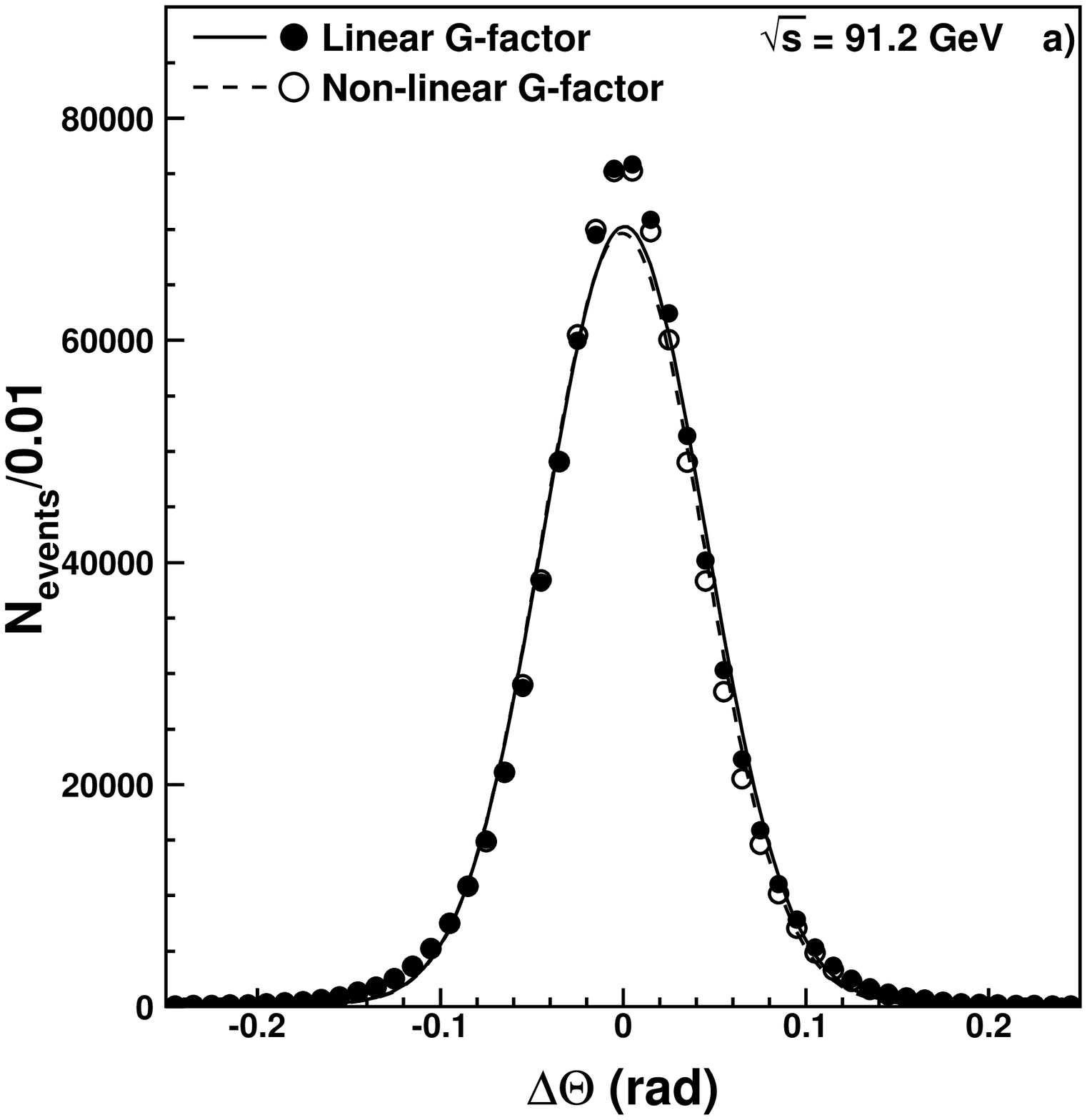}
  \includegraphics[width=.5\figwidth]{\mydirfig 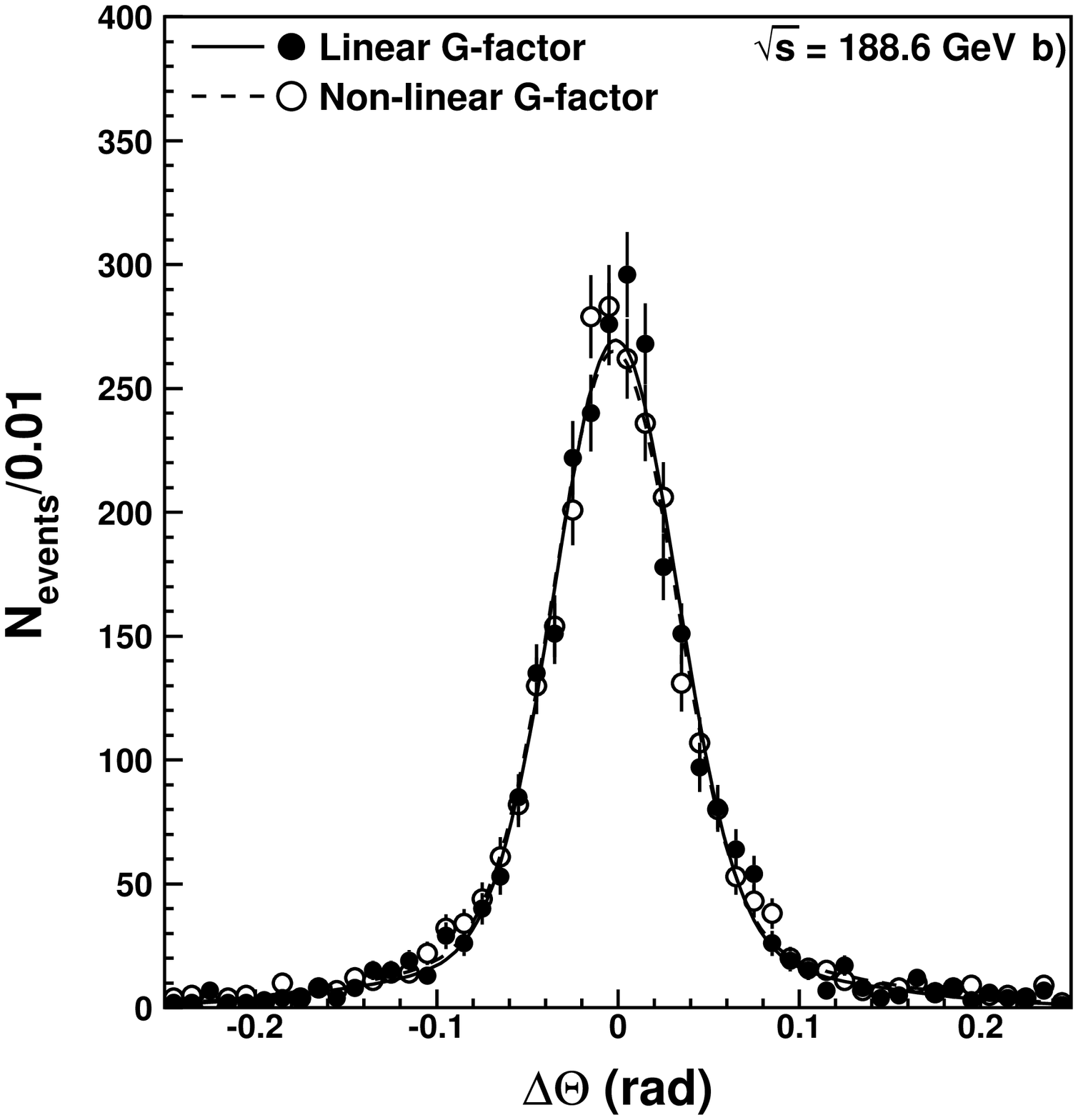}
  \includegraphics[width=.5\figwidth]{\mydirfig 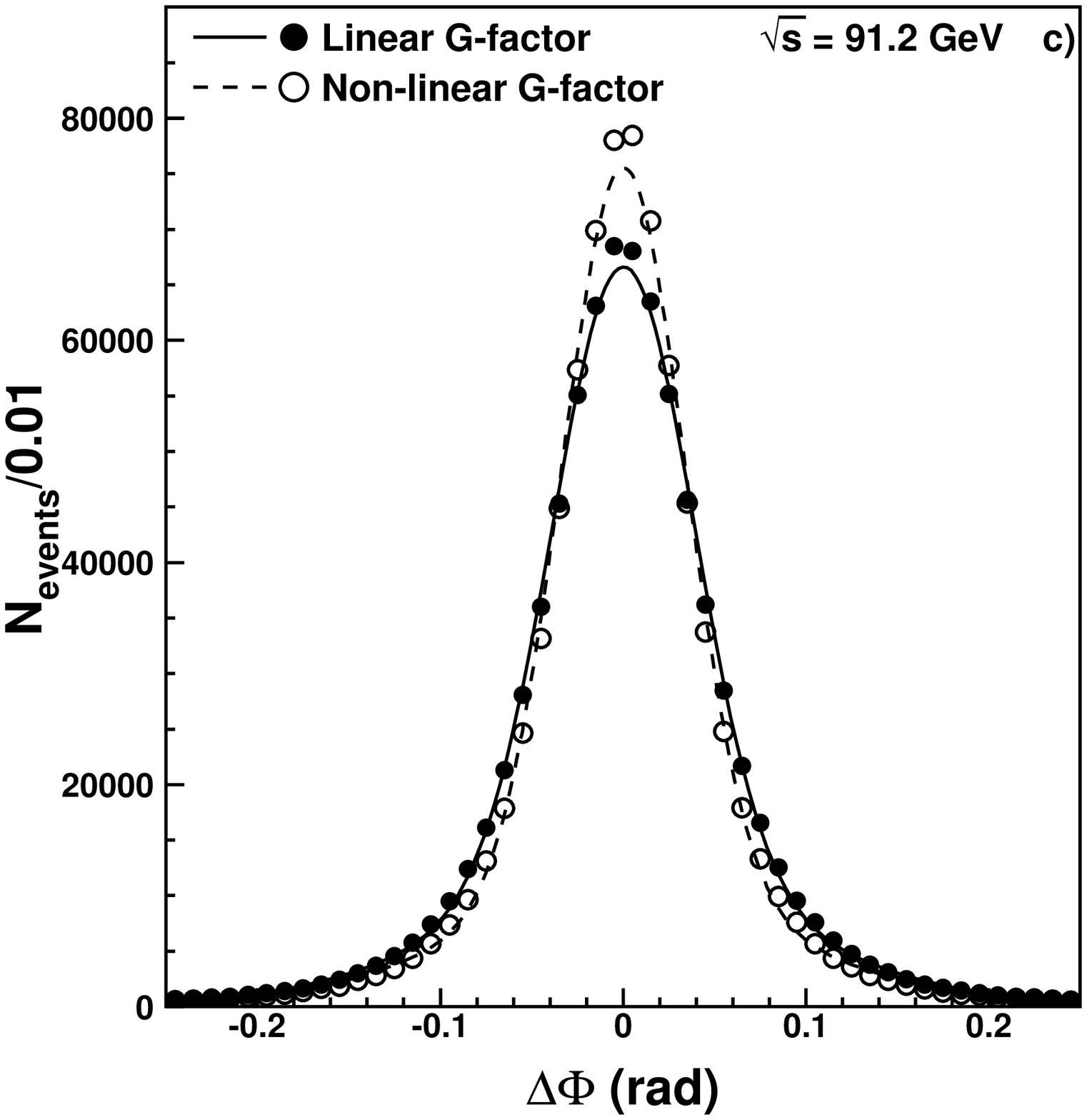}
  \includegraphics[width=.5\figwidth]{\mydirfig 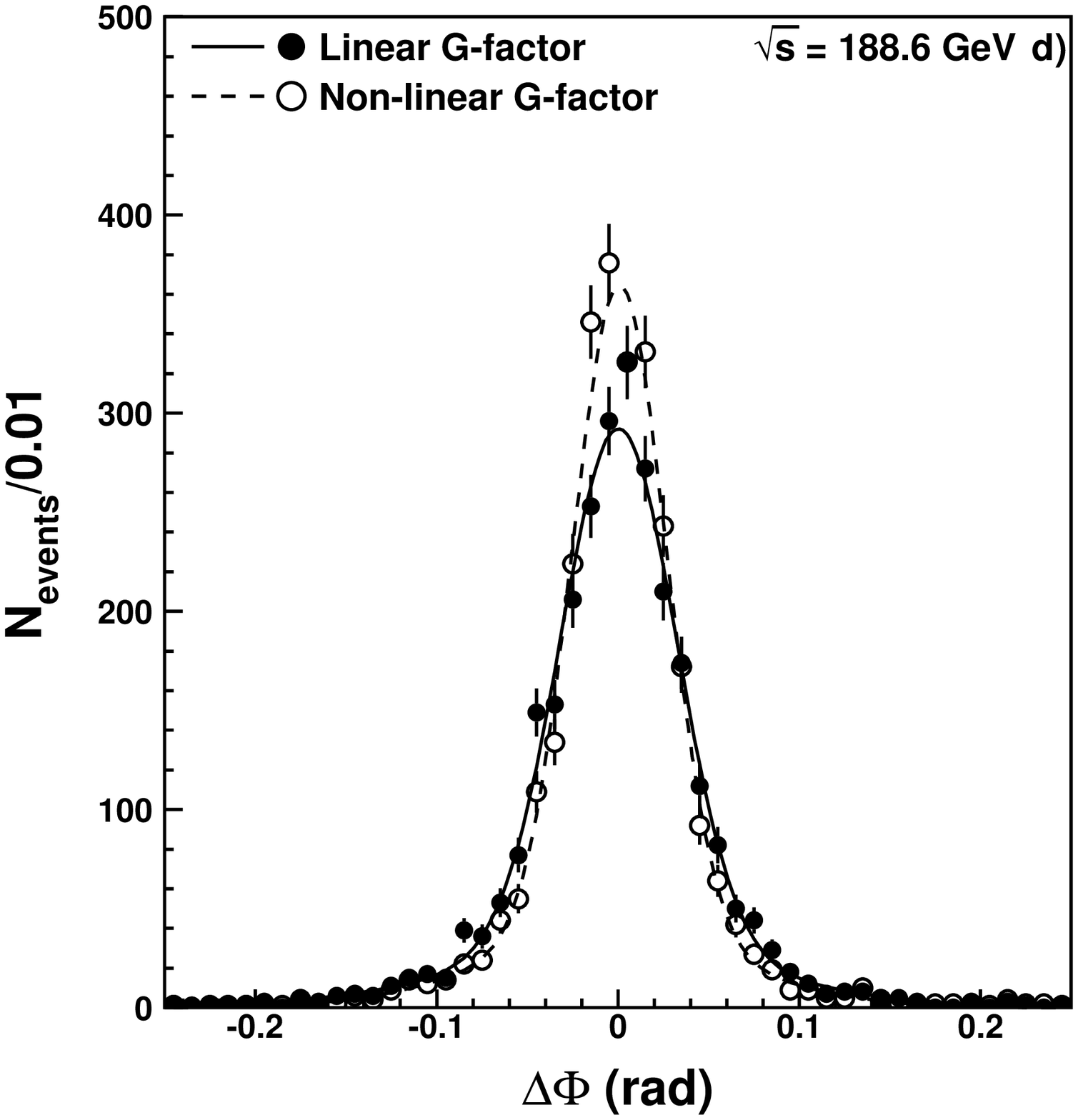}
\end{center}
  \caption{Jet angular resolutions obtained from the differences of
  (a,b) polar
  ($\Delta\Theta=\abs{\Theta_2-\Theta_1}-\pi$)
  and (c,d) azimuthal
  ($\Delta\Phi=\abs{\Phi_2-\Phi_1}-\pi$)
  angles  of the two jets in
          two-jet events at (a,c)  $\rs=91.2\,\GeV$ and (b,d) $\rs=188.6\,\GeV$
          with non-linear \Gfactors.}
 \label{fig:resth}
\end{figure}
 
 
 
\begin{figure}[htbp]
\begin{center}
  \includegraphics*[width=\figwidth]{\mydirfig 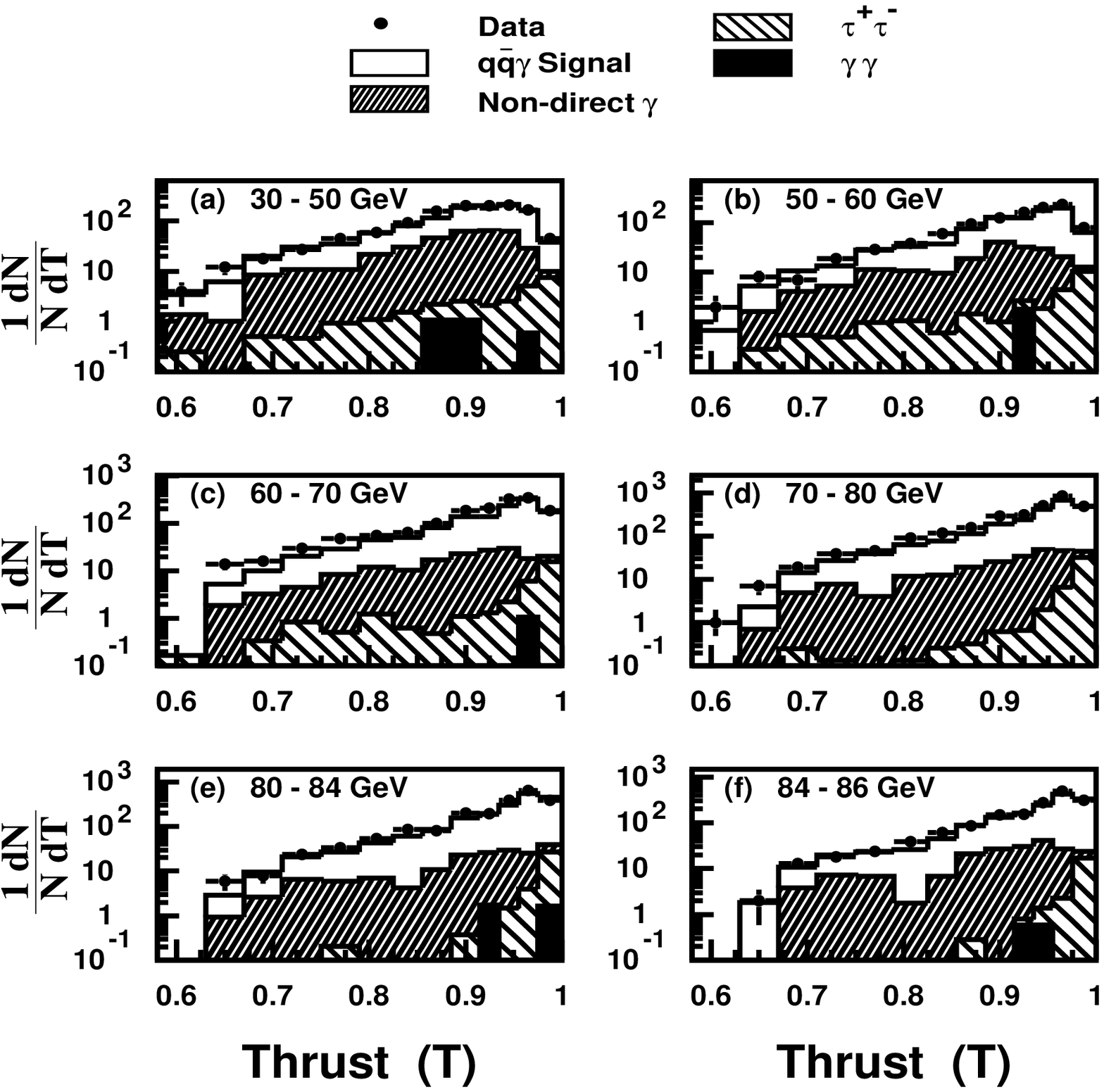}
\end{center}
\caption{Measured thrust distributions at different reduced centre-of-mass energies
           (a) 30--50\,\GeV, (b) 50--60\,\GeV, (c) 60--70\,\GeV, (d) 70--80\,\GeV,
           (e) 80--84\,\GeV, (f) 84--86\,\GeV.
           The solid lines correspond to the overall expectations from theory.
           The shaded areas refer to different backgrounds and the clear
           area refers to the signal predicted by \textsc{Jetset}.}
\label{fig:raw1}
\end{figure}

\clearpage
 
\begin{figure}[htbp]
\begin{center}
  \includegraphics*[width=.5\figwidth,bbllx=5,bblly=30,bburx=525,bbury=560]{\mydirfig 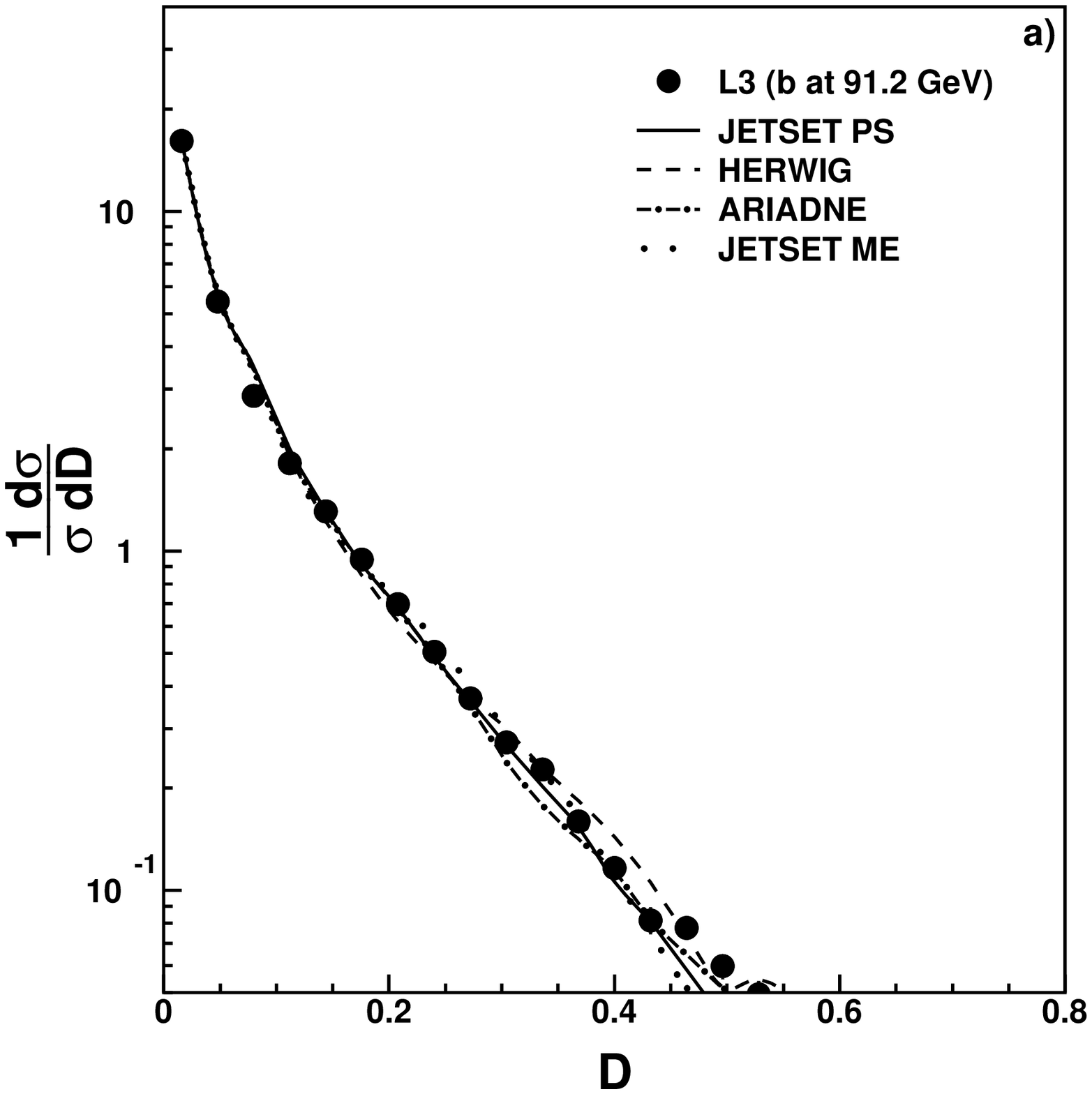}
  \includegraphics*[width=.5\figwidth,bbllx=5,bblly=30,bburx=525,bbury=560]{\mydirfig 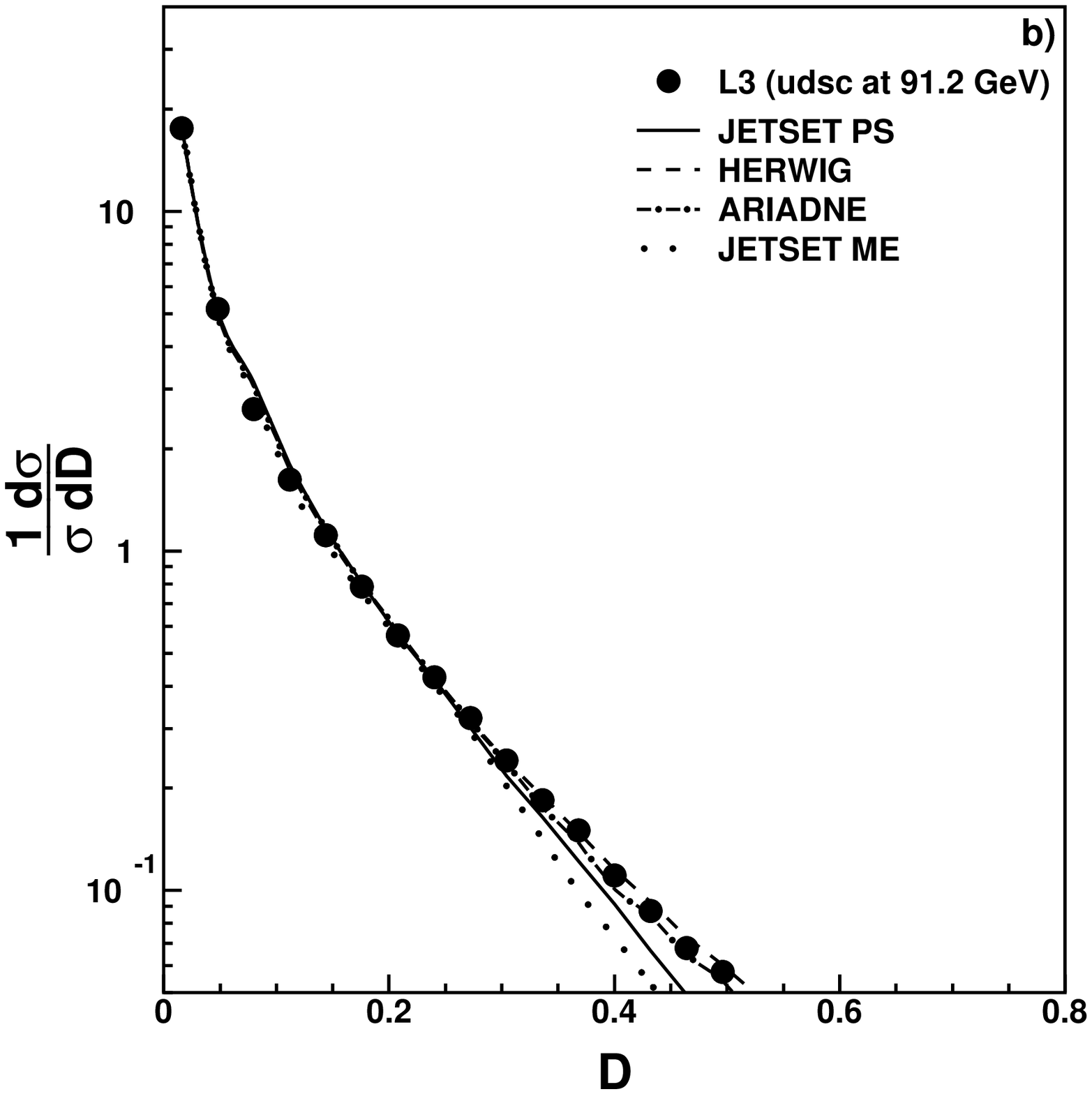}
  \includegraphics*[width=.5\figwidth,bbllx=5,bblly=30,bburx=525,bbury=560]{\mydirfig 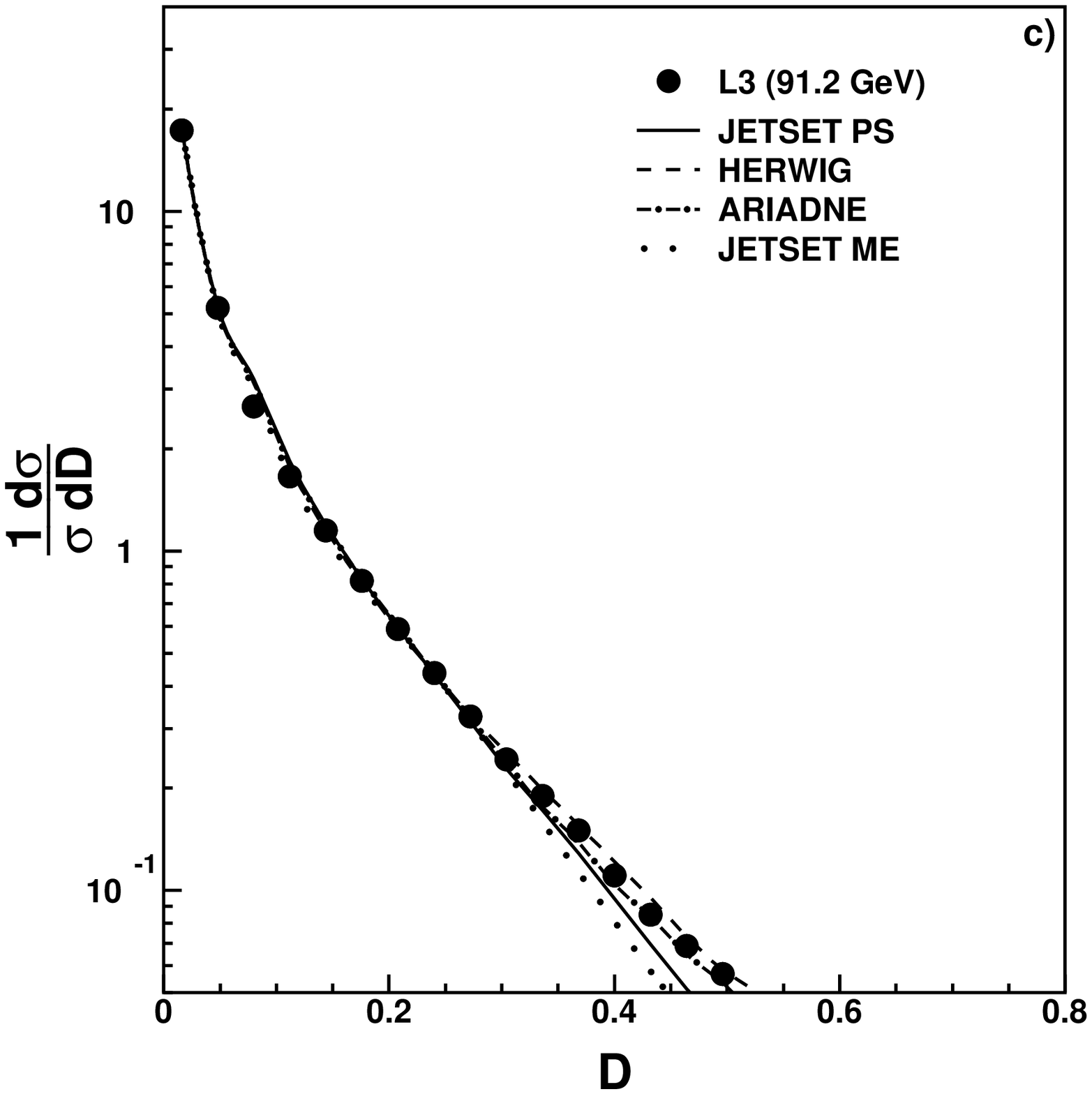}
  \includegraphics*[width=.5\figwidth,bbllx=5,bblly=30,bburx=525,bbury=560]{\mydirfig 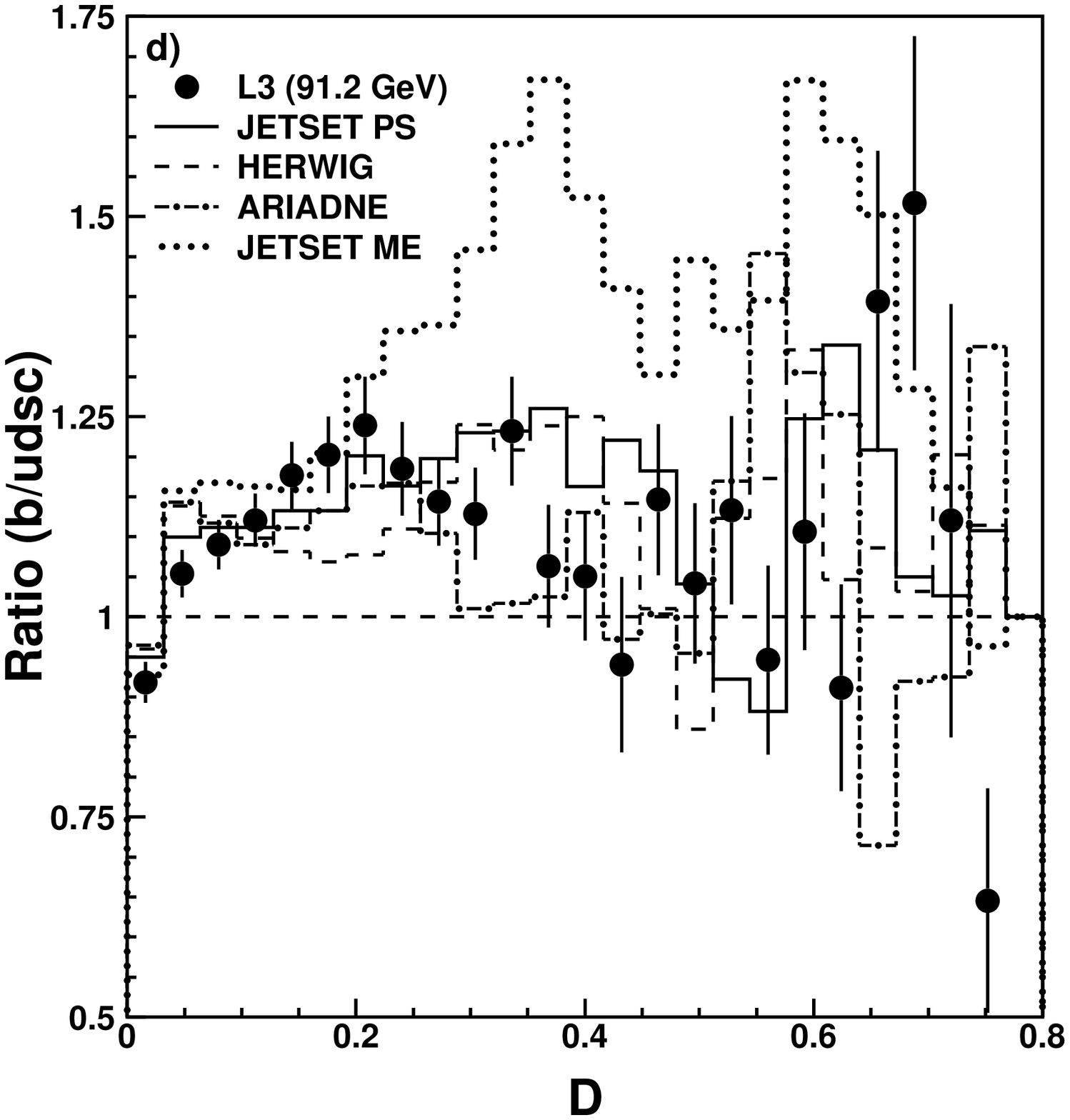}
\end{center}
\caption{$D$-parameter distributions at
         $\rs=91.2\,\GeV$
         for b, udsc, and all quark flavours
         and the ratio b/udsc
         compared to several \QCD\ models.
         }
\label{fig:part-d}
\end{figure}
 
\clearpage
 
\begin{figure}[htbp]
\begin{center}
    \includegraphics[width=\figwidth]{\mydirfig 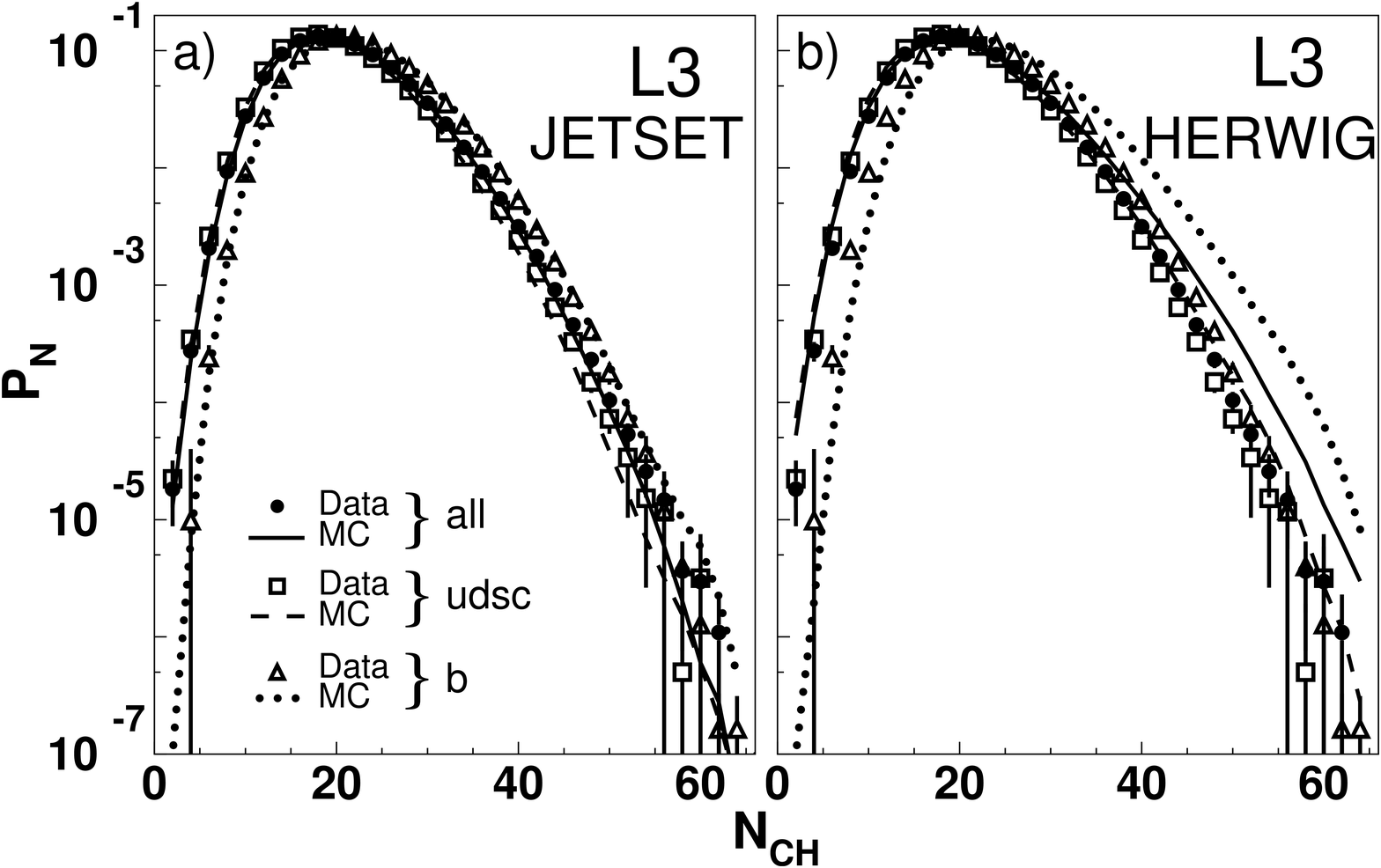}
    \includegraphics[width=\figwidth]{\mydirfig 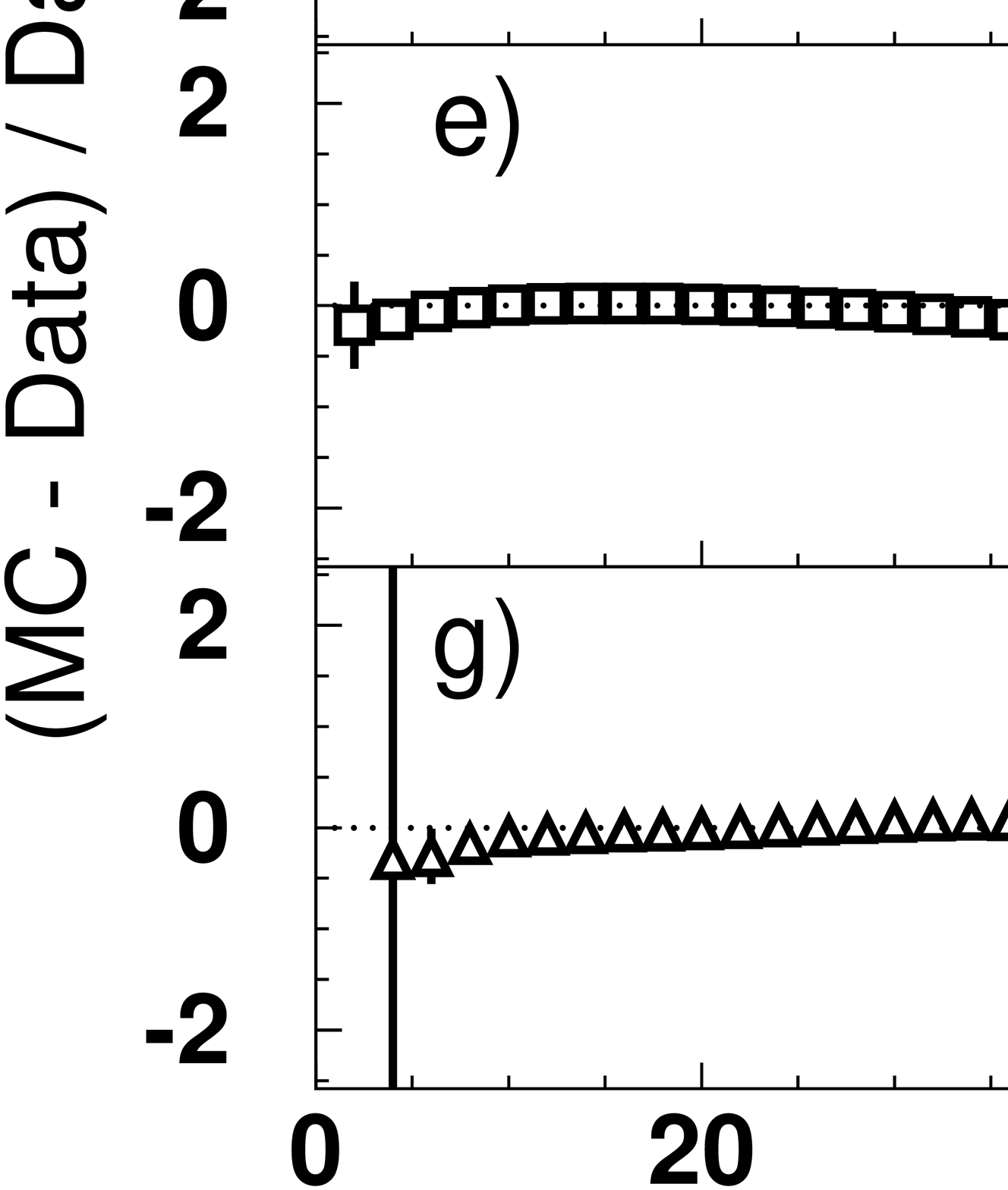}
\end{center}
\caption{Charged particle multiplicity distributions, normalised to unity, at
         $\rs=91.2\,\GeV$
       compared to (a, c, e, g) \textsc{Jetset PS} and (b, d, f, h) \textsc{Herwig}.}
\label{fig:chmulZ}
\end{figure}

\begin{figure}[htbp]
\begin{center}
   \includegraphics*[width=0.705\figwidth]{\mydirfig 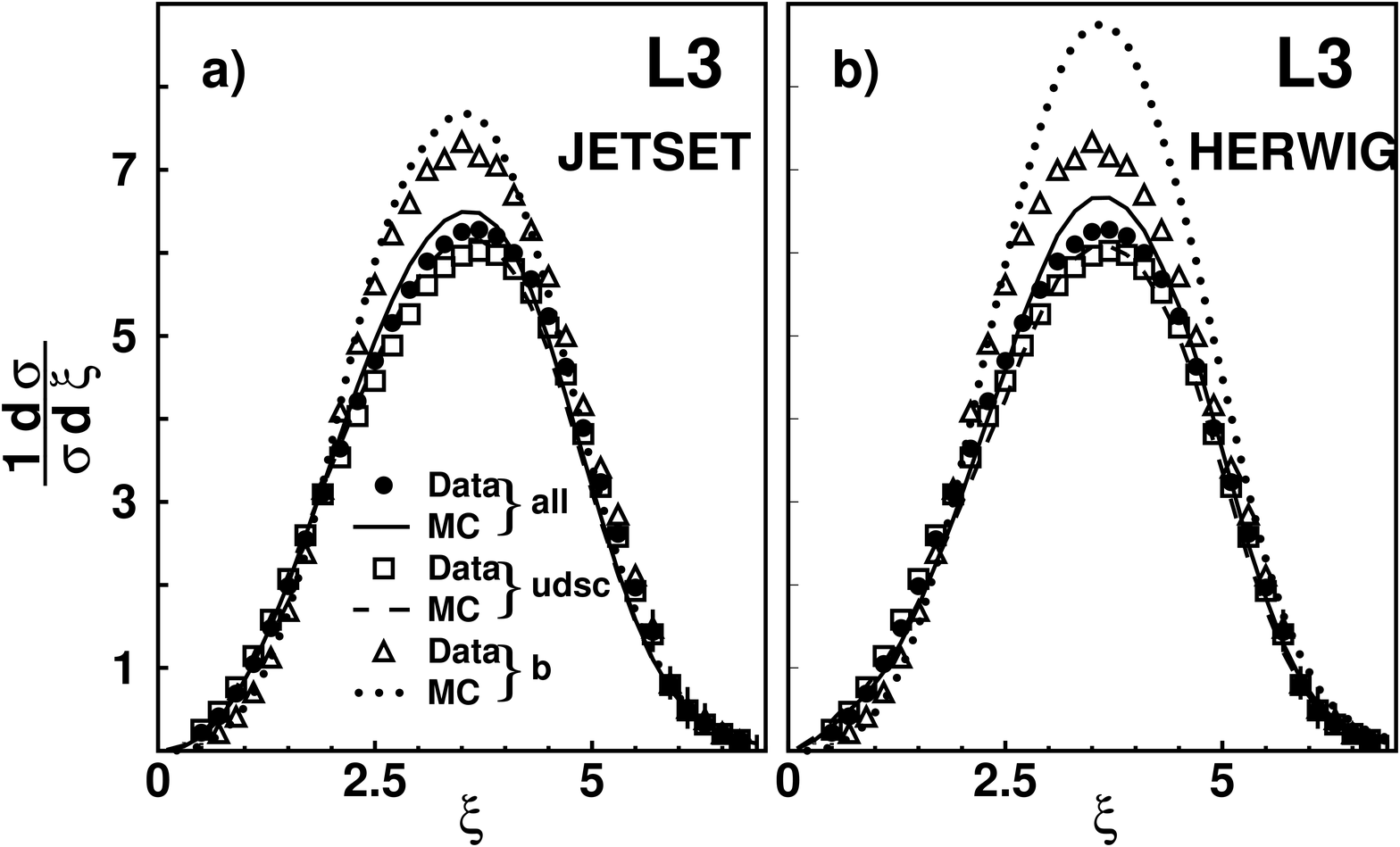}
 
   \includegraphics*[width=\figwidth]{\mydirfig 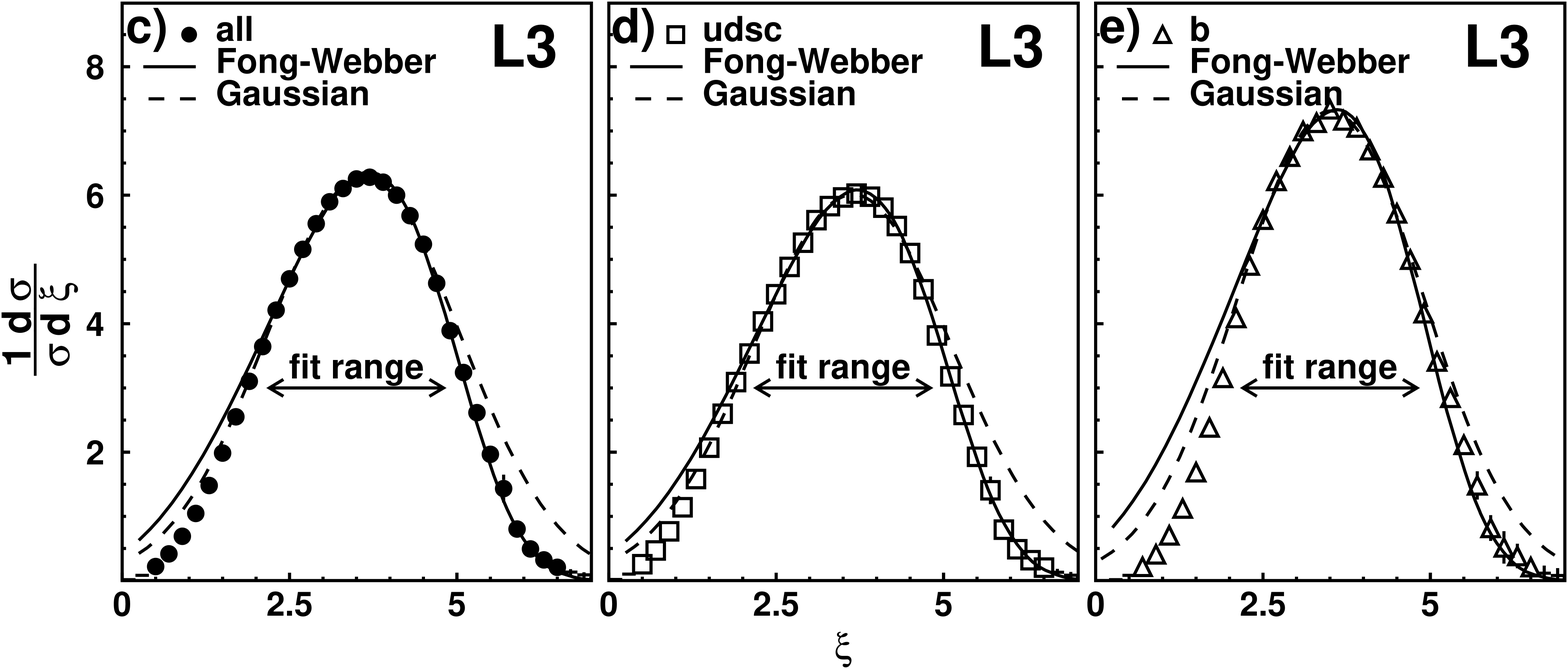}
\end{center}
\caption{Corrected $\xi$ distributions at $\rs= 91.2\,\GeV$ compared to
           (a) \textsc{Jetset PS} and (b) \textsc{Herwig} and
           together with the results of fits to Gaussian and Fong-Webber parametrisations
           for the (c) all-flavour, (d) udsc- and (e) b-quark samples.
           }
\label{fig:ksiZ}
\end{figure}
 
\newpage
 
 
\begin{thebibliography}{100}
  \bibitem{qcd_qqgth}
     M. Gell-Mann, Acta Phys. Austriaca Suppl. {\bf IX} (1972) 733.
  \bibitem{qcd_qqgth1}
     H. Fritzsch and M. Gell-Mann, 16th International Conference
       on High Energy Physics, Batavia, 1972; editors J. D. Jackson and
  \bibitem{qcd_qqgth2}
     A. Roberts, National Accelerator Laboratory (1972).
  \bibitem{qcd_qqgth3}
     H. Fritzsch, M. Gell-Mann and H. Leytwyler, Phys. Lett. {\bf  B47} (1973) 365.
  \bibitem{qcd_qqgth4}
     D.J. Gross and F. Wilczek, Phys. Rev. Lett. {\bf 30} (1973) 1343.
  \bibitem{qcd_qqgth5}
     D.J. Gross and F. Wilczek, Phys. Rev. {\bf D8} (1973) 3633.
  \bibitem{qcd_qqgth6}
     H.D. Politzer, Phys. Rev. Lett. {\bf  30} (1973) 1346.
  \bibitem{qcd_qqgth7}
     G. 't Hooft, Nucl. Phys. {\bf B33} (1971) 173.
  \bibitem{qcd_qqgth8}
     S. Weinberg, Phys. Rev. Lett. {\bf 31} (1973) 494.
 
  \bibitem{l3:det}
     L3 Collaboration, B. Adeva \etal, Nucl. Inst. Meth. {\bf A289} (1990) 35.
  \bibitem{l3:det1}
     M. Chemarin \etal, Nucl. Inst. Meth. {\bf A349} (1994) 345.
  \bibitem{l3:det2}
     I.C. Brock \etal, Nucl. Inst. Meth. {\bf A381} (1996) 236.
  \bibitem{l3:det3}
     H. Anderhub \etal, Nucl. Inst. Meth. {\bf A515} (2003) 31.
 
  \bibitem{l3:smd}
     M. Acciarri \etal, Nucl. Inst. Meth. {\bf A351} (1994) 300.
  \bibitem{fbmuon}
     A. Adam     \etal, Nucl. Inst. Meth. {\bf A383} (1996) 342.
  \bibitem{spaghetti}
     G. Basti    \etal, Nucl. Inst. Meth. {\bf A374} (1996) 293.
 
  \bibitem{qcdpower}
     B.R. Webber, Phys. Lett. {\bf B339} (1994) 148.
  \bibitem{qcdpower1}
     Yu.L. Dokshitzer and B.R. Webber, Phys. Lett. {\bf B352} (1995) 451.
  \bibitem{qcdpower2}
     Yu.L. Dokshitzer \etal, Nucl. Phys. {\bf B511} (1997) 396.
  \bibitem{qcdpower3}
     Yu.L. Dokshitzer \etal, J. High Energy Phys. {\bf 05} (1998) 3.
  \bibitem{milan}
     Yu.L. Dokshitzer \etal, Eur. Phys. J. Direct {\bf C3} (1999) 1; {\it Erratum} {\bf C3} (2001) 1.
 
  \bibitem{kunznas}
     Z. Kunzst and P. Nason, {\it Z Physics at LEP1}, CERN Yellow Report 89-08
     (1989) Vol.~I, p.\,373.
 
  \bibitem{ert-me}
     R.K. Ellis, D.A. Ross and A.E. Terrano,  Nucl. Phys. {\bf B178} (1981) 421.
 
  \bibitem{qcd-t}
     S. Catani \etal, Phys. Lett. {\bf B263} (1991) 491.
 
  \bibitem{qcd-hjm}
     S. Catani \etal, Phys. Lett. {\bf B272} (1991) 368.
 
  \bibitem{qcd-jbm}
     S. Catani \etal, Phys. Lett. {\bf B295} (1992) 269.
  \bibitem{qcd-jbm2}
     Yu.L. Dokshitzer \etal, J. High Energy Phys. {\bf 01} (1998) 11.
 
  \bibitem{qcd-oth}
     S. Catani \etal, Nucl. Phys. {\bf B407} (1993) 3.
 
  \bibitem{qcd-c}
     S. Catani \etal, Phys. Lett. {\bf B427} (1998) 377.
 
  \bibitem{l3qcd91z}
     L3 Collaboration, B. Adeva \etal, Z. Phys. {\bf C55} (1992) 39.
 
  \bibitem{l3qcd91}
     L3 Collaboration, B. Adeva \etal, Phys. Lett. {\bf B284} (1992) 471.
  \bibitem{l3qcd91a}
     L3 Collaboration, O. Adriani \etal, Phys. Rep. {\bf 236} (1993) 1.
 
  \bibitem{l3qcd133}
     L3 Collaboration, M. Acciarri \etal, Phys. Lett. {\bf B371} (1996) 137.
 
  \bibitem{l3qcdlep2}
     L3 Collaboration, M. Acciarri \etal, Phys. Lett. {\bf B404} (1997) 390.
  \bibitem{l3qcdlep2a}
     L3 Collaboration, M. Acciarri \etal, Phys. Lett. {\bf B444} (1998) 569.
  \bibitem{l3qcdlep2b}
     L3 Collaboration, M. Acciarri \etal, Phys. Lett. {\bf B489} (2000) 65.
  \bibitem{l3qcdlep2c}
     L3 Collaboration, P. Achard \etal, Phys. Lett. {\bf B536} (2002) 217.
 
  \bibitem{l3qqg}
     L3 Collaboration, M. Acciarri \etal, Phys. Lett. {\bf B411} (1997) 339.
 
  \bibitem{as3}
     Z. Nagy and Z. Tr\'ocs\'anyi, Nucl. Phys. Proc. Suppl. {\bf 64} (1998) 63.
  \bibitem{as3a}
     Z. Nagy and Z. Tr\'ocs\'anyi, Phys. Rev. {\bf D59} (1999) 014020;
     {\it Erratum} Phys. Rev. {\bf D62} (2000) 099902.
 
  \bibitem{lla}
     K. Konishi, A. Ukawa and G. Veneziano, Nucl. Phys. {\bf B157} (1979) 45.
  \bibitem{lla1}
     R. Odorico, Nucl. Phys. {\bf B172} (1980) 157.
  \bibitem{lla2}
     G.C. Fox and S. Wolfram, Nucl. Phys. {\bf B168} (1980) 285.
  \bibitem{lla3}
     T.D. Gottschalk, Nucl. Phys. {\bf B214} (1983) 201.
  \bibitem{lla4}
     G. Marchesini and B.R. Webber, Nucl. Phys. {\bf B238} (1984) 1.
 
  \bibitem{nlla}
     J. Kalinowski, K. Konishi and T.R. Taylor, Nucl. Phys. {\bf B181} (1981) 221.
  \bibitem{nlla1}
     J. Kalinowski \etal, Nucl. Phys. {\bf B181} (1981) 253.
  \bibitem{nlla2}
     J.F. Gunion and J. Kalinowski, Phys. Rev. {\bf D29} (1984) 1545.
  \bibitem{nlla3}
     J.F. Gunion, J. Kalinowski and L. Szymanowski, Phys. Rev. {\bf D32} (1985) 2303.
 
  \bibitem{mlla1}
     A.H. Mueller, Nucl. Phys. {\bf B213} (1983) 85.
  \bibitem{mlla2}
     Yu.L. Dokshitzer \etal, Rev. Mod. Phys. {\bf60} (1988) 373.
  \bibitem{mlla3}
     Yu.L. Dokshitzer \etal, {\it Basics of Perturbative \QCD}, Editions Fronti\`eres, Gif-sur-Yvette, 1991.
 
  \bibitem{power}
     Yu.L. Dokshitzer and B.R. Webber, Phys. Lett. {\bf B352} (1995) 451;
     preprint Cavendish HEP-97-2 (1997).
 
  \bibitem{lphd}
   Ya.I. Azimov \etal, Z. Phys. {\bf C27} (1985) 65.
  \bibitem{lphd2}
   Ya.I. Azimov \etal, Z. Phys. {\bf C31} (1986) 213.
 
  \bibitem{quark} MARK-I Collaboration, G. Hanson \etal, Phys. Rev. Lett.  {\bf 35} (1975) 1609.
 
  \bibitem{obla}  MARK-J Collaboration, D.P. Barber \etal, Phys. Rev. Lett. {\bf 43} (1979) 830.
 
  \bibitem{gluonT}
   TASSO Collaboration, R. Brandelik \etal, Phys. Lett.  {\bf B86} (1979) 243.
  \bibitem{gluonP}
   PLUTO Collaboration, C. Berger \etal, Phys. Lett.  {\bf B86} (1979) 418.
  \bibitem{gluonJ}
   JADE W. Bartel \etal, Phys. Lett. {\bf B91} (1980) 142.
 
  \bibitem{jadefl}
    JADE Collaboration, W. Bartel \etal, Phys. Lett. {\bf B101} (1981) 129.
  \bibitem{jadefl1}
    JADE Collaboration, W. Bartel \etal, Z. Phys. {\bf C21} (1983) 37.
 
  \bibitem{amy}
    AMY Collaboration, I.H. Park \etal, Phys. Rev. Lett.  {\bf 62} (1989) 1713.
 
  \bibitem{qcdee}
    P. M\"{a}ttig, Phys. Rep.  {\bf 177} (1989) 141.
 
  \bibitem{qcdleprev}
    T. Hebbeker, Phys. Rep.  {\bf 217} (1992) 69.
  \bibitem{qcdleprev1}
    S. Bethke and J.E. Pilcher, Ann. Rev. Nucl. Part. Sci. {\bf 42} (1992) 251.
  \bibitem{qcdleprev2}
    M. Schmelling, Phys. Scr. {\bf51} (1995) 683.
  \bibitem{qcdleprev3}
    P.N. Burrows, Proc. of XXIV SLAC Summer Institute on Particle Physics (1996) 101.
  \bibitem{qcdleprev4}
    D. Duchesneau, J.H. Field and H. Jeremie, C. R. Physique {\bf 3} (2002) 1211.
 
  \bibitem{herarunning}
     H1 Collaboration, C. Adloff \etal, Eur. Phys. J. {\bf C19} (2001) 289.
  \bibitem{herarunning1}
     ZEUS Collaboration, S. Chekanov \etal, Phys. Lett. {\bf B547} (2002) 164.
 
  \bibitem{fnalrunning}
     CDF  Collaboration, C. Affolder \etal, Phys. Rev. Lett. {\bf 88} (2002) 042001.
 
  \bibitem{lepgen}
     For a review see:
     T.~Sj\"ostrand \etal, ``Z Physics at LEP 1", eds. G.~Altarelli \etal,
     CERN Report CERN-89-08, Vol. 3 (1989) 143.
 
  \bibitem{coh}
    A. Bassetto, M. Ciafaloni and G. Marchesini, Phys. Rep. {\bf 100} (1983) 201.
  \bibitem{coh1}
    Yu.L. Dokshitzer \etal,  Rev. Mod. Phys. {\bf 60} (1988) 373.
  \bibitem{coh2}
    B. R. Webber, Ann. Rev. Nucl. Part. Sci. {\bf 36} (1986) 253.
 
  \bibitem{dglap}
     V.N. Gribov and L.N. Lipatov, Sov. J. Nucl. Phys. {\bf15} (1972) 438.
  \bibitem{dglap1}
     V.N. Gribov and L.N. Lipatov, Sov. J. Nucl. Phys. {\bf15} (1972) 675.
  \bibitem{dglap2}
     Yu. L. Dokshitzer, Sov. J. Phys. JETP {\bf46} (1977) 641.
  \bibitem{dglap3}
     G. Altarelli and G. Parisi, Nucl. Phys. {\bf B126} (1977) 298.
 
  \bibitem{if}
     A. Krzywicki and B. Petersson, Phys. Rev. {\bf D6} (1972) 924.
  \bibitem{if1}
     J. Finkelstein and R. Peccei, Phys. Rev. {\bf D6} (1972) 2606.
  \bibitem{if2}
     F. Niedermayer, Nucl. Phys. {\bf B79} (1974) 355.
  \bibitem{if3}
     A. Casher, J. Kogut and L. Susskind, Phys. Rev. {\bf D10} (1974) 732.
 
  \bibitem{ifff}
     R.D. Field and R.P. Feynman, Nucl. Phys. {\bf B136} (1978) 1.
 
  \bibitem{ifalihoyer}
     P. Hoyer \etal, Nucl. Phys. {\bf B161} (1979) 349.
  \bibitem{ifalihoyer1}
     A. Ali   \etal, Nucl. Phys. {\bf B168} (1980) 409.
  \bibitem{ifalihoyer2}
     A. Ali   \etal, Phys. Lett. {\bf B93} (1980) 155.
 
  \bibitem{sf}
     X. Artru and G. Mennessier, Nucl. Phys. {\bf B70} (1974) 93.
  \bibitem{sf1}
     B. Andersson, G. Gustafson and T. Sj\"ostrand, Z. Phys. {\bf C6} (1980) 235.
  \bibitem{sf2}
     B. Andersson \etal, Phys. Rep. {\bf97} (1983) 31.
  \bibitem{sf3}
     X. Artru,                  Phys. Rep. {\bf97} (1983) 147.
 
  \bibitem{cf}
     R.D. Field and S. Wolfram, Nucl. Phys. {\bf B213} (1983) 65.
  \bibitem{cf1}
     B.R. Webber, Nucl. Phys. {\bf B238} (1984) 492.
 
  \bibitem{jetset-pythia}
     \textsc{Jetset} 7.4 and \textsc{Pythia} 5.7 Monte Carlo Programs:
     T. Sj\"ostrand, Comp. Phys. Comm. {\bf 82} (1994) 74;
     CERN-TH-7112/93 (1993), revised August 1995.
 
  \bibitem{ariadne}
     \textsc{Ariadne}      Monte Carlo Program:
     L. L{\"{o}}nnblad, Comp. Phys. Comm. {\bf71} (1992) 15.
 
  \bibitem{herwig}
     G. Marchesini and B. Webber, Nucl. Phys. {\bf B310} (1988) 461.
  \bibitem{herwig1}
     I.G. Knowles, Nucl. Phys. {\bf B310} (1988) 571.
  \bibitem{herwig59}
     \textsc{Herwig} 5.9 Monte Carlo Program:
     G. Marchesini \etal, Comp. Phys. Comm. {\bf 67} (1992) 465.
 
  \bibitem{cojets6}
     \textsc{Cojets} 6.23 Monte Carlo Program:
     R. Odorico, Comp. Phys. Comm. {\bf 32} (1984) 139;
      {\it Erratum} Comp. Phys. Comm. {\bf 34} (1985) 43.
  \bibitem{cojets}
     R. Odorico, Nucl. Phys. {\bf B228} (1983) 381.
  \bibitem{cojets1}
     R. Mazzanti and R. Odorico, Nucl. Phys. {\bf B370} (1992) 23 and
     Bologna preprint DFUB 92/1.
 
 
  \bibitem{pythiasix}
     \textsc{Pythia} 6.2 Monte Carlo Program:
     T. Sj\"ostrand \etal, Comp. Phys. Comm. {\bf 135}(2001) 238.
 
  \bibitem{jetewc}
     F.A. Berends, R. Kleiss and S. Jadach, Nucl. Phys. {\bf B202} (1982) 63.
  \bibitem{jetewc1}
     F.A. Berends, R. Kleiss and S. Jadach, Comp. Phys. Comm. {\bf 29} (1983) 185.
 
  \bibitem{peterson}
     C. Peterson \etal, Phys. Rev. {\bf D27} (1983) 105.
 
  \bibitem{lundsym}
     B. Andersson, G. Gustafson and B. S\"oderberg. Z. Phys. {\bf C20} (1983) 317.
 
  \bibitem{azi}
     Ya.I. Azimov, Phys. Lett. {\bf B165} (1985) 147.
 
  \bibitem{opt}
     P.M. Stevenson, Phys. Rev. {\bf D23} (1981) 2916.
  \bibitem{opt1}
     J.H. Field, Ann. Phys. {\bf226} (1993) 209.
 
  \bibitem{scd}
     S. Bethke, Z. Phys. {\bf C43} (1989) 331.
 
  \bibitem{zhu}
     R. Y. Zhu, Ph.D. thesis, MIT (1983).
 
  \bibitem{preconfine}
     D. Amati and G. Veneziano, Phys. Lett. {\bf B83} (1979) 87.
 
  \bibitem{kk2f}
     \textsc{KK2}f 4.14 Monte Carlo Program:
     S. Jadach, B.F.L. Ward and Z. W\c{a}s,  Comp. Phys. Comm. {\bf 130} (2000) 260.
  \bibitem{kk2f1}
     S. Jadach, B.F.L. Ward and Z. W\c{a}s,  Phys. Rev.        {\bf D63} (2001) 113009.
 
  \bibitem{geant}
     \textsc{Geant} 3.15 detector simulation program:
     R. Brun \etal, ``GEANT 3'', CERN DD/EE/84-1 (Revised), September 1987.
  \bibitem{gheisha}
     \textsc{Gheisha}  program: H. Fesefeldt, RWTH Aachen Report PITHA 85/02 (1985).
 
  \bibitem{phojet}
     \textsc{Phojet} Monte Carlo Program:
     R. Engel, Z. Phys. {\bf C66} (1995) 203.
  \bibitem{phojet1}
     R. Engel, J. Ranft and S. Roesler, Phys. Rev. {\bf D52} (1995) 1459.
 
  \bibitem{koralz}
     \textsc{KoralZ} Monte Carlo Program:
     S. Jadach, B.F.L. Ward and Z. W\c{a}s, Comp. Phys. Comm. {\bf 79} (1994) 503.
 
  \bibitem{bhagene}
     J.H. Field, Phys. Lett. {\bf B323} (1994) 432.
  \bibitem{bhageneMC}
     \textsc{Bhagene} Monte Carlo Program:
     J.H. Field and T. Riemann, Comp. Phys. Comm. {\bf 94} (1996) 53.
 
  \bibitem{bhwide}
     \textsc{Bhwide} Monte Carlo Program:
     S. Jadach \etal, Phys. Lett. {\bf B390} (1997) 298.
 
  \bibitem{koralw}
     \textsc{KoralW} Monte Carlo Program:
     M. Skrzypek \etal, Comp. Phys. Comm. {\bf 94} (1996) 216.
  \bibitem{koralw1}
     M. Skrzypek \etal, Phys. Lett. {\bf B372} (1996) 289.
 
  \bibitem{kt}
     Yu.L. Dokshitzer, Contribution to the Workshop on Jets at LEP and HERA (1990).
  \bibitem{kt1}
     N. Brown and W. J. Stirling, Rutherford Preprint RAL-91-049.
  \bibitem{kt2}
     S. Catani \etal, Phys. Lett. {\bf B269} (1991) 432.
  \bibitem{kt3}
     S. Bethke \etal, Nucl. Phys. {\bf B370} (1992) 310.
 
  \bibitem{l3btag}
     L3 Collaboration, M. Acciarri \etal, Phys. Lett. {\bf B411} (1997) 373.
 
  \bibitem{thrust}
     S. Brandt \etal, Phys. Lett. {\bf 12} (1964) 57.
  \bibitem{thrust1}
     E. Fahri, Phys. Rev. Lett. {\bf 39} (1977) 1587.
 
  \bibitem{hjm}
     T. Chandramohan and L. Clavelli, Nucl. Phys. {\bf B184} (1981) 365.
  \bibitem{hjm1}
     MARK-II Collaboration, A. Peterson \etal, Phys. Rev. {\bf D37} (1988) 1.
  \bibitem{hjm2}
     TASSO Collaboration, W. Braunschweig \etal, Z. Phys. {\bf C45} (1989) 11.
 
  \bibitem{tmatx}
     G. Parisi, Phys. Lett. {\bf B74} (1978) 65.
  \bibitem{tmatx1}
     J.F. Donoghue, F.E. Low, and S.Y. Pi, Phys. Rev. {\bf D20} (1979) 2759.
 
  \bibitem{mns}
     MARK-J Collaboration, D.P. Barber \etal, Phys. Lett. {\bf B89} (1979) 139.
 
  \bibitem{jade}
     JADE Collaboration, W. Bartel \etal,  Z. Phys. {\bf C33} (1986) 23.
  \bibitem{jade1}
     JADE Collaboration, S. Bethke \etal, Phys. Lett. {\bf B213} (1988) 235.
 
  \bibitem{fwm}
     G.C. Fox and F. Wolfram, Phys. Rev. Lett. {\bf 41} (1978) 1581.
  \bibitem{fwm1}
     G.C. Fox and F. Wolfram, Nucl. Phys. {\bf B149} (1979) 413.
  \bibitem{fwm2}
     G.C. Fox and F. Wolfram, Phys. Lett. {\bf B82} (1979) 134.
 
  \bibitem{smatx}
     J.D. Bjorken and S.J. Brodsky, Phys. Rev. {\bf D1} (1970) 1416.
 
  \bibitem{spher}
     H. Georgi \etal, Phys. Rev. Lett. {\bf 39} (1977) 1237.
  \bibitem{spher1}
     S. Brandt \etal, Z. Phys. {\bf C1} (1979) 61.
 
  \bibitem{scalepi0}
     L3 Collaboration, O. Adriani \etal, Phys. Lett. {\bf B292} (1992) 472
 
  \bibitem{minuit}
     MINUIT, CERN program library D506:
     F. James and M. Roos, Comp. Phys. Comm. {\bf 10} (1975) 343.
 
  \bibitem{herwigsix}
     \textsc{Herwig} 6.2 Monte Carlo Program:
     Gennaro Corcella \etal, J. High Energy Phys.  {\bf 01}(2001) 10.
 
  \bibitem{hfwg}
     The LEP heavy flavour group, preprint LEPHF/96-01, ALEPH Note 96-099, DELPHI
     96-67 PHYS 627, L3 Note 1969, OPAL Technical Note TN391 (1996).
 
  \bibitem{lundbe}
     L. L\"onnblad and T. Sj\"ostrand, Phys. Lett. {\bf B351} (1995) 293.
  \bibitem{lundbe1}
     L. L\"onnblad and T. Sj\"ostrand, Eur. Phys. J. {\bf C2} (1998) 165.
 
  \bibitem{cambridge}
     Yu.L. Dokshitzer \etal,  J. High Energy Phys.  {\bf 08}(1997) 001; preprint hep-ph/9707323 (1997).
 
  \bibitem{mark2}
     MARK-II Collaboration, S. Bethke \etal, Z. Phys. {\bf C43} (1989) 325.
 
  \bibitem{r3-tasso}
     TASSO Collaboration, W. Braunschweig \etal, Phys. Lett. {\bf B214} (1988) 286.
 
  \bibitem{r3-venus}
     VENUS Collaboration, K. Abe \etal, Phys. Lett. {\bf B240} (1990) 232.
 
  \bibitem{qcd-event2}
     S. Catani and M. Seymour, Phys. Lett. {\bf B378} (1996) 287.
 
  \bibitem{dpar}
     G.P. Salam and Z. Tr\'ocs\'anyi, private communication.
 
  \bibitem{salamDW}
     Yu.L. Dokshitzer and B.R. Webber, Phys. Lett. {\bf B404} (1997) 321.
  \bibitem{salamW}
     B.R. Webber, Nucl. Phys. Proc. Suppl. {\bf 71} (1999) 66.
  \bibitem{salam}
     G. Salam, private communication.
 
  \bibitem{pdg}
     Particle Data Group, K. Hagiwara \etal, Phys. Rev.  {\bf D15} (2002) 010001.
 
  \bibitem{alephas}
   ALEPH Coll., D. Buskulic \etal, Z. Phys.  {\bf C73} (1997) 409.
  \bibitem{alephas1}
   ALEPH Coll., R. Barate \etal, Phys. Rep. {\bf 294} (1998) 1.
  \bibitem{alephas2}
   ALEPH Coll., A. Heister \etal, preprint CERN-EP/2003-084, submitted to E. Phys. J.
 
  \bibitem{delphias}
   DELPHI Coll., P. Abreu \etal, Phys. Lett. {\bf B456} (1999) 322.
  \bibitem{delphias1}
   DELPHI Coll., P. Abreu \etal, E. Phys. J. {\bf C14} (2000) 557.
  \bibitem{delphias2}
   DELPHI Coll., J. Abdallah \etal, preprint CERN-EP-2004-007, accepted by E. Phys. J.
 
 
  \bibitem{opalas}
   OPAL Coll., P.D. Acton et al., Z. Phys. {\bf C59} (1993) 1.
  \bibitem{opalas1}
   OPAL Coll., G. Alexander \etal, Z. Phys. {\bf C72} (1996) 191.
  \bibitem{opalas2}
   OPAL Coll., K. Ackerstaff \etal, Z. Phys. {\bf C75} (1997) 193.
  \bibitem{opalas3}
   OPAL Coll., G. Abbiendi \etal,  Eur. Phys. J {\bf C16} (2000) 185.
 
  \bibitem{sldas}
   SLD Coll., K. Abe \etal., Phys. Rev. {\bf D51} (1995) 962.
 
 
 
 
  \bibitem{l3tau}
     L3 Collaboration, M. Acciarri \etal, Phys. Lett. {\bf B507} (2001) 47.
 
  \bibitem{l3lineshape}
     L3 Collaboration, M. Acciarri \etal, Eur. Phys. J. {\bf C16} (2000) 1.
 
  \bibitem{l3mult}
     L3 Collaboration, P.~Achard \etal, Phys. Lett. {\bf B577} (2003) 109.
 
  \bibitem{dagost}
     G. D'Agostini, Nucl. Inst. Meth. {\bf A362} (1995) 487.
 
  \bibitem{3nlo1}
     I.M. Dremin and J.W. Gary, Phys. Lett. {\bf B459} (1999) 341.
 
  \bibitem{3nlo2}
     A. Capella \etal, Phys. Rev. {\bf D61} (2000) 074009.
 
 
 
  \bibitem{mean:JADE}
     JADE Collaboration, W. Bartel \etal, Z. Phys. {\bf C20} (1983) 187.
  \bibitem{mean:TASSO1}
     TASSO Collaboration, W. Braunschweig \etal, Z. Phys. {\bf C45} (1989) 193.
  \bibitem{mean:TASSO2}
     TASSO Collaboration, W. Braunschweig \etal, Z. Phys. {\bf C47} (1990) 187.
  \bibitem{mean:AMY1}
     AMY Collaboration, Y.K. Li \etal, Phys. Rev. {\bf D41} (1990) 2675.
  \bibitem{mean:AMY2}
     AMY Collaboration, H.W. Zheng \etal, Phys. Rev. {\bf D42} (1990) 737.
 
  \bibitem{gaus1}
     A.H. Mueller, in {\em Proc. 1981 International Symposium on Lepton and
     Photon Interactions at High Energies}, ed. W. Pfeil (Bonn, 1981), p.\,689.
  \bibitem{gaus1a}
     Yu.L. Dokshitzer, V.S. Fadin and V.A. Khoze, Phys. Lett. {\bf B115} (1982) 242.
 
  \bibitem{gaus2}
     A.H. Mueller, Nucl. Phys. {\bf B241} (1984) 141.
 
  \bibitem{sgaus}
     C.P. Fong and B.R. Webber, Phys. Lett. {\bf B229} (1989) 289.
 
  \bibitem{ksi-l3}
     L3 Collaboration, B. Adeva \etal,  Phys. Lett. {\bf B259} (1991) 199.
  \bibitem{ksi-l3a}
     L3 Collaboration, M. Acciarri \etal, Phys. Lett. {\bf B444} (1998) 569.
 
  \bibitem{ksi-tasso}
    TASSO Collaboration, W. Braunschweig \etal, Z. Phys. {\bf C47} (1990) 187.
 
 
 
 
\end{thebibliography}
\end{document}